%% file: main.tex
\documentclass[10pt,twocolumn,letterpaper]{article}

\usepackage[pagenumbers]{cvpr} %

\input{preamble}

\definecolor{cvprblue}{rgb}{0.21,0.49,0.74}
\usepackage[pagebackref,breaklinks,colorlinks,citecolor=cvprblue]{hyperref}

\def\confYear{2024}

\title{Deep Video Codec Control for Vision Models}

\newcommand{\authorstep}{\hspace{0.3cm}}
\newcommand{\affiliationstep}{\hspace{0.25cm}}

\author{Christoph Reich\textsuperscript{\normalfont{}1,2,3,4} 
\authorstep Biplob Debnath\textsuperscript{\normalfont{}1}
\authorstep Deep Patel\textsuperscript{\normalfont{}1}
\authorstep Tim Prangemeier\textsuperscript{\normalfont{}3}
\\ Daniel Cremers\textsuperscript{\normalfont{}2,4}
\authorstep Srimat Chakradhar\textsuperscript{\normalfont{}1}
\and \textsuperscript{1}NEC Laboratories America, Inc.\affiliationstep \textsuperscript{2}TU Munich \affiliationstep \textsuperscript{3}TU Darmstadt \\\textsuperscript{4}Munich Center for Machine Learning (MCML)
}

\usepackage{fancyhdr}
\usepackage{setspace}

\fancypagestyle{firststyle}
{

    \fancyhf{}
    \lfoot{{\footnotesize\begin{spacing}{.5}\parbox{\linewidth}{\vspace{1.85em}
    To appear in \emph{Proceedings of the IEEE/CVF Conference on Computer Vision and Pattern Recognition}, 1\textsuperscript{st} Workshop on AI for Streaming @ CVPR, 2024.%
    \\\hrule\vspace{\baselineskip}
    \copyright~2024 IEEE. Personal use of this material is permitted. Permission from IEEE must be obtained for all other uses, in any current or future media, including reprinting/republishing this material for advertising or promotional purposes, creating new collective works, for resale or redistribution to servers or lists, or reuse of any copyrighted component of this work in other works. 
    }\end{spacing}}}
}

\begin{document}
\maketitle

\begin{abstract}
Standardized lossy video coding is at the core of almost all real-world video processing pipelines. Rate control is used to enable standard codecs to adapt to different network bandwidth conditions or storage constraints. However, standard video codecs (\eg, H.264) and their rate control modules aim to minimize video distortion \wrt{} human quality assessment. We demonstrate empirically that standard-coded videos vastly deteriorate the performance of deep vision models. To overcome the deterioration of vision performance, this paper presents the first end-to-end learnable \emph{deep video codec control} that considers both \emph{bandwidth constraints} and \emph{downstream deep vision performance}, while \emph{adhering to existing standardization}. We demonstrate that our approach better preserves downstream deep vision performance than traditional standard video coding.
\end{abstract}
\thispagestyle{firststyle}

\input{content/introduction.tex}
\input{content/related_work.tex}
\input{content/method.tex}
\input{content/experiments.tex}
\input{content/discussion.tex}
\input{content/conclusion.tex}

{
    \small
    \bibliographystyle{ieeenat_fullname}
    \bibliography{main}
}

\input{content/appendix}

\end{document}

%% file: preamble.tex
\usepackage[dvipsnames]{xcolor}

\usepackage{graphicx}
\usepackage{booktabs}
\usepackage{pifont}
\usepackage{tabularx}
\usepackage{float}
\usepackage{multirow}
\usepackage{makecell}
\usepackage{cuted}
\usepackage{tikz}
\usepackage[export]{adjustbox}
\usepackage{pgfplots}
\usepackage{fontawesome}
\usepackage{bbding}
\usetikzlibrary{spy}
\usetikzlibrary{shapes.arrows}
\usetikzlibrary{arrows.meta, arrows}
\usetikzlibrary{patterns, shapes.arrows}
\usetikzlibrary{positioning}
\usetikzlibrary{matrix}
\usetikzlibrary{pgfplots.groupplots}
\usepgfplotslibrary{groupplots,dateplot}
\usepgfplotslibrary{fillbetween}
\tikzset{every picture/.style={/utils/exec={\sffamily}}}
\DeclareUnicodeCharacter{2212}{−}
\pgfplotsset{compat=newest}

\usepackage[per-mode=symbol, group-digits=false, detect-all=true,input-symbols={()}]{siunitx}
\DeclareSIUnit\flop{FLOP}
\DeclareSIUnit\fps{fps}
\newcommand{\qp}{\mathrm{QP}}
\newcommand{\gop}{\mathrm{GOP}}

\usepackage{algorithm}
\usepackage{algorithmic}
\usepackage{newfloat}
\usepackage{listings}
\floatstyle{ruled}
\newfloat{listing}{tb}{lst}{}
\floatname{listing}{Listing}

\usepackage{xcolor}

\definecolor{tud0d}{RGB}{83,83,83}
\definecolor{tud0c}{RGB}{137,137,137}
\definecolor{tud0b}{RGB}{181,181,181}
\definecolor{tud0a}{RGB}{220,220,220}
\definecolor{tud1a}{RGB}{93,133,195}
\definecolor{tud2a}{RGB}{0,156,218}
\definecolor{tud3a}{RGB}{80,182,149}
\definecolor{tud4a}{RGB}{175,204,80}
\definecolor{tud5a}{RGB}{221,223,72}
\definecolor{tud6a}{RGB}{255,224,92}
\definecolor{tud7a}{RGB}{248,186,60}
\definecolor{tud8a}{RGB}{238,122,52}
\definecolor{tud9a}{RGB}{233,80,62}
\definecolor{tud10a}{RGB}{201,48,142}
\definecolor{tud11a}{RGB}{128,69,151}
\definecolor{tud1b}{RGB}{0,90,169}
\definecolor{tud2b}{RGB}{0,131,204}
\definecolor{tud3b}{RGB}{0,157,129}
\definecolor{tud4b}{RGB}{153,192,0}
\definecolor{tud5b}{RGB}{201,212,0}
\definecolor{tud6b}{RGB}{253,202,0}
\definecolor{tud7b}{RGB}{245,163,0}
\definecolor{tud8b}{RGB}{236,101,0}
\definecolor{tud9b}{RGB}{230,0,26}
\definecolor{tud10b}{RGB}{166,0,132}
\definecolor{tud11b}{RGB}{114,16,133}
\definecolor{tud1c}{RGB}{0,78,138}
\definecolor{tud2c}{RGB}{0,104,157}
\definecolor{tud3c}{RGB}{0,136,119}
\definecolor{tud4c}{RGB}{127,171,22}
\definecolor{tud5c}{RGB}{177,189,0}
\definecolor{tud6c}{RGB}{215,172,0}
\definecolor{tud7c}{RGB}{210,135,0}
\definecolor{tud8c}{RGB}{204,76,3}
\definecolor{tud9c}{RGB}{185,15,34}
\definecolor{tud10c}{RGB}{149,17,105}
\definecolor{tud11c}{RGB}{97,28,115}
\definecolor{tud1d}{RGB}{36,53,114}
\definecolor{tud2d}{RGB}{0,78,115}
\definecolor{tud3d}{RGB}{0,113,94}
\definecolor{tud4d}{RGB}{106,139,55}
\definecolor{tud5d}{RGB}{153,166,4}
\definecolor{tud6d}{RGB}{174,142,0}
\definecolor{tud7d}{RGB}{190,111,0}
\definecolor{tud8d}{RGB}{169,73,19}
\definecolor{tud9d}{RGB}{156,28,38}
\definecolor{tud10d}{RGB}{115,32,84}
\definecolor{tud11d}{RGB}{76,34,106}

\definecolor{necgray}{HTML}{7d7d7d}
\definecolor{necblue}{HTML}{2c328f}
\definecolor{necbluedark}{HTML}{00285e}
\definecolor{necorange}{HTML}{eb6e00}
\definecolor{necyellow}{HTML}{fff484}
\definecolor{green}{HTML}{388a73}

\definecolor{qp51}{rgb}{0.5,0,0}
\definecolor{qp0}{rgb}{0,0,0.5}

\newcommand{\cmark}{\textcolor{tud3c}{\ding{51}}}
\newcommand{\xmark}{\textcolor{tud9c}{\ding{55}}}
\newcommand{\ok}{\textcolor{tud7b}{$\sim$}{ }}

%% file: content/introduction.tex
\section{Introduction}
\label{sec:introduction}

Video data is a major source of internet traffic~\cite{Barnett2018}. A significant and increasing amount of these videos is consumed and analyzed by deep vision models~\cite{Hu2023}. Streaming videos over a network or storing these requires lossy video codecs and rate control to meet dynamic bandwidth or storage constraints, preventing video corruption or dropping~\cite{Itsumi2022}. 

Almost all real-world video processing pipelines utilize standardized video coding (\eg, H.264~\cite{Wiegand2003}) to ensure interoperability and low-costs~\cite{Lederer2019}. While deep video codecs have demonstrated promising results and can be optimized for deep vision models, they find minimal to no application in real-world~\cite{Lederer2019, Wood2022}. This is due to the lack of standardization (ISO) and limited support for rate control (\cf{}~\cref{tab:overview})~\cite{Zhang2023}.

\begin{table}[t!]
    \centering
    \caption{\textbf{High-level comparison to existing approaches.} Our deep video codec control offers both rate control and standardization while being optimized for deep vision models}
    \vspace{-4pt}
    \small
    \setlength\tabcolsep{0.5pt}
    \renewcommand\arraystretch{0.87}
        \begin{tabular}{>{\raggedright\arraybackslash}p{4.55cm}>{\centering\arraybackslash}p{1.675cm}>{\centering\arraybackslash}p{0.95cm}>{\centering\arraybackslash}p{0.85cm}}
    	\toprule
    	 & Optimize vision perf. & Rate control & $\;\;\;\;\;$ ISO \\
    	\midrule
    	Deep video codecs & \cmark & \ok & \xmark \\
    	Standard video codecs (\eg, H.264) & \xmark & \cmark & \cmark \\
    	\textbf{Deep video codec control} & \cmark & \cmark & \cmark \\
    	\bottomrule
    \end{tabular}
    \label{tab:overview}
\end{table}

Standard video codecs are developed to minimize image distortion \wrt{} human quality assessment (\cf{} \cref{tab:overview})~\cite{Richardson2004}. We demonstrate empirically that this is suboptimal for current deep vision models. More specifically, using H.264 coded videos during inference leads to a vast deterioration in downstream deep vision performance~\cite{Otani2022}.

In this paper, we aim to optimize standard video codecs for deep vision models (\eg, a semantic segmentation network) by learning a deep video codec control (\cf{} \cref{fig:generalarchitecture}). For a given video clip $\mathbf{V}$, we formulate the codec control task as a constrained optimization problem
\begin{equation}\label{eq:problem}
  \begin{aligned}
       \max_{\qp}\,&\mathrm{M}\!\left(\mathrm{DNN}\!\left(\operatorname{H.\!264}\!\left(\mathbf{V},\mathrm{C}_{\theta}\!\left(\mathbf{V},b\right)\right)\right)\right)\\[-3.5pt]
       \text{s.t.}\,&\tilde{b}\leq b.
   \end{aligned}
\end{equation}
Our lightweight control network $\mathrm{C}_{\theta}$ consumes both the video clip $\mathbf{V}$ and the (dynamic) target bitrate $b$ to predict high-dimensional codec parameters $\qp$. We aim to learn a content and bandwidth-aware prediction of $\qp$, controlling the H.264 coding, such that we maximized the performance of a deep vision model $\mathrm{DNN}$, measured by a task-specific metric $\mathrm{M}$ (\eg, accuracy). Additionally, the resulting video bitrate $\tilde{b}$ should not exceed the target bitrate $b$. To the best of our knowledge, we present the first end-to-end learnable codec control taking vision performance, bandwidth conditions, and existing standardizations into account (\cf{} \cref{tab:overview}). Making our deep video codec control the first standardized approach to support real-world bandwidth conditions as well as deep vision models.

Although approaches for optimizing standard video codecs for vision models within the scope of standardization have been proposed, they entail significant limitations impeding their real-world application~\cite{Galteri2018, Du2022, Itsumi2022}. First, existing approaches heavily rely on the use of saliency maps and hand-crafted rules~\cite{Galteri2018, Du2022}. However, saliency maps often fail to express highly relevant regions for deep vision models due to saturating gradients~\cite{Sundararajan2017}. Second, existing approaches only consider the preservation of vision performance and do not consider rate control~\cite{Galteri2018, Du2022, Itsumi2022}. In contrast, our deep video codec control is not limited by the concept of saliency maps, does not rely on hand-crafted rules, and considers both the preservation of downstream vision performance as well as rate control, offering a greater potential for practical application.

\paragraph{Contributions.} Motivated by the \emph{potential performance gains} and the \emph{need for standardization}, we make the following contributions: \emph{(1)} We showcase that downstream deep vision performance vastly deteriorates when standard video coding is employed. \emph{(2)} We propose the first end-to-end learnable codec control, directly optimizing \cref{eq:problem}, while maintaining compliance with standardization. \emph{(3)} To enable end-to-end learning, we introduce a conditional differentiable surrogate model of a standard codec, allowing gradient propagation through the non-differentiable standard codec \wrt{} high-dimensional codec parameters. We showcase the effectiveness of our deep codec control by controlling H.264 for two classical vision tasks (semantic segmentation and optical flow estimation) and on two datasets (Cityscapes~\cite{Cordts2016} and CamVid~\cite{Brostow2009}).

%% file: content/related_work.tex
\section{Related Work}
\label{sec:relatedwork}

\paragraph{Video compression.} The inherent memory complexity and temporal redundancy of video data have motivated the development of numerous algorithms for compressing videos \cite{Wu2001, Richardson2011, Liu2020}. Significant time has been dedicated to the development of standardized video compression algorithms, examples include H.264~\cite{Wiegand2003}, H.265~\cite{Sullivan2012}, and MPEG-5~\cite{Choi2020}. Recently, deep learning-based video compression algorithms have been introduced~\cite{Wood2022, Zhang2023}. Despite offering better preservation of the perceptual quality in the coded video and supporting custom quality objectives (\eg, preserving downstream performance), deep video compression algorithms find minimal applicability, as they lack widely supported standardization, are computationally heavy, and offer limited customizability as well as control over the compression process for rate control~\cite{Zhang2023}.

\paragraph{Video compression for machines.} With an increasing amount of video data being analyzed by machines, such as deep vision models, a new line of research on video compression for machines has emerged~\cite{Duan2020}. A common approach is to extract vision feature maps from an image or video and perform compression at the feature level~\cite{Emmons2019, Matsubara2019, Duan2020, Xia2020}. Another approach is to train a deep compression algorithm for a specific vision task, such as object detection~\cite{Beye2022}. While some efforts have been made towards standardizing video compression for machines, no general standard has emerged yet~\cite{Duan2020, Gao2021, Wood2022}, and the applicability of such approaches remains very limited. We do not aim to develop a new video codec for machines, rather our aim is to control widely used standard codecs for deep vision models as content and network bandwidth changes.

\paragraph{Video codec control for vision models.} A common approach for identifying regions of interest for vision tasks is to utilize local and simple heuristics~\cite{Zhang2018b, Li2020b}. Hard-coded rules are used to decide how to set the codec control~\cite{Zhang2018b, Li2020b}. These approaches implicitly assume static scenes and trivially fail on dynamic scenes. Other more sophisticated approaches utilize feedback from a server-side vision model during inference for controlling the encoding~\cite{Liu2019, Du2020, Zhang2021}. Feedback loop-based approaches entail complicated architectures, add additional points of failure, can break standardization, and often only support a specific vision task. Mandhane~\etal~\cite{Mandhane2022} use the MuZero reinforcement learning algorithm to learn a codec control. This control predicts a single codec parameter for each frame under a bandwidth constraint and aims to optimize perceptual quality but does not target vision models. Itsumi~\etal~\cite{Itsumi2022} propose an RL approach for finding regions relevant to a vision model. Paired with hard-coded rules they presented a codec control for H.265. While this approach considers relevant regions for a deep vision model, Itsumi \etal{} do not consider a dynamic bandwidth constraint. AccMPEG~\cite{Du2022} and Galteri~\etal~\cite{Galteri2018} also estimate regions relevant for a deep vision model by learning to predict saliency maps~\cite{Simonyan2013}. These approaches also do not consider a bandwidth condition, use hard-coded rules for the control, and are limited by the concept of saliency maps~\cite{Sundararajan2017}.

\paragraph{Video codec surrogate.} Standard video codecs, such as H.264 and H.265, are not differentiable \wrt{} the input video and codec parameters. This non-differentiability has motivated the use of deep neural networks to approximate standard codecs in a differentiable manner~\cite{Zhao2019, Qiu2021, Klopp2021, Tian2021, Isik2023}. For instance, Tian~\etal~\cite{Tian2021} proposed a differentiable surrogate of H.265 that predicts the coded video, for a small subset of clip-wise compression strengths. While we are not the first to learn a differentiable approximation of a standard video codec, we present the first conditional surrogate with support for fine-grain macroblock-wise quantization and offer a differentiable file size prediction.

%% file: content/method.tex
\section{H.264 and Deep Vision Performance}
\label{subsec:h264}

H.264/AVC performs efficient lossy video compression by exploiting image compression techniques and temporal redundancies \cite{Wiegand2003}. The predictive coding architecture of the H.264 codec utilizes sophisticated hand-crafted transformations to analyze redundancy within videos. A macroblock-wise motion-compensated discrete cosine transform (DCT) followed by quantization and lossless coding is used to archive effective video compression. 

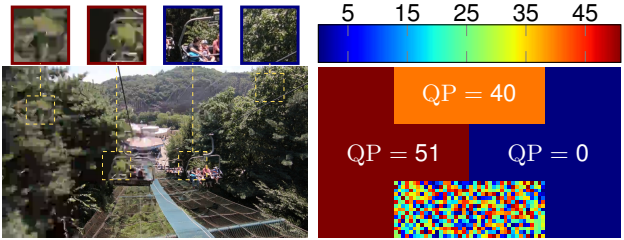
\begin{figure}[b]
    \centering
    \vspace{-0.5em}
    \input{artwork/qp_example_small_2.tex}%
    \vspace{-0.5em}
    \caption{\textbf{H.264 macroblock-wise encoding example.} Effect of different macroblock-wise quantization parameters on the visual quality. Video frame (left) is encoded with $\qp$ map (right) as an \textsc{I}-frame. Video data from REDS dataset~\cite{Nah2019}.}
    \label{fig:h264}
\end{figure}

\paragraph{Quantization parameter.} The H.264 codec allows for a variety of different customizations to the encoding process~\cite{Richardson2011}. The quantization parameter ($\qp$) controls the quantization strength and is the key parameter to control the compression strength. $\qp$ ranges from $0$ to $51$ (integer range), with high values leading to stronger compression. Strong compression reduces bitrate at the cost of video distortion, measured by SSIM (structural similarity index measure)~\cite{Wang2004}. Note that, for a given set of codec parameters, the bitrate remains non-trivially dependent on the video content. For example, a video with entirely black frames requires vastly fewer bits than a natural video, while being encoded with the same codec parameters.

\paragraph{Macroblock-wise quantization.} H.264 offers support for macroblock-wise quantization, in which regions of the video, in this case, $16\times 16$ frame patches (macroblocks), are compressed with varying $\qp$ values. Thus, irrelevant regions can be compressed with a high $\qp$ value (strong compression) and relevant regions with a lower $\qp$ value (less compression). An example of macroblock-wise coding is given in \cref{fig:h264}. We employ macroblock-wise quantization to facilitate a fine-grain spatial and temporal control of the video distortion and bitrate~\cite{Lampert2006}.

\paragraph{Group of pictures.} The group of pictures ($\gop$) size further influences the encoding, by controlling which frames are encoded as an \textsc{I}, \textsc{B}, or \textsc{P}-frame. \textsc{I}-frames (intra-coded frames) are only compressed by utilizing spatial redundancies (similar to image coding), whereas \textsc{B}-frames (bidirectional predicted frames) and \textsc{P}-frames (predicted frames) also use information from adjacent frames. In particular, \textsc{B}-frames are compressed by utilizing a previous and a subsequent \textsc{I}- or \textsc{P}-frame. For compressing \textsc{P}-frames only a single previous \textsc{I}- or \textsc{P}-frame is used.

\begin{figure}[t]
    \centering
    \input{artwork/cityscapes/acc_vs_bw_cityscapes_ss_of_subplot_small.tex}
    \vspace{-0.5em}
    \caption{\textbf{Vision performance \vs{} compression.} Cityscapes segmentation accuracy and optical flow estimation performance, measured by the average endpoint error (AEPE), for different H.264 quantization parameters between the raw clip predictions (pseudo label) and the coded clip predictions. $\qp$ is applied uniformly.}
    \label{fig:accvsqp}
\end{figure}
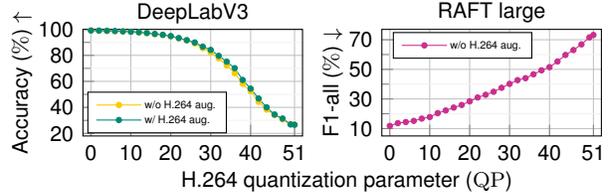

\begin{figure*}[t]
    \centering
    \input{artwork/general_architecture_new.tex}
    \vspace{-0.5em}
    \caption{\textbf{Deep video codec control pipeline.} The control network predicts high-dimensional codec parameters for an input clip and a given dynamic network bandwidth (BW) condition. The video clip is encoded using the predicted codec parameters, sent over the network to the server-side, decoded, and analyzed by a deep vision model (\eg{}, segmentation model). At training, the pre-trained server-side model is fixed and a differentiable surrogate model of the standard codec is used to propagate gradients from the server-side model and the file size prediction to the control network. During inference, the surrogate model is not used. Video frames from~\cite{Nah2019}.}
    \label{fig:generalarchitecture}
\end{figure*}
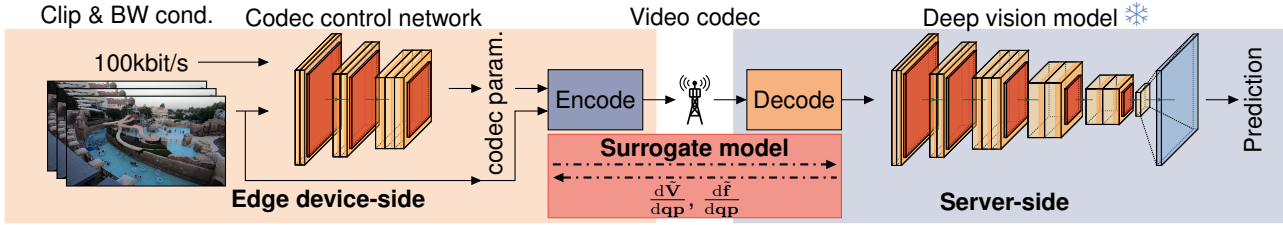

\paragraph{Deep vision performance on H.264 coded videos.} We empirically showcase that lossy compressing videos with H.264 can severely deteriorate downstream performance. When uniformly (for all macroblocks) increasing $\qp$, both segmentation accuracy (DeepLabV3~\cite{Chen2017}) and optical flow estimation (RAFT~\cite{Teed2020}) performance decreased, with reference to the prediction obtained on the original uncompressed video. Otani~\etal~\cite{Otani2022} observed the same for action recognition. Using H.264 augmented frames during training cannot help to overcome this issue. We trained a DeepLabV3 model (w/ ResNet-18~\cite{He2016} backbone) on H.264 coded frames (H.264 aug.). While consuming coded frames during training performance is still highly affected by the compression. \cref{fig:accvsqp} visualizes the results of these experiments. We aim to learn the allocation of $\qp$ such that downstream performance is better preserved and a target bandwidth is met. Additional experiments and experimental details are in the supplement.

\section{Method}
\label{sec:method}

Here we introduce our deep video codec control (\cf{} \cref{fig:generalarchitecture}). Before we describe our novel deep video codec control we present our H.264 codec surrogate model. Surrogate modeling the H.264 in a differentiable manner is required to enable our end-to-end codec control training.

\subsection{Differentiable H.264 surrogate model}
\label{subsec:surrogate}

The H.264 video codec is not differentiable due to discreet operations (non-differentiable) and quantizations (gradient zero or undefined). To enable a gradient flow from the vision model and the generated bandwidth to the codec control network we aim to build a conditional differentiable surrogate model of H.264.

We consider H.264 coding (encoding \& decoding) as a continuous\footnote{Standard H.264 maps from discreet input to discreet outputs. To enable continuous differentiation we extend the H.264 surrogate mapping to real-valued input and output videos.} black-box function mapping the original (raw) video $\mathbf{V}$ conditioned on the macroblock-wise quantization parameters $\mathbf{QP}$ to the encoded and decoded video $\hat{\mathbf{V}}$ as well as the encoded per-frame file size $\mathbf{f}$
\begin{equation}\label{eq:h264}
    \begin{aligned}
        \operatorname{H.\!264}&\!\left(\mathbf{V}, \mathbf{QP}\right)=\bigl(\hat{\mathbf{V}}, \mathbf{f}\bigr),\;\mathbf{V},\hat{\mathbf{V}}\in\mathbb{R}^{3\times {\rm T}\times {\rm H} \times {\rm W}},\\[-3pt]
        \mathbf{QP}&\in\{0, \ldots, 51\}^{{\rm T}\times {\rm H} / 16 \times {\rm W} / 16},\;\mathbf{f}\in\mathbb{R}_{+}^{\rm T}.
    \end{aligned}
\end{equation}
${\rm T}$ indicates the number of frames and ${\rm H}\times {\rm W}$ the spatial dimensions of the RGB video. Other H.264 parameters are considered to be constant. In particular, we consider a $\gop$ of $8$ (thus, ${\rm T}=8$) and a default preset~\cite{Tomar2006}.

\begin{figure}[b!]
    \centering
    \includegraphics{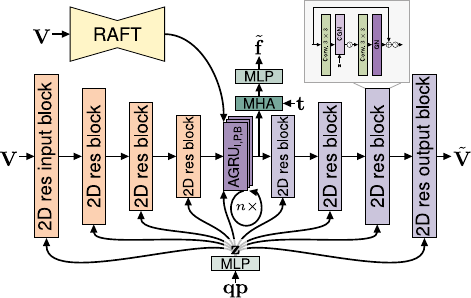}
    \vspace{-0.2em}
    \caption{\textbf{Surrogate model architecture}. Our model is composed of a 2D encoder (orange), a 2D decoder (blue), an MHA-based file size head, and three AGRU bottleneck blocks. We use RAFT to compute the optical flows for the AGRU blocks. For embedding the $\mathbf{qp}$ we use an MLP. Skip-connections omitted for simplicity.}
    \label{fig:surrogatearchitecture}
\end{figure}

Intuitively, our surrogate model fulfills two tasks during the codec control training (\cf{} \cref{fig:generalarchitecture}). First, it allows our codec control network to consume gradient-based feedback from the downstream model regarding its performance. Second, the codec control can also get gradient-based feedback \wrt{} the generated and required bandwidth/file size through our differentiable file size prediction.

\paragraph{Surrogate model architecture.} Our H.264 surrogate model architecture (\cf{} \cref{fig:surrogatearchitecture}) is motivated by computational and memory efficiency and encodes inductive biases from the original (non-diff) H.264 coding approach. In general, our surrogate model entails an encoder-decoder architecture with a bottleneck stage~\cite{Ronneberger2015}. For computational efficiency, we constrain the encoder and decoder to the frame level. To efficiently learn temporal interactions, we present an aligned convolutional gated recurrent unit (AGRU) for each frame type (\textsc{I}-, \textsc{P}-, and \textsc{B}-frame)~\cite{Ballas2016}. By using the optical flow prediction of a pre-trained (small) RAFT model~\cite{Teed2020}, we align adjacent frames used to compress \textsc{P}- and \textsc{B}-frames in each AGRU iteration. Based on the output features of the AGRU we regress the file size on a per-frame level. Our file size head utilizes Multi-Head Attention~\cite{Vaswani2017} to perform cross-attention between learnable query tokens $\mathbf{t}\in\mathbb{R}^{3\times \rm C_{\mathbf{t}}}$ and the AGRU output features to regress the per-frame file size. We utilize individual tokens for each frame type (\textsc{I}-, \textsc{P}-, and \textsc{B}-frame) and repeat the query tokens for each frame.

To condition both the encoder, decoder, and bottleneck blocks on $\qp$ we use a multilayer perceptron (MLP) to embed $\qp$ into a latent vector $\mathbf{z}\in\mathbb{R}^{{\rm C}_{\mathbf{z}} \times {\rm T} \times {\rm H}/16 \times {\rm W}/16}$. To incorporate $\mathbf{z}$ into the surrogate we used conditional group normalization (CGN)~\cite{Devries2017}. The CGN layer combines a spatial feature transform layer~\cite{Wang2018} followed by a group normalization layer~\cite{Wu2018} (w/o learnable parameters). Note our surrogate model uses one-hot encoded quantization parameters, denoted as $\mathbf{qp}\in \{0, 1\}^{52 \times {\rm T} \times {\rm H}/16 \times {\rm W}/16}$. This allows us to later formulate the prediction of the integer-valued $\qp$ as a classification problem. Our resulting surrogate is fully differentiable \wrt{} both the input video clip $\mathbf{V}$ and $\mathbf{qp}$.

\paragraph{Aligned convolutional gated recurrent unit.} Taking inspiration from the motion-compensated and $\gop$-based compression performed by H.264 (and other standard codecs) we propose to utilize AGRUs in the bottleneck stage of our surrogate model. Our AGRU aims to approximate H.264 compression in an iterative fashion in latent space. Through an alignment in the latent space temporal interactions between the frame to be compressed and the reference frames are efficiently modeled. In particular, we utilize separate AGRUs for each frame type. 

The \textsc{B}-frame AGRU is described by
\begin{equation}\label{eq:agru}
    \small
    \begin{aligned}
       \mathbf{Z}_{t+1}&=\sigma(\operatorname{CGN}(\operatorname{C}_{3\times3}(\mathbf{H}_t)+\operatorname{C}_{3\times3}([\hat{\mathbf{A}}_t, \tilde{\mathbf{A}}_t]),\mathbf{z}))\\[-1pt]
       \mathbf{R}_{t+1}&=\sigma(\operatorname{GN}(\operatorname{C}_{1\times1}(\mathbf{H}_t)+\operatorname{C}_{1\times1}([\hat{\mathbf{A}}_t, \tilde{\mathbf{A}}_t])))\\[-1pt]
       \overline{\mathbf{H}}_{t+1}&=\tanh(\operatorname{GN}(\operatorname{C}_{1\times1}(\mathbf{R}_{t+1}\odot\mathbf{H}_t)+\operatorname{C}_{1\times1}([\hat{\mathbf{A}}_t, \tilde{\mathbf{A}}_t])))\\[-1pt]
       \mathbf{H}_{t+1}&=(1-\mathbf{Z}_{t+1})\odot\mathbf{H}_{t}+\mathbf{Z}_{t+1}\odot\overline{\mathbf{H}}_{t+1},
    \end{aligned}
\end{equation}
where $\operatorname{C}_{3\times 3}$ and $\operatorname{C}_{1\times 1}$ denote a $3\times 3$ and a $1\times 1$ 2D convolution, respectively. We do not share parameters between convolutions. $\mathbf{H}_t$ are the latent features of the \textsc{B}-frame. $\hat{\mathbf{A}}_t$ and $\tilde{\mathbf{A}}_t$ are the aligned previous and subsequent frame features used for compression based on the $\gop$ structure. We align the features of the frames used for compression by $\hat{\mathbf{A}}_t=\overleftarrow{w}(\hat{\mathbf{H}}_t, \operatorname{RAFT}(\mathbf{V}_t, \hat{\mathbf{V}}_t))$. We backward warp ($\overleftarrow{w}$) the unaligned features $\hat{\mathbf{H}}_t$ based on the optical flow between the frame to be compressed $\mathbf{V}_i$ and the reference frame $\hat{\mathbf{V}}_j$, using RAFT. We downsample the optical flow to match the spatial dimension of the latent features. For \textsc{P}-frames, we utilize one reference frame. In the case of an \textsc{I}-frame we fully omit the conditioning in the AGRU. Note the reference frames for \textsc{B}- and \textsc{P}-frame compression can be obtained by the known $\gop$ structure~\cite{Wiegand2003}.

\paragraph{Surrogate model training.} In general, we are interested that our surrogate model approximates both the H.264 function ($\hat{\mathbf{V}}\!\approx\!\tilde{\mathbf{V}}$, $\mathbf{f}\!\approx\!\tilde{\mathbf{f}}$, \cf{} Eq.~\eqref{eq:h264}) and provide smooth gradients \wrt{} the quantization parameters ($\frac{\mathrm{d}\tilde{\mathbf{V}}}{\mathrm{d}\mathbf{qp}}$, $\frac{\mathrm{d}\tilde{\mathbf{f}}}{\mathrm{d}\mathbf{qp}}$). Based on the control variates theory, the surrogate model can become a low-variance and continuous gradient estimator of \cref{eq:h264} if \emph{(1)} the difference between the output of the surrogate and the true H.264 function is minimized and \emph{(2)} the two output distributions are maximizing the correlation coefficients $\rho$ \cite{Glynn2002, Grathwohl2018}. We enforce both requirements for $\tilde{\mathbf{V}}$ and $\tilde{f}$ by minimizing $\mathcal{L}_{\mathrm{s}}= \mathcal{L}_{\mathrm{s}_{v}} + \mathcal{L}_{\mathrm{s}_{f}}$ during training. $\mathcal{L}_{\mathrm{s}_{v}}$ supervises the per-frame file size prediction and is defined as
\begin{equation}
    \mathcal{L}_{\mathrm{s}_{f}}=\alpha_{\rho_{f}}\,\mathcal{L}_{\rho_{f}} + \alpha_{\mathrm{L1}}\,\mathcal{L}_{\mathrm{L1}}.
    \label{eq:surrogatefsloss}
\end{equation}%
$\mathcal{L}_{\rho_{f}}$ is the correlation coefficient loss~\cite{Tian2021} between the true file size $\mathbf{f}$ and the predicted file size $\tilde{\mathbf{f}}$. $\mathcal{L}_{\mathrm{L1}}$ is used to minimize the distance between $\mathbf{f}$ and $\tilde{\mathbf{f}}$. Note that we learn $\tilde{\mathbf{f}}$ in $\log_{10}$-space due to the large range of file sizes.

To learn the prediction of the coded video $\tilde{\mathbf{V}}$ we minimize
\begin{equation}
    \mathcal{L}_{\mathrm{s}_{v}}=\alpha_{\rho_{v}}\,\mathcal{L}_{\rho_{v}} + \alpha_{\mathrm{SSIM}}\,\mathcal{L}_{\mathrm{SSIM}} + \alpha_{\mathrm{FF}}\,\mathcal{L}_{\mathrm{FF}}.
    \label{eq:surrogatevideoloss}
\end{equation}
$\mathcal{L}_{\rho_{f}}$ is the correlation coefficient loss for the coded video prediction. We minimize the distance between $\hat{\mathbf{V}}$ and $\tilde{\mathbf{V}}$ both in pixel space and frequency space. In particular, we use the structural similarity measure (SSIM) loss~\cite{Zhao2016} for minimizing the error in pixel space. Taking inspiration from the frequency-based compression utilized by H.264, we utilize the focal frequency loss~\cite{Jiang2021}, minimizing prediction error in the frequency space. $\alpha$ is used to denote the different weighting parameters.

\subsection{Deep video codec control}
\label{subsec:videocodeccontrol}

We aim to learn a deep codec control solving our constrained optimization (\cf{} \cref{eq:problem}). In particular, given a video clip and a dynamic target bandwidth our deep codec control should predict high-dimensional codec parameters ($\qp$), such that downstream performance is preserved while staying within the target bandwidth. Our general architecture is depicted in \cref{fig:generalarchitecture}. Note, while we demonstrate our general codec control on the task of semantic segmentation and optical flow estimation in this paper, the general training formulation (\cf{} \cref{eq:loss}) is agnostic to the vision task to be performed. Our approach only assumes the differentiability of the server-side vision model.

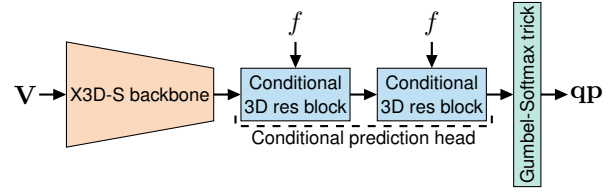
\begin{figure}[t!]
    \centering
    \input{artwork/control_network}
    \vspace{-0.5em}
    \caption{\textbf{Control network architecture.} We use a pre-trained X3D-S followed by two conditional 3D residual blocks. The $\mathbf{qp}$ one-hot vector is obtained by using the Gumbel-Softmax trick.}
    \label{fig:control_network}
\end{figure}

\paragraph{Control network.} Our control network (\cf{} \cref{fig:control_network}) consumes a original video clip $\mathbf{V}$ as well as a target bandwidth $f$ and predicts the macroblock-wise $\qp$. To facilitate the deployment on a standard edge device, such as the NVIDIA Jetson Nano, we use a lightweight architecture. In particular, we utilize X3D-S~\cite{Feichtenhofer2020} as our control network. To input the bandwidth condition to the network we omit the X3D classification head and use two residual blocks with conditional normalization~\cite{Devries2017}.

Due to the discreet nature of $\qp$ (integer-valued), we formulate the $\qp$ prediction as a classification. Our control network learns to predict a logit vector over all possible $\qp$ values. During end-to-end training, the Gumbel-Softmax trick~\cite{Jang2017} is used to produce a differentiable one-hot $\mathbf{qp}$ vector based on the predicted logits. During inference, when used as an input to the original H.264 codec and not to the surrogate, we apply the $\arg\max$ function.

\paragraph{Self-supervised control training.} We propose a self-supervised training strategy to train our control network. To utilize end-to-end gradient-based learning we reformulate the constrained optimization problem (\cf{} \cref{eq:problem}) as a continuous optimization task. In particular, our control loss $\mathcal{L}_{\mathrm{c}}$ consists of three terms -  the bandwidth loss $\mathcal{L}_{\mathrm{b}}$, a performance loss $\mathcal{L}_{\mathrm{p}}$, and a bandwidth regularizer $\mathcal{L}_{\mathrm{r}}$,
\begin{equation}
    \mathcal{L}_{\mathrm{c}} = \alpha_{\mathrm{b}}\,\mathcal{L}_{\mathrm{b}} + 
    \alpha_{\mathrm{p}}\,\mathcal{L}_{\mathrm{p}} + \alpha_{\mathrm{r}}\,\mathcal{L}_{\mathrm{r}}.
    \label{eq:loss}
\end{equation}
where each individual loss is weighted by a separate regularization parameter $\alpha$.

The bandwidth loss $\mathcal{L}_{\mathrm{b}}$ is used to enforce that the deep codec control satisfies the network bandwidth condition (\cf{} \cref{eq:problem}) and is defined as
\begin{equation}
    \mathcal{L}_{\mathrm{b}}=\max\bigl(0, \tilde{b} - b\, (1 - \epsilon_{\mathrm{p}})\bigr),
    \label{eq:lossbandwidth}
\end{equation}
where $b$ is the maximum available bandwidth (bandwidth condition). $\tilde{b}$ denotes the bandwidth prediction computed based on the surrogate model file size prediction $\tilde{\mathbf{f}}$. We convert the per-frame file size (in $\si{\byte}$) to the bandwidth (in $\si{\bit\per\second}$), with the known frame rate ($\mathrm{fps}$), the number of video frames $\rm T$, and the temporal stride $\Delta t$, by $\tilde{b}=\frac{8\,\sum_{i=1}^{\rm T}\tilde{f}_i\,\mathrm{fps}}{{\rm T}\,\Delta t}$, assuming a constant stream for the duration of the input clip. We use a small $\epsilon_{\mathrm{B}}$ in order to enforce the generated bandwidth to be just below the available bandwidth.

The performance loss enforces the preservation of downstream vision performance. In the case of optical flow estimation, we use a scaled absolute error loss ${\operatorname{H}(b - \tilde{b}\,(1 + \epsilon_{\mathrm{p}}))\|\mathbf{O}-\tilde{\mathbf{O}}\|_1}$. $\mathbf{O}$ denotes the optical flow prediction for the coded video clip and $\tilde{\mathbf{O}}$ represents the optical flow prediction based on the original (raw) video clip, used as a pseudo label. We scale the absolute error with a Heaviside function $\operatorname{H}$, only considering the server-side model's performance if the target bandwidth is met with a small tolerance $\epsilon_{\mathrm{p}}$. For semantic segmentation, we replace the absolute error with the Kullback–Leibler divergence. Note, using a different server-side model (\eg{} object detection model) can require adapting $\mathcal{L}_{\mathrm{p}}$ to the new task.

During preliminary experiments, we observed the control network can struggle to use the whole $\qp$ value range. Motivated by this observation, we regularize the control network toward generating a bitrate close to the target by
\begin{equation}
    \mathcal{L}_{\mathrm{r}}=\bigl|\min\bigl(0, \tilde{b} - b\, (1 - \epsilon_{\mathrm{r}})\bigr)\bigr|.
\end{equation}
This regularization loss penalizes the control network if the bandwidth prediction $\tilde{b}$ is far away from the target bandwidth $b$. We utilize ${\epsilon_{\mathrm{B}}<\epsilon_{\mathrm{r}}}$ to not push the generated bandwidth above the target bandwidth.

\paragraph{Training schedule} We train both our deep codec control and surrogate in an alternating fashion. To ensure a stable training of our control network from the beginning on, we pre-train the surrogate model before fine-tuning it during the control training. The codec control training is depicted in pseudocode in the supplement. Note, that our H.264 surrogate is only required for training, not for inference. For inference, the standard H.264 codec is used. We maintain an exponential moving average of the control network (w/ a decay of $0.99$) to combat the noise in our learning signal introduced by our surrogate and the Gumbel-Softmax trick.

\paragraph{Discussion.} We propose to learn high-dimensional codec parameters using end-to-end self-supervised learning, directly optimizing our control objective (\cref{eq:problem}). However, learning codec parameters could theoretically also be achieved using reinforcement learning (RL)~\cite{Mandhane2022}. We argue that using RL is infeasible due to the high-dimensional action space (codec parameters) and the complex loss surface of downstream vision model~\cite{Bucsoniu2018, Li2018}. Note that our deep codec control training can be also viewed as a kind of knowledge distillation~\cite{Gao2021}. We learn our control network with knowledge distilled from the downstream model.

%% file: artwork/qp_example_small_2.tex
\begin{tikzpicture}[every node/.style={font=\small}, spy using outlines={tud6a, line width=0.80mm, dashed, dash pattern=on 1.5pt off 1.5pt, magnification=2, size=0.75cm, connect spies}]

\begin{groupplot}[group style={group size=2 by 1, horizontal sep=4pt, vertical sep=4pt}]

\nextgroupplot[
width=0.6761\columnwidth,
ticks=none,
axis lines=none, 
xtick=\empty, 
ytick=\empty,
axis equal image,
xmin=-0.5, xmax=1279.5,
ymin=-0.5, ymax=719.5,
]
\addplot graphics [includegraphics cmd=\pgfimage, xmin=-0.5, xmax=1279.5, ymin=719.5, ymax=-0.5] {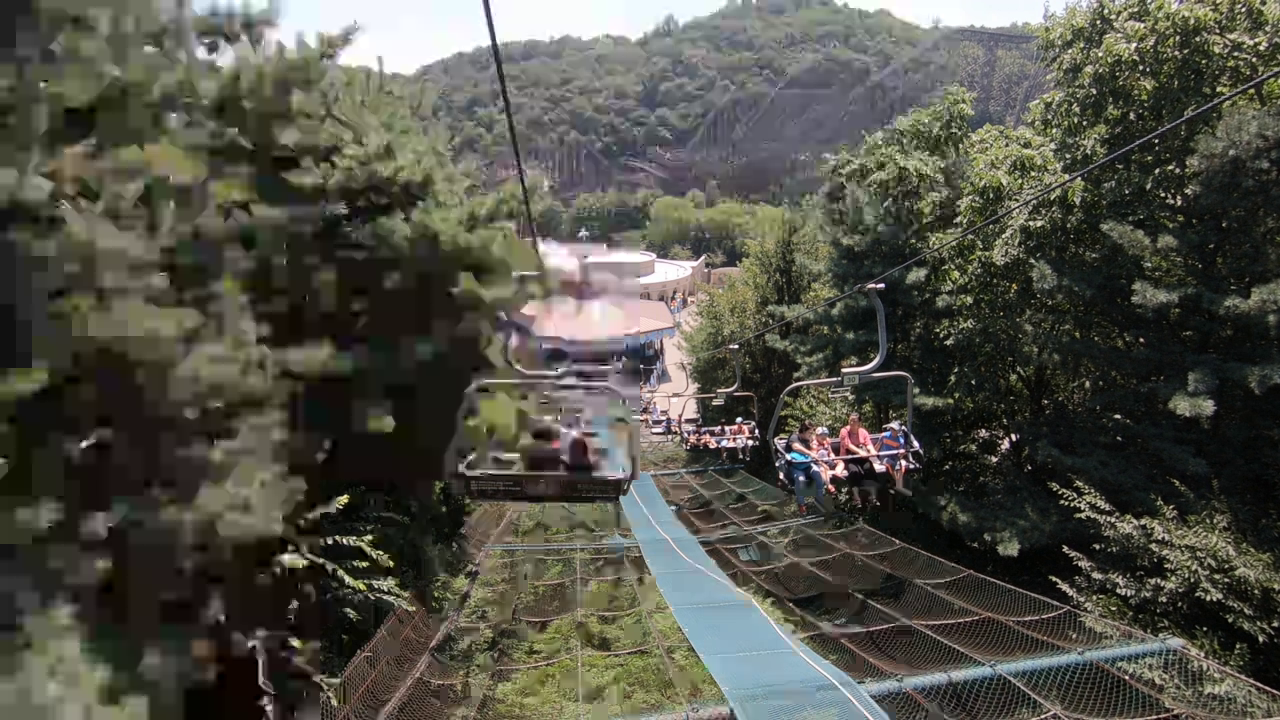};

\nextgroupplot[
width=0.6725\columnwidth,
ticks=none,
axis lines=none, 
xtick=\empty, 
ytick=\empty,
axis equal image,
xmin=-0.5, xmax=79.5,
ymin=-0.5, ymax=44.5,
colorbar,
colorbar horizontal,
colorbar style={
xlabel={},
xtick={5, 15, 25, 35, 45},
xticklabels={5, 15, 25, 35, 45},
xticklabel pos=upper,
xticklabel style={yshift=-2.25pt},
at={(0.0,1.0275)},
anchor=south west,
},
colormap={mymap}{[1pt]
  rgb(0pt)=(0,0,0.5);
  rgb(22pt)=(0,0,1);
  rgb(25pt)=(0,0,1);
  rgb(68pt)=(0,0.86,1);
  rgb(70pt)=(0,0.9,0.967741935483871);
  rgb(75pt)=(0.0806451612903226,1,0.887096774193548);
  rgb(128pt)=(0.935483870967742,1,0.0322580645161291);
  rgb(130pt)=(0.967741935483871,0.962962962962963,0);
  rgb(132pt)=(1,0.925925925925926,0);
  rgb(178pt)=(1,0.0740740740740741,0);
  rgb(182pt)=(0.909090909090909,0,0);
  rgb(200pt)=(0.5,0,0)
},
point meta max=51,
point meta min=0,
]
\addplot graphics [includegraphics cmd=\pgfimage, xmin=-0.5, xmax=79.5, ymin=44.5, ymax=-0.5] {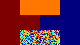};
\end{groupplot}
\node[anchor=center, text=white] at (7.25, 1.1) {$\qp=\text{0}$};
\node[anchor=center, text=white] at (5.2, 1.1) {$\qp=\text{51}$};
\node[anchor=center, text=white] at (6.2, 1.9) {$\qp=\text{40}$};
\spy on (0.5, 1.7) in node [above, dash pattern=, line width=0.4mm, qp51] at (0.5, 2.32);
\spy on (1.5125, 0.96) in node [above, dash pattern=, line width=0.4mm, qp51] at (1.5125, 2.32);
\spy on (2.525, 0.96) in node [above, dash pattern=, line width=0.4mm, qp0] at (2.525, 2.32);
\spy on (3.55, 2.0) in node [above, dash pattern=, line width=0.4mm, qp0] at (3.55, 2.32);

\end{tikzpicture}

%% file: artwork/cityscapes/acc_vs_bw_cityscapes_ss_of_subplot_small.tex
\begin{tikzpicture}[every node/.style={font=\footnotesize}]
    \node[anchor=center] at (0.421\columnwidth, -0.60) {H.264 quantization parameter ($\qp$)};
	\begin{groupplot}[
        group style={
            group name=accvsbw,
            group size=2 by 2,
            ylabels at=edge left,
            horizontal sep=30pt,
            vertical sep=0pt,
        },
        legend style={nodes={scale=0.5}},
        height=3.0cm,
        width=0.54\columnwidth,
        grid=both,
        grid style={line width=.1pt, draw=gray!10},
        major grid style={line width=.2pt,draw=gray!50},
        minor tick num=1,
        xtick pos=bottom,
        ytick pos=left,
        xmin=-2,
        xmax=53,
	    ymin=18,   
        ymax=100,
        ylabel shift=-4.5pt,
        ytick={
            20, 40, 60, 80, 100
        },
        yticklabels={
            20, 40, 60, 80, 100
        },
        xtick={
            0, 10, 20, 30, 40, 51
        },
        xticklabels={
            0, 10, 20, 30, 40, 51
        },
        ticklabel style = {font=\footnotesize},
        legend style={at={(0.02,0.22)}, anchor=west},
        legend cell align={left},
        ]
        \nextgroupplot[title={DeepLabV3}, title style={yshift=-7pt, font=\footnotesize}, ylabel=Accuracy ($\%$) $\uparrow$]
    	\addplot[color=tud6b, mark=*, mark size=1.0pt] coordinates {
    		(0, 99.12)
            (2, 98.97)
            (4, 98.85)
            (6, 98.63)
            (8, 98.48)
            (10, 98.14)
            (12, 97.71)
            (14, 97.17)
            (16, 96.39)
            (18, 95.71)
            (20, 94.71)
            (22, 93.17)
            (24, 91.72)
            (26, 89.19)
            (28, 85.60)
            (30, 82.30)
            (32, 77.75)
            (34, 72.30)
            (36, 66.33)
            (38, 58.12)
            (40, 52.46)
            (42, 44.25)
            (44, 38.27)
            (46, 34.16)
            (48, 30.73)
            (50, 27.16)
            (51, 26.71)
    	};
        \addlegendentry{w/o H.264 aug.}
        \addplot[color=tud3c, mark=*, mark size=1.0pt] coordinates {
    		(0, 99.15)
            (2, 99.06)
            (4, 98.93)
            (6, 98.72)
            (8, 98.49)
            (10, 98.21)
            (12, 97.80)
            (14, 97.22)
            (16, 96.48)
            (18, 95.82)
            (20, 94.84)
            (22, 93.31)
            (24, 91.68)
            (26, 89.95)
            (28, 86.73)
            (30, 84.20)
            (32, 79.65)
            (34, 75.17)
            (36, 70.16)
            (38, 61.10)
            (40, 54.39)
            (42, 46.82)
            (44, 40.01)
            (46, 34.48)
            (48, 31.41)
            (50, 26.84)
            (51, 26.61)
    	};
        \addlegendentry{w/ H.264 aug.}
        \nextgroupplot[
        title={RAFT large},
        title style={yshift=-7pt, font=\footnotesize},
        legend style={at={(0.05,0.85)}, anchor=west},
        ymin=5, ymax=77,
        ytick={
            10, 30, 50, 70
        },
        yticklabels={
            10, 30, 50, 70
        }, ylabel=F1-all ($\%$) $\downarrow$,]
        \addplot[color=tud10a, mark=*, mark size=1.0pt] coordinates {
    		(0, 11.8073)
            (2, 13.6577)
            (4, 14.4103)
            (6, 15.2006)
            (8, 16.7328)
            (10, 17.7773)
            (12, 20.4003)
            (14, 22.2215)
            (16, 24.1599)
            (18, 25.9383)
            (20, 28.3980)
            (22, 30.9835)
            (24, 32.8198)
            (26, 34.8446)
            (28, 37.5349)
            (30, 40.1892)
            (32, 42.7077)
            (34, 44.0429)
            (36, 46.6707)
            (38, 49.3368)
            (40, 51.4957)
            (42, 55.3088)
            (44, 59.7154)
            (46, 63.0313)
            (48, 66.8569)
            (50, 71.6018)
            (51, 73.2276)
    	};
     \addlegendentry{w/o H.264 aug.}
	\end{groupplot}
\end{tikzpicture}

%% file: artwork/general_architecture_new.tex
\begin{tikzpicture}[xscale=1.258, yscale=1, >={Stealth[inset=0pt,length=4.0pt,angle'=45]}, every node/.style={font=\small}]
    \draw[white, fill=tud8b!20] (-3.8, 0.95) rectangle++ (6.9, -2.6);
    \draw[white, fill=tud1d!20] (3.9, 0.95) rectangle++ (5.85, -2.6);
    \begin{scope}[xshift=-0.5cm]
        \node[] at (-2.0, 1.1) {Clip \& BW cond.};
        \begin{scope}[yshift=-0.3cm, xshift=0.1cm]
            \begin{scope}[yshift=-0.05cm]
                \node[] at (-2.1, 0.0) {\includegraphics[width=2.1cm, frame]{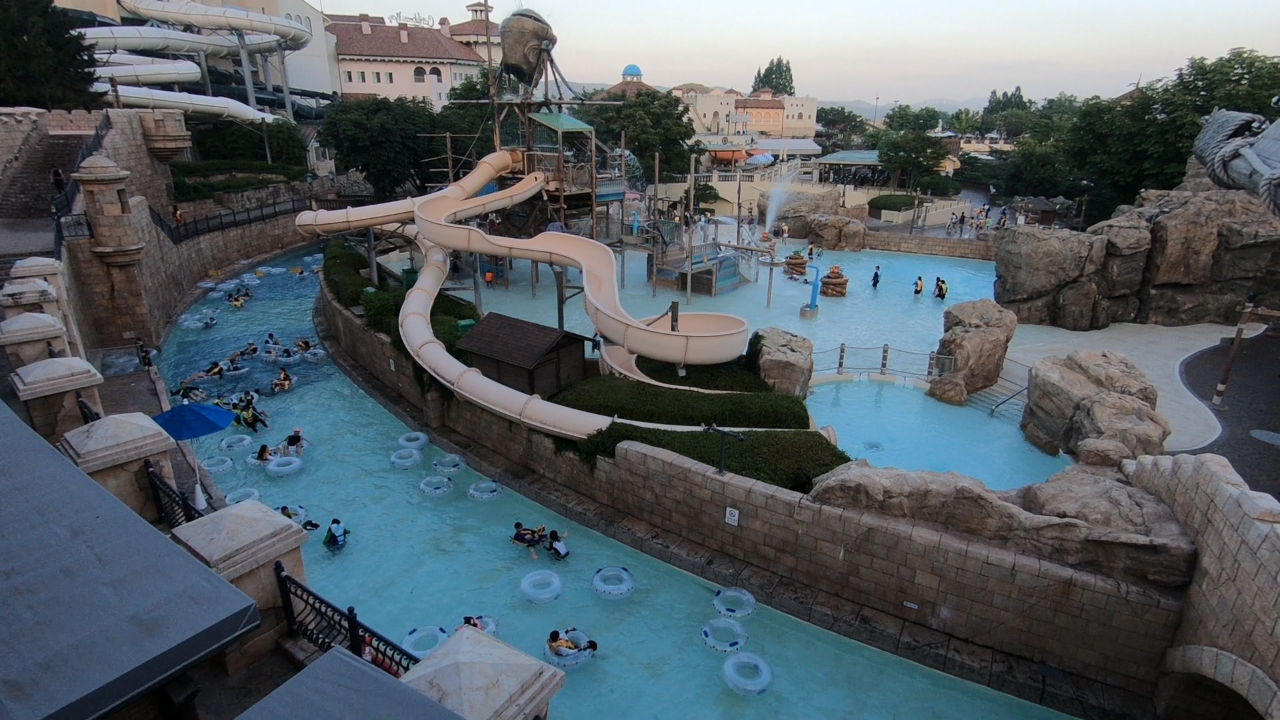}};
                \node[] at (-2.0, -0.1) {\includegraphics[width=2.1cm, frame]{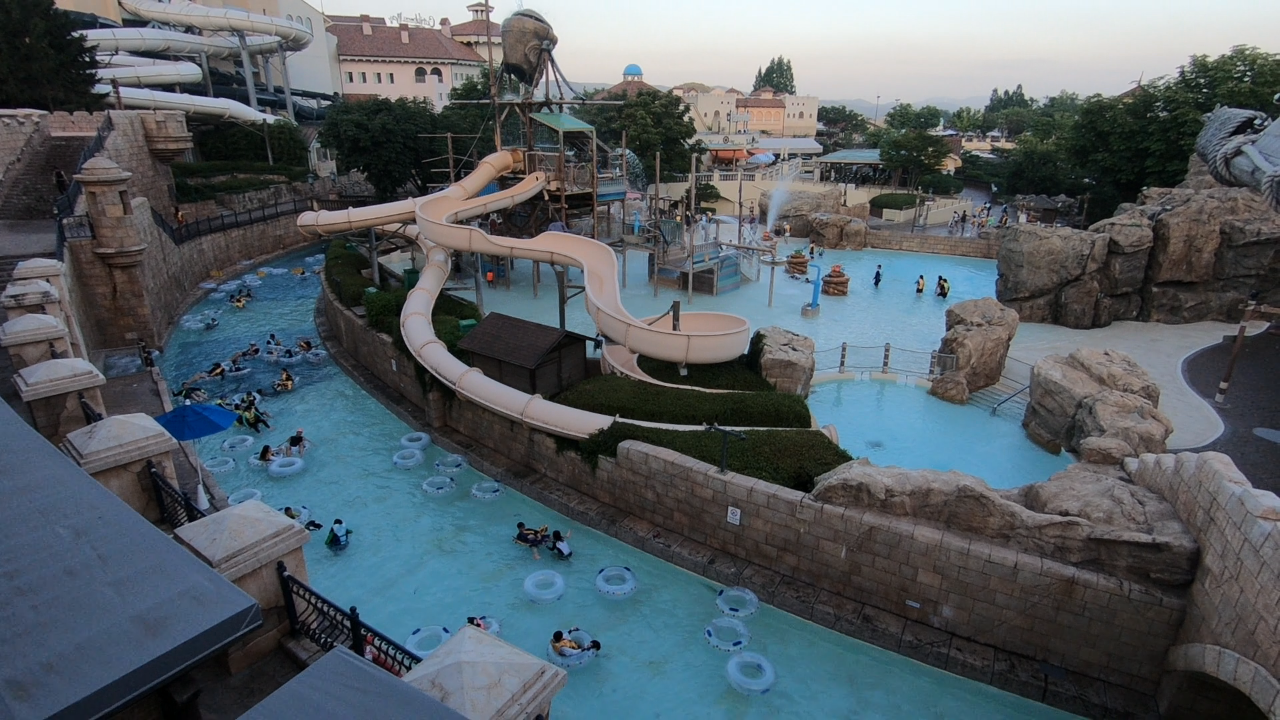}};
                \node[] at (-1.9, -0.2) {\includegraphics[width=2.1cm, frame]{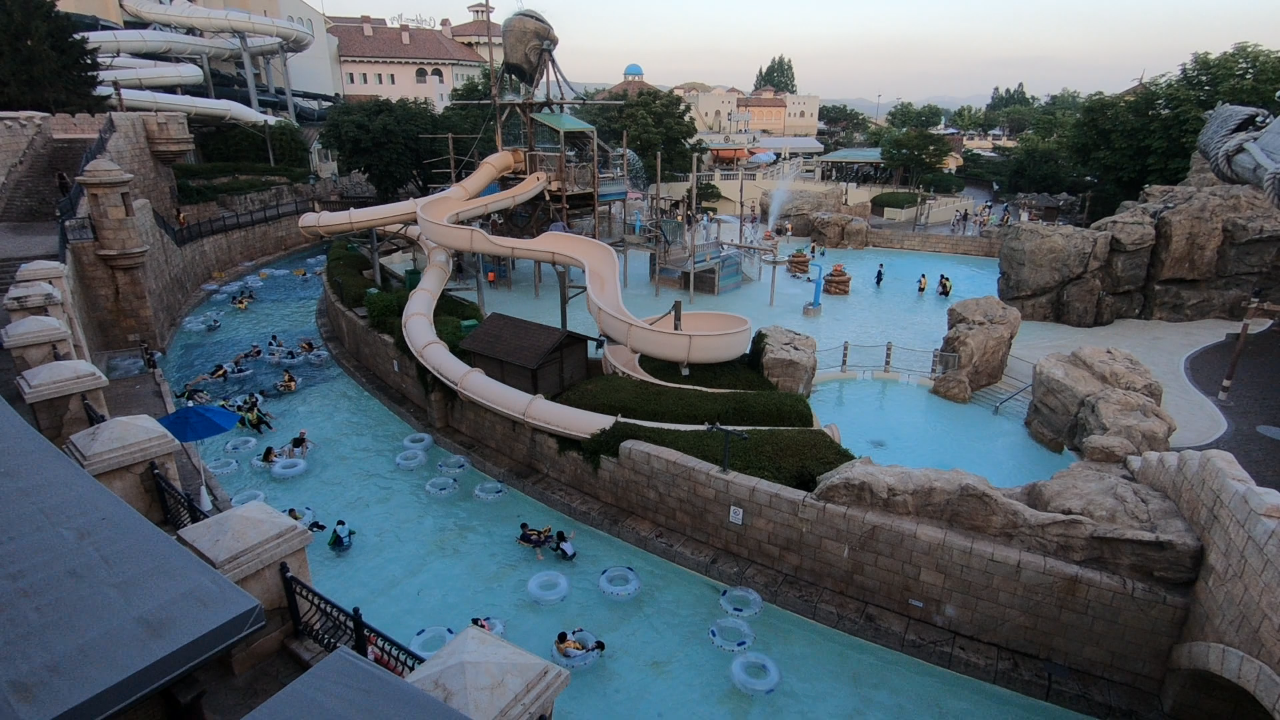}};
            \end{scope}
        \end{scope}
    \end{scope}
    \draw[->] (-1.25, -0.15) -- (-1.25, -1.075) -- (1.60, -1.075) -- (1.60, -0.15) -- (1.95, -0.15);
    \draw[->] (1.15, -1.075) -- (1.25, -1.075);
    \draw[->] (-1.75, 0.5) node[left] {100\si{\kilo\bit\per\second}} -- (-1.0, 0.5);
    \draw[->] (-1.35, -0.15) -- (-1.0, -0.15);
    \node[] at (0.0, 1.1) {Codec control network};
    \node[] at (0.0, 0.0) {\includegraphics[scale=0.63]{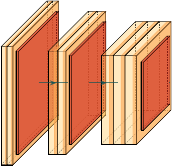}};
    \node[] at (-0.375, -1.35) {\textbf{Edge device-side}};
    \draw[->] (0.9, 0.15) -- (1.25, 0.15);
    \node[rotate=90] at (1.4, -0.04) {codec param.};
    \draw[->] (1.6, 0.15) -- (1.95, 0.15);
    \draw[black, fill=tud1d!50] (1.95, 0.4) rectangle++ (1.0, -0.8);
    \node[] at (2.45, 0.0) {Encode};
    \draw[->] (2.95, 0.0) -- (3.3, 0.0);
    \node[] at (3.5, 1.1) {Video codec};
    \node[] at (3.5, 0.0) {\includegraphics[scale=0.3]{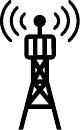}};
    \draw[->] (3.7, 0.0) -- (4.05, 0.0);
    \draw[black, fill=tud8b!50] (4.05, 0.4) rectangle++ (1.0, -0.8);
    \node[] at (4.55, 0.0) {Decode};
    \draw[->] (5.05, 0.0) -- (5.4, 0.0);
    \begin{scope}[yshift=0.04cm]
        \draw[tud9a, fill=tud9a!60] (1.95, -0.5) rectangle (5.05, -1.61);
        \draw[thick, dashdotted, ->] (2.0, -0.9) -- (5.0, -0.9);
        \draw[thick, dashdotted, <-] (2.0, -1.075) -- (5.0, -1.075);
        \node[] at (3.5, -0.7) {\textbf{Surrogate model}};
        \node[] at (3.5, -1.36) {$\frac{\mathrm{d}\tilde{\mathbf{V}}}{\mathrm{d}\mathbf{qp}}$, $\frac{\mathrm{d}\tilde{\mathbf{f}}}{\mathrm{d}\mathbf{qp}}$};
    \end{scope}
    \node[] at (7.1, 1.1) {Deep vision model \textcolor{tud1a}{\SnowflakeChevron}};
    \node[] at (7.2, 0.0) {\includegraphics[scale=0.158]{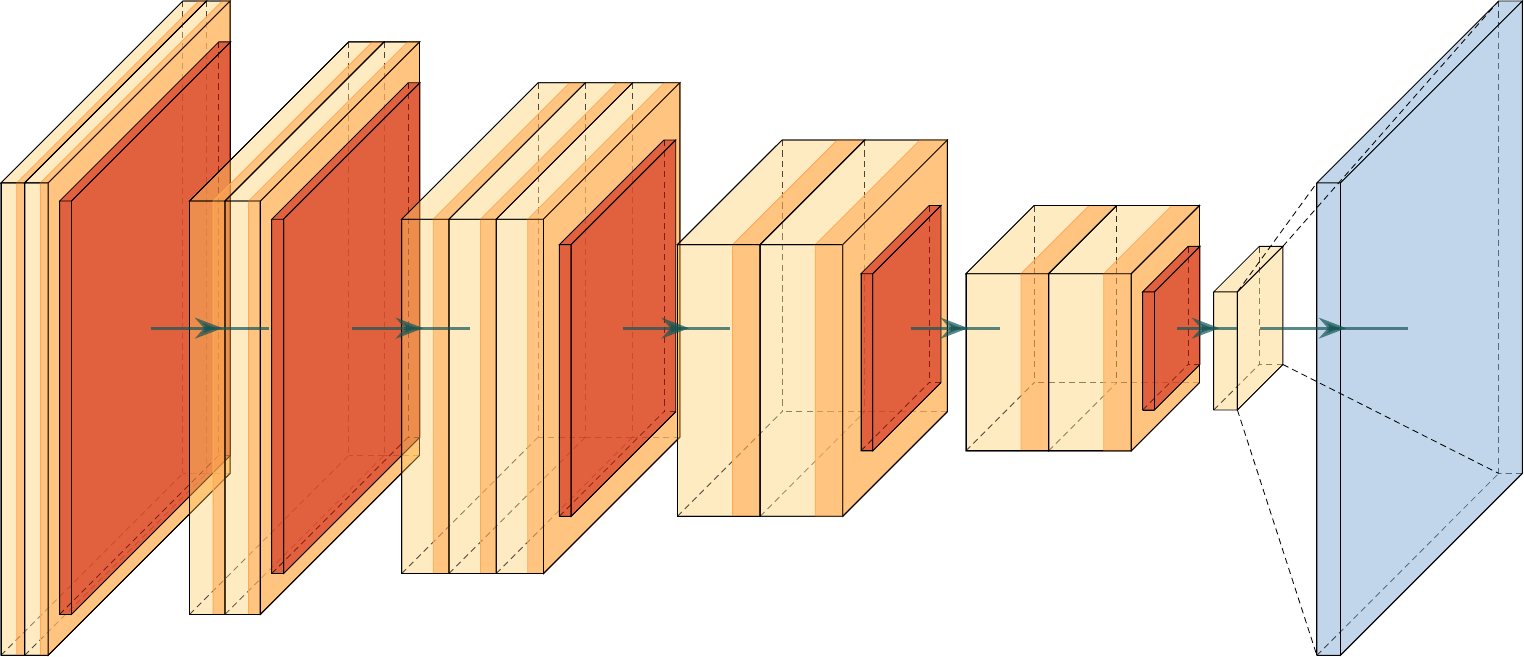}};
    \draw[->] (8.9, 0.0) -- (9.25, 0.0);
    \node[rotate=90] at (9.4, 0.0) {Prediction};
    \node[] at (6.775, -1.35) {\textbf{Server-side}};
\end{tikzpicture}

%% file: artwork/control_network.tex
\begin{tikzpicture}[xscale=0.725, yscale=0.7, >={Stealth[inset=0pt,length=5pt,angle'=45]}]

    \draw[thick, black, ->] (-0.2, 1.0) node[left=-3pt] {$\mathbf{V}$} -- (0.3, 1.0); 
	\draw[black, fill=tud8b!25] (0.3, 2.0) -- (3.0, 1.5) -- (3.0, 0.5) -- (0.3, 0.0) -- cycle;
    \node[anchor=center] at (1.65, 1.0) {\scriptsize X3D-S backbone};
    \draw[thick, black, ->] (3.0, 1.0) -- (3.5, 1.0);
    \draw[black, fill=tud2b!25] (3.5, 1.5) rectangle (5.5, 0.5);
    \node[anchor=center, align=center, text width=1.9cm] at (4.5, 1.0) {\scriptsize Conditional\\[-3pt] 3D$\!$ res$\!$ block};
    \draw[thick, black, ->] (5.5, 1.0) -- (6.0, 1.0);
    \draw[thick, black, ->] (4.5, 2.0) node[above=-1pt] {$f$} -- (4.5, 1.5);
    \begin{scope}[xshift=2.5cm]
        \draw[black, fill=tud2b!25] (3.5, 1.5) rectangle (5.5, 0.5);
        \node[anchor=center, align=center, text width=1.9cm] at (4.5, 1.0) {\scriptsize Conditional\\[-3pt] 3D$\!$ res$\!$ block};
        \draw[thick, black, ->] (5.5, 1.0) -- (6.0, 1.0); %
        \draw[thick, black, ->] (4.5, 2.0) node[above=-1pt] {$f$} -- (4.5, 1.5);
        \draw[black, fill=tud3b!25] (6.0, 2.75) rectangle (6.5, -0.75);
        \node[rotate=90, anchor=center] at (6.25, 1.0) {\scriptsize Gumbel-Softmax trick};
        \draw[thick, black, ->] (6.5, 1.0) -- (7.0, 1.0) node [right=-3pt] {$\mathbf{qp}$};
    \end{scope}
    
    \draw[black, dashed, thick] (3.4, 0.6) --  (3.4, 0.4) -- (8.1, 0.4) node[midway, below=-2pt] {\scriptsize Conditional prediction head} -- (8.1, 0.6);
 
\end{tikzpicture}

%% file: content/experiments.tex
\section{Experiments}
\label{sec:experiments}

\paragraph{Dataset.} We perform experiments on Cityscapes~\cite{Cordts2016} and CamVid~\cite{Brostow2009}. We utilize the unlabeled sequences of Cityscapes including 30-frame video clips ($\sim\!10\si{\kilo\relax}$ frames total) with $2967$ training and $498$ validation clips. The frame rate is $17\si{\fps}$. Similarly for Camvid, we use the available videos (at $15\si{\fps}$) composed of $\sim\!29\si{\kilo\relax}$ frames in total. We use three videos for training and one video for validation. For codec control training \& validation, we sample clips of $8$ frames with a temporal stride of $3$. For the surrogate model pre-training, we vary the temporal stride randomly between $1$, $2$, and $3$.

\paragraph{Baseline.} Since we are not aware of an existing general or task-specific approach to directly solve our constrained optimization problem (\cf{} \cref{eq:problem}) within the scope of standardization, we utilize the generic H.264 rate control. In particular, we follow similar work and use 2-pass average bitrate (ABR) control~\cite{Tomar2006, Mandhane2022}. 2-pass ABR is a competitive baseline, since, in contrast to our control, 2-pass ABR consumes the video clip twice. The first pass gathers information about the motion and prediction error. The second pass uses this information to set $\qp$ for meeting a target bandwidth while minimizing distortion. Note that 2-pass ABR does not guarantee that the target bandwidth is met.

\paragraph{Control validation.} We aim to validate the codec controls directly on the constrained optimization problem (\cf{} \cref{eq:problem}) by measuring a task-specific metric, considering clip dropping. In a real-world deployment, exceeding the available network bandwidth can result in block noises, frame skipping, or even clip dropping (worst case)~\cite{Itsumi2022}. For validating segmentation performance we use the pixel-wise accuracy $\mathrm{acc}_{\mathrm{seg}}$ (in \%) considering clip-dropping if bandwidth is exceeded. If a clip is dropped $\mathrm{acc}_{\mathrm{seg}}$ is considered to be zero for the respective clip. For optical flow estimation, we use the F1-all metric, reporting the outlier ratio (in \%) of pixels exceeding a threshold of $3$ pixels or $5\%$ \wrt{} the optical flow (pseudo) label. If bandwidth is exceeded we set F1-all to $100\%$ for the respective clip. Note we employ the prediction on the original (uncomp.) clip as a pseudo label. In addition to the task-specific metric, we also measure the bandwidth condition accuracy $\mathrm{acc}_{\mathrm{bw}}$ (in \%). Following similar work, we measure both the task-specific metric ($\mathrm{acc}_{\mathrm{seg}}$ or F1-all) and $\mathrm{acc}_{\mathrm{bw}}$ with a tolerance of zero ($\Delta0\%$), five ($\Delta2\%$), and ten ($\Delta5\%$) percent on the bandwidth condition~\cite{Mandhane2022}.

\paragraph{Implementation details.} We implement our surrogate model and codec control pipeline using PyTorch~\cite{Paszke2019} and Lightning~\cite{Falcon2019}. For the macroblock-wise H.264 compression, we rely on a modified FFmpeg implementation of AccMPEG~\cite{Tomar2006, Du2022}. We pre-train our surrogate model for $45\si{\kilo\relax}$ iterations. Our deep codec control is trained for just $6.0\si{\kilo\relax}$ iterations. For both, we utilize two NVIDIA A6000 ($48\si{\giga\byte}$) GPUs. Surrogate pre-training takes approximately one day, whereas the codec control training requires $16\si{\hour}$ to complete. We sample the network bandwidth condition $b$ from a log uniform distribution $\mathcal{U}_{\log}\!\left(30\si{\kilo\bit\per\second}, 0.9\si{\mega\bit\per\second}\right)$ capturing the working range of the H.264 codec. We sample from a log-uniform distribution to explore $\qp$ uniformly during training. For more implementation details and hyperparameters please refer to the supplement.

\begin{figure}[t!]
    \centering
    \input{artwork/surrogate_results/camvid/surrogate_results_camvid}%
    \vspace{-0.5em}
    \caption{\textbf{Qualitative surrogate model results.} On the left frame coded by H.264 for the given $\qp$ map (right). In the middle is the coded video prediction of our surrogate model. We show the first frame of the clip. CamVid dataset used.}
    \label{fig:surrogate_results}
\end{figure}
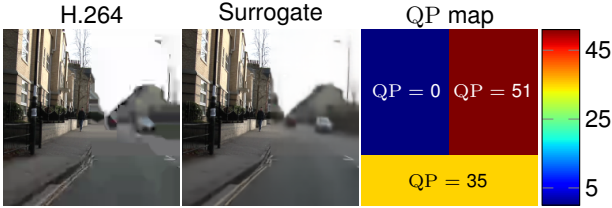

\paragraph{Surrogate model results.} In \cref{fig:surrogate_results}, we showcase qualitative results of our codec surrogate model in approximating H.264 video distortion. Our conditional surrogate can adapt to different macroblock-wise quantization parameters and accurately approximate video distortion introduced by H.264. For low $\qp$ regions, our conditional surrogate maintains details of the video frame. For high $\qp$ regions, details and high-frequencies are discarded and the video is distorted similarly to the original coded video. 

\begin{table}[t]
    \setlength\tabcolsep{0.75pt}
    \renewcommand\arraystretch{0.88}
    \caption{\textbf{Conditional surrogate model validation results.} We report SSIM and L1 scores for the distorted video prediction and the relative error for the file size prediction. Note a pixel range of $[0, 255]$ is used for validation. We show results averaged over the full $\qp$ range. We validate the surrogate model once on each possible $\qp$ value using a uniform $\qp$ (for all macroblocks).}
    \vspace{-4pt}
    \input{table/surrogate_results}
    \label{table:surrogate_results_summary}
\end{table}

We also validate the effectiveness of our conditional H.264 surrogate model qualitatively (\cf{} \cref{table:surrogate_results_summary}). In addition, we also ablate the effect of our AGRU bottleneck block. For comparison, we train a conditional surrogate model with three standard 3D residual blocks, instead of our AGRU, in the bottleneck stage. Our conditional surrogate model is able to approximate the codec video of standard H.264 well. Additionally, our surrogate model also accurately predicts the generated file size. Using our AGRU improves over the 3D convolutional baseline. Note we offer additional results of our surrogate in the supplement.

\begin{table}[b]
    \setlength\tabcolsep{0.75pt}
    \renewcommand\arraystretch{0.88}
    \caption{\textbf{Semantic segmentation validation results.} BW ($\mathrm{acc}_{\mathrm{bw}}$) \& segmentation accuracies ($\mathrm{acc}_{\mathrm{seg}}$) for difference BW tolerances reported. Metrics averaged over ten BW conditions.}
    \vspace{-4pt}
    \input{table/results_summary_ss}
    \label{table:results_summary_ss}
\end{table}

\begin{table}[t]
    \setlength\tabcolsep{0.75pt}
    \renewcommand\arraystretch{0.88}
    \caption{\textbf{Optical flow validation results.} We report bandwidth accuracy ($\mathrm{acc}_{\mathrm{bw}}$) and F1-all scores for difference tolerances on the BW conditions. Metrics averaged over ten BW conditions.}
    \vspace{-4pt}
    \input{table/results_summary_of}
    \label{table:results_summary_of}
\end{table}

\paragraph{Codec control results: semantic segmentation.} We compare our deep codec control to 2-pass ABR on the downstream task of semantic segmentation. As the downstream model, we utilize a DeepLabV3 model (w/ ResNet-18 backbone) trained on the respective dataset. Our obtained results are depicted in \cref{table:results_summary_ss}. On both Cityscapes and CamVid, our deep codec control strongly outperforms the 2-pass ABR control. Our deep codec control better preserves downstream performance (with no bandwidth tolerance) and predicts codec parameters such that the dynamic bandwidth condition is met. In contrast, 2-pass ABR tends to overshoot the bandwidth condition. This can be observed by the vastly increasing $\mathrm{acc}_{\mathrm{bw}}$ if a bandwidth tolerance is permitted. This behavior is also reflected in an increased segmentation accuracy (considering drops) of 2-pass ABR if a bandwidth tolerance is permitted. However, in the most representative case, if no bandwidth tolerance is permitted, our deep codec control leads to vastly better downstream results. In particular, on Cityscapes our deep codec control leads to improvements in $\mathrm{acc}_{\mathrm{seg}}$ of $20.5\%$ and on CamVid to $11.6\%$ over 2-pass ABR considering no BW tolerance.

\paragraph{Codec control results: optical flow estimation.} We also analyze the performance of deep codec control and our 2-pass ABR baseline on optical flow estimation as the downstream vision task. As the downstream model, we utilize RAFT large, trained on synthetic data and driving data from KITTI~\cite{Geiger2013, Teed2020}. \cref{table:results_summary_of} presents our codec control results for optical flow estimation. Similar to semantic segmentation, our codec control also outperforms 2-pass ABR in controlling H.264 for optical flow estimation. Our deep codec control better preserves downstream performance (w/ no bandwidth tolerance) and better follows dynamic bandwidth conditions. Since the 2-pass ABR control often overshoots the bandwidth condition downstream performance is increased if a bandwidth tolerance is permitted. If no bandwidth tolerance is permitted, the most representative setting, our deep codec control leads to more than $10\%$ fewer outliers (F1-all) in the optical flow prediction than 2-pass ABR.

\paragraph{Codec control transfer results} In \cref{table:results_transfer}, we report results when transferring our deep codec control between downstream tasks during inference. When transferring our deep codec control trained to preserve optical flow performance to semantic segmentation we observe a drop in performance. This demonstrates both the ability of our codec control to learn a task-specific behavior and showcases the effectiveness of the surrogate model's gradients.

\paragraph{Additional results.} We refer the reader to the supplement for additional (qualitative \& quantitative) results of both our deep video codec control and conditional surrogate model. 

\begin{table}[t]
    \setlength\tabcolsep{0.75pt}
    \renewcommand\arraystretch{0.88}
    \caption{\textbf{Transfer results of our codec control from optical flow estimation to semantic segmentation on Cityscapes.} For reference, we also report results when directly trained on semantic segmentation.}
    \vspace{-4pt}
    \input{table/transfer_results}
    \label{table:results_transfer}
\end{table}

%% file: artwork/surrogate_results/camvid/surrogate_results_camvid.tex
\begin{tikzpicture}[every node/.style={font=\small}, spy using outlines={tud6a, line width=0.80mm, dashed, dash pattern=on 1.5pt off 1.5pt, magnification=2, size=0.75cm, connect spies}]

\begin{groupplot}[group style={group size=3 by 1, horizontal sep=1.25pt, vertical sep=1.25pt}]

\nextgroupplot[
width=0.545\columnwidth,
ticks=none,
axis lines=none, 
xtick=\empty, 
ytick=\empty,
axis equal image,
xmin=0, xmax=224,
ymin=0, ymax=224,
title={H.264\vphantom{g}},
title style={yshift=-9pt},
]
\addplot graphics [includegraphics cmd=\pgfimage, xmin=0, xmax=224, ymin=0, ymax=224] {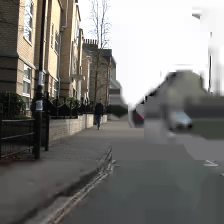};

\nextgroupplot[
width=0.545\columnwidth,
ticks=none,
axis lines=none, 
xtick=\empty, 
ytick=\empty,
axis equal image,
xmin=0, xmax=224,
ymin=0, ymax=224,
title={Surrogate},
title style={yshift=-8pt},
]
\addplot graphics [includegraphics cmd=\pgfimage, xmin=0, xmax=224, ymin=0, ymax=224] {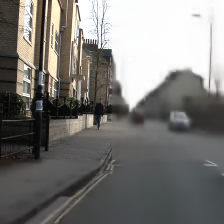};

\nextgroupplot[
width=0.545\columnwidth,
ticks=none,
axis lines=none, 
xtick=\empty, 
ytick=\empty,
axis equal image,
xmin=0, xmax=14,
ymin=0, ymax=14,
colorbar,
colorbar style={
    ylabel={},
    ytick={5, 25, 45},
    yticklabels={5, 25, 45},
    at={(1.025,1)},
    yticklabel style={xshift=-1.5pt},
},
colormap={mymap}{[1pt]
    rgb(0pt)=(0,0,0.5);
    rgb(22pt)=(0,0,1);
    rgb(25pt)=(0,0,1);
    rgb(68pt)=(0,0.86,1);
    rgb(70pt)=(0,0.9,0.967741935483871);
    rgb(75pt)=(0.0806451612903226,1,0.887096774193548);
    rgb(128pt)=(0.935483870967742,1,0.0322580645161291);
    rgb(130pt)=(0.967741935483871,0.962962962962963,0);
    rgb(132pt)=(1,0.925925925925926,0);
    rgb(178pt)=(1,0.0740740740740741,0);
    rgb(182pt)=(0.909090909090909,0,0);
    rgb(200pt)=(0.5,0,0)
},
point meta max=51,
point meta min=0,
title={$\qp$ map\vphantom{g}},
title style={yshift=-9pt},
]
\addplot graphics [includegraphics cmd=\pgfimage, xmin=0, xmax=14, ymin=0, ymax=14] {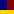};
\end{groupplot}
\node[anchor=center, text=white] at (6.5, 1.5) {\scriptsize $\qp=\text{51}$};
\node[anchor=center, text=white] at (5.35, 1.5) {\scriptsize $\qp=\text{0}$};
\node[anchor=center] at (5.9, 0.3) {\scriptsize $\qp=\text{35}$};

\end{tikzpicture}

%% file: table/surrogate_results.tex
\small
\begin{tabular*}{\columnwidth}{@{\extracolsep{\fill}}l@{\hskip 0.25em}c@{\hskip 0.25em}S[table-format=1.3]S[table-format=1.3]S[table-format=1.3]@{}}
	\toprule
	&  & \multicolumn{2}{c}{$\tilde{\mathbf{V}}$ } & {$\tilde{\mathbf{f}}$}  \\
    \cmidrule{3-4}\cmidrule{5-5}
	{Method} & {AGRU} & {SSIM $\uparrow$} & {L1 $\downarrow$} & {Rel. error $(\%)$ $\downarrow$} \\
	\midrule
    \multicolumn{5}{@{}c}{\textit{Cityscapes}~\cite{Cordts2016}} \\
    \midrule
    Conditional surrogate & \xmark & 0.949 & 3.892 & 6.719 \\ 
    Conditional surrogate & \cmark & 0.964 & 2.552 & 5.150 \\
    \midrule
    \multicolumn{5}{@{}c}{\textit{CamVid}~\cite{Brostow2009}} \\
    \midrule
    Conditional surrogate & \cmark & 0.958 & 2.876 & 2.246 \\
	\bottomrule
\end{tabular*}

%% file: table/results_summary_ss.tex
\small
\begin{tabular*}{\columnwidth}{@{\extracolsep{\fill}}l@{\hskip 0.25em}S[table-format=2.2]S[table-format=2.2]S[table-format=2.2]S[table-format=2.2]S[table-format=2.2]S[table-format=2.2]@{}}
	\toprule
	& \multicolumn{3}{c}{$\mathrm{acc}_{\mathrm{bw}}$ $(\%)$ $\uparrow$} & \multicolumn{3}{c}{$\mathrm{acc}_{\mathrm{seg}}$ $(\%)$ $\uparrow$}  \\
    \cmidrule{2-4}\cmidrule{5-7}
	{Method} & {$\Delta0\%$} & {$\Delta2\%$} & {$\Delta5\%$} & {$\Delta0\%$} & {$\Delta2\%$} & {$\Delta5\%$} \\
	\midrule
    \multicolumn{7}{@{}c}{\textit{Cityscapes}~\cite{Cordts2016}} \\
    \midrule
	2-pass ABR (H.264) & 68.13 & 74.98 & 82.31 & 64.29 & 70.57 & 77.07 \\
    \textbf{Deep codec control} & \textbf{96.22} & \textbf{97.05} & \textbf{97.91} & \textbf{84.79} & \textbf{85.50} & \textbf{86.28} \\
    \midrule
    \multicolumn{7}{@{}c}{\textit{CamVid}~\cite{Brostow2009}} \\
    \midrule
    2-pass ABR (H.264) & 63.91 & 74.43 & 85.36 & 54.06 & 62.49 & \textbf{71.53} \\
    \textbf{Deep codec control} & \textbf{94.64} & \textbf{95.61} & \textbf{96.46} & \textbf{65.70} & \textbf{62.52} & 59.01 \\
	\bottomrule
\end{tabular*}

%% file: table/results_summary_of.tex
\small
\begin{tabular*}{\columnwidth}{@{\extracolsep{\fill}}l@{\hskip 0.25em}S[table-format=2.2]S[table-format=2.2]S[table-format=2.2]S[table-format=2.2]S[table-format=2.2]S[table-format=2.2]@{}}
	\toprule
	& \multicolumn{3}{c}{$\mathrm{acc}_{\mathrm{bw}}$ $(\%)$ $\uparrow$} & \multicolumn{3}{c}{F1-all $(\%)$ $\downarrow$}  \\
    \cmidrule{2-4}\cmidrule{5-7}
	{Method} & {$\Delta0\%$} & {$\Delta2\%$} & {$\Delta5\%$} & {$\Delta0\%$} & {$\Delta2\%$} & {$\Delta5\%$} \\
	\midrule
    \multicolumn{7}{@{}c}{\textit{Cityscapes}~\cite{Cordts2016}} \\
    \midrule
	2-pass ABR (H.264) & 69.60 & 76.48 & 83.09 & 41.99 & 36.37 & 31.18 \\
    \textbf{Deep codec control} & \textbf{98.05} & \textbf{98.29} & \textbf{98.86} & \textbf{27.57} & \textbf{27.42} & \textbf{27.03} \\
    \midrule
    \multicolumn{7}{@{}c}{\textit{CamVid}~\cite{Brostow2009}} \\
    \midrule
    2-pass ABR (H.264) & 63.99 & 73.93 & 85.34 & 43.54 & 34.73 & 26.31 \\
    \textbf{Deep codec control} & \textbf{97.41} & \textbf{98.73} & \textbf{98.09} & \textbf{21.06} & \textbf{20.55} & \textbf{20.08} \\
	\bottomrule
\end{tabular*}

%% file: table/transfer_results.tex
\small
\begin{tabular*}{\columnwidth}{@{\extracolsep{\fill}}l@{\hskip 0.25em}S[table-format=2.2]S[table-format=2.2]S[table-format=2.2]S[table-format=2.2]S[table-format=2.2]S[table-format=2.2]@{}}
	\toprule
	& \multicolumn{3}{c}{$\mathrm{acc}_{\mathrm{bw}}$ $\uparrow$} & \multicolumn{3}{c}{$\mathrm{acc}_{\mathrm{seg}}$ $\uparrow$}  \\
    \cmidrule{2-4}\cmidrule{5-7}
	{Training task} & {$\Delta0\%$} & {$\Delta2\%$} & {$\Delta5\%$} & {$\Delta0\%$} & {$\Delta2\%$} & {$\Delta5\%$} \\
    \midrule
    \textbf{Optical flow estimation} & 97.79 & 98.31 & 98.90 & 75.03 & 75.37 & 75.76 \\
    Semantic segmentation & 96.22 & 97.05 & 97.91 & 84.79 & 85.50 & 86.28 \\
	\bottomrule
\end{tabular*}

%% file: content/discussion.tex
\section{Discussion}
\label{sec:discussion}

We demonstrated the general feasibility of our deep codec control in overcoming the limitations of the standard H.264 codec and conserving downstream performance. Our experiments demonstrate compelling results for two classical downstream tasks: semantic segmentation and optical flow estimation. However, currently, we train a deep codec control for a specific downstream task and model. Training a deep control for every task and model might not be practical. An avenue for future investigation lies in exploring how control networks can be transferred across different downstream tasks and models. Showcasing the potential application to control other video codecs, such as H.265, also represents an interesting direction for future research. Our deep codec control pipeline offers a solid foundation for future research toward cross-task support, a generalized deep codec control, and support for other standard video codecs. We hope that our contribution opens up new avenues in video coding and enables the effective use of standard codecs in state-of-the-art computer vision pipelines.

%% file: content/conclusion.tex
\section{Conclusion}
\label{sec:conclusion}

We presented the first fully end-to-end learnable deep codec control for standard video codecs, to conserve downstream deep vision performance in the face of dynamic bandwidth conditions. Our novel conditional differentiable codec surrogate model enables us to learn a content and network bandwidth-dependent codec control using self-supervised learning. We empirically demonstrate that our deep codec control can control the standard H.264 codec and is able to meet dynamic bandwidth conditions, while better preserving the downstream performance of a deep vision model than a standard baseline (H.264 2-pass ABR control). Our deep video codec control not only offers an alternative approach for optimizing video codecs for deep vision models but we can also conserve current standards.

{\small
\paragraph{Acknowledgements.} We thank Simone Schaub-Meyer for feedback on the initial version of the paper. We also thank Heinz Koeppl for the general support.
}

%% file: content/appendix.tex
\onecolumn
\appendix
{\centering
\Large
\textbf{\thetitle}\\
\vspace{0.5em}Supplementary Material \\
\vspace{1.0em}}

\section*{Supplement}

In this supplement, we provide additional details of our deep video codec control and conditional surrogate model. In particular, we present additional material on our methodology and also provide additional experimental results.

\section{Method}

This section provides additional details about our novel self-supervised control training, the surrogate model pre-training, the conditional group normalization layer, and the encoder/decoder residual block used in our conditional surrogate model. Finally, we provide a more detailed discussion of the novelty of our conditional surrogate model supporting the main paper's related work section.

\subsection{Self-supervised deep codec control}

Here we provide our self-supervised control training in PyTorch-like pseudo-code (\cref{alg:training}). Note, we train our codec control and conditional surrogate model in an alternating fashion. We utilize the prediction on the original uncompressed clip as a pseudo label in our deep codec control training (line 14). More advanced pseudo-labeling approaches (\eg, \cite{Araslanov2021} or \cite{Jeong2023}) might offer an improved learning signal. However, for the sake of simplicity, we refrain from using advanced pseudo-labeling approaches.

\begin{algorithm}[ht!]
    \caption{Our end-to-end deep codec control training.}
    \input{algorithm/training.tex}
    \label{alg:training}
\end{algorithm}

\subsection{Surrogate model pre-training}

During preliminary surrogate model pre-training runs, we observed that the surrogate model entails a tendency to be biased toward the identity function. In particular, the surrogate model solved the task of predicting the distorted video by predicting the original video. We combat this behavior by first learning on high quantization parameter ($\qp$) values and gradually introducing a larger range of   $\qp$ during training. We start by just sampling a macroblock-wise $\qp$ of $51$ before linearly increasing the $\qp$ sampling range to the full range between $0$ and $51$ until half of the pre-training. We observed that first learning strong compression rates, including strong video distortion, prevents the surrogate from just learning the identity mapping. We believe the strong video distortion at the beginning of the training lets the conditional surrogate model diverge from the shortcut solution (identity mapping).

To artificially enlarge the number of available video clips, we use data augmentation during surrogate pre-training. In particular, we randomly (with a probability of $0.1$) convert the RGB frames to grayscale frames. We also utilize temporal augmentations. We randomly reverse the order of frames (with a probability of $0.5$) and repeat frames (with a probability of $0.1$).

We sample the $\qp$ parameters at different resolution stages to mimic regions with uniform $\qp$ values. Additionally, with a probability of $0.4$, we utilize the same $\qp$ map for all frames in the video clip, mimicking uniform $\qp$ values in the spatial dimension. $\qp$ sample generated by our technique can be seen in \cref{fig:surrogate_results_camvid_full} (and subsequent Figures).

\subsection{Conditional group normalization} 

To incorporate the one-hot macroblock-wise quantization parameters $\mathbf{qp}$ into our conditional surrogate model, we utilize conditional group normalization (CGN). Before applying CGN, we embed $\mathbf{qp}$ using a two-layer feed-forward neural network to the latent vector $\mathbf{z}$. Similar to conditional batch normalization~\cite{Devries2017}, we predict the affine parameters of group normalization~\cite{Wu2018} based on the condition latent vector $\mathbf{z}$. In particular, we use the spatial feature transform layer introduced by Wang~\etal~\cite{Wang2018} to predict macroblock-wise affine parameters. Formally, our CGN later is described as
\begin{equation}
    \hat{\mathbf{X}}=\operatorname{Softplus}\!\left(\operatorname{\xi}_{\mu}\!\left(\mathbf{z}\right)\right) \operatorname{GroupNorm}\!\left(\mathbf{X}\right) + \operatorname{\xi}_{\sigma}\!\left(\mathbf{z}\right),
\end{equation}
where $\mathbf{X}$ is the 4D spatio-temporal input feature map and $\hat{\mathbf{X}}$ is the output feature map of the same shape. $\mathbf{z}$ is the 4D condition embedding generated from $\mathbf{qp}$. $\operatorname{\xi}$ denotes a learnable linear mapping transforming the condition embedding. A Softplus activation~\cite{Nair2010} is used to ensure a positive scaling. $\operatorname{GroupNorm}$ denotes the standard group normalization layer without affine parameters. To ensure matching spatial dimensions between the feature map and macroblock-wise affine parameters, nearest-neighbor interpolation is used to the output of $\operatorname{\xi}_{\mu}$ and $\operatorname{\xi}_{\sigma}$.

\subsection{Encoder/decoder 2D residual block}

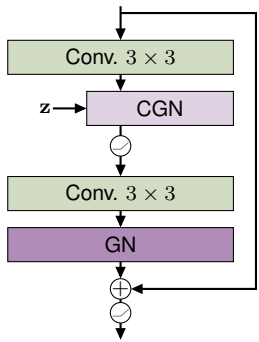
\begin{figure}[ht!]
    \centering
    \input{artwork/res_net_block_new}
    \vspace{-0.5em}
    \caption{Residual encoder and decoder block conditioned on the latent vector $\mathbf{z}$.}
    \label{fig:block}
\end{figure}

Both our surrogate model encoder and decoder utilize 2D residual building blocks~\cite{He2016}. In particular, our residual block is composed of two 2D $3\times 3$ convolutions, two leaky ReLU activations~\cite{Maas2013}, a CGN layer, and a standard GN layer~\cite{Wu2018}. In the encoder, we utilize a strided convolution to reduce the spatial dimensions. To upsample the feature maps in the decoder, we employ a $4\times 4$ transposed convolution instead of the first $3\times 3$ convolution. The full block is visualized in \cref{fig:block}.

\subsection{Conditional surrogate model discussion.} 

Note that we are not the first to propose a differentiable surrogate model of a non-differentiable standard video codec (\eg, H.264 or H.265)~\cite{Zhao2019, Qiu2021, Klopp2021, Tian2021, Isik2023}. However, to the best of our knowledge, we are proposing the first surrogate model offering support for fine-grain conditioning (macroblock-wise quantization). Additionally, our conditional surrogate model offers a differentiable prediction of the file size of the encoded video. While we are the first to offer a differentiable file size prediction for a standard video codec, existing work on surrogate modeling image codecs (\eg, JPEG~\cite{Wallace1992}) also offers differentiable file size predictions~\cite{Luo2021, Xie2022}. In summary, our novel conditional surrogate model is the first to approximate the H.264 standard video codec while offering macroblock-wise conditioning (over the whole $\qp$ range) and a differentiable file size prediction. This conditional surrogate model enables us the learn our deep video codec control in a fully end-to-end fashion, while existing surrogate models are unable to facilitate the end-to-end learning.

\section{Experiments}

First, we provide additional implementation details, before introducing our downstream models and their downstream performance. Next, we showcase additional results on how H.264 coding affects the downstream performance of deep vision models. Finally, we provide additional deep codec control and surrogate model results.

\subsection{Further implementation details} 

We implement our surrogate model and codec control pipeline using PyTorch~\cite{Paszke2019}, PyTorch Lightning~\cite{Falcon2019}, and Kornia~\cite{Riba2020}. For the macroblock-wise H.264 compression, we rely on the modified FFmpeg implementation of AccMPEG~\cite{Du2022, Tomar2006}. We pre-train our conditional surrogate model for $45\si{\kilo\relax}$ iterations with a base learning rate of $4\cdot 10^{-4}$ and a weight decay of $10^{-5}$. We employ a cosine learning rate schedule with linear annealing~\cite{Loshchilov2017}. Our deep codec control is trained for just $4.5\si{\kilo\relax}$ iterations. For training the control network, we also use the AdamW optimizer with a cosine learning rate schedule (without annealing). The base learning rates are set to $10^{-4}$ for the prediction head and to $10^{-5}$ for the pre-trained backbone blocks of the control network. We use a weight decay of $10^{-3}$. For fine-tuning the conditional surrogate model, we use a fixed learning rate of $1\cdot 10^{-4}$. For both surrogate pre-training and codec control training, we utilize two NVIDIA A6000 ($48\si{\giga\byte}$) GPUs. Surrogate pre-training takes approximately one day, whereas our codec control training requires approximately $12$ hours to complete. For both surrogate pre-training and codec control training, we use a batch size of $8$ per GPU and half-precision training.

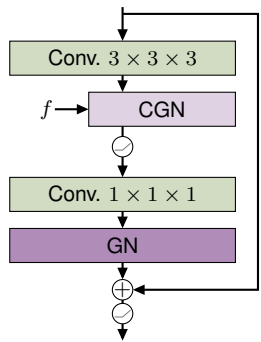
\begin{figure}[ht!]
    \centering
    \input{artwork/cond_3d_res_block}
    \vspace{-0.5em}
    \caption{Conditional 3D residual block composed of two convolutions, two leaky ReLU activations, a GN layer, and a CGN layer. For the first convolution, we use a group size equal to the number of channels.}
    \label{fig:cond_3d_block}
\end{figure}

The surrogate model's encoder uses four residual blocks with 64, 128, 256, and 1024 convolutional filters, respectively. The decoder is composed of four residual blocks with 512, 256, 128, and 64 convolutional filters, respectively. A conditional embedding dimension ${\rm C}_{\mathbf{z}}$ of $256$ is used. The AGRU utilizes a channel dimension of $1024$ for each convolution and eight recurrent iterations. For computing the optical flow used to align features before each AGRU iteration, we use a pre-trained RAFT small model from TorchVision~\cite{Torchvision2016, Teed2020}. We train the RAFT weights together with the other surrogate parameters. For the backbone blocks (X3D-S~\cite{Feichtenhofer2020}) of the control network, we utilize pre-trained weights obtained from action recognition on Kinetics-400~\cite{Fan2021b}. We freeze the backbones batch normalization layer during training. For the prediction head, we use two conditional 3D residual blocks (\cf{} \cref{fig:cond_3d_block}). During codec control training we start with a Gumbel-Softmax temperature of $2.0$ and decrease the temperature in a cosine schedule to $0.1$ at the end of the training. In total, the control network is composed of only $3\si{\mega\relax}$ parameters, making it easier to deploy the control network on an edge device, such as the NVIDIA Jetson Nano. 

As the segmentation model, we utilize our trained DeepLabV3 models (Cityscapes \& CamVid) trained with the MMSegmentation~\cite{Mmseg2020} protocol and H.264 augmentation. For optical flow estimation, we utilize a RAFT large model from Torchvision~\cite{Torchvision2016, Teed2020}.

We utilize both Cityscapes~\cite{Cordts2016} and CamVid~\cite{Brostow2009} with a resolution of $224\times 224$. On Cityscapes, this resolution is achieved by downsampling the frame by a factor of $4$ followed by cropping. On CamVid, we downsample the frames by a factor of $3$ followed by clipping. We use a resolution of $224\times 224$ to combat the large memory requirements of our codec control pipeline introduced by our surrogate model.

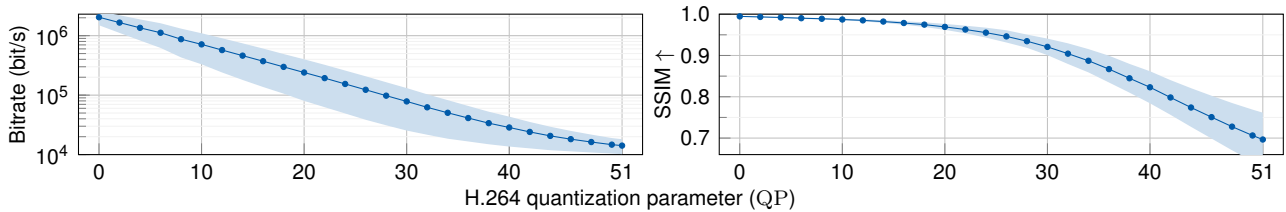
\begin{figure}[ht!]
    \centering
    \input{artwork/cityscapes/qp_vs_bw_and_ssim_cityscapes.tex}
    \vspace{-0.5em}
    \caption{\textbf{Rate-distortion tradeoff.} Bandwidth and SSIM (to raw video) with two standard dev. for different quantization parameters ($\qp$). A uniform $\qp$ over all macroblocks is used. Cityscapes (val seq.) with a resolution of $224\times 224$, a clip length (and $\gop$) of $8$, and a temporal stride of $3$ is utilized. Frame rate is $17\si{\fps}$.}
    \label{fig:qpvsbw}
\end{figure}

We set the surrogate loss weights to $\alpha_{\rho_{v}}=10^{-4}$, $\alpha_{\mathrm{SSIM}}=2$, $\alpha_{\mathrm{FF}}=200$, $\alpha_{\rho_{f}}=10^{-4}$, and $\alpha_{\mathrm{L1}}=0.1$. We use $\alpha_{\mathrm{p}}=2$, $\alpha_{\mathrm{b}}=6$, $\alpha_{\mathrm{r}}=1$, $\epsilon_{\mathrm{b}}=\epsilon_{\mathrm{p}}=0.02$, and $\epsilon_{\mathrm{r}}=0.05$ for the control loss. We sample the bandwidth condition $\tilde{b}$ from a log uniform distribution $\mathcal{U}_{\log}\!\left(30\si{\kilo\bit\per\second}, 0.9\si{\mega\bit\per\second}\right)$ capturing the working range of the H.264 codec (\cf{} \cref{fig:qpvsbw}). We use a log-uniform distribution since the bandwidth correlates logarithmically with $\qp$ and we want to explore $\qp$ uniformly during training. For validation, we sample $10$ bandwidth conditions equally spaced in $\log_{10}$-space.

\subsection{Downstream models}

For our semantic segmentation codec control experiments, we utilize a DeepLabV3~\cite{Chen2017} with ResNet-18~\cite{He2016} backbone (w/ H.264 aug.). We trained this model on the resolution ($224\times 224$) used by our codec control. Additionally, we also trained different variants of the model for analyzing the downstream performance of these models on H.264 codec video. To showcase the effect of H.264 as a data augmentation during downstream training, we train each model variant with and without H.264 as data augmentation. In our H.264 data aug. implementation, we use H.264 coded frames (with random $\qp$) in $50\%$ of the case. Except for the optional addition of the H.264 aug., we follow the Cityscapes training protocol of MMSegmentation~\cite{Mmseg2020} for both the Cityscapes and the CamVid datasets. Downstream validation results on non-coded frames are shown in \cref{table:downstream_ss_results}.

\begin{table}[ht!]
    \small
    \footnotesize
    \setlength\tabcolsep{0.75pt}
    \renewcommand\arraystretch{0.9}
    \caption{\textbf{Downstream model results (semantic segmentation).} We report the mean IoU on the respective validation set using the MMSegmentation validation pipeline.}
    \vspace{-4pt}
    \input{table/semantic_segmentation_mmseg_results}
    \label{table:downstream_ss_results}
\end{table}

For our optical flow codec control experiments, we utilize a RAFT~\cite{Teed2020} large model, trained supervised on synthetic data and KITTI~\cite{Geiger2013}. In particular, we utilize the public checkpoint from Torchvision~\cite{Torchvision2016}.

\subsection{Deep vision performance on H.264 coded videos}

\begin{figure}[ht!]
    \centering
    \input{artwork/cityscapes/acc_vs_bw_cityscapes_ss_of_subplot_small_2.tex}
    \vspace{-0.5em}
    \caption{\textbf{Downstream performance \vs{} compression trade-off.} Cityscapes (val sequences) segmentation accuracy and optical flow estimation performance, measured by average endpoint error (AEPE), for different H.264 quantization parameters between the raw clip predictions (pseudo label) and the compressed clip predictions. Temporal stride is $3$ and clip length (and $\gop$) is $8$. $\qp$ is applied uniformly.}
    \label{fig:accvsqp_cs}
\end{figure}
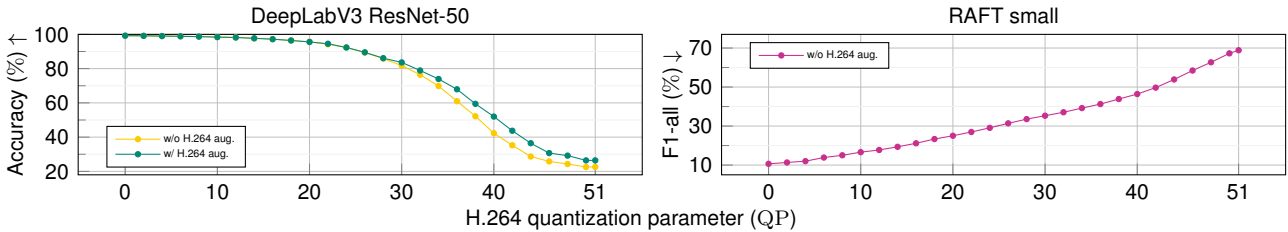

We analyze the performance of our downstream models trained for semantic segmentation on H.264 coded videos. We also analyze the optical flow estimation performance of RAFT (large \& small) on coded videos. Cityscapes results are shown in \cref{fig:accvsqp_cs}. If $\qp$ is increased downstream performance on Cityscapes is vastly deteriorated.

Downstream performance results on CamVid using H.264 coded videos are shown in \cref{fig:accvsqp_cv}. Consistent with the results on Cityscapes, downstream performance on CamVid also deteriorates as $\qp$ is increased. These findings also align with the findings by Otani~\etal~\cite{Otani2022} on the deterioration of action recognition performance on H.264 (and JPEG~\cite{Wallace1992}) coded videos/frames.

{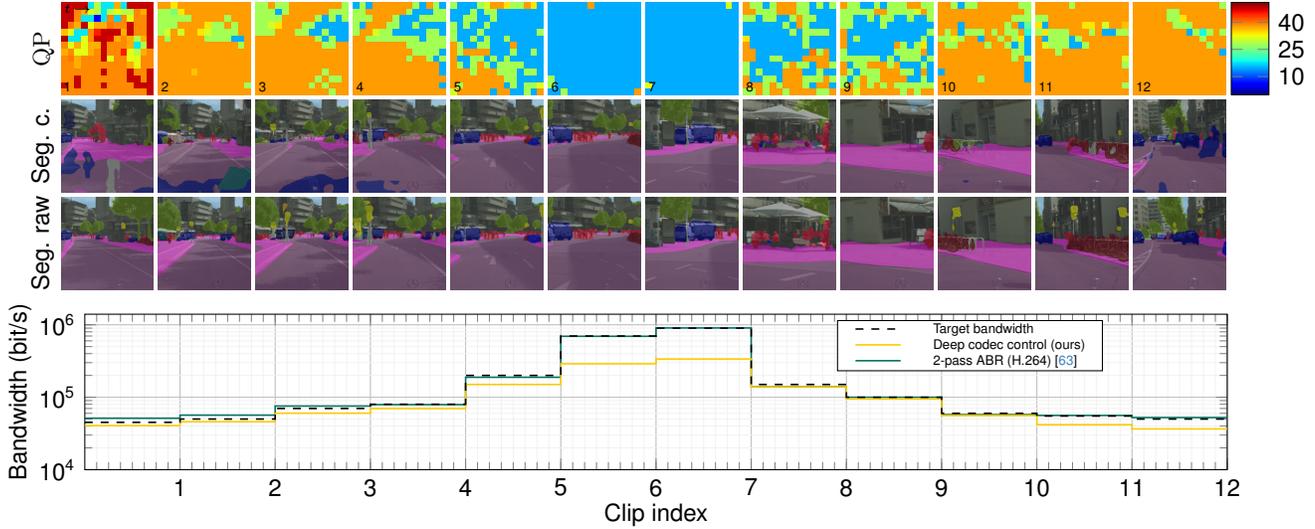
\begin{figure*}[t!]
    \centering
    \input{artwork/cityscapes/long_video}
    \vspace{-0.5em}
    \caption{\textbf{Qualitative results.} Cityscapes Stuttgart demo~\cite{Cordts2016} with clip-wise varying bandwidth condition. $12$ clips with $8$ frames each (temp. stride 3) used. $\qp$ prediction in the top row, segmentation prediction on the compressed video middle row, and segmentation prediction on the raw video bottom row. For clarity, we only visualize the $\qp$ prediction and segmentation for the first frame of each clip. DeepLabV3 (ResNet18 and H.264 aug. training) utilized. The graph shows the generated clip-wise bandwidths by H.264 and our approach. The segmentation accuracy (considering drops) over all clips is $45.71\%$ for H.264 and $82.98\%$ for our deep codec control. Zoom in for details; best viewed in color.}
    \label{fig:longvideoappendix}
\end{figure*}}

\begin{figure}[ht!]
    \centering
    \input{artwork/camvid/acc_vs_bw_camvid_ss_of_subplot_small.tex}
    \vspace{-0.5em}
    \caption{\textbf{Downstream performance \vs{} compression trade-off.} CamVid (val) segmentation accuracy and optical flow estimation performance, measured by average endpoint error (AEPE), for different H.264 quantization parameters between the raw clip predictions (pseudo label) and the compressed clip predictions. Temporal stride is $3$ and clip length (and $\gop$) is $8$. $\qp$ is applied uniformly.}
    \label{fig:accvsqp_cv}
\end{figure}
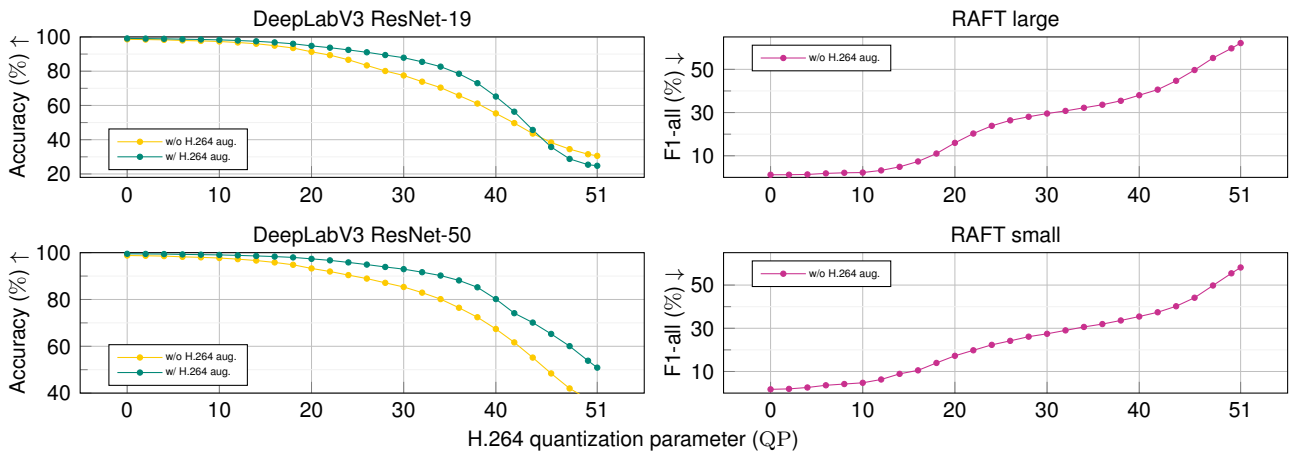

Using H.264 coded frames during supervised training only leads to minor residents to coded frames during inference. Still, downstream segmentation performance is highly affected. Interestingly, H.264 augmentation seems to have a better impact if the backbone network has a higher capacity.

\subsection{Codec control results: semantic segmentation}

We present additional qualitative and quantitative results of our deep video codec control on semantic segmentation as a downstream task. In \cref{fig:longvideoappendix}, we showcase our deep codec control on a sequence of $12$ video clips of the Cityscapes Stuttgart demo. Our codec control network adapts the macroblock-wise QP to the bandwidth constraints and captures interesting regions. Typically, the control network assigns lower QP values (less compression) to crowded frame regions. Still, the segmentation is affected if only limited bandwidth is available. We also observe the tendency of 2-pass ABR to overshoot the target bandwidth. In contrast, our deep codec control better follows the target bandwidth.

In \cref{fig:acc_bw_ss_cityscapes} \& \cref{fig:acc_p_ss_cityscapes}, we provide fine-grain results to Tab.~1 of the main paper on the Cityscapes dataset. Our deep video codec control largely outperforms 2-pass ABR control. However, for some bandwidth conditions, we perform on par with 2-pass ABR. Surprisingly, we observe that 2-pass ABR not only struggles to meet the bandwidth conditioning for low bandwidth values but also tends to overshoot the target bandwidth for larger bandwidth values.

\begin{figure}[th!]
    \centering
    \input{artwork/cityscapes/acc_b_ss_cs}
    \vspace{-0.5em}
    \caption{\textbf{Bandwidth condition results.} Bandwidth accuracy ($\mathrm{acc}_{\mathrm{bw}}$), \ie, how often the bandwidth constraint has been satisfied, for different network bandwidth conditions on Cityscapes (val sequences).}
    \label{fig:acc_bw_ss_cityscapes}
\end{figure}
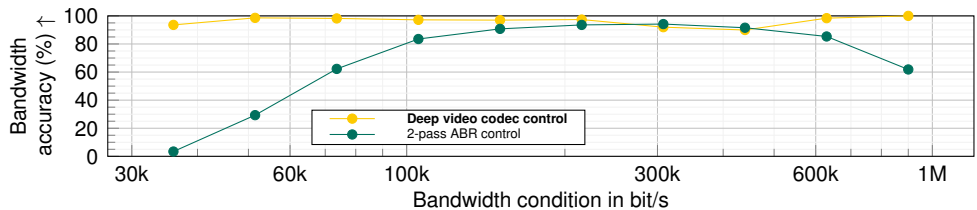

\begin{figure}[th!]
    \centering
    \input{artwork/cityscapes/acc_p_ss_cs}
    \vspace{-0.5em}
    \caption{\textbf{Downstream results.} Segmentation accuracy ($\mathrm{acc}_{\mathrm{seg}}$), considering clip dropping, for different network bandwidth conditions on Cityscapes (val sequences).}
    \label{fig:acc_p_ss_cityscapes}
\end{figure}
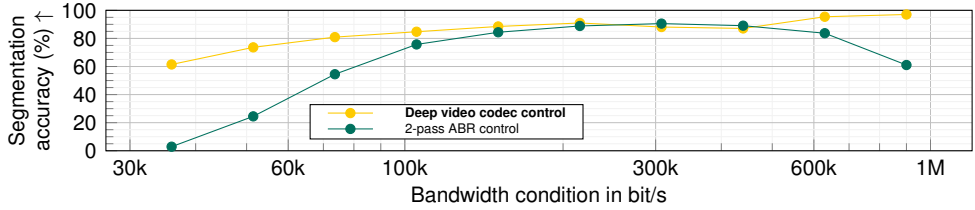

In \cref{table:ema_ablation}, we ablate the use of EMA on the control networks parameters. While scoring similar bandwidth condition accuracies EMA helps in downstream performance. In the most relevant setting, with no bandwidth tolerance, using EMA improves $\mathrm{acc}_{\mathrm{seg}}$ by approximately a half percent.

\begin{table}[ht!]
    \small
    \footnotesize
    \caption{\textbf{EMA ablation results on Cityscapes (semantic segmentation).} Here we compare our learn control with and without EMA (exponential moving average) on the parameters. We report bandwidth ($\mathrm{acc}_{\mathrm{bw}}$) and segmentation accuracies ($\mathrm{acc}_{\mathrm{seg}}$) for difference bandwidth tolerances. Metrics averaged over ten different bandwidth conditions.}
    \setlength\tabcolsep{0.75pt}
    \renewcommand\arraystretch{0.9}
    \vspace{-4pt}
    \input{table/ema_ablation}
    \label{table:ema_ablation}
\end{table}

\subsection{Codec control results: optical flow estimation}

In \cref{fig:acc_bw_ss_cityscapes} \& \cref{fig:acc_p_ss_cityscapes}, we provide fine-grain results to Tab.~2 of the main paper on the CamVid dataset. Similar to semantic segmentation, our deep codec control better follows the bandwidth conditioning and also better preserves downstream performance over most of the bandwidth range. Interestingly, for large bandwidth conditions, 2-pass ABR control tends to lead to slightly better downstream results. However, in general, our deep codec control leads to superior results. Especially for low network bandwidth conditions, our deep codec control vastly outperforms 2-pass ABR control.

\begin{figure}[th!]
    \centering
    \input{artwork/camvid/acc_b_of_cv}
    \vspace{-0.5em}
    \caption{\textbf{Bandwidth condition results.} Bandwidth accuracy ($\mathrm{acc}_{\mathrm{bw}}$), \ie, how often the bandwidth constraint has been satisfied, for different network bandwidth conditions on CamVid (val sequences).}
    \label{fig:acc_bw_ss_camvid}
\end{figure}
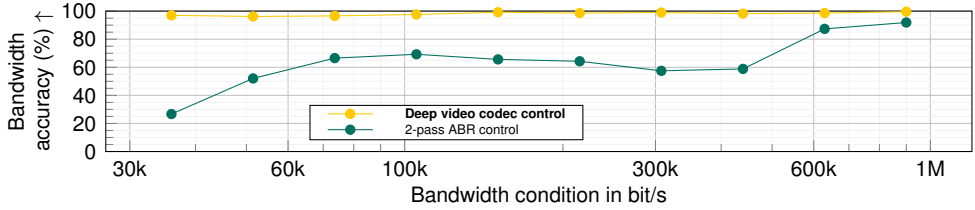

\begin{figure}[th!]
    \centering
    \input{artwork/camvid/f1_of_cv}
    \vspace{-0.5em}
    \caption{\textbf{Downstream results.} F1-all, considering clip dropping, for different network bandwidth conditions on CamVid (val sequences).}
    \label{fig:f1_of_camvid}
\end{figure}
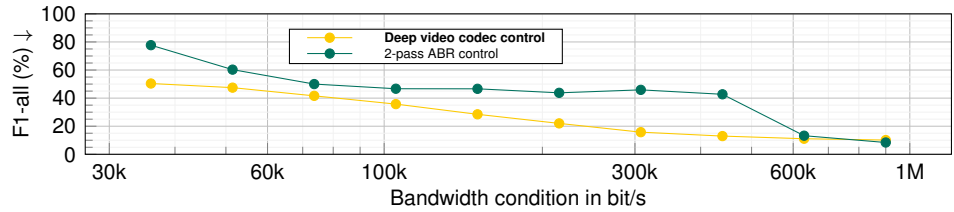

\subsection{Surrogate model results}

We provide fine-grain results of our conditional surrogate model on CamVid (\cref{fig:surrogate_camvid_video_distortion} \& \cref{fig:surrogate_camvid_file_size}) and Cityscapes (\cref{fig:surrogate_cityscapes_video_distortion} \& \cref{fig:surrogate_cityscapes_file_size}). Note our conditional surrogate model is considerably better than using the naive solution, the identity function. This can be observed by comparing \cref{fig:surrogate_camvid_video_distortion} and Fig.~2 of the main paper.

In the following \cref{fig:surrogate_results_camvid_full}-\cref{fig:surrogate_results_cityscapes_low}, we provide additional qualitative results of our conditional surrogate in approximating H.264 video distortion for different macroblock-wise quantization parameters. Note we present standard video clips but also show video clips with repeated frames and in reversed order, showcasing the generalization ability of our conditional surrogate model. We observe our surrogate model is able to accurately approximate the video distortion of H.264 for different macroblock-wise quantization parameters.

\begin{figure}[ht!]
    \centering
    \input{artwork/surrogate_results/camvid/video_distortion_results}
    \vspace{-0.5em}
    \caption{\textbf{Surrogate model coded video prediction results on CamVid.} We utilize a uniform $\qp$ (for all macroblocks).}
    \label{fig:surrogate_camvid_video_distortion}
\end{figure}
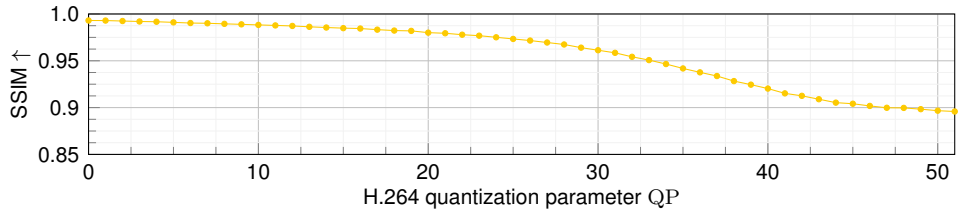

\begin{figure}[ht!]
    \centering
    \input{artwork/surrogate_results/camvid/file_size_results}
    \vspace{-0.5em}
    \caption{\textbf{Surrogate model file size prediction results on CamVid.} We utilize a uniform $\qp$ (for all macroblocks).}
    \label{fig:surrogate_camvid_file_size}
\end{figure}
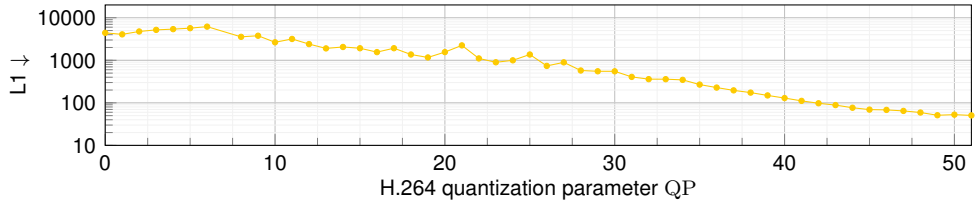

\begin{figure}[ht!]
    \centering
    \input{artwork/surrogate_results/cityscapes/video_distortion_results}
    \vspace{-0.5em}
    \caption{\textbf{Surrogate model coded video prediction results on CamVid.} We utilize a uniform $\qp$ (for all macroblocks).}
    \label{fig:surrogate_cityscapes_video_distortion}
\end{figure}
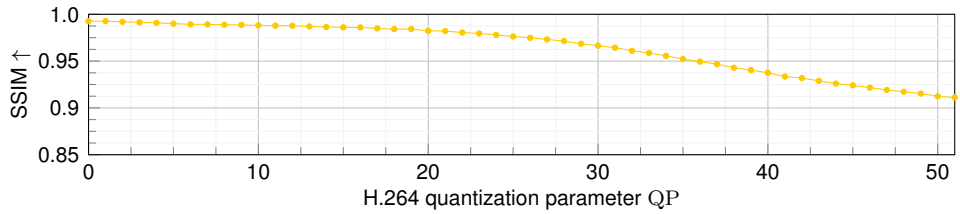

\begin{figure}[ht!]
    \centering
    \input{artwork/surrogate_results/cityscapes/file_size_results}
    \vspace{-0.5em}
    \caption{\textbf{Surrogate model file size prediction results on Cityscapes.} We utilize a uniform $\qp$ (for all macroblocks).}
    \label{fig:surrogate_cityscapes_file_size}
\end{figure}
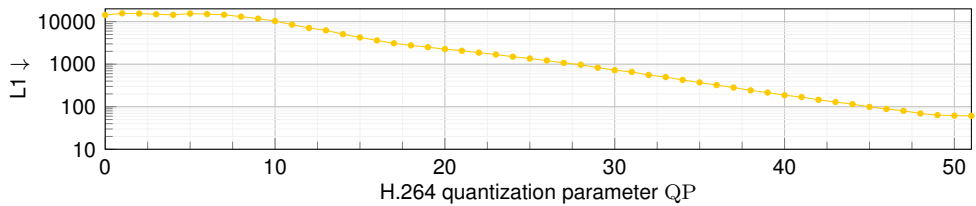

\vfill

{\begin{figure*}[t]
    \centering
    \input{artwork/surrogate_results/camvid/supplement/surrogate_results_camvid_full_1}\\[1cm]
    \input{artwork/surrogate_results/camvid/supplement/surrogate_results_camvid_full_2}\\[1cm]
    \input{artwork/surrogate_results/camvid/supplement/surrogate_results_camvid_full_3}
    \vspace{-0.5em}
    \caption{\textbf{Qualitative surrogate model results.} The first row shows the H.264 coded clip, the middle row shows the prediction of our conditional surrogate model, and the bottom row shows the macroblock-wise $\qp$ map. $\qp$ sampled over the full range. CamVid dataset used.}
    \label{fig:surrogate_results_camvid_full}
\end{figure*}
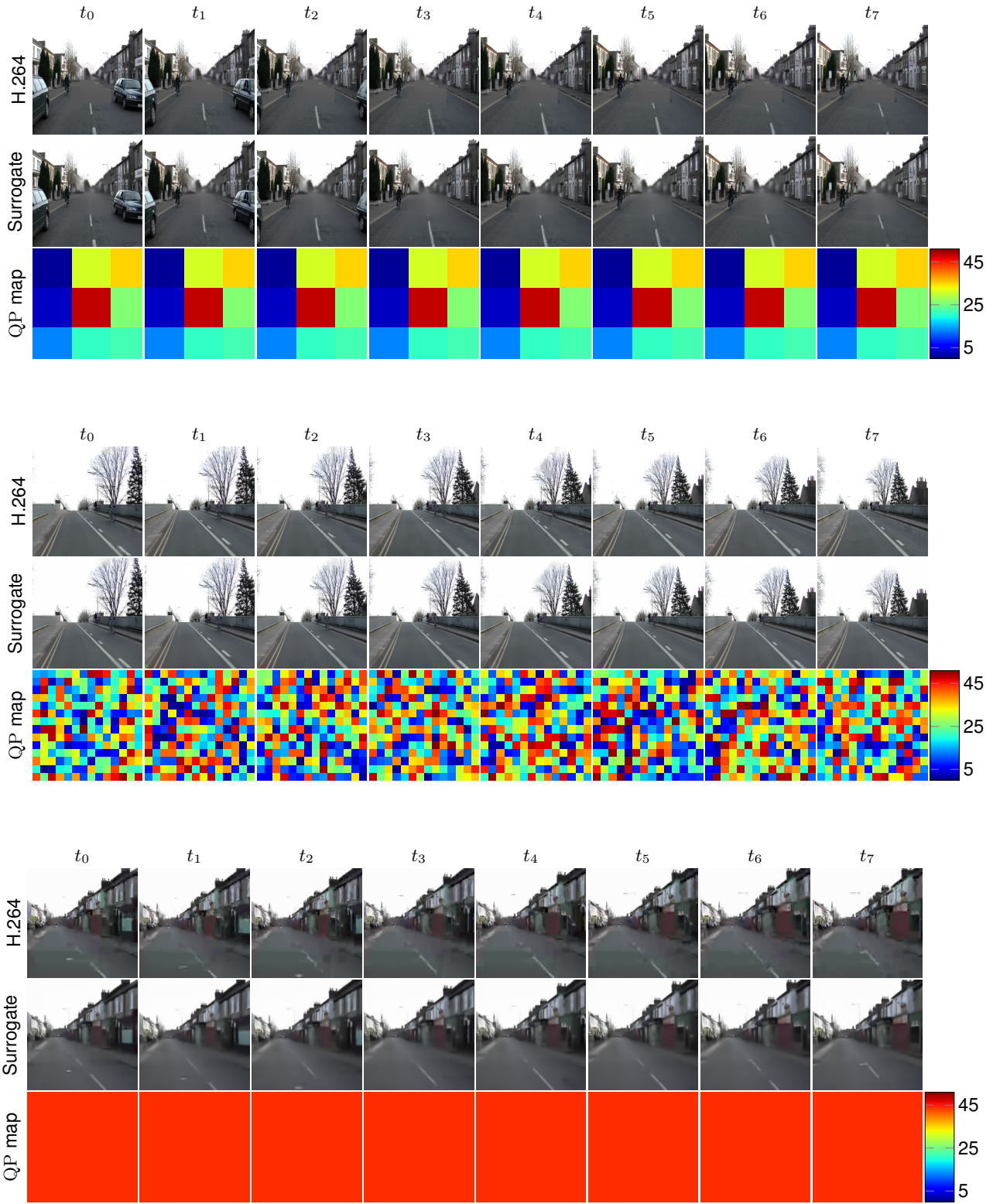}

{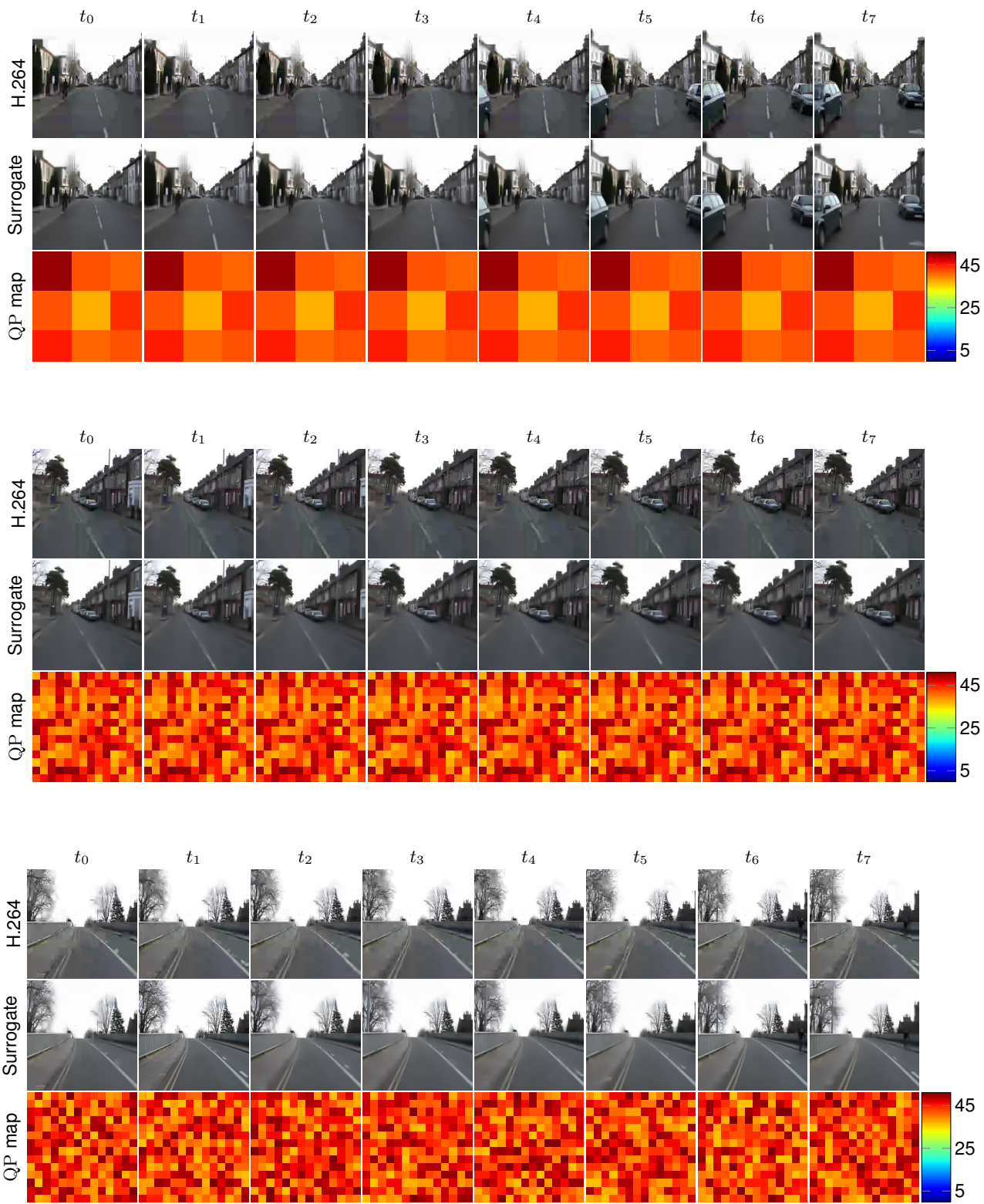
\begin{figure*}[t]
    \centering
    \input{artwork/surrogate_results/camvid/supplement/surrogate_results_camvid_high_1}\\[1cm]
    \input{artwork/surrogate_results/camvid/supplement/surrogate_results_camvid_high_2}\\[1cm]
    \input{artwork/surrogate_results/camvid/supplement/surrogate_results_camvid_high_3}
    \vspace{-0.5em}
    \caption{\textbf{Qualitative surrogate model results.} The first row shows the H.264 coded clip, the middle row shows the prediction of our conditional surrogate model, and the bottom row shows the macroblock-wise $\qp$ map. $\qp$ sampled from a range between $36$ and $51$. CamVid dataset used.}
    \label{fig:surrogate_results_camvid_high}
\end{figure*}}

{\begin{figure*}[t]
    \centering
    \input{artwork/surrogate_results/camvid/supplement/surrogate_results_camvid_low_1}\\[1cm]
    \input{artwork/surrogate_results/camvid/supplement/surrogate_results_camvid_low_2}\\[1cm]
    \input{artwork/surrogate_results/camvid/supplement/surrogate_results_camvid_low_3}
    \vspace{-0.5em}
    \caption{\textbf{Qualitative surrogate model results.} The first row shows the H.264 coded clip, the middle row shows the prediction of our conditional surrogate model, and the bottom row shows the macroblock-wise $\qp$ map. $\qp$ sampled from a range between $0$ and $15$. CamVid dataset used.}
    \label{fig:surrogate_results_camvid_low}
\end{figure*}
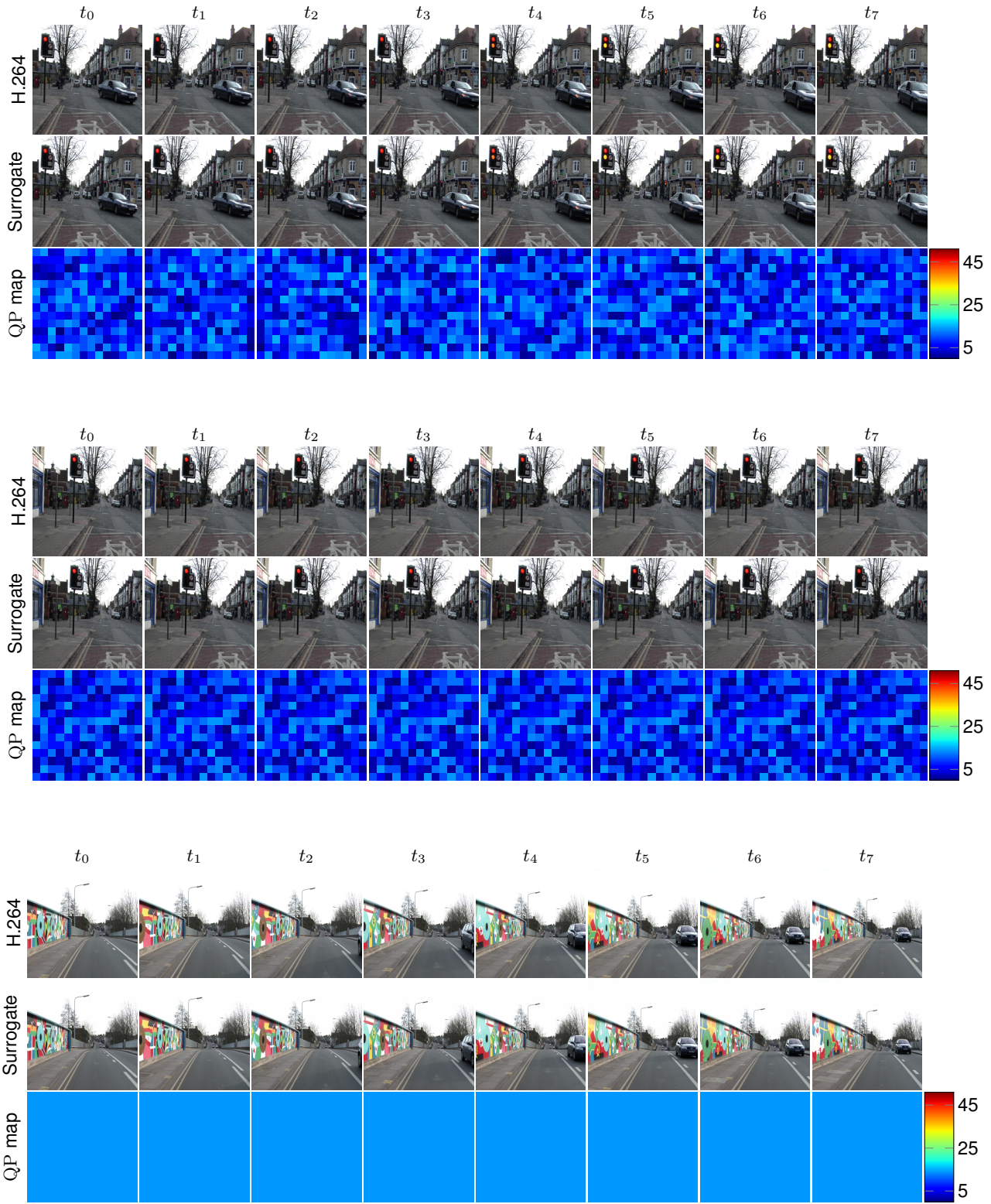}

{\begin{figure*}[t]
    \centering
    \input{artwork/surrogate_results/cityscapes/supplement/surrogate_results_cityscapes_full_1}\\[1cm]
    \input{artwork/surrogate_results/cityscapes/supplement/surrogate_results_cityscapes_full_2}\\[1cm]
    \input{artwork/surrogate_results/cityscapes/supplement/surrogate_results_cityscapes_full_3}
    \vspace{-0.5em}
    \caption{\textbf{Qualitative surrogate model results.} The first row shows the H.264 coded clip, the middle row shows the prediction of our conditional surrogate model, and the bottom row shows the macroblock-wise $\qp$ map. $\qp$ sampled over the full range. Cityscapes dataset used.}
    \label{fig:surrogate_results_cityscapes_full}
\end{figure*}}

{\begin{figure*}[t]
    \centering
    \input{artwork/surrogate_results/cityscapes/supplement/surrogate_results_cityscapes_high_1}\\[1cm]
    \input{artwork/surrogate_results/cityscapes/supplement/surrogate_results_cityscapes_high_2}\\[1cm]
    \input{artwork/surrogate_results/cityscapes/supplement/surrogate_results_cityscapes_high_3}
    \vspace{-0.5em}
    \caption{\textbf{Qualitative surrogate model results.} The first row shows the H.264 coded clip, the middle row shows the prediction of our conditional surrogate model, and the bottom row shows the macroblock-wise $\qp$ map. $\qp$ sampled from a range between $36$ and $51$. Cityscapes dataset used.}
    \label{fig:surrogate_results_cityscapes_high}
\end{figure*}}

{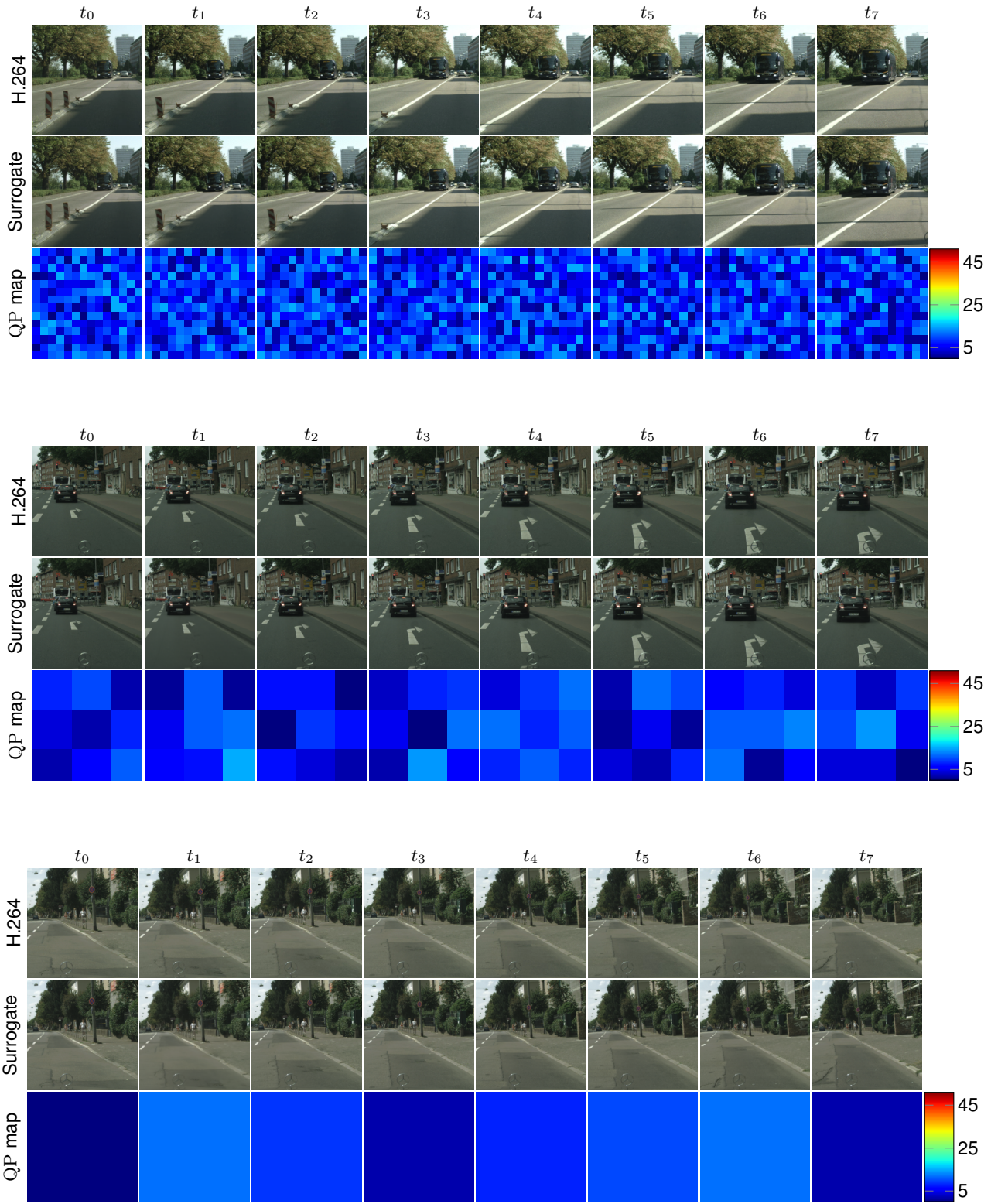
\begin{figure*}[t]
    \centering
    \input{artwork/surrogate_results/cityscapes/supplement/surrogate_results_cityscapes_low_1}\\[1cm]
    \input{artwork/surrogate_results/cityscapes/supplement/surrogate_results_cityscapes_low_2}\\[1cm]
    \input{artwork/surrogate_results/cityscapes/supplement/surrogate_results_cityscapes_low_3}
    \vspace{-0.5em}
    \caption{\textbf{Qualitative surrogate model results.} The first row shows the H.264 coded clip, the middle row shows the prediction of our conditional surrogate model, and the bottom row shows the macroblock-wise $\qp$ map. $\qp$ sampled from a range between $0$ and $15$. Cityscapes dataset used.}
    \label{fig:surrogate_results_cityscapes_low}
\end{figure*}}

%% file: algorithm/training.tex
\lstset{
backgroundcolor=\color{white},
basicstyle=\normalsize\ttfamily\selectfont,
breaklines=true,
captionpos=b,
commentstyle=\normalsize\color{tud1b},
keywordstyle=\normalsize\color{tud8c},
numbers=left,
xleftmargin=2.55em,
framexleftmargin=0.0em,
framexrightmargin=-0.65em,
columns=fullflexible,
}
\begin{lstlisting}[language=python]
for clip in data_loader:
    # Sample BW condition
    bw_c = log_uniform(bw_min, bw_max)
    # Make one-hot QP prediction
    qp_oh = control_network(clip, bw_c)
    # Forward pass surrogate model
    clip_c, fs = h264_sg(clip, qp_oh)
    # Convert file size to BW
    bw = file_size_to_bandwidth(fs)
    # Make downstream prediction
    pred = downstream_model(clip_c)
    # Downstream prediction raw clip
    with no_grad():
        label_pseudo = downstream_model(clip)
    # Compute loss
    loss = a_b * l_b(bw, bw_c) \ 
        + a_r * regularizer_b(bw, bw_c) \
        + a_p * h(bw_c - bw) * l_p(pred, label_pseudo)
    # Backward pass
    loss.backward()
    # Optimization step
    optimizer_control_network.step()
    # Update EMA control network
    ema(control_network, control_network_ema, decay=0.99)
    # Next: Surrogate model training step
\end{lstlisting}

%% file: artwork/res_net_block_new.tex
\begin{tikzpicture}[>={Stealth[inset=0pt,length=4.5pt,angle'=45]}, every node/.style={font=\footnotesize}, xscale=1.5, yscale=0.90625]
    \draw[->, thick] (-2.0, 3.4) -- (-0.8, 3.4) -- (-0.8, -0.65) -- (-1.91, -0.65);
    \draw[->, thick] (-2.0, 3.5) -- (-2.0, 3.0);
    \draw[black, fill=tud4d!30] (-3.0, 3.0) rectangle (-1.0, 2.5);
    \node[] at (-2.0, 2.75) {Conv. $3 \times 3$};
    \draw[->, thick] (-2.0, 2.5) -- (-2.0, 2.25);
    \draw[black, fill=tud11c!20] (-2.3, 2.25) rectangle (-1.0, 1.75);
    \node[] at (-1.65, 2.00) {CGN};
    \draw[->, thick] (-2.6, 2.0) node[left=-3pt] {$\mathbf{z}$} -- (-2.3, 2.0);
    \draw[->, thick] (-2.0, 1.75) -- (-2.0, 1.0);
    \begin{scope}[yshift=-0.75cm]
        \draw[fill=white] (-2.0, 2.2) ellipse (0.090625cm and 0.15cm);
        \draw[gray] (-2.07, 2.15) -- (-2.0, 2.15) -- (-1.93, 2.25);
    \end{scope}
    \begin{scope}[yshift=-2.0cm]
        \draw[black, fill=tud4d!30] (-3.0, 3.0) rectangle (-1.0, 2.5);
        \node[] at (-2.0, 2.75) {Conv. $3 \times 3$};
        \draw[->, thick] (-2.0, 2.5) -- (-2.0, 2.25);
        \draw[black, fill=tud11c!50] (-3.0, 2.25) rectangle (-1.0, 1.75);
        \node[] at (-2.0, 2.00) {GN};
        \draw[->, thick] (-2.0, 1.4) -- (-2.0, 0.6);
        \draw[->, thick] (-2.0, 1.75) -- (-2.0, 1.5);
        \begin{scope}[yshift=-0.85cm]
            \draw[fill=white] (-2.0, 2.2) ellipse (0.090625cm and 0.15cm);
            \node[] at (-2.0, 2.2) {\footnotesize $+$};
        \end{scope}
        \begin{scope}[yshift=-1.2cm]
            \draw[fill=white] (-2.0, 2.2) ellipse (0.090625cm and 0.15cm);
            \draw[gray] (-2.07, 2.15) -- (-2.0, 2.15) -- (-1.93, 2.25);
        \end{scope}
    \end{scope}
        
\end{tikzpicture}

%% file: artwork/cond_3d_res_block.tex
\begin{tikzpicture}[>={Stealth[inset=0pt,length=4.5pt,angle'=45]}, every node/.style={font=\footnotesize}, xscale=1.5, yscale=0.90625]
    \draw[->, thick] (-2.0, 3.4) -- (-0.8, 3.4) -- (-0.8, -0.65) -- (-1.91, -0.65);
    \draw[->, thick] (-2.0, 3.5) -- (-2.0, 3.0);
    \draw[black, fill=tud4d!30] (-3.0, 3.0) rectangle (-1.0, 2.5);
    \node[] at (-2.0, 2.75) {Conv. $3 \times 3 \times 3$};
    \draw[->, thick] (-2.0, 2.5) -- (-2.0, 2.25);
    \draw[black, fill=tud11c!20] (-2.3, 2.25) rectangle (-1.0, 1.75);
    \node[] at (-1.65, 2.00) {CGN};
    \draw[->, thick] (-2.6, 2.0) node[left=-3pt] {$f$} -- (-2.3, 2.0);
    \draw[->, thick] (-2.0, 1.75) -- (-2.0, 1.0);
    \begin{scope}[yshift=-0.75cm]
        \draw[fill=white] (-2.0, 2.2) ellipse (0.090625cm and 0.15cm);
        \draw[gray] (-2.07, 2.15) -- (-2.0, 2.15) -- (-1.93, 2.25);
    \end{scope}
    \begin{scope}[yshift=-2.0cm]
        \draw[black, fill=tud4d!30] (-3.0, 3.0) rectangle (-1.0, 2.5);
        \node[] at (-2.0, 2.75) {Conv. $1 \times 1 \times 1$};
        \draw[->, thick] (-2.0, 2.5) -- (-2.0, 2.25);
        \draw[black, fill=tud11c!50] (-3.0, 2.25) rectangle (-1.0, 1.75);
        \node[] at (-2.0, 2.00) {GN};
        \draw[->, thick] (-2.0, 1.4) -- (-2.0, 0.6);
        \draw[->, thick] (-2.0, 1.75) -- (-2.0, 1.5);
        \begin{scope}[yshift=-0.85cm]
            \draw[fill=white] (-2.0, 2.2) ellipse (0.090625cm and 0.15cm);
            \node[] at (-2.0, 2.2) {\footnotesize $+$};
        \end{scope}
        \begin{scope}[yshift=-1.2cm]
            \draw[fill=white] (-2.0, 2.2) ellipse (0.090625cm and 0.15cm);
            \draw[gray] (-2.07, 2.15) -- (-2.0, 2.15) -- (-1.93, 2.25);
        \end{scope}
    \end{scope}

\end{tikzpicture}

%% file: artwork/cityscapes/qp_vs_bw_and_ssim_cityscapes.tex
\begin{tikzpicture}[every node/.style={font=\footnotesize}]
    \node[anchor=center] at (0.42\columnwidth, -0.6) {H.264 quantization parameter ($\qp$)};
	\begin{groupplot}[
        group style={
            group name=accvsbw,
            group size=2 by 2,
            ylabels at=edge left,
            horizontal sep=29pt,
            vertical sep=9.5pt,
        },
        legend style={nodes={scale=0.5}},
        height=3.45cm,
        width=0.52\columnwidth,
        grid=both,
        xtick pos=bottom,
        ytick pos=left,
        grid style={line width=.1pt, draw=gray!10},
        major grid style={line width=.2pt,draw=gray!50},
        minor tick num=1,
        xmin=-2,
        xmax=53,
        ylabel shift=-4.5pt,
        xtick={
            0, 10, 20, 30, 40, 51
        },
        xticklabels={
            0, 10, 20, 30, 40, 51
        },
        ticklabel style = {font=\footnotesize},
        ]
        \nextgroupplot[
        ylabel=Bitrate (\si{\bit\per\second}),
        ymode=log,
        ymin=10000.00,
        ymax=2300000,
        ytickten={4, 5, 6, 7},
        yticklabels={10\textsuperscript{4}, 10\textsuperscript{5}, 10\textsuperscript{6}, 10\textsuperscript{7}},
        ]
    	\addplot[color=tud1b, mark=*, mark size=1.0pt] table[x=x,y=y] {h264_bw_c.dat};
        \addplot[name path=upper,draw=none] table[x=x,y expr=\thisrow{y}+2*\thisrow{std}] {h264_bw_c.dat};
        \addplot[name path=lower,draw=none] table[x=x,y expr=\thisrow{y}-2*\thisrow{std}] {h264_bw_c.dat};
        \addplot[fill=tud1b!20] fill between[of=upper and lower];
        \nextgroupplot[
        ylabel=SSIM $\uparrow$,
        ytick={0.7, 0.8, 0.9, 1.0},
        yticklabels={0.7, 0.8, 0.9, 1.0},
        ymin=0.66,   
        ymax=1,
        ]
        \addplot[color=tud1b, mark=*, mark size=1.0pt] table[x=x,y=y] {h264_ssim_c.dat};
        \addplot[name path=upper,draw=none] table[x=x,y expr=\thisrow{y}+2*\thisrow{std}] {h264_ssim_c.dat};
        \addplot[name path=lower,draw=none] table[x=x,y expr=\thisrow{y}-2*\thisrow{std}] {h264_ssim_c.dat};
        \addplot[fill=tud1b!20] fill between[of=upper and lower];
	\end{groupplot}
\end{tikzpicture}

%% file: table/semantic_segmentation_mmseg_results.tex
\begin{tabular*}{\columnwidth}{@{\extracolsep{\fill}}l@{\hskip 0.25em}c@{\hskip 0.25em}S[table-format=2.2]@{}}
	\toprule
	{Method} & {H.264 aug.} & {mIoU ($\%$) $\uparrow$} \\
	\midrule
    \multicolumn{3}{@{}c}{\textit{Cityscapes}~\cite{Cordts2016}} \\
    \midrule
	DeepLabV3 (ResNet-18) & \xmark & 64.47 \\
    DeepLabV3 (ResNet-18) & \cmark & 64.96 \\
    DeepLabV3 (ResNet-50) & \xmark & 67.62 \\
    DeepLabV3 (ResNet-50) & \cmark & 66.74 \\
    \midrule
    \multicolumn{3}{@{}c}{\textit{CamVid}~\cite{Brostow2009}} \\
    \midrule
    DeepLabV3 (ResNet-18) & \xmark & 50.68 \\
    DeepLabV3 (ResNet-18) & \cmark & 50.48 \\
    DeepLabV3 (ResNet-50) & \xmark & 56.65 \\
    DeepLabV3 (ResNet-50) & \cmark & 55.35 \\
	\bottomrule
\end{tabular*}

%% file: artwork/cityscapes/acc_vs_bw_cityscapes_ss_of_subplot_small_2.tex
\begin{tikzpicture}[every node/.style={font=\footnotesize}]
    \node[anchor=center] at (0.421\columnwidth, -0.60) {H.264 quantization parameter ($\qp$)};
	\begin{groupplot}[
        group style={
            group name=accvsbw,
            group size=2 by 4,
            ylabels at=edge left,
            horizontal sep=30pt,
            vertical sep=28.5pt,
        },
        legend style={nodes={scale=0.5}},
        height=3.45cm,
        width=0.52\columnwidth,
        grid=both,
        grid style={line width=.1pt, draw=gray!10},
        major grid style={line width=.2pt,draw=gray!50},
        minor tick num=1,
        xtick pos=bottom,
        ytick pos=left,
	    ymin=18,   
        ymax=100,
        ylabel shift=-4.5pt,
        ytick={
            20, 40, 60, 80, 100
        },
        yticklabels={
            20, 40, 60, 80, 100
        },
        xtick={
            0, 10, 20, 30, 40, 51
        },
        xticklabels={
            0, 10, 20, 30, 40, 51
        },
        ticklabel style = {font=\footnotesize},
        legend style={at={(0.05,0.2)}, anchor=west},
        legend cell align={left},
        ]
        \nextgroupplot[title={DeepLabV3 ResNet-50}, title style={yshift=-7pt, font=\footnotesize}, ylabel=Accuracy ($\%$) $\uparrow$]
    	\addplot[color=tud6b, mark=*, mark size=1.0pt] coordinates {
    		(0, 99.24)
            (2, 99.15)
            (4, 99.03)
            (6, 98.84)
            (8, 98.68)
            (10, 98.48)
            (12, 98.26)
            (14, 97.73)
            (16, 97.23)
            (18, 96.43)
            (20, 95.64)
            (22, 94.22)
            (24, 92.41)
            (26, 89.43)
            (28, 85.87)
            (30, 81.78)
            (32, 76.45)
            (34, 69.84)
            (36, 60.98)
            (38, 52.18)
            (40, 42.28)
            (42, 35.25)
            (44, 28.69)
            (46, 25.83)
            (48, 24.32)
            (50, 22.57)
            (51, 22.62)
    	};
        \addlegendentry{w/o H.264 aug.}
        \addplot[color=tud3c, mark=*, mark size=1.0pt] coordinates {
    		(0, 99.24)
            (2, 99.14)
            (4, 99.01)
            (6, 98.87)
            (8, 98.70)
            (10, 98.39)
            (12, 98.17)
            (14, 97.69)
            (16, 97.18)
            (18, 96.44)
            (20, 95.59)
            (22, 94.52)
            (24, 92.33)
            (26, 89.50)
            (28, 86.10)
            (30, 83.63)
            (32, 78.91)
            (34, 73.95)
            (36, 67.93)
            (38, 59.44)
            (40, 52.00)
            (42, 43.74)
            (44, 36.45)
            (46, 30.69)
            (48, 29.21)
            (50, 26.41)
            (51, 26.42)
    	};
        \addlegendentry{w/ H.264 aug.}
        \nextgroupplot[
        title={RAFT small\vphantom{p}},
        title style={yshift=-7pt, font=\footnotesize},
        legend style={at={(0.05,0.85)}, anchor=west},
        ymin=5, ymax=77,
        ytick={
            10, 30, 50, 70
        },
        yticklabels={
            10, 30, 50, 70
        }, ylabel=F1-all ($\%$) $\downarrow$,]
        \addplot[color=tud10a, mark=*, mark size=1.0pt] coordinates {
    		(0, 10.6500)
            (2, 11.3175)
            (4, 11.9947)
            (6, 13.8510)
            (8, 15.0113)
            (10, 16.6122)
            (12, 17.7385)
            (14, 19.3612)
            (16, 21.1556)
            (18, 23.3428)
            (20, 25.0265)
            (22, 26.9530)
            (24, 29.0873)
            (26, 31.3834)
            (28, 33.5527)
            (30, 35.2991)
            (32, 37.0659)
            (34, 39.2056)
            (36, 41.2646)
            (38, 43.8459)
            (40, 46.4035)
            (42, 49.7029)
            (44, 53.8914)
            (46, 58.4632)
            (48, 62.6836)
            (50, 67.2630)
            (51, 68.8548)
    	};
     \addlegendentry{w/o H.264 aug.}
	\end{groupplot}
\end{tikzpicture}

%% file: artwork/cityscapes/long_video.tex
\raggedleft
\begin{tikzpicture}[every node/.style={font=\small}]

\newlength\imgW
\setlength{\imgW}{0.1864\textwidth}

\begin{groupplot}[group style={group size=12 by 3, horizontal sep=2pt, vertical sep=2pt, ylabels at=edge left}]

\nextgroupplot[
width=\imgW,
ticks=none,
axis line style={white, line width=0.0pt},
xtick=\empty,
ytick=\empty,
ylabel shift=-1pt,
ylabel=$\qp$,
axis equal image,
xmin=-0.5, xmax=13.5, 
ymin=-0.5, ymax=13.5,
]
\addplot graphics [includegraphics cmd=\pgfimage, xmin=-0.5, xmax=13.5, ymax=13.5, ymin=-0.5] {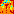};
\node[anchor=south west] at (axis cs:-1.5, -1.5) {\tiny 1};
\node[anchor=south west] at (axis cs:-1.5, 10.5) {\tiny $t\!\to$};

\nextgroupplot[
width=\imgW,
ticks=none,
axis line style={white, line width=0.0pt},
xtick=\empty, 
ytick=\empty,
axis equal image,
xmin=-0.5, xmax=13.5, 
ymin=-0.5, ymax=13.5,
]
\addplot graphics [includegraphics cmd=\pgfimage, xmin=-0.5, xmax=13.5, ymax=13.5, ymin=-0.5] {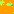};
\node[anchor=south west] at (axis cs:-1.5, -1.5) {\tiny 2};

\nextgroupplot[
width=\imgW,
ticks=none,
axis line style={white, line width=0.0pt},
xtick=\empty, 
ytick=\empty,
axis equal image,
xmin=-0.5, xmax=13.5, 
ymin=-0.5, ymax=13.5,
]
\addplot graphics [includegraphics cmd=\pgfimage, xmin=-0.5, xmax=13.5, ymax=13.5, ymin=-0.5] {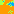};
\node[anchor=south west] at (axis cs:-1.5, -1.5) {\tiny 3};

\nextgroupplot[
width=\imgW,
ticks=none,
axis line style={white, line width=0.0pt},
xtick=\empty, 
ytick=\empty,
axis equal image,
xmin=-0.5, xmax=13.5, 
ymin=-0.5, ymax=13.5,
]
\addplot graphics [includegraphics cmd=\pgfimage, xmin=-0.5, xmax=13.5, ymax=13.5, ymin=-0.5] {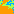};
\node[anchor=south west] at (axis cs:-1.5, -1.5) {\tiny 4};

\nextgroupplot[
width=\imgW,
ticks=none,
axis line style={white, line width=0.0pt},
xtick=\empty, 
ytick=\empty,
axis equal image,
xmin=-0.5, xmax=13.5, 
ymin=-0.5, ymax=13.5,
]
\addplot graphics [includegraphics cmd=\pgfimage, xmin=-0.5, xmax=13.5, ymax=13.5, ymin=-0.5] {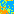};
\node[anchor=south west] at (axis cs:-1.5, -1.5) {\tiny 5};

\nextgroupplot[
width=\imgW,
ticks=none,
axis line style={white, line width=0.0pt},
xtick=\empty, 
ytick=\empty,
axis equal image,
xmin=-0.5, xmax=13.5, 
ymin=-0.5, ymax=13.5,
]
\addplot graphics [includegraphics cmd=\pgfimage, xmin=-0.5, xmax=13.5, ymax=13.5, ymin=-0.5] {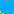};
\node[anchor=south west] at (axis cs:-1.5, -1.5) {\tiny 6};

\nextgroupplot[
width=\imgW,
ticks=none,
axis line style={white, line width=0.0pt}, 
xtick=\empty, 
ytick=\empty,
axis equal image,
xmin=-0.5, xmax=13.5, 
ymin=-0.5, ymax=13.5,
]
\addplot graphics [includegraphics cmd=\pgfimage, xmin=-0.5, xmax=13.5, ymax=13.5, ymin=-0.5] {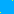};
\node[anchor=south west] at (axis cs:-1.5, -1.5) {\tiny 7};

\nextgroupplot[
width=\imgW,
ticks=none,
axis line style={white, line width=0.0pt},
xtick=\empty, 
ytick=\empty,
axis equal image,
xmin=-0.5, xmax=13.5, 
ymin=-0.5, ymax=13.5,
]
\addplot graphics [includegraphics cmd=\pgfimage, xmin=-0.5, xmax=13.5, ymax=13.5, ymin=-0.5] {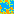};
\node[anchor=south west] at (axis cs:-1.5, -1.5) {\tiny 8};

\nextgroupplot[
width=\imgW,
ticks=none,
axis line style={white, line width=0.0pt},
xtick=\empty, 
ytick=\empty,
axis equal image,
xmin=-0.5, xmax=13.5, 
ymin=-0.5, ymax=13.5,
]
\addplot graphics [includegraphics cmd=\pgfimage, xmin=-0.5, xmax=13.5, ymax=13.5, ymin=-0.5] {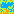};
\node[anchor=south west] at (axis cs:-1.5, -1.5) {\tiny 9};

\nextgroupplot[
width=\imgW,
ticks=none,
axis line style={white, line width=0.0pt},
xtick=\empty, 
ytick=\empty,
axis equal image,
xmin=-0.5, xmax=13.5, 
ymin=-0.5, ymax=13.5,
]
\addplot graphics [includegraphics cmd=\pgfimage, xmin=-0.5, xmax=13.5, ymax=13.5, ymin=-0.5] {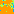};
\node[anchor=south west] at (axis cs:-1.5, -1.5) {\tiny 10};

\nextgroupplot[
width=\imgW,
ticks=none,
axis line style={white, line width=0.0pt},
xtick=\empty, 
ytick=\empty,
axis equal image,
xmin=-0.5, xmax=13.5, 
ymin=-0.5, ymax=13.5,
]
\addplot graphics [includegraphics cmd=\pgfimage, xmin=-0.5, xmax=13.5, ymax=13.5, ymin=-0.5] {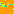};
\node[anchor=south west] at (axis cs:-1.5, -1.5) {\tiny 11};

\nextgroupplot[
width=\imgW,
ticks=none,
axis line style={white, line width=0.0pt}, 
xtick=\empty, 
ytick=\empty,
axis equal image,
xmin=-0.5, xmax=13.5, 
ymin=-0.5, ymax=13.5,
colorbar,
colorbar style={
ylabel={},
ytick={10, 25, 40},
yticklabels={10, 25, 40},
at={(1.06,1)},
},
colormap={mymap}{[1pt]
  rgb(0pt)=(0,0,0.5);
  rgb(22pt)=(0,0,1);
  rgb(25pt)=(0,0,1);
  rgb(68pt)=(0,0.86,1);
  rgb(70pt)=(0,0.9,0.967741935483871);
  rgb(75pt)=(0.0806451612903226,1,0.887096774193548);
  rgb(128pt)=(0.935483870967742,1,0.0322580645161291);
  rgb(130pt)=(0.967741935483871,0.962962962962963,0);
  rgb(132pt)=(1,0.925925925925926,0);
  rgb(178pt)=(1,0.0740740740740741,0);
  rgb(182pt)=(0.909090909090909,0,0);
  rgb(200pt)=(0.5,0,0)
},
point meta max=51,
point meta min=0,
]
\addplot graphics [includegraphics cmd=\pgfimage, xmin=-0.5, xmax=13.5, ymax=13.5, ymin=-0.5] {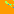};
\node[anchor=south west] at (axis cs:-1.5, -1.5) {\tiny 12};

\nextgroupplot[
width=\imgW,
ticks=none,
axis line style={white, line width=0.0pt},
xtick=\empty, 
ytick=\empty,
ylabel shift=-1pt,
ylabel=Seg. c.,
axis equal image,
xmin=-0.5, xmax=223.5, 
ymin=-0.5, ymax=223.5,
]
\addplot graphics [includegraphics cmd=\pgfimage, xmin=-0.5, xmax=223.5, ymax=223.5, ymin=-0.5] {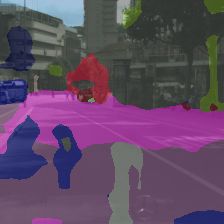};

\nextgroupplot[
width=\imgW,
ticks=none,
axis line style={white, line width=0.0pt}, 
xtick=\empty, 
ytick=\empty,
axis equal image,
xmin=-0.5, xmax=223.5, 
ymin=-0.5, ymax=223.5,
]
\addplot graphics [includegraphics cmd=\pgfimage, xmin=-0.5, xmax=223.5, ymax=223.5, ymin=-0.5] {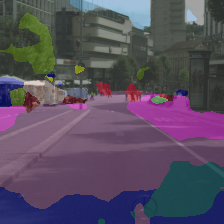};

\nextgroupplot[
width=\imgW,
ticks=none,
axis line style={white, line width=0.0pt},
xtick=\empty, 
ytick=\empty,
axis equal image,
xmin=-0.5, xmax=223.5, 
ymin=-0.5, ymax=223.5,
]
\addplot graphics [includegraphics cmd=\pgfimage, xmin=-0.5, xmax=223.5, ymax=223.5, ymin=-0.5] {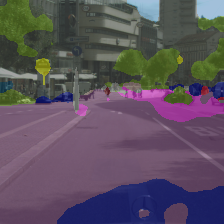};

\nextgroupplot[
width=\imgW,
ticks=none,
axis line style={white, line width=0.0pt},
xtick=\empty, 
ytick=\empty,
axis equal image,
xmin=-0.5, xmax=223.5, 
ymin=-0.5, ymax=223.5,
]
\addplot graphics [includegraphics cmd=\pgfimage, xmin=-0.5, xmax=223.5, ymax=223.5, ymin=-0.5] {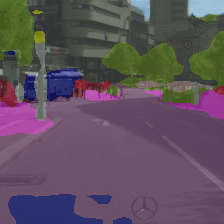};

\nextgroupplot[
width=\imgW,
ticks=none,
axis line style={white, line width=0.0pt},
xtick=\empty, 
ytick=\empty,
axis equal image,
xmin=-0.5, xmax=223.5, 
ymin=-0.5, ymax=223.5,
]
\addplot graphics [includegraphics cmd=\pgfimage, xmin=-0.5, xmax=223.5, ymax=223.5, ymin=-0.5] {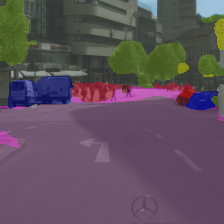};

\nextgroupplot[
width=\imgW,
ticks=none,
axis line style={white, line width=0.0pt},
xtick=\empty, 
ytick=\empty,
axis equal image,
xmin=-0.5, xmax=223.5, 
ymin=-0.5, ymax=223.5,
]
\addplot graphics [includegraphics cmd=\pgfimage, xmin=-0.5, xmax=223.5, ymax=223.5, ymin=-0.5] {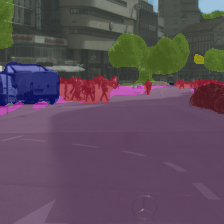};

\nextgroupplot[
width=\imgW,
ticks=none,
axis line style={white, line width=0.0pt},
xtick=\empty, 
ytick=\empty,
axis equal image,
xmin=-0.5, xmax=223.5, 
ymin=-0.5, ymax=223.5,
]
\addplot graphics [includegraphics cmd=\pgfimage, xmin=-0.5, xmax=223.5, ymax=223.5, ymin=-0.5] {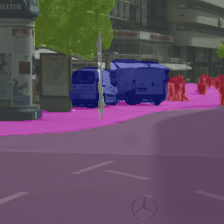};

\nextgroupplot[
width=\imgW,
ticks=none,
axis line style={white, line width=0.0pt},
xtick=\empty, 
ytick=\empty,
axis equal image,
xmin=-0.5, xmax=223.5, 
ymin=-0.5, ymax=223.5,
]
\addplot graphics [includegraphics cmd=\pgfimage, xmin=-0.5, xmax=223.5, ymax=223.5, ymin=-0.5] {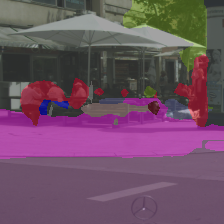};

\nextgroupplot[
width=\imgW,
ticks=none,
axis line style={white, line width=0.0pt},
xtick=\empty, 
ytick=\empty,
axis equal image,
xmin=-0.5, xmax=223.5, 
ymin=-0.5, ymax=223.5,
]
\addplot graphics [includegraphics cmd=\pgfimage, xmin=-0.5, xmax=223.5, ymax=223.5, ymin=-0.5] {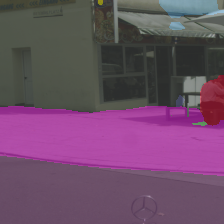};

\nextgroupplot[
width=\imgW,
ticks=none,
axis line style={white, line width=0.0pt},
xtick=\empty, 
ytick=\empty,
axis equal image,
xmin=-0.5, xmax=223.5, 
ymin=-0.5, ymax=223.5,
]
\addplot graphics [includegraphics cmd=\pgfimage, xmin=-0.5, xmax=223.5, ymax=223.5, ymin=-0.5] {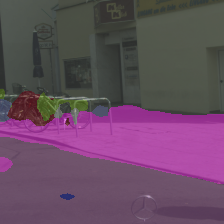};

\nextgroupplot[
width=\imgW,
ticks=none,
axis line style={white, line width=0.0pt},
xtick=\empty, 
ytick=\empty,
axis equal image,
xmin=-0.5, xmax=223.5, 
ymin=-0.5, ymax=223.5,
]
\addplot graphics [includegraphics cmd=\pgfimage, xmin=-0.5, xmax=223.5, ymax=223.5, ymin=-0.5] {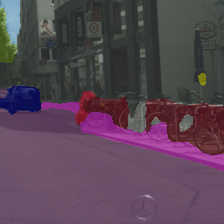};

\nextgroupplot[
width=\imgW,
ticks=none,
axis line style={white, line width=0.0pt},
xtick=\empty, 
ytick=\empty,
axis equal image,
xmin=-0.5, xmax=223.5, 
ymin=-0.5, ymax=223.5,
]
\addplot graphics [includegraphics cmd=\pgfimage, xmin=-0.5, xmax=223.5, ymax=223.5, ymin=-0.5] {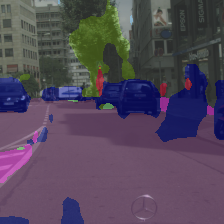};

\nextgroupplot[
width=\imgW,
ticks=none,
axis line style={white, line width=0.0pt},
xtick=\empty, 
ytick=\empty,
ylabel shift=-1pt,
ylabel=Seg. raw,
axis equal image,
xmin=-0.5, xmax=223.5, 
ymin=-0.5, ymax=223.5,
]
\addplot graphics [includegraphics cmd=\pgfimage, xmin=-0.5, xmax=223.5, ymax=223.5, ymin=-0.5] {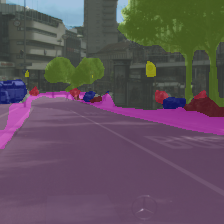};

\nextgroupplot[
width=\imgW,
ticks=none,
axis line style={white, line width=0.0pt}, 
xtick=\empty, 
ytick=\empty,
axis equal image,
xmin=-0.5, xmax=223.5, 
ymin=-0.5, ymax=223.5,
]
\addplot graphics [includegraphics cmd=\pgfimage, xmin=-0.5, xmax=223.5, ymax=223.5, ymin=-0.5] {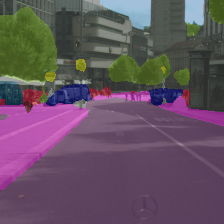};

\nextgroupplot[
width=\imgW,
ticks=none,
axis line style={white, line width=0.0pt},
xtick=\empty, 
ytick=\empty,
axis equal image,
xmin=-0.5, xmax=223.5, 
ymin=-0.5, ymax=223.5,
]
\addplot graphics [includegraphics cmd=\pgfimage, xmin=-0.5, xmax=223.5, ymax=223.5, ymin=-0.5] {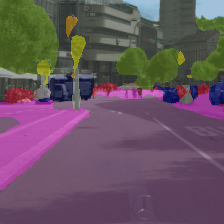};

\nextgroupplot[
width=\imgW,
ticks=none,
axis line style={white, line width=0.0pt},
xtick=\empty, 
ytick=\empty,
axis equal image,
xmin=-0.5, xmax=223.5, 
ymin=-0.5, ymax=223.5,
]
\addplot graphics [includegraphics cmd=\pgfimage, xmin=-0.5, xmax=223.5, ymax=223.5, ymin=-0.5] {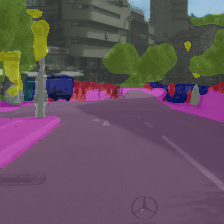};

\nextgroupplot[
width=\imgW,
ticks=none,
axis line style={white, line width=0.0pt},
xtick=\empty, 
ytick=\empty,
axis equal image,
xmin=-0.5, xmax=223.5, 
ymin=-0.5, ymax=223.5,
]
\addplot graphics [includegraphics cmd=\pgfimage, xmin=-0.5, xmax=223.5, ymax=223.5, ymin=-0.5] {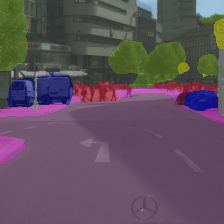};

\nextgroupplot[
width=\imgW,
ticks=none,
axis line style={white, line width=0.0pt},
xtick=\empty, 
ytick=\empty,
axis equal image,
xmin=-0.5, xmax=223.5, 
ymin=-0.5, ymax=223.5,
]
\addplot graphics [includegraphics cmd=\pgfimage, xmin=-0.5, xmax=223.5, ymax=223.5, ymin=-0.5] {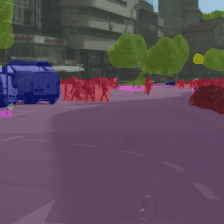};

\nextgroupplot[
width=\imgW,
ticks=none,
axis line style={white, line width=0.0pt},
xtick=\empty, 
ytick=\empty,
axis equal image,
xmin=-0.5, xmax=223.5, 
ymin=-0.5, ymax=223.5,
]
\addplot graphics [includegraphics cmd=\pgfimage, xmin=-0.5, xmax=223.5, ymax=223.5, ymin=-0.5] {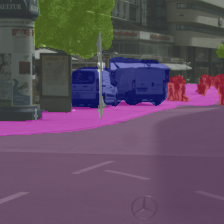};

\nextgroupplot[
width=\imgW,
ticks=none,
axis line style={white, line width=0.0pt},
xtick=\empty, 
ytick=\empty,
axis equal image,
xmin=-0.5, xmax=223.5, 
ymin=-0.5, ymax=223.5,
]
\addplot graphics [includegraphics cmd=\pgfimage, xmin=-0.5, xmax=223.5, ymax=223.5, ymin=-0.5] {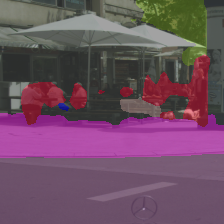};

\nextgroupplot[
width=\imgW,
ticks=none,
axis line style={white, line width=0.0pt},
xtick=\empty, 
ytick=\empty,
axis equal image,
xmin=-0.5, xmax=223.5, 
ymin=-0.5, ymax=223.5,
]
\addplot graphics [includegraphics cmd=\pgfimage, xmin=-0.5, xmax=223.5, ymax=223.5, ymin=-0.5] {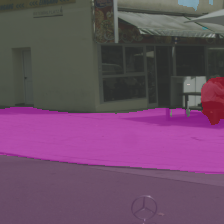};

\nextgroupplot[
width=\imgW,
ticks=none,
axis line style={white, line width=0.0pt},
xtick=\empty, 
ytick=\empty,
axis equal image,
xmin=-0.5, xmax=223.5, 
ymin=-0.5, ymax=223.5,
]
\addplot graphics [includegraphics cmd=\pgfimage, xmin=-0.5, xmax=223.5, ymax=223.5, ymin=-0.5] {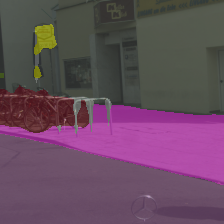};

\nextgroupplot[
width=\imgW,
ticks=none,
axis line style={white, line width=0.0pt},
xtick=\empty, 
ytick=\empty,
axis equal image,
xmin=-0.5, xmax=223.5, 
ymin=-0.5, ymax=223.5,
]
\addplot graphics [includegraphics cmd=\pgfimage, xmin=-0.5, xmax=223.5, ymax=223.5, ymin=-0.5] {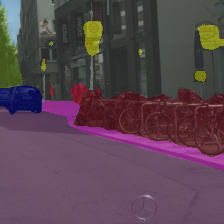};

\nextgroupplot[
width=\imgW,
ticks=none,
axis line style={white, line width=0.0pt},
xtick=\empty, 
ytick=\empty,
axis equal image,
xmin=-0.5, xmax=223.5, 
ymin=-0.5, ymax=223.5,
]
\addplot graphics [includegraphics cmd=\pgfimage, xmin=-0.5, xmax=223.5, ymax=223.5, ymin=-0.5] {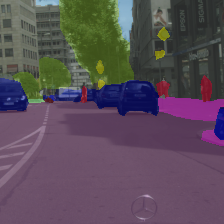};

\end{groupplot}
\end{tikzpicture}%
\vspace{-0.1cm}

\raggedright
\begin{tikzpicture}[every node/.style={font=\small}]
	\begin{semilogyaxis}[
        legend style={nodes={scale=0.5}},
        ylabel shift=-4pt,
        xlabel shift=-4pt,
        height=3.65cm,
        width=0.9603\textwidth,
        grid=both,
        grid style={line width=.1pt, draw=gray!10},
        major grid style={line width=.2pt,draw=gray!50},
	    ymin=10000.00,
        ymax=1400000,
        xmin=0.0,
        xmax=12.0,
        xlabel=Clip index,
        ylabel=Bandwidth (\si{\bit\per\second}),
        ytickten={4, 5, 6, 7},
        yticklabels={10\textsuperscript{4}, 10\textsuperscript{5}, 10\textsuperscript{6}},
        xtick={1, 2, 3, 4, 5, 6, 7, 8, 9, 10, 11, 12},
        xticklabels={1, 2, 3, 4, 5, 6, 7, 8, 9, 10, 11, 12},
        minor xtick={0.125, 0.25, 0.375, 0.5, 0.625, 0.75, 0.875, 1.0, 1.125, 1.25, 1.375, 1.5, 1.625, 1.75, 1.875, 2.0, 2.125, 2.25, 2.375, 2.5, 2.625, 2.75, 2.875, 3.0, 3.125, 3.25, 3.375, 3.5, 3.625, 3.75, 3.875, 4.0, 4.125, 4.25, 4.375, 4.5, 4.625, 4.75, 4.875, 5.0, 5.125, 5.25, 5.375, 5.5, 5.625, 5.75, 5.875, 6.0, 6.125, 6.25, 6.375, 6.5, 6.625, 6.75, 6.875, 7.0, 7.125, 7.25, 7.375, 7.5, 7.625, 7.75, 7.875, 8.0, 8.125, 8.25, 8.375, 8.5, 8.625, 8.75, 8.875, 9.0, 9.125, 9.25, 9.375, 9.5, 9.625, 9.75, 9.875, 10.0, 10.125, 10.25, 10.375, 10.5, 10.625, 10.75, 10.875, 11.0, 11.125, 11.25, 11.375, 11.5, 11.625, 11.75, 11.875, 12.0},
        set layers
        ]
        \addplot[color=tud3d, mark=none, semithick] coordinates {
            (0, 51328.66796875)
            (1, 51328.66796875)
            (1, 56610.0)
            (2, 56610.0)
            (2, 75548.0)
            (3, 75548.0)
            (3, 79084.0)
            (4, 79084.0)
            (4, 188643.328125)
            (5, 188643.328125)
            (5, 694121.3125)
            (6, 694121.3125)
            (6, 904479.3125)
            (7, 904479.3125)
            (7, 139524.671875)
            (8, 139524.671875)
            (8, 99903.3359375)
            (9, 99903.3359375)
            (9, 57465.66796875)
            (10, 57465.66796875)
            (10, 56088.66796875)
            (11, 56088.66796875)
            (11, 52450.66796875)
            (12, 52450.66796875)
    	};
        \label{pgfplots:longvideoh264}
        \addplot[color=tud6b, mark=none, semithick] coordinates {
            (0, 40601.66796875)
            (1, 40601.66796875)
            (1, 45911.33203125)
            (2, 45911.33203125)
            (2, 60219.66796875)
            (3, 60219.66796875)
            (3, 69802.0)
            (4, 69802.0)
            (4, 149991.0)
            (5, 149991.0)
            (5, 289561.0)
            (6, 289561.0)
            (6, 336838.0)
            (7, 336838.0)
            (7, 139377.328125)
            (8, 139377.328125)
            (8, 94848.6640625)
            (9, 94848.6640625)
            (9, 55907.33203125)
            (10, 55907.33203125)
            (10, 41808.66796875)
            (11, 41808.66796875)
            (11, 36521.66796875)
            (12, 36521.66796875)
    	};
        \label{pgfplots:longvideocc}
        \addplot[color=black, mark=none, dashed, semithick] coordinates {
            (0, 45000)
            (1, 45000)
            (1, 50000)
            (2, 50000)
            (2, 70000)
            (3, 70000)
            (3, 80000)
            (4, 80000)
            (4, 200000)
            (5, 200000)
            (5, 700000)
            (6, 700000)
            (6, 900000)
            (7, 900000)
            (7, 150000)
            (8, 150000)
            (8, 100000)
            (9, 100000)
            (9, 60000)
            (10, 60000)
            (10, 55000)
            (11, 55000)
            (11, 50000)
            (12, 50000)
    	};
        \label{pgfplots:longvideotarget}
	\end{semilogyaxis}
    \node[draw,fill=white, inner sep=0.5pt, anchor=west] at (10.0, 1.65) {\tiny
    \begin{tabular}{cl}
    \ref*{pgfplots:longvideotarget} & Target bandwidth \\
    \ref*{pgfplots:longvideocc} & Deep codec control (ours) \\
    \ref*{pgfplots:longvideoh264} & 2-pass ABR (H.264)~\cite{Wiegand2003} \\
    \end{tabular}};
\end{tikzpicture}

%% file: artwork/camvid/acc_vs_bw_camvid_ss_of_subplot_small.tex
\begin{tikzpicture}[every node/.style={font=\footnotesize}]
    \node[anchor=center] at (0.421\columnwidth, -3.50) {H.264 quantization parameter ($\qp$)};
	\begin{groupplot}[
        group style={
            group name=accvsbw,
            group size=2 by 4,
            ylabels at=edge left,
            horizontal sep=30pt,
            vertical sep=28.5pt,
        },
        legend style={nodes={scale=0.5}},
        height=3.45cm,
        width=0.52\columnwidth,
        grid=both,
        grid style={line width=.1pt, draw=gray!10},
        major grid style={line width=.2pt,draw=gray!50},
        minor tick num=1,
        xtick pos=bottom,
        ytick pos=left,
	    ymin=18,   
        ymax=100,
        ylabel shift=-4.5pt,
        ytick={
            20, 40, 60, 80, 100
        },
        yticklabels={
            20, 40, 60, 80, 100
        },
        xtick={
            0, 10, 20, 30, 40, 51
        },
        xticklabels={
            0, 10, 20, 30, 40, 51
        },
        ticklabel style = {font=\footnotesize},
        legend style={at={(0.05,0.2)}, anchor=west},
        legend cell align={left},
        ]
        \nextgroupplot[title={DeepLabV3 ResNet-19}, title style={yshift=-7pt, font=\footnotesize}, ylabel=Accuracy ($\%$) $\uparrow$]
    	\addplot[color=tud6b, mark=*, mark size=1.0pt] coordinates {
            (0, 98.60)
            (2, 98.47)
            (4, 98.30)
            (6, 97.94)
            (8, 97.68)
            (10, 97.34)
            (12, 96.76)
            (14, 96.03)
            (16, 94.91)
            (18, 93.53)
            (20, 91.32)
            (22, 89.37)
            (24, 86.66)
            (26, 83.37)
            (28, 80.11)
            (30, 77.50)
            (32, 73.88)
            (34, 70.41)
            (36, 65.77)
            (38, 61.12)
            (40, 55.35)
            (42, 49.70)
            (44, 43.58)
            (46, 38.36)
            (48, 34.49)
            (50, 31.51)
            (51, 30.58)
    	};
        \addlegendentry{w/o H.264 aug.}
        \addplot[color=tud3c, mark=*, mark size=1.0pt] coordinates {
    		(0, 99.11)
            (2, 99.03)
            (4, 98.91)
            (6, 98.68)
            (8, 98.51)
            (10, 98.29)
            (12, 97.93)
            (14, 97.46)
            (16, 96.81)
            (18, 96.00)
            (20, 94.78)
            (22, 93.72)
            (24, 92.43)
            (26, 91.04)
            (28, 89.43)
            (30, 87.87)
            (32, 85.46)
            (34, 82.66)
            (36, 78.49)
            (38, 73.00)
            (40, 65.20)
            (42, 56.35)
            (44, 45.70)
            (46, 35.75)
            (48, 28.79)
            (50, 25.41)
            (51, 24.78)
    	};
        \addlegendentry{w/ H.264 aug.}
        \nextgroupplot[
        title={RAFT large},
        title style={yshift=-7pt, font=\footnotesize},
        legend style={at={(0.05,0.85)}, anchor=west},
        ymin=0, ymax=65,
        ytick={
            10, 30, 50
        },
        yticklabels={
            10, 30, 50
        }, ylabel=F1-all ($\%$) $\downarrow$,]
        \addplot[color=tud10a, mark=*, mark size=1.0pt] coordinates {
    		(0, 1.2135)
            (2, 1.2320)
            (4, 1.3586)
            (6, 1.8426)
            (8, 2.1205)
            (10, 2.2289)
            (12, 3.2271)
            (14, 4.8793)
            (16, 7.3639)
            (18, 11.0327)
            (20, 15.9708)
            (22, 20.2911)
            (24, 23.8577)
            (26, 26.3856)
            (28, 28.0547)
            (30, 29.5496)
            (32, 30.7670)
            (34, 32.2151)
            (36, 33.6238)
            (38, 35.4435)
            (40, 38.0235)
            (42, 40.6130)
            (44, 44.7080)
            (46, 49.6858)
            (48, 55.2609)
            (50, 59.7110)
            (51, 62.1554)
    	};
     \addlegendentry{w/o H.264 aug.}
     \nextgroupplot[title={DeepLabV3 ResNet-50}, title style={yshift=-7pt, font=\footnotesize}, ylabel=Accuracy ($\%$) $\uparrow$,
     ymin=40,   
    ymax=100,
    ylabel shift=-4.5pt,
    ytick={
        40, 60, 80, 100
    },
    yticklabels={
        40, 60, 80, 100
    }
    ]
    	\addplot[color=tud6b, mark=*, mark size=1.0pt] coordinates {
            (0, 98.79)
            (2, 98.67)
            (4, 98.54)
            (6, 98.22)
            (8, 98.00)
            (10, 97.71)
            (12, 97.23)
            (14, 96.64)
            (16, 95.83)
            (18, 94.82)
            (20, 93.26)
            (22, 91.95)
            (24, 90.41)
            (26, 88.93)
            (28, 87.11)
            (30, 85.34)
            (32, 82.92)
            (34, 80.15)
            (36, 76.46)
            (38, 72.41)
            (40, 67.40)
            (42, 61.66)
            (44, 55.18)
            (46, 48.40)
            (48, 42.00)
            (50, 37.44)
            (51, 36.27)
    	};
        \addlegendentry{w/o H.264 aug.}
        \addplot[color=tud3c, mark=*, mark size=1.0pt] coordinates {
    				(0, 99.51)
            (2, 99.46)
            (4, 99.40)
            (6, 99.26)
            (8, 99.16)
            (10, 99.04)
            (12, 98.84)
            (14, 98.60)
            (16, 98.33)
            (18, 97.97)
            (20, 97.36)
            (22, 96.71)
            (24, 95.84)
            (26, 94.91)
            (28, 93.89)
            (30, 92.96)
            (32, 91.65)
            (34, 90.22)
            (36, 88.14)
            (38, 85.20)
            (40, 80.18)
            (42, 74.14)
            (44, 70.13)
            (46, 65.28)
            (48, 60.07)
            (50, 53.83)
            (51, 50.88)
    	};
        \addlegendentry{w/ H.264 aug.}
        \nextgroupplot[
        title={RAFT small\vphantom{p}},
        title style={yshift=-7pt, font=\footnotesize},
        legend style={at={(0.05,0.85)}, anchor=west},
        ymin=0, ymax=65,
        ytick={
            10, 30, 50
        },
        yticklabels={
            10, 30, 50
        }, ylabel=F1-all ($\%$) $\downarrow$,]
        \addplot[color=tud10a, mark=*, mark size=1.0pt] coordinates {
    		(0, 1.7371)
            (2, 1.9482)
            (4, 2.5926)
            (6, 3.6127)
            (8, 4.1928)
            (10, 4.7633)
            (12, 6.2966)
            (14, 8.9047)
            (16, 10.5253)
            (18, 13.9595)
            (20, 17.2517)
            (22, 19.8084)
            (24, 22.3260)
            (26, 24.1816)
            (28, 26.1145)
            (30, 27.4458)
            (32, 29.0208)
            (34, 30.6204)
            (36, 31.9456)
            (38, 33.6149)
            (40, 35.4356)
            (42, 37.4069)
            (44, 40.1907)
            (46, 44.1358)
            (48, 49.8009)
            (50, 55.4400)
            (51, 58.1419)
    	};
     \addlegendentry{w/o H.264 aug.}
	\end{groupplot}
\end{tikzpicture}

%% file: artwork/cityscapes/acc_b_ss_cs.tex
\begin{tikzpicture}[every node/.style={font=\footnotesize}]
	\begin{semilogxaxis}[
        ylabel shift=-4pt,
        xlabel shift=-3pt,
        legend style={nodes={scale=0.5}},
        height=3.45cm,
        width=0.75\columnwidth,
        grid=both,
        xtick pos=bottom,
        ytick pos=left,
        grid style={line width=.1pt, draw=gray!10},
        major grid style={line width=.2pt,draw=gray!50},
        minor tick num=3,
	    ymin=0,   
        ymax=100,
        xmin=27000,   
        xmax=1200000,
        ylabel style={align=center},
        ylabel=Bandwidth\\ accuracy (\%) $\uparrow$,
        xlabel=Bandwidth condition in \si{\bit\per\second},
        ytick={
            0, 20, 40, 60, 80, 100
        },
        yticklabels={
            0, 20, 40, 60, 80, 100
        },
        xtick={
            30000, 60000, 100000, 300000, 600000, 1000000
        },
        xticklabels={
            30k, 60k, 100k, 300k, 600k, 1M
        },
        minor xtick={40000, 50000, 70000, 80000, 90000, 200000, 400000, 500000, 700000, 800000, 900000},
        ticklabel style = {font=\footnotesize},
        legend style={at={(0.05,0.25)},anchor=west}
        ]
        \addplot[color=tud6b, mark=*, mark size=1.75pt] coordinates {
            (36000, 93.57429718875502)
            (51479, 98.59437751004017)
            (73613, 98.19277108433735)
            (105265, 97.18875502008032)
            (150525, 96.98795180722891)
            (215246, 97.38955823293173)
            (307796, 91.96787148594378)
            (440138, 89.95983935742972)
            (629384, 98.39357429718876)
            (900000, 100)
    	};
        \label{pgfplots:dccaccbwcs}
        \addplot[color=tud3d, mark=*, mark size=1.75pt] coordinates {
            (36000, 3.413654618473896)
            (51479, 29.31726907630522)
            (73613, 62.24899598393574)
            (105265, 83.53413654618473)
            (150525, 90.76305220883534)
            (215246, 93.57429718875502)
            (307796, 94.17670682730924)
            (440138, 91.56626506024096)
            (629384, 85.34136546184738)
            (900000, 61.84738955823293)
    	};
        \label{pgfplots:abraccbwcs}
	\end{semilogxaxis}
    \node[draw,fill=white, inner sep=0.5pt, anchor=south east] at (6.35, 0.15) {\tiny
    \begin{tabular}{cl}
    \ref*{pgfplots:dccaccbwcs} & \textbf{Deep video codec control}\\
    \ref*{pgfplots:abraccbwcs} & 2-pass ABR control\\
    \end{tabular}};
\end{tikzpicture}

%% file: artwork/cityscapes/acc_p_ss_cs.tex
\begin{tikzpicture}[every node/.style={font=\footnotesize}]
	\begin{semilogxaxis}[
        ylabel shift=-4pt,
        xlabel shift=-3pt,
        legend style={nodes={scale=0.5}},
        height=3.45cm,
        width=0.75\columnwidth,
        grid=both,
        xtick pos=bottom,
        ytick pos=left,
        grid style={line width=.1pt, draw=gray!10},
        major grid style={line width=.2pt,draw=gray!50},
        minor tick num=3,
	    ymin=0,   
        ymax=100,
        xmin=27000,   
        xmax=1200000,
        ylabel style={align=center},
        ylabel=Segmentation\\ accuracy (\%) $\uparrow$,
        xlabel=Bandwidth condition in \si{\bit\per\second},
        ytick={
            0, 20, 40, 60, 80, 100
        },
        yticklabels={
            0, 20, 40, 60, 80, 100
        },
        xtick={
            30000, 60000, 100000, 300000, 600000, 1000000
        },
        xticklabels={
            30k, 60k, 100k, 300k, 600k, 1M
        },
        minor xtick={40000, 50000, 70000, 80000, 90000, 200000, 400000, 500000, 700000, 800000, 900000},
        ticklabel style = {font=\footnotesize},
        legend style={at={(0.05,0.25)},anchor=west}
        ]
        \addplot[color=tud6b, mark=*, mark size=1.75pt] coordinates {
            (36000, 61.4527702331543)
            (51479, 73.63640666007996)
            (73613, 80.92343807220459)
            (105265, 84.75565314292908)
            (150525, 88.51199746131897)
            (215246, 90.98583459854126)
            (307796, 88.11136484146118)
            (440138, 87.09592223167419)
            (629384, 95.42379379272461)
            (900000, 97.04542756080627)
    	};
        \label{pgfplots:dccaccpcs}
        \addplot[color=tud3d, mark=*, mark size=1.75pt] coordinates {
            (36000, 2.8865477070212364)
            (51479, 24.493250250816345)
            (73613, 54.49244379997253)
            (105265, 75.71422457695007)
            (150525, 84.37908887863159)
            (215246, 88.86411190032959)
            (307796, 90.54772853851318)
            (440138, 89.06646966934204)
            (629384, 83.70336294174194)
            (900000, 61.05825901031494)
    	};
        \label{pgfplots:abraccpcs}
	\end{semilogxaxis}
    \node[draw,fill=white, inner sep=0.5pt, anchor=south east] at (6.35, 0.15) {\tiny
    \begin{tabular}{cl}
    \ref*{pgfplots:dccaccpcs} & \textbf{Deep video codec control}\\
    \ref*{pgfplots:abraccpcs} & 2-pass ABR control\\
    \end{tabular}};
\end{tikzpicture}

%% file: table/ema_ablation.tex
\begin{tabular*}{\columnwidth}{@{\extracolsep{\fill}}c@{\hskip 0.25em}S[table-format=2.2]S[table-format=2.2]S[table-format=2.2]S[table-format=2.2]S[table-format=2.2]S[table-format=2.2]@{}}
	\toprule
	& \multicolumn{3}{c}{$\mathrm{acc}_{\mathrm{bw}}$ $\uparrow$} & \multicolumn{3}{c}{$\mathrm{acc}_{\mathrm{seg}}$ $\uparrow$}  \\
    \cmidrule{2-4}\cmidrule{5-7}
	{Control network EMA} & {$\Delta0\%$} & {$\Delta5\%$} & {$\Delta10\%$} & {$\Delta0\%$} & {$\Delta5\%$} & {$\Delta10\%$} \\
    \midrule
    \cmark & 96.22 & 97.05 & 97.91 & 84.79 & 85.50 & 86.28 \\
    \xmark & 96.12 & 97.01  & 96.89 & 84.36 & 85.15 & 86.12 \\
	\bottomrule
\end{tabular*}

%% file: artwork/camvid/acc_b_of_cv.tex
\begin{tikzpicture}[every node/.style={font=\footnotesize}]
	\begin{semilogxaxis}[
        ylabel shift=-4pt,
        xlabel shift=-3pt,
        legend style={nodes={scale=0.5}},
        height=3.45cm,
        width=0.75\columnwidth,
        grid=both,
        xtick pos=bottom,
        ytick pos=left,
        grid style={line width=.1pt, draw=gray!10},
        major grid style={line width=.2pt,draw=gray!50},
        minor tick num=3,
	    ymin=0,   
        ymax=100,
        xmin=27000,   
        xmax=1200000,
        ylabel style={align=center},
        ylabel=Bandwidth\\ accuracy (\%) $\uparrow$,
        xlabel=Bandwidth condition in \si{\bit\per\second},
        ytick={
            0, 20, 40, 60, 80, 100
        },
        yticklabels={
            0, 20, 40, 60, 80, 100
        },
        xtick={
            30000, 60000, 100000, 300000, 600000, 1000000
        },
        xticklabels={
            30k, 60k, 100k, 300k, 600k, 1M
        },
        minor xtick={40000, 50000, 70000, 80000, 90000, 200000, 400000, 500000, 700000, 800000, 900000},
        ticklabel style = {font=\footnotesize},
        legend style={at={(0.05,0.25)},anchor=west}
        ]
        \addplot[color=tud6b, mark=*, mark size=1.75pt] coordinates {
            (36000, 96.98795180722891)
            (51479, 96.18473895582329)
            (73613, 96.58634538152611)
            (105265, 97.59036144578314)
            (150525, 99.19678714859438)
            (215246, 98.59437751004017)
            (307796, 98.99598393574297)
            (440138, 98.19277108433735)
            (629384, 98.59437751004017)
            (900000, 99.59839357429718)
    	};
        \label{pgfplots:dccaccbwcv}
        \addplot[color=tud3d, mark=*, mark size=1.75pt] coordinates {
            (36000, 26.700212)
            (51479, 52.042786)
            (73613, 66.524257)
            (105265, 69.239533)
            (150525, 65.619165)
            (215246, 64.261528)
            (307796, 57.473338)
            (440138, 58.830976)
            (629384, 87.341372)
            (900000, 91.866832)
    	};
        \label{pgfplots:abraccbwcv}
	\end{semilogxaxis}
    \node[draw,fill=white, inner sep=0.5pt, anchor=south east] at (6.35, 0.15) {\tiny
    \begin{tabular}{cl}
    \ref*{pgfplots:dccaccbwcv} & \textbf{Deep video codec control}\\
    \ref*{pgfplots:abraccbwcv} & 2-pass ABR control\\
    \end{tabular}};
\end{tikzpicture}

%% file: artwork/camvid/f1_of_cv.tex
\begin{tikzpicture}[every node/.style={font=\footnotesize}]
	\begin{semilogxaxis}[
        ylabel shift=-4pt,
        xlabel shift=-3pt,
        legend style={nodes={scale=0.5}},
        height=3.45cm,
        width=0.75\columnwidth,
        grid=both,
        xtick pos=bottom,
        ytick pos=left,
        grid style={line width=.1pt, draw=gray!10},
        major grid style={line width=.2pt,draw=gray!50},
        minor tick num=3,
	    ymin=0,   
        ymax=100,
        xmin=27000,   
        xmax=1200000,
        ylabel style={align=center},
        ylabel=F1-all (\%) $\downarrow$,
        xlabel=Bandwidth condition in \si{\bit\per\second},
        ytick={
            0, 20, 40, 60, 80, 100
        },
        yticklabels={
            0, 20, 40, 60, 80, 100
        },
        xtick={
            30000, 60000, 100000, 300000, 600000, 1000000
        },
        xticklabels={
            30k, 60k, 100k, 300k, 600k, 1M
        },
        minor xtick={40000, 50000, 70000, 80000, 90000, 200000, 400000, 500000, 700000, 800000, 900000},
        ticklabel style = {font=\footnotesize},
        legend style={at={(0.05,0.25)},anchor=west}
        ]
        \addplot[color=tud6b, mark=*, mark size=1.75pt] coordinates {
            (36000, 50.40447825858422)
            (51479, 47.45974190708205)
            (73613, 41.62391044478396)
            (105265, 35.74635819123355)
            (150525, 28.472495053666663)
            (215246, 22.006215955843274)
            (307796, 15.75969054975199)
            (440138, 12.970684418375871)
            (629384, 11.088450355915331)
            (900000, 10.146342042517515)
    	};
        \label{pgfplots:dccf1cv}
        \addplot[color=tud3d, mark=*, mark size=1.75pt] coordinates {
            (36000, 77.70990656249818)
            (51479, 60.292401881846175)
            (73613, 50.00665062797171)
            (105265, 46.68733065797132)
            (150525, 46.63093700798927)
            (215246, 43.770591124090295)
            (307796, 45.88789928218466)
            (440138, 42.72859144317388)
            (629384, 13.242987276310503)
            (900000, 8.401948520512352)
    	};
        \label{pgfplots:abrf1cv}
	\end{semilogxaxis}
    \node[draw,fill=white, inner sep=0.5pt, anchor=south east] at (6.35, 1.2) {\tiny
    \begin{tabular}{cl}
    \ref*{pgfplots:dccf1cv} & \textbf{Deep video codec control}\\
    \ref*{pgfplots:abrf1cv} & 2-pass ABR control\\
    \end{tabular}};
\end{tikzpicture}

%% file: artwork/surrogate_results/camvid/video_distortion_results.tex
\begin{tikzpicture}[every node/.style={font=\footnotesize}]
	\begin{axis}[
        ylabel shift=-4pt,
        xlabel shift=-3pt,
        legend style={nodes={scale=0.5}},
        height=3.45cm,
        width=0.75\columnwidth,
        xtick pos=bottom,
        ytick pos=left,
        grid=both,
        grid style={line width=.1pt, draw=gray!10},
        major grid style={line width=.2pt,draw=gray!50},
        minor tick num=3,
	    ymin=0.85,   
        ymax=1,
        xmin=0,   
        xmax=51,
        ylabel style={align=center},
        ylabel=SSIM $\uparrow$,
        xlabel=H.264 quantization parameter $\qp$,
        ytick={
            0.85, 0.9, 0.95, 1.0
        },
        yticklabels={
            0.85, 0.9, 0.95, 1.0
        },
        xtick={
            0, 10, 20, 30, 40, 50
        },
        xticklabels={
            0, 10, 20, 30, 40, 50
        },
        ticklabel style = {font=\footnotesize},
        legend style={at={(0.05,0.25)}, anchor=west}
        ]
        \addplot[color=tud6b, mark=*, mark size=1pt] coordinates {
            (0, 0.9930420517921448)
            (1, 0.9929826855659485)
            (2, 0.9924761652946472)
            (3, 0.9919763803482056)
            (4, 0.9917157292366028)
            (5, 0.9910449981689453)
            (6, 0.9903277158737183)
            (7, 0.9900669455528259)
            (8, 0.9894698262214661)
            (9, 0.9889363050460815)
            (10, 0.9883257746696472)
            (11, 0.9877781271934509)
            (12, 0.9872408509254456)
            (13, 0.9862929582595825)
            (14, 0.9855368137359619)
            (15, 0.9848710298538208)
            (16, 0.98445725440979)
            (17, 0.9832679629325867)
            (18, 0.9822978973388672)
            (19, 0.9819504618644714)
            (20, 0.9800617098808289)
            (21, 0.9793978929519653)
            (22, 0.9778733253479004)
            (23, 0.9768207669258118)
            (24, 0.9751002192497253)
            (25, 0.9733517169952393)
            (26, 0.9716167449951172)
            (27, 0.9695338606834412)
            (28, 0.9674190282821655)
            (29, 0.964041531085968)
            (30, 0.9612835645675659)
            (31, 0.95851069688797)
            (32, 0.9542566537857056)
            (33, 0.9507468342781067)
            (34, 0.9465392231941223)
            (35, 0.9418298602104187)
            (36, 0.9376657605171204)
            (37, 0.9337335824966431)
            (38, 0.9283332228660583)
            (39, 0.9245188236236572)
            (40, 0.9203258752822876)
            (41, 0.9152200222015381)
            (42, 0.9125767350196838)
            (43, 0.9089648127555847)
            (44, 0.9053136110305786)
            (45, 0.9041290283203125)
            (46, 0.9017984867095947)
            (47, 0.899826169013977)
            (48, 0.899685800075531)
            (49, 0.8983321785926819)
            (50, 0.896716296672821)
            (51, 0.8958501219749451)
    	};
	\end{axis}
\end{tikzpicture}

%% file: artwork/surrogate_results/camvid/file_size_results.tex
\begin{tikzpicture}[every node/.style={font=\footnotesize}]
	\begin{semilogyaxis}[
        ylabel shift=-4pt,
        xlabel shift=-3pt,
        legend style={nodes={scale=0.5}},
        height=3.45cm,
        width=0.75\columnwidth,
        xtick pos=bottom,
        ytick pos=left,
        grid=both,
        grid style={line width=.1pt, draw=gray!10},
        major grid style={line width=.2pt,draw=gray!50},
        minor tick num=3,
	    ymin = 10,
        ymax = 20000,
        xmin=0,
        xmax=51,
        ylabel style={align=center},
        ylabel=L1 $\downarrow$,
        xlabel=H.264 quantization parameter $\qp$,
        ytick={
            10, 100, 1000, 10000
        },
        yticklabels={
            10, 100, 1000, 10000
        },
        xtick={
            0, 10, 20, 30, 40, 50
        },
        xticklabels={
            0, 10, 20, 30, 40, 50
        },
        ticklabel style = {font=\footnotesize},
        legend style={at={(0.05,0.25)}, anchor=west}
        ]
        \addplot[color=tud6b, mark=*, mark size=1pt] coordinates {
            (0, 4415.35986328125)
            (1, 4070.22607421875)
            (2, 4757.736328125)
            (3, 5187.82958984375)
            (4, 5400.30224609375)
            (5, 5686.0126953125)
            (6, 6184.8466796875)
            (7, 0)
            (8, 3557.12890625)
            (9, 3777.2705078125)
            (10, 2649.449951171875)
            (11, 3173.621826171875)
            (12, 2405.251953125)
            (13, 1903.6470947265625)
            (14, 2055.179931640625)
            (15, 1919.6041259765625)
            (16, 1556.741455078125)
            (17, 1922.898681640625)
            (18, 1368.2266845703125)
            (19, 1165.823974609375)
            (20, 1558.8221435546875)
            (21, 2240.105224609375)
            (22, 1105.9820556640625)
            (23, 897.6957397460938)
            (24, 998.2044677734375)
            (25, 1367.601318359375)
            (26, 737.6646118164062)
            (27, 894.2593383789062)
            (28, 575.4366455078125)
            (29, 551.722900390625)
            (30, 550.843017578125)
            (31, 407.14324951171875)
            (32, 360.4018859863281)
            (33, 359.50689697265625)
            (34, 346.6927185058594)
            (35, 268.73199462890625)
            (36, 228.449951171875)
            (37, 196.3734893798828)
            (38, 174.04161071777344)
            (39, 149.30320739746094)
            (40, 129.46910095214844)
            (41, 111.2187271118164)
            (42, 97.67758178710938)
            (43, 88.50192260742188)
            (44, 76.67709350585938)
            (45, 69.32530975341797)
            (46, 68.265625)
            (47, 64.60723876953125)
            (48, 59.19744873046875)
            (49, 51.08645248413086)
            (50, 52.74142074584961)
            (51, 50.63410568237305)
    	};
	\end{semilogyaxis}
\end{tikzpicture}

%% file: artwork/surrogate_results/cityscapes/video_distortion_results.tex
\begin{tikzpicture}[every node/.style={font=\footnotesize}]
	\begin{axis}[
        ylabel shift=-4pt,
        xlabel shift=-3pt,
        legend style={nodes={scale=0.5}},
        height=3.45cm,
        width=0.75\columnwidth,
        xtick pos=bottom,
        ytick pos=left,
        grid=both,
        grid style={line width=.1pt, draw=gray!10},
        major grid style={line width=.2pt,draw=gray!50},
        minor tick num=3,
	    ymin=0.85,   
        ymax=1,
        xmin=0,   
        xmax=51,
        ylabel style={align=center},
        ylabel=SSIM $\uparrow$,
        xlabel=H.264 quantization parameter $\qp$,
        ytick={
            0.85, 0.9, 0.95, 1.0
        },
        yticklabels={
            0.85, 0.9, 0.95, 1.0
        },
        xtick={
            0, 10, 20, 30, 40, 50
        },
        xticklabels={
            0, 10, 20, 30, 40, 50
        },
        ticklabel style = {font=\footnotesize},
        legend style={at={(0.05,0.25)}, anchor=west}
        ]
        \addplot[color=tud6b, mark=*, mark size=1pt] coordinates {
            (0, 0.9926615357398987)
            (1, 0.9926746487617493)
            (2, 0.991992175579071)
            (3, 0.9913950562477112)
            (4, 0.9909408688545227)
            (5, 0.9900208115577698)
            (6, 0.9893037676811218)
            (7, 0.9892423152923584)
            (8, 0.9888352751731873)
            (9, 0.9886770844459534)
            (10, 0.9881774187088013)
            (11, 0.987903892993927)
            (12, 0.9877846837043762)
            (13, 0.9869989156723022)
            (14, 0.9864421486854553)
            (15, 0.9860401153564453)
            (16, 0.985963761806488)
            (17, 0.9849522709846497)
            (18, 0.9842058420181274)
            (19, 0.984208881855011)
            (20, 0.9823741912841797)
            (21, 0.9820764064788818)
            (22, 0.9805131554603577)
            (23, 0.9795517921447754)
            (24, 0.9779638051986694)
            (25, 0.9763618111610413)
            (26, 0.9748175144195557)
            (27, 0.9731575846672058)
            (28, 0.971369743347168)
            (29, 0.9685648679733276)
            (30, 0.9664642214775085)
            (31, 0.9643315076828003)
            (32, 0.9610609412193298)
            (33, 0.9586877226829529)
            (34, 0.9555799961090088)
            (35, 0.9522244930267334)
            (36, 0.9494631886482239)
            (37, 0.9467443227767944)
            (38, 0.9428926706314087)
            (39, 0.9404300451278687)
            (40, 0.937524139881134)
            (41, 0.9335266947746277)
            (42, 0.9318516850471497)
            (43, 0.9289624094963074)
            (44, 0.925991415977478)
            (45, 0.9241498112678528)
            (46, 0.9216626286506653)
            (47, 0.9192006587982178)
            (48, 0.9172995090484619)
            (49, 0.9153491854667664)
            (50, 0.9124337434768677)
            (51, 0.9111444354057312)
    	};
	\end{axis}
\end{tikzpicture}

%% file: artwork/surrogate_results/cityscapes/file_size_results.tex
\begin{tikzpicture}[every node/.style={font=\footnotesize}]
	\begin{semilogyaxis}[
        ylabel shift=-4pt,
        xlabel shift=-3pt,
        legend style={nodes={scale=0.5}},
        height=3.45cm,
        width=0.75\columnwidth,
        xtick pos=bottom,
        ytick pos=left,
        grid=both,
        grid style={line width=.1pt, draw=gray!10},
        major grid style={line width=.2pt,draw=gray!50},
        minor tick num=3,
	    ymin = 10,
        ymax = 20000,
        xmin=0,
        xmax=51,
        ylabel style={align=center},
        ylabel=L1 $\downarrow$,
        xlabel=H.264 quantization parameter $\qp$,
        ytick={
            10, 100, 1000, 10000
        },
        yticklabels={
            10, 100, 1000, 10000
        },
        xtick={
            0, 10, 20, 30, 40, 50
        },
        xticklabels={
            0, 10, 20, 30, 40, 50
        },
        ticklabel style = {font=\footnotesize},
        legend style={at={(0.05,0.25)}, anchor=west}
        ]
        \addplot[color=tud6b, mark=*, mark size=1pt] coordinates {
            (0, 14334.3017578125)
            (1, 15657.060546875)
            (2, 15472.80078125)
            (3, 14918.076171875)
            (4, 14448.716796875)
            (5, 15468.0146484375)
            (6, 15056.181640625)
            (7, 14509.216796875)
            (8, 13136.4306640625)
            (9, 11763.142578125)
            (10, 10341.64453125)
            (11, 8535.24609375)
            (12, 7081.94384765625)
            (13, 6240.85107421875)
            (14, 5068.2353515625)
            (15, 4252.27783203125)
            (16, 3626.620849609375)
            (17, 3094.9375)
            (18, 2767.915771484375)
            (19, 2510.66162109375)
            (20, 2259.973876953125)
            (21, 2068.8388671875)
            (22, 1869.7435302734375)
            (23, 1684.2640380859375)
            (24, 1503.236572265625)
            (25, 1359.66552734375)
            (26, 1219.64404296875)
            (27, 1071.7288818359375)
            (28, 972.6589965820312)
            (29, 828.7520141601562)
            (30, 723.5234985351562)
            (31, 658.35302734375)
            (32, 558.302734375)
            (33, 498.2701721191406)
            (34, 426.27691650390625)
            (35, 372.47283935546875)
            (36, 321.955078125)
            (37, 281.9697570800781)
            (38, 242.87725830078125)
            (39, 215.5325927734375)
            (40, 186.33612060546875)
            (41, 169.17552185058594)
            (42, 145.68553161621094)
            (43, 129.02073669433594)
            (44, 115.13391876220703)
            (45, 99.33672332763672)
            (46, 88.94337463378906)
            (47, 80.14433288574219)
            (48, 69.4856948852539)
            (49, 63.478858947753906)
            (50, 61.79032516479492)
            (51, 61.27623748779297)
    	};
	\end{semilogyaxis}
\end{tikzpicture}

%% file: artwork/surrogate_results/camvid/supplement/surrogate_results_camvid_full_1.tex
\begin{tikzpicture}[every node/.style={font=\small}, spy using outlines={tud6a, line width=0.80mm, dashed, dash pattern=on 1.5pt off 1.5pt, magnification=2, size=0.75cm, connect spies}]

\begin{groupplot}[group style={group size=8 by 3, horizontal sep=1.25pt, vertical sep=1.25pt}]

\nextgroupplot[
width=0.23\columnwidth,
ticks=none,
axis line style={draw=none}, 
xtick=\empty, 
ytick=\empty,
axis equal image,
xmin=0, xmax=224,
ymin=0, ymax=224,
ylabel={H.264},
title={$t_0$},
title style={yshift=-7pt},
]
\addplot graphics [includegraphics cmd=\pgfimage, xmin=0, xmax=224, ymin=0, ymax=224] {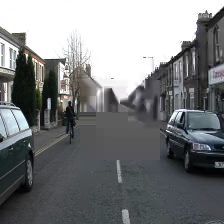};

\nextgroupplot[
width=0.23\columnwidth,
ticks=none,
axis lines=none, 
xtick=\empty, 
ytick=\empty,
axis equal image,
xmin=0, xmax=224,
ymin=0, ymax=224,
title={$t_1$},
title style={yshift=-7pt},
]
\addplot graphics [includegraphics cmd=\pgfimage, xmin=0, xmax=224, ymin=0, ymax=224] {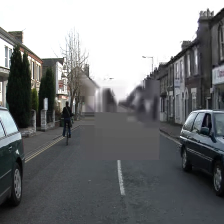};

\nextgroupplot[
width=0.23\columnwidth,
ticks=none,
axis lines=none, 
xtick=\empty, 
ytick=\empty,
axis equal image,
xmin=0, xmax=224,
ymin=0, ymax=224,
title={$t_2$},
title style={yshift=-7pt},
]
\addplot graphics [includegraphics cmd=\pgfimage, xmin=0, xmax=224, ymin=0, ymax=224] {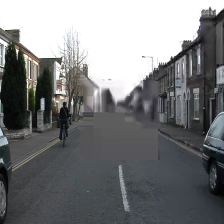};

\nextgroupplot[
width=0.23\columnwidth,
ticks=none,
axis lines=none, 
xtick=\empty, 
ytick=\empty,
axis equal image,
xmin=0, xmax=224,
ymin=0, ymax=224,
title={$t_3$},
title style={yshift=-7pt},
]
\addplot graphics [includegraphics cmd=\pgfimage, xmin=0, xmax=224, ymin=0, ymax=224] {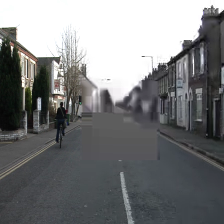};

\nextgroupplot[
width=0.23\columnwidth,
ticks=none,
axis lines=none, 
xtick=\empty, 
ytick=\empty,
axis equal image,
xmin=0, xmax=224,
ymin=0, ymax=224,
title={$t_4$},
title style={yshift=-7pt},
]
\addplot graphics [includegraphics cmd=\pgfimage, xmin=0, xmax=224, ymin=0, ymax=224] {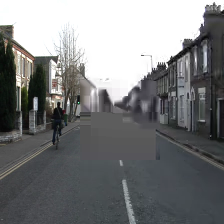};

\nextgroupplot[
width=0.23\columnwidth,
ticks=none,
axis lines=none, 
xtick=\empty, 
ytick=\empty,
axis equal image,
xmin=0, xmax=224,
ymin=0, ymax=224,
title={$t_5$},
title style={yshift=-7pt},
]
\addplot graphics [includegraphics cmd=\pgfimage, xmin=0, xmax=224, ymin=0, ymax=224] {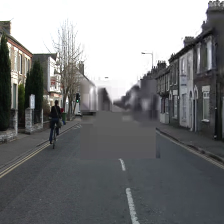};

\nextgroupplot[
width=0.23\columnwidth,
ticks=none,
axis lines=none, 
xtick=\empty, 
ytick=\empty,
axis equal image,
xmin=0, xmax=224,
ymin=0, ymax=224,
title={$t_6$},
title style={yshift=-7pt},
]
\addplot graphics [includegraphics cmd=\pgfimage, xmin=0, xmax=224, ymin=0, ymax=224] {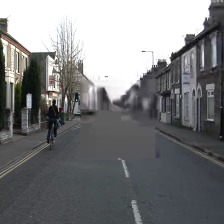};

\nextgroupplot[
width=0.23\columnwidth,
ticks=none,
axis lines=none, 
xtick=\empty, 
ytick=\empty,
axis equal image,
xmin=0, xmax=224,
ymin=0, ymax=224,
title={$t_7$},
title style={yshift=-7pt},
]
\addplot graphics [includegraphics cmd=\pgfimage, xmin=0, xmax=224, ymin=0, ymax=224] {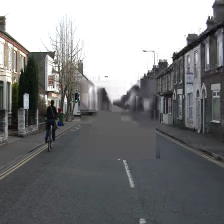};

\nextgroupplot[
width=0.23\columnwidth,
ticks=none,
axis line style={draw=none}, 
xtick=\empty, 
ytick=\empty,
axis equal image,
xmin=0, xmax=224,
ymin=0, ymax=224,
ylabel={Surrogate},
title style={yshift=-7pt},
]
\addplot graphics [includegraphics cmd=\pgfimage, xmin=0, xmax=224, ymin=0, ymax=224] {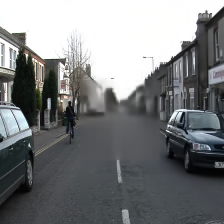};

\nextgroupplot[
width=0.23\columnwidth,
ticks=none,
axis lines=none, 
xtick=\empty, 
ytick=\empty,
axis equal image,
xmin=0, xmax=224,
ymin=0, ymax=224,
title style={yshift=-7pt},
]
\addplot graphics [includegraphics cmd=\pgfimage, xmin=0, xmax=224, ymin=0, ymax=224] {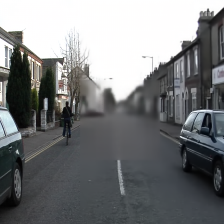};

\nextgroupplot[
width=0.23\columnwidth,
ticks=none,
axis lines=none, 
xtick=\empty, 
ytick=\empty,
axis equal image,
xmin=0, xmax=224,
ymin=0, ymax=224,
title style={yshift=-7pt},
]
\addplot graphics [includegraphics cmd=\pgfimage, xmin=0, xmax=224, ymin=0, ymax=224] {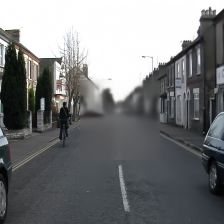};

\nextgroupplot[
width=0.23\columnwidth,
ticks=none,
axis lines=none, 
xtick=\empty, 
ytick=\empty,
axis equal image,
xmin=0, xmax=224,
ymin=0, ymax=224,
title style={yshift=-7pt},
]
\addplot graphics [includegraphics cmd=\pgfimage, xmin=0, xmax=224, ymin=0, ymax=224] {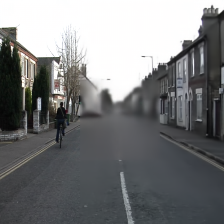};

\nextgroupplot[
width=0.23\columnwidth,
ticks=none,
axis lines=none, 
xtick=\empty, 
ytick=\empty,
axis equal image,
xmin=0, xmax=224,
ymin=0, ymax=224,
title style={yshift=-7pt},
]
\addplot graphics [includegraphics cmd=\pgfimage, xmin=0, xmax=224, ymin=0, ymax=224] {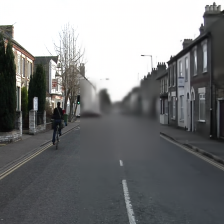};

\nextgroupplot[
width=0.23\columnwidth,
ticks=none,
axis lines=none, 
xtick=\empty, 
ytick=\empty,
axis equal image,
xmin=0, xmax=224,
ymin=0, ymax=224,
title style={yshift=-7pt},
]
\addplot graphics [includegraphics cmd=\pgfimage, xmin=0, xmax=224, ymin=0, ymax=224] {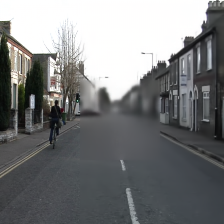};

\nextgroupplot[
width=0.23\columnwidth,
ticks=none,
axis lines=none, 
xtick=\empty, 
ytick=\empty,
axis equal image,
xmin=0, xmax=224,
ymin=0, ymax=224,
title style={yshift=-7pt},
]
\addplot graphics [includegraphics cmd=\pgfimage, xmin=0, xmax=224, ymin=0, ymax=224] {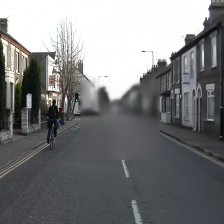};

\nextgroupplot[
width=0.23\columnwidth,
ticks=none,
axis lines=none, 
xtick=\empty, 
ytick=\empty,
axis equal image,
xmin=0, xmax=224,
ymin=0, ymax=224,
title style={yshift=-7pt},
]
\addplot graphics [includegraphics cmd=\pgfimage, xmin=0, xmax=224, ymin=0, ymax=224] {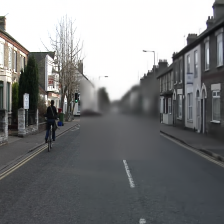};

\nextgroupplot[
width=0.23\columnwidth,
ticks=none,
axis line style={draw=none}, 
xtick=\empty, 
ytick=\empty,
axis equal image,
xmin=0, xmax=224,
ymin=0, ymax=224,
ylabel={$\qp$ map},
title style={yshift=-7pt},
]
\addplot graphics [includegraphics cmd=\pgfimage, xmin=0, xmax=224, ymin=0, ymax=224] {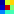};

\nextgroupplot[
width=0.23\columnwidth,
ticks=none,
axis lines=none, 
xtick=\empty, 
ytick=\empty,
axis equal image,
xmin=0, xmax=224,
ymin=0, ymax=224,
title style={yshift=-7pt},
]
\addplot graphics [includegraphics cmd=\pgfimage, xmin=0, xmax=224, ymin=0, ymax=224] {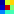};

\nextgroupplot[
width=0.23\columnwidth,
ticks=none,
axis lines=none, 
xtick=\empty, 
ytick=\empty,
axis equal image,
xmin=0, xmax=224,
ymin=0, ymax=224,
title style={yshift=-7pt},
]
\addplot graphics [includegraphics cmd=\pgfimage, xmin=0, xmax=224, ymin=0, ymax=224] {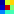};

\nextgroupplot[
width=0.23\columnwidth,
ticks=none,
axis lines=none, 
xtick=\empty, 
ytick=\empty,
axis equal image,
xmin=0, xmax=224,
ymin=0, ymax=224,
title style={yshift=-7pt},
]
\addplot graphics [includegraphics cmd=\pgfimage, xmin=0, xmax=224, ymin=0, ymax=224] {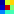};

\nextgroupplot[
width=0.23\columnwidth,
ticks=none,
axis lines=none, 
xtick=\empty, 
ytick=\empty,
axis equal image,
xmin=0, xmax=224,
ymin=0, ymax=224,
title style={yshift=-7pt},
]
\addplot graphics [includegraphics cmd=\pgfimage, xmin=0, xmax=224, ymin=0, ymax=224] {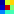};

\nextgroupplot[
width=0.23\columnwidth,
ticks=none,
axis lines=none, 
xtick=\empty, 
ytick=\empty,
axis equal image,
xmin=0, xmax=224,
ymin=0, ymax=224,
title style={yshift=-7pt},
]
\addplot graphics [includegraphics cmd=\pgfimage, xmin=0, xmax=224, ymin=0, ymax=224] {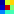};

\nextgroupplot[
width=0.23\columnwidth,
ticks=none,
axis lines=none, 
xtick=\empty, 
ytick=\empty,
axis equal image,
xmin=0, xmax=224,
ymin=0, ymax=224,
title style={yshift=-7pt},
]
\addplot graphics [includegraphics cmd=\pgfimage, xmin=0, xmax=224, ymin=0, ymax=224] {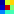};

\nextgroupplot[
width=0.23\columnwidth,
ticks=none,
axis lines=none, 
xtick=\empty, 
ytick=\empty,
axis equal image,
xmin=0, xmax=224,
ymin=0, ymax=224,
title style={yshift=-7pt},
colorbar,
colorbar style={
    ylabel={},
    ytick={5, 25, 45},
    yticklabels={5, 25, 45},
    at={(1.025,1)},
    yticklabel style={xshift=-1.5pt},
},
colormap={mymap}{[1pt]
    rgb(0pt)=(0,0,0.5);
    rgb(22pt)=(0,0,1);
    rgb(25pt)=(0,0,1);
    rgb(68pt)=(0,0.86,1);
    rgb(70pt)=(0,0.9,0.967741935483871);
    rgb(75pt)=(0.0806451612903226,1,0.887096774193548);
    rgb(128pt)=(0.935483870967742,1,0.0322580645161291);
    rgb(130pt)=(0.967741935483871,0.962962962962963,0);
    rgb(132pt)=(1,0.925925925925926,0);
    rgb(178pt)=(1,0.0740740740740741,0);
    rgb(182pt)=(0.909090909090909,0,0);
    rgb(200pt)=(0.5,0,0)
},
point meta max=51,
point meta min=0,
]
\addplot graphics [includegraphics cmd=\pgfimage, xmin=0, xmax=224, ymin=0, ymax=224] {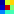};

\end{groupplot}

\end{tikzpicture}

%% file: artwork/surrogate_results/camvid/supplement/surrogate_results_camvid_full_2.tex
\begin{tikzpicture}[every node/.style={font=\small}, spy using outlines={tud6a, line width=0.80mm, dashed, dash pattern=on 1.5pt off 1.5pt, magnification=2, size=0.75cm, connect spies}]

\begin{groupplot}[group style={group size=8 by 3, horizontal sep=1.25pt, vertical sep=1.25pt}]

\nextgroupplot[
width=0.23\columnwidth,
ticks=none,
axis line style={draw=none}, 
xtick=\empty, 
ytick=\empty,
axis equal image,
xmin=0, xmax=224,
ymin=0, ymax=224,
ylabel={H.264},
title={$t_0$},
title style={yshift=-7pt},
]
\addplot graphics [includegraphics cmd=\pgfimage, xmin=0, xmax=224, ymin=0, ymax=224] {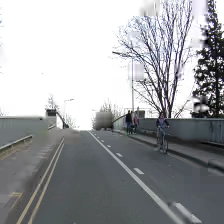};

\nextgroupplot[
width=0.23\columnwidth,
ticks=none,
axis lines=none, 
xtick=\empty, 
ytick=\empty,
axis equal image,
xmin=0, xmax=224,
ymin=0, ymax=224,
title={$t_1$},
title style={yshift=-7pt},
]
\addplot graphics [includegraphics cmd=\pgfimage, xmin=0, xmax=224, ymin=0, ymax=224] {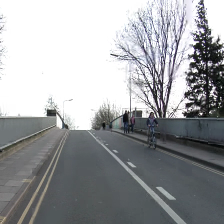};

\nextgroupplot[
width=0.23\columnwidth,
ticks=none,
axis lines=none, 
xtick=\empty, 
ytick=\empty,
axis equal image,
xmin=0, xmax=224,
ymin=0, ymax=224,
title={$t_2$},
title style={yshift=-7pt},
]
\addplot graphics [includegraphics cmd=\pgfimage, xmin=0, xmax=224, ymin=0, ymax=224] {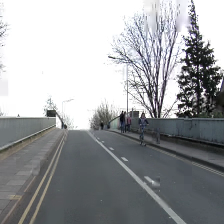};

\nextgroupplot[
width=0.23\columnwidth,
ticks=none,
axis lines=none, 
xtick=\empty, 
ytick=\empty,
axis equal image,
xmin=0, xmax=224,
ymin=0, ymax=224,
title={$t_3$},
title style={yshift=-7pt},
]
\addplot graphics [includegraphics cmd=\pgfimage, xmin=0, xmax=224, ymin=0, ymax=224] {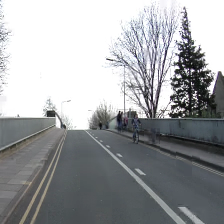};

\nextgroupplot[
width=0.23\columnwidth,
ticks=none,
axis lines=none, 
xtick=\empty, 
ytick=\empty,
axis equal image,
xmin=0, xmax=224,
ymin=0, ymax=224,
title={$t_4$},
title style={yshift=-7pt},
]
\addplot graphics [includegraphics cmd=\pgfimage, xmin=0, xmax=224, ymin=0, ymax=224] {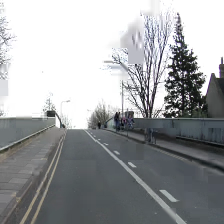};

\nextgroupplot[
width=0.23\columnwidth,
ticks=none,
axis lines=none, 
xtick=\empty, 
ytick=\empty,
axis equal image,
xmin=0, xmax=224,
ymin=0, ymax=224,
title={$t_5$},
title style={yshift=-7pt},
]
\addplot graphics [includegraphics cmd=\pgfimage, xmin=0, xmax=224, ymin=0, ymax=224] {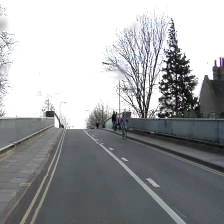};

\nextgroupplot[
width=0.23\columnwidth,
ticks=none,
axis lines=none, 
xtick=\empty, 
ytick=\empty,
axis equal image,
xmin=0, xmax=224,
ymin=0, ymax=224,
title={$t_6$},
title style={yshift=-7pt},
]
\addplot graphics [includegraphics cmd=\pgfimage, xmin=0, xmax=224, ymin=0, ymax=224] {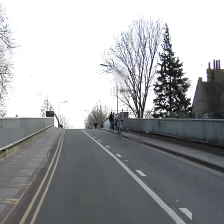};

\nextgroupplot[
width=0.23\columnwidth,
ticks=none,
axis lines=none, 
xtick=\empty, 
ytick=\empty,
axis equal image,
xmin=0, xmax=224,
ymin=0, ymax=224,
title={$t_7$},
title style={yshift=-7pt},
]
\addplot graphics [includegraphics cmd=\pgfimage, xmin=0, xmax=224, ymin=0, ymax=224] {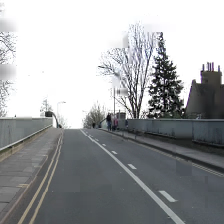};

\nextgroupplot[
width=0.23\columnwidth,
ticks=none,
axis line style={draw=none}, 
xtick=\empty, 
ytick=\empty,
axis equal image,
xmin=0, xmax=224,
ymin=0, ymax=224,
ylabel={Surrogate},
title style={yshift=-7pt},
]
\addplot graphics [includegraphics cmd=\pgfimage, xmin=0, xmax=224, ymin=0, ymax=224] {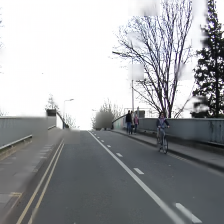};

\nextgroupplot[
width=0.23\columnwidth,
ticks=none,
axis lines=none, 
xtick=\empty, 
ytick=\empty,
axis equal image,
xmin=0, xmax=224,
ymin=0, ymax=224,
title style={yshift=-7pt},
]
\addplot graphics [includegraphics cmd=\pgfimage, xmin=0, xmax=224, ymin=0, ymax=224] {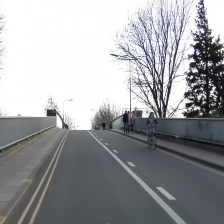};

\nextgroupplot[
width=0.23\columnwidth,
ticks=none,
axis lines=none, 
xtick=\empty, 
ytick=\empty,
axis equal image,
xmin=0, xmax=224,
ymin=0, ymax=224,
title style={yshift=-7pt},
]
\addplot graphics [includegraphics cmd=\pgfimage, xmin=0, xmax=224, ymin=0, ymax=224] {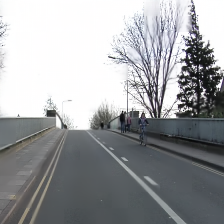};

\nextgroupplot[
width=0.23\columnwidth,
ticks=none,
axis lines=none, 
xtick=\empty, 
ytick=\empty,
axis equal image,
xmin=0, xmax=224,
ymin=0, ymax=224,
title style={yshift=-7pt},
]
\addplot graphics [includegraphics cmd=\pgfimage, xmin=0, xmax=224, ymin=0, ymax=224] {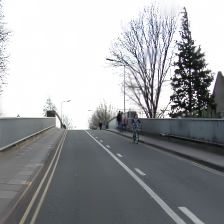};

\nextgroupplot[
width=0.23\columnwidth,
ticks=none,
axis lines=none, 
xtick=\empty, 
ytick=\empty,
axis equal image,
xmin=0, xmax=224,
ymin=0, ymax=224,
title style={yshift=-7pt},
]
\addplot graphics [includegraphics cmd=\pgfimage, xmin=0, xmax=224, ymin=0, ymax=224] {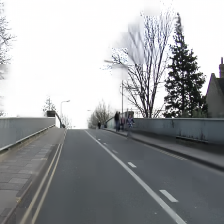};

\nextgroupplot[
width=0.23\columnwidth,
ticks=none,
axis lines=none, 
xtick=\empty, 
ytick=\empty,
axis equal image,
xmin=0, xmax=224,
ymin=0, ymax=224,
title style={yshift=-7pt},
]
\addplot graphics [includegraphics cmd=\pgfimage, xmin=0, xmax=224, ymin=0, ymax=224] {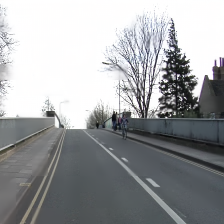};

\nextgroupplot[
width=0.23\columnwidth,
ticks=none,
axis lines=none, 
xtick=\empty, 
ytick=\empty,
axis equal image,
xmin=0, xmax=224,
ymin=0, ymax=224,
title style={yshift=-7pt},
]
\addplot graphics [includegraphics cmd=\pgfimage, xmin=0, xmax=224, ymin=0, ymax=224] {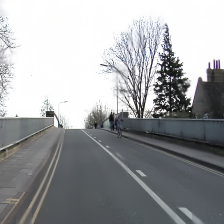};

\nextgroupplot[
width=0.23\columnwidth,
ticks=none,
axis lines=none, 
xtick=\empty, 
ytick=\empty,
axis equal image,
xmin=0, xmax=224,
ymin=0, ymax=224,
title style={yshift=-7pt},
]
\addplot graphics [includegraphics cmd=\pgfimage, xmin=0, xmax=224, ymin=0, ymax=224] {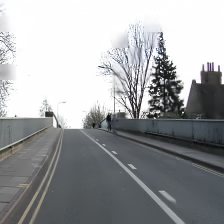};

\nextgroupplot[
width=0.23\columnwidth,
ticks=none,
axis line style={draw=none}, 
xtick=\empty, 
ytick=\empty,
axis equal image,
xmin=0, xmax=224,
ymin=0, ymax=224,
ylabel={$\qp$ map},
title style={yshift=-7pt},
]
\addplot graphics [includegraphics cmd=\pgfimage, xmin=0, xmax=224, ymin=0, ymax=224] {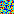};

\nextgroupplot[
width=0.23\columnwidth,
ticks=none,
axis lines=none, 
xtick=\empty, 
ytick=\empty,
axis equal image,
xmin=0, xmax=224,
ymin=0, ymax=224,
title style={yshift=-7pt},
]
\addplot graphics [includegraphics cmd=\pgfimage, xmin=0, xmax=224, ymin=0, ymax=224] {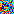};

\nextgroupplot[
width=0.23\columnwidth,
ticks=none,
axis lines=none, 
xtick=\empty, 
ytick=\empty,
axis equal image,
xmin=0, xmax=224,
ymin=0, ymax=224,
title style={yshift=-7pt},
]
\addplot graphics [includegraphics cmd=\pgfimage, xmin=0, xmax=224, ymin=0, ymax=224] {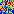};

\nextgroupplot[
width=0.23\columnwidth,
ticks=none,
axis lines=none, 
xtick=\empty, 
ytick=\empty,
axis equal image,
xmin=0, xmax=224,
ymin=0, ymax=224,
title style={yshift=-7pt},
]
\addplot graphics [includegraphics cmd=\pgfimage, xmin=0, xmax=224, ymin=0, ymax=224] {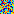};

\nextgroupplot[
width=0.23\columnwidth,
ticks=none,
axis lines=none, 
xtick=\empty, 
ytick=\empty,
axis equal image,
xmin=0, xmax=224,
ymin=0, ymax=224,
title style={yshift=-7pt},
]
\addplot graphics [includegraphics cmd=\pgfimage, xmin=0, xmax=224, ymin=0, ymax=224] {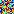};

\nextgroupplot[
width=0.23\columnwidth,
ticks=none,
axis lines=none, 
xtick=\empty, 
ytick=\empty,
axis equal image,
xmin=0, xmax=224,
ymin=0, ymax=224,
title style={yshift=-7pt},
]
\addplot graphics [includegraphics cmd=\pgfimage, xmin=0, xmax=224, ymin=0, ymax=224] {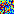};

\nextgroupplot[
width=0.23\columnwidth,
ticks=none,
axis lines=none, 
xtick=\empty, 
ytick=\empty,
axis equal image,
xmin=0, xmax=224,
ymin=0, ymax=224,
title style={yshift=-7pt},
]
\addplot graphics [includegraphics cmd=\pgfimage, xmin=0, xmax=224, ymin=0, ymax=224] {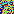};

\nextgroupplot[
width=0.23\columnwidth,
ticks=none,
axis lines=none, 
xtick=\empty, 
ytick=\empty,
axis equal image,
xmin=0, xmax=224,
ymin=0, ymax=224,
title style={yshift=-7pt},
colorbar,
colorbar style={
    ylabel={},
    ytick={5, 25, 45},
    yticklabels={5, 25, 45},
    at={(1.025,1)},
    yticklabel style={xshift=-1.5pt},
},
colormap={mymap}{[1pt]
    rgb(0pt)=(0,0,0.5);
    rgb(22pt)=(0,0,1);
    rgb(25pt)=(0,0,1);
    rgb(68pt)=(0,0.86,1);
    rgb(70pt)=(0,0.9,0.967741935483871);
    rgb(75pt)=(0.0806451612903226,1,0.887096774193548);
    rgb(128pt)=(0.935483870967742,1,0.0322580645161291);
    rgb(130pt)=(0.967741935483871,0.962962962962963,0);
    rgb(132pt)=(1,0.925925925925926,0);
    rgb(178pt)=(1,0.0740740740740741,0);
    rgb(182pt)=(0.909090909090909,0,0);
    rgb(200pt)=(0.5,0,0)
},
point meta max=51,
point meta min=0,
]
\addplot graphics [includegraphics cmd=\pgfimage, xmin=0, xmax=224, ymin=0, ymax=224] {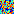};

\end{groupplot}

\end{tikzpicture}

%% file: artwork/surrogate_results/camvid/supplement/surrogate_results_camvid_full_3.tex
\begin{tikzpicture}[every node/.style={font=\small}, spy using outlines={tud6a, line width=0.80mm, dashed, dash pattern=on 1.5pt off 1.5pt, magnification=2, size=0.75cm, connect spies}]

\begin{groupplot}[group style={group size=8 by 3, horizontal sep=1.25pt, vertical sep=1.25pt}]

\nextgroupplot[
width=0.23\columnwidth,
ticks=none,
axis line style={draw=none}, 
xtick=\empty, 
ytick=\empty,
axis equal image,
xmin=0, xmax=224,
ymin=0, ymax=224,
ylabel={H.264},
title={$t_0$},
title style={yshift=-7pt},
]
\addplot graphics [includegraphics cmd=\pgfimage, xmin=0, xmax=224, ymin=0, ymax=224] {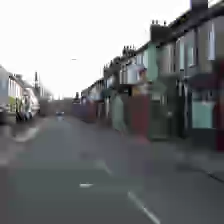};

\nextgroupplot[
width=0.23\columnwidth,
ticks=none,
axis lines=none, 
xtick=\empty, 
ytick=\empty,
axis equal image,
xmin=0, xmax=224,
ymin=0, ymax=224,
title={$t_1$},
title style={yshift=-7pt},
]
\addplot graphics [includegraphics cmd=\pgfimage, xmin=0, xmax=224, ymin=0, ymax=224] {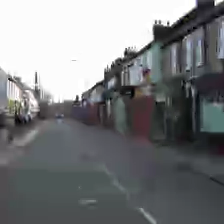};

\nextgroupplot[
width=0.23\columnwidth,
ticks=none,
axis lines=none, 
xtick=\empty, 
ytick=\empty,
axis equal image,
xmin=0, xmax=224,
ymin=0, ymax=224,
title={$t_2$},
title style={yshift=-7pt},
]
\addplot graphics [includegraphics cmd=\pgfimage, xmin=0, xmax=224, ymin=0, ymax=224] {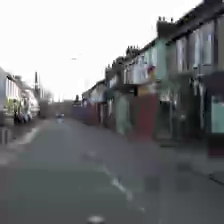};

\nextgroupplot[
width=0.23\columnwidth,
ticks=none,
axis lines=none, 
xtick=\empty, 
ytick=\empty,
axis equal image,
xmin=0, xmax=224,
ymin=0, ymax=224,
title={$t_3$},
title style={yshift=-7pt},
]
\addplot graphics [includegraphics cmd=\pgfimage, xmin=0, xmax=224, ymin=0, ymax=224] {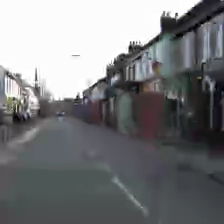};

\nextgroupplot[
width=0.23\columnwidth,
ticks=none,
axis lines=none, 
xtick=\empty, 
ytick=\empty,
axis equal image,
xmin=0, xmax=224,
ymin=0, ymax=224,
title={$t_4$},
title style={yshift=-7pt},
]
\addplot graphics [includegraphics cmd=\pgfimage, xmin=0, xmax=224, ymin=0, ymax=224] {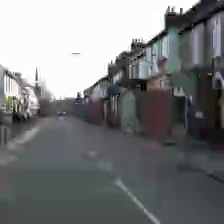};

\nextgroupplot[
width=0.23\columnwidth,
ticks=none,
axis lines=none, 
xtick=\empty, 
ytick=\empty,
axis equal image,
xmin=0, xmax=224,
ymin=0, ymax=224,
title={$t_5$},
title style={yshift=-7pt},
]
\addplot graphics [includegraphics cmd=\pgfimage, xmin=0, xmax=224, ymin=0, ymax=224] {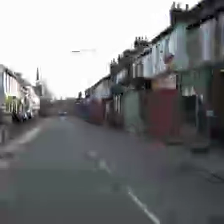};

\nextgroupplot[
width=0.23\columnwidth,
ticks=none,
axis lines=none, 
xtick=\empty, 
ytick=\empty,
axis equal image,
xmin=0, xmax=224,
ymin=0, ymax=224,
title={$t_6$},
title style={yshift=-7pt},
]
\addplot graphics [includegraphics cmd=\pgfimage, xmin=0, xmax=224, ymin=0, ymax=224] {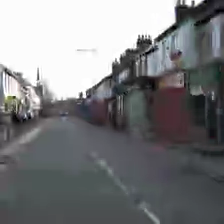};

\nextgroupplot[
width=0.23\columnwidth,
ticks=none,
axis lines=none, 
xtick=\empty, 
ytick=\empty,
axis equal image,
xmin=0, xmax=224,
ymin=0, ymax=224,
title={$t_7$},
title style={yshift=-7pt},
]
\addplot graphics [includegraphics cmd=\pgfimage, xmin=0, xmax=224, ymin=0, ymax=224] {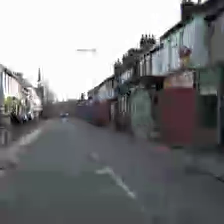};

\nextgroupplot[
width=0.23\columnwidth,
ticks=none,
axis line style={draw=none}, 
xtick=\empty, 
ytick=\empty,
axis equal image,
xmin=0, xmax=224,
ymin=0, ymax=224,
ylabel={Surrogate},
title style={yshift=-7pt},
]
\addplot graphics [includegraphics cmd=\pgfimage, xmin=0, xmax=224, ymin=0, ymax=224] {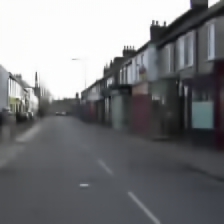};

\nextgroupplot[
width=0.23\columnwidth,
ticks=none,
axis lines=none, 
xtick=\empty, 
ytick=\empty,
axis equal image,
xmin=0, xmax=224,
ymin=0, ymax=224,
title style={yshift=-7pt},
]
\addplot graphics [includegraphics cmd=\pgfimage, xmin=0, xmax=224, ymin=0, ymax=224] {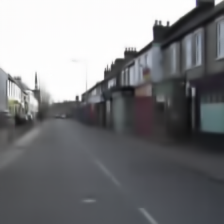};

\nextgroupplot[
width=0.23\columnwidth,
ticks=none,
axis lines=none, 
xtick=\empty, 
ytick=\empty,
axis equal image,
xmin=0, xmax=224,
ymin=0, ymax=224,
title style={yshift=-7pt},
]
\addplot graphics [includegraphics cmd=\pgfimage, xmin=0, xmax=224, ymin=0, ymax=224] {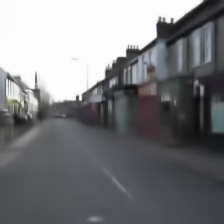};

\nextgroupplot[
width=0.23\columnwidth,
ticks=none,
axis lines=none, 
xtick=\empty, 
ytick=\empty,
axis equal image,
xmin=0, xmax=224,
ymin=0, ymax=224,
title style={yshift=-7pt},
]
\addplot graphics [includegraphics cmd=\pgfimage, xmin=0, xmax=224, ymin=0, ymax=224] {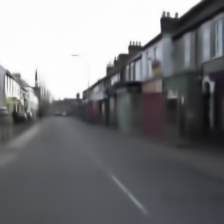};

\nextgroupplot[
width=0.23\columnwidth,
ticks=none,
axis lines=none, 
xtick=\empty, 
ytick=\empty,
axis equal image,
xmin=0, xmax=224,
ymin=0, ymax=224,
title style={yshift=-7pt},
]
\addplot graphics [includegraphics cmd=\pgfimage, xmin=0, xmax=224, ymin=0, ymax=224] {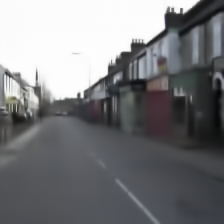};

\nextgroupplot[
width=0.23\columnwidth,
ticks=none,
axis lines=none, 
xtick=\empty, 
ytick=\empty,
axis equal image,
xmin=0, xmax=224,
ymin=0, ymax=224,
title style={yshift=-7pt},
]
\addplot graphics [includegraphics cmd=\pgfimage, xmin=0, xmax=224, ymin=0, ymax=224] {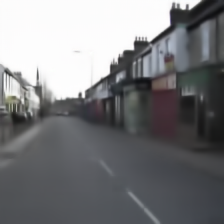};

\nextgroupplot[
width=0.23\columnwidth,
ticks=none,
axis lines=none, 
xtick=\empty, 
ytick=\empty,
axis equal image,
xmin=0, xmax=224,
ymin=0, ymax=224,
title style={yshift=-7pt},
]
\addplot graphics [includegraphics cmd=\pgfimage, xmin=0, xmax=224, ymin=0, ymax=224] {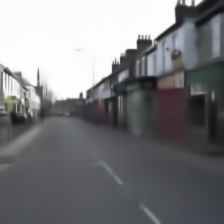};

\nextgroupplot[
width=0.23\columnwidth,
ticks=none,
axis lines=none, 
xtick=\empty, 
ytick=\empty,
axis equal image,
xmin=0, xmax=224,
ymin=0, ymax=224,
title style={yshift=-7pt},
]
\addplot graphics [includegraphics cmd=\pgfimage, xmin=0, xmax=224, ymin=0, ymax=224] {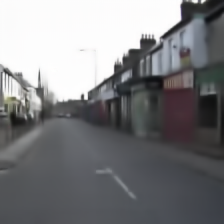};

\nextgroupplot[
width=0.23\columnwidth,
ticks=none,
axis line style={draw=none}, 
xtick=\empty, 
ytick=\empty,
axis equal image,
xmin=0, xmax=224,
ymin=0, ymax=224,
ylabel={$\qp$ map},
title style={yshift=-7pt},
]
\addplot graphics [includegraphics cmd=\pgfimage, xmin=0, xmax=224, ymin=0, ymax=224] {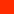};

\nextgroupplot[
width=0.23\columnwidth,
ticks=none,
axis lines=none, 
xtick=\empty, 
ytick=\empty,
axis equal image,
xmin=0, xmax=224,
ymin=0, ymax=224,
title style={yshift=-7pt},
]
\addplot graphics [includegraphics cmd=\pgfimage, xmin=0, xmax=224, ymin=0, ymax=224] {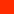};

\nextgroupplot[
width=0.23\columnwidth,
ticks=none,
axis lines=none, 
xtick=\empty, 
ytick=\empty,
axis equal image,
xmin=0, xmax=224,
ymin=0, ymax=224,
title style={yshift=-7pt},
]
\addplot graphics [includegraphics cmd=\pgfimage, xmin=0, xmax=224, ymin=0, ymax=224] {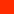};

\nextgroupplot[
width=0.23\columnwidth,
ticks=none,
axis lines=none, 
xtick=\empty, 
ytick=\empty,
axis equal image,
xmin=0, xmax=224,
ymin=0, ymax=224,
title style={yshift=-7pt},
]
\addplot graphics [includegraphics cmd=\pgfimage, xmin=0, xmax=224, ymin=0, ymax=224] {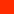};

\nextgroupplot[
width=0.23\columnwidth,
ticks=none,
axis lines=none, 
xtick=\empty, 
ytick=\empty,
axis equal image,
xmin=0, xmax=224,
ymin=0, ymax=224,
title style={yshift=-7pt},
]
\addplot graphics [includegraphics cmd=\pgfimage, xmin=0, xmax=224, ymin=0, ymax=224] {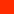};

\nextgroupplot[
width=0.23\columnwidth,
ticks=none,
axis lines=none, 
xtick=\empty, 
ytick=\empty,
axis equal image,
xmin=0, xmax=224,
ymin=0, ymax=224,
title style={yshift=-7pt},
]
\addplot graphics [includegraphics cmd=\pgfimage, xmin=0, xmax=224, ymin=0, ymax=224] {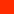};

\nextgroupplot[
width=0.23\columnwidth,
ticks=none,
axis lines=none, 
xtick=\empty, 
ytick=\empty,
axis equal image,
xmin=0, xmax=224,
ymin=0, ymax=224,
title style={yshift=-7pt},
]
\addplot graphics [includegraphics cmd=\pgfimage, xmin=0, xmax=224, ymin=0, ymax=224] {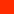};

\nextgroupplot[
width=0.23\columnwidth,
ticks=none,
axis lines=none, 
xtick=\empty, 
ytick=\empty,
axis equal image,
xmin=0, xmax=224,
ymin=0, ymax=224,
title style={yshift=-7pt},
colorbar,
colorbar style={
    ylabel={},
    ytick={5, 25, 45},
    yticklabels={5, 25, 45},
    at={(1.025,1)},
    yticklabel style={xshift=-1.5pt},
},
colormap={mymap}{[1pt]
    rgb(0pt)=(0,0,0.5);
    rgb(22pt)=(0,0,1);
    rgb(25pt)=(0,0,1);
    rgb(68pt)=(0,0.86,1);
    rgb(70pt)=(0,0.9,0.967741935483871);
    rgb(75pt)=(0.0806451612903226,1,0.887096774193548);
    rgb(128pt)=(0.935483870967742,1,0.0322580645161291);
    rgb(130pt)=(0.967741935483871,0.962962962962963,0);
    rgb(132pt)=(1,0.925925925925926,0);
    rgb(178pt)=(1,0.0740740740740741,0);
    rgb(182pt)=(0.909090909090909,0,0);
    rgb(200pt)=(0.5,0,0)
},
point meta max=51,
point meta min=0,
]
\addplot graphics [includegraphics cmd=\pgfimage, xmin=0, xmax=224, ymin=0, ymax=224] {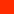};

\end{groupplot}

\end{tikzpicture}

%% file: artwork/surrogate_results/camvid/supplement/surrogate_results_camvid_high_1.tex
\begin{tikzpicture}[every node/.style={font=\small}, spy using outlines={tud6a, line width=0.80mm, dashed, dash pattern=on 1.5pt off 1.5pt, magnification=2, size=0.75cm, connect spies}]

\begin{groupplot}[group style={group size=8 by 3, horizontal sep=1.25pt, vertical sep=1.25pt}]

\nextgroupplot[
width=0.23\columnwidth,
ticks=none,
axis line style={draw=none}, 
xtick=\empty, 
ytick=\empty,
axis equal image,
xmin=0, xmax=224,
ymin=0, ymax=224,
ylabel={H.264},
title={$t_0$},
title style={yshift=-7pt},
]
\addplot graphics [includegraphics cmd=\pgfimage, xmin=0, xmax=224, ymin=0, ymax=224] {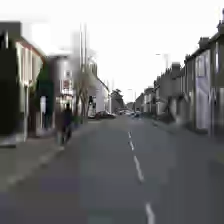};

\nextgroupplot[
width=0.23\columnwidth,
ticks=none,
axis lines=none, 
xtick=\empty, 
ytick=\empty,
axis equal image,
xmin=0, xmax=224,
ymin=0, ymax=224,
title={$t_1$},
title style={yshift=-7pt},
]
\addplot graphics [includegraphics cmd=\pgfimage, xmin=0, xmax=224, ymin=0, ymax=224] {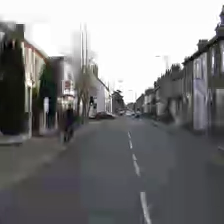};

\nextgroupplot[
width=0.23\columnwidth,
ticks=none,
axis lines=none, 
xtick=\empty, 
ytick=\empty,
axis equal image,
xmin=0, xmax=224,
ymin=0, ymax=224,
title={$t_2$},
title style={yshift=-7pt},
]
\addplot graphics [includegraphics cmd=\pgfimage, xmin=0, xmax=224, ymin=0, ymax=224] {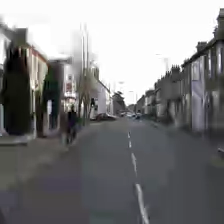};

\nextgroupplot[
width=0.23\columnwidth,
ticks=none,
axis lines=none, 
xtick=\empty, 
ytick=\empty,
axis equal image,
xmin=0, xmax=224,
ymin=0, ymax=224,
title={$t_3$},
title style={yshift=-7pt},
]
\addplot graphics [includegraphics cmd=\pgfimage, xmin=0, xmax=224, ymin=0, ymax=224] {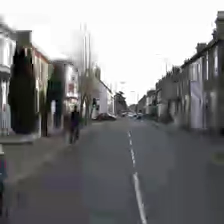};

\nextgroupplot[
width=0.23\columnwidth,
ticks=none,
axis lines=none, 
xtick=\empty, 
ytick=\empty,
axis equal image,
xmin=0, xmax=224,
ymin=0, ymax=224,
title={$t_4$},
title style={yshift=-7pt},
]
\addplot graphics [includegraphics cmd=\pgfimage, xmin=0, xmax=224, ymin=0, ymax=224] {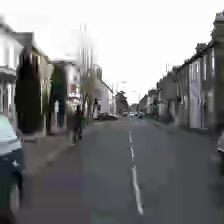};

\nextgroupplot[
width=0.23\columnwidth,
ticks=none,
axis lines=none, 
xtick=\empty, 
ytick=\empty,
axis equal image,
xmin=0, xmax=224,
ymin=0, ymax=224,
title={$t_5$},
title style={yshift=-7pt},
]
\addplot graphics [includegraphics cmd=\pgfimage, xmin=0, xmax=224, ymin=0, ymax=224] {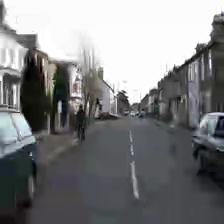};

\nextgroupplot[
width=0.23\columnwidth,
ticks=none,
axis lines=none, 
xtick=\empty, 
ytick=\empty,
axis equal image,
xmin=0, xmax=224,
ymin=0, ymax=224,
title={$t_6$},
title style={yshift=-7pt},
]
\addplot graphics [includegraphics cmd=\pgfimage, xmin=0, xmax=224, ymin=0, ymax=224] {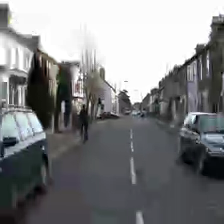};

\nextgroupplot[
width=0.23\columnwidth,
ticks=none,
axis lines=none, 
xtick=\empty, 
ytick=\empty,
axis equal image,
xmin=0, xmax=224,
ymin=0, ymax=224,
title={$t_7$},
title style={yshift=-7pt},
]
\addplot graphics [includegraphics cmd=\pgfimage, xmin=0, xmax=224, ymin=0, ymax=224] {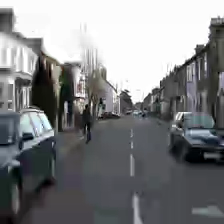};

\nextgroupplot[
width=0.23\columnwidth,
ticks=none,
axis line style={draw=none}, 
xtick=\empty, 
ytick=\empty,
axis equal image,
xmin=0, xmax=224,
ymin=0, ymax=224,
ylabel={Surrogate},
title style={yshift=-7pt},
]
\addplot graphics [includegraphics cmd=\pgfimage, xmin=0, xmax=224, ymin=0, ymax=224] {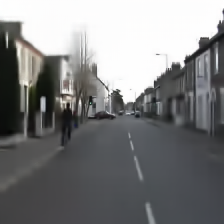};

\nextgroupplot[
width=0.23\columnwidth,
ticks=none,
axis lines=none, 
xtick=\empty, 
ytick=\empty,
axis equal image,
xmin=0, xmax=224,
ymin=0, ymax=224,
title style={yshift=-7pt},
]
\addplot graphics [includegraphics cmd=\pgfimage, xmin=0, xmax=224, ymin=0, ymax=224] {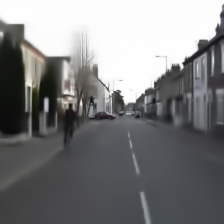};

\nextgroupplot[
width=0.23\columnwidth,
ticks=none,
axis lines=none, 
xtick=\empty, 
ytick=\empty,
axis equal image,
xmin=0, xmax=224,
ymin=0, ymax=224,
title style={yshift=-7pt},
]
\addplot graphics [includegraphics cmd=\pgfimage, xmin=0, xmax=224, ymin=0, ymax=224] {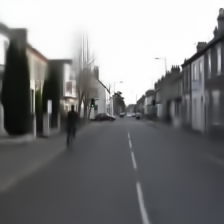};

\nextgroupplot[
width=0.23\columnwidth,
ticks=none,
axis lines=none, 
xtick=\empty, 
ytick=\empty,
axis equal image,
xmin=0, xmax=224,
ymin=0, ymax=224,
title style={yshift=-7pt},
]
\addplot graphics [includegraphics cmd=\pgfimage, xmin=0, xmax=224, ymin=0, ymax=224] {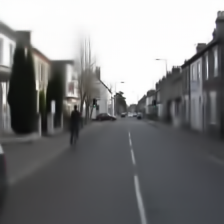};

\nextgroupplot[
width=0.23\columnwidth,
ticks=none,
axis lines=none, 
xtick=\empty, 
ytick=\empty,
axis equal image,
xmin=0, xmax=224,
ymin=0, ymax=224,
title style={yshift=-7pt},
]
\addplot graphics [includegraphics cmd=\pgfimage, xmin=0, xmax=224, ymin=0, ymax=224] {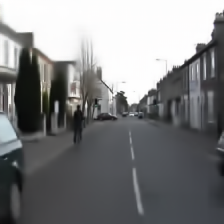};

\nextgroupplot[
width=0.23\columnwidth,
ticks=none,
axis lines=none, 
xtick=\empty, 
ytick=\empty,
axis equal image,
xmin=0, xmax=224,
ymin=0, ymax=224,
title style={yshift=-7pt},
]
\addplot graphics [includegraphics cmd=\pgfimage, xmin=0, xmax=224, ymin=0, ymax=224] {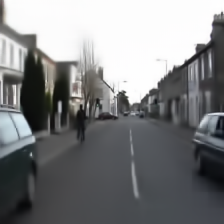};

\nextgroupplot[
width=0.23\columnwidth,
ticks=none,
axis lines=none, 
xtick=\empty, 
ytick=\empty,
axis equal image,
xmin=0, xmax=224,
ymin=0, ymax=224,
title style={yshift=-7pt},
]
\addplot graphics [includegraphics cmd=\pgfimage, xmin=0, xmax=224, ymin=0, ymax=224] {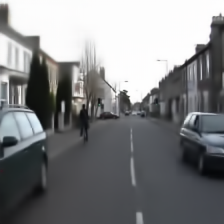};

\nextgroupplot[
width=0.23\columnwidth,
ticks=none,
axis lines=none, 
xtick=\empty, 
ytick=\empty,
axis equal image,
xmin=0, xmax=224,
ymin=0, ymax=224,
title style={yshift=-7pt},
]
\addplot graphics [includegraphics cmd=\pgfimage, xmin=0, xmax=224, ymin=0, ymax=224] {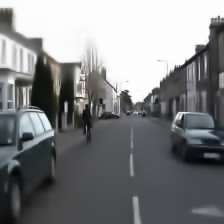};

\nextgroupplot[
width=0.23\columnwidth,
ticks=none,
axis line style={draw=none}, 
xtick=\empty, 
ytick=\empty,
axis equal image,
xmin=0, xmax=224,
ymin=0, ymax=224,
ylabel={$\qp$ map},
title style={yshift=-7pt},
]
\addplot graphics [includegraphics cmd=\pgfimage, xmin=0, xmax=224, ymin=0, ymax=224] {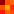};

\nextgroupplot[
width=0.23\columnwidth,
ticks=none,
axis lines=none, 
xtick=\empty, 
ytick=\empty,
axis equal image,
xmin=0, xmax=224,
ymin=0, ymax=224,
title style={yshift=-7pt},
]
\addplot graphics [includegraphics cmd=\pgfimage, xmin=0, xmax=224, ymin=0, ymax=224] {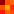};

\nextgroupplot[
width=0.23\columnwidth,
ticks=none,
axis lines=none, 
xtick=\empty, 
ytick=\empty,
axis equal image,
xmin=0, xmax=224,
ymin=0, ymax=224,
title style={yshift=-7pt},
]
\addplot graphics [includegraphics cmd=\pgfimage, xmin=0, xmax=224, ymin=0, ymax=224] {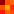};

\nextgroupplot[
width=0.23\columnwidth,
ticks=none,
axis lines=none, 
xtick=\empty, 
ytick=\empty,
axis equal image,
xmin=0, xmax=224,
ymin=0, ymax=224,
title style={yshift=-7pt},
]
\addplot graphics [includegraphics cmd=\pgfimage, xmin=0, xmax=224, ymin=0, ymax=224] {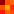};

\nextgroupplot[
width=0.23\columnwidth,
ticks=none,
axis lines=none, 
xtick=\empty, 
ytick=\empty,
axis equal image,
xmin=0, xmax=224,
ymin=0, ymax=224,
title style={yshift=-7pt},
]
\addplot graphics [includegraphics cmd=\pgfimage, xmin=0, xmax=224, ymin=0, ymax=224] {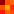};

\nextgroupplot[
width=0.23\columnwidth,
ticks=none,
axis lines=none, 
xtick=\empty, 
ytick=\empty,
axis equal image,
xmin=0, xmax=224,
ymin=0, ymax=224,
title style={yshift=-7pt},
]
\addplot graphics [includegraphics cmd=\pgfimage, xmin=0, xmax=224, ymin=0, ymax=224] {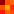};

\nextgroupplot[
width=0.23\columnwidth,
ticks=none,
axis lines=none, 
xtick=\empty, 
ytick=\empty,
axis equal image,
xmin=0, xmax=224,
ymin=0, ymax=224,
title style={yshift=-7pt},
]
\addplot graphics [includegraphics cmd=\pgfimage, xmin=0, xmax=224, ymin=0, ymax=224] {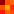};

\nextgroupplot[
width=0.23\columnwidth,
ticks=none,
axis lines=none, 
xtick=\empty, 
ytick=\empty,
axis equal image,
xmin=0, xmax=224,
ymin=0, ymax=224,
title style={yshift=-7pt},
colorbar,
colorbar style={
    ylabel={},
    ytick={5, 25, 45},
    yticklabels={5, 25, 45},
    at={(1.025,1)},
    yticklabel style={xshift=-1.5pt},
},
colormap={mymap}{[1pt]
    rgb(0pt)=(0,0,0.5);
    rgb(22pt)=(0,0,1);
    rgb(25pt)=(0,0,1);
    rgb(68pt)=(0,0.86,1);
    rgb(70pt)=(0,0.9,0.967741935483871);
    rgb(75pt)=(0.0806451612903226,1,0.887096774193548);
    rgb(128pt)=(0.935483870967742,1,0.0322580645161291);
    rgb(130pt)=(0.967741935483871,0.962962962962963,0);
    rgb(132pt)=(1,0.925925925925926,0);
    rgb(178pt)=(1,0.0740740740740741,0);
    rgb(182pt)=(0.909090909090909,0,0);
    rgb(200pt)=(0.5,0,0)
},
point meta max=51,
point meta min=0,
]
\addplot graphics [includegraphics cmd=\pgfimage, xmin=0, xmax=224, ymin=0, ymax=224] {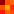};

\end{groupplot}

\end{tikzpicture}

%% file: artwork/surrogate_results/camvid/supplement/surrogate_results_camvid_high_2.tex
\begin{tikzpicture}[every node/.style={font=\small}, spy using outlines={tud6a, line width=0.80mm, dashed, dash pattern=on 1.5pt off 1.5pt, magnification=2, size=0.75cm, connect spies}]

\begin{groupplot}[group style={group size=8 by 3, horizontal sep=1.25pt, vertical sep=1.25pt}]

\nextgroupplot[
width=0.23\columnwidth,
ticks=none,
axis line style={draw=none}, 
xtick=\empty, 
ytick=\empty,
axis equal image,
xmin=0, xmax=224,
ymin=0, ymax=224,
ylabel={H.264},
title={$t_0$},
title style={yshift=-7pt},
]
\addplot graphics [includegraphics cmd=\pgfimage, xmin=0, xmax=224, ymin=0, ymax=224] {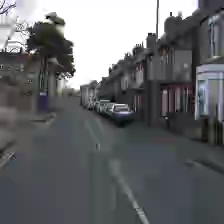};

\nextgroupplot[
width=0.23\columnwidth,
ticks=none,
axis lines=none, 
xtick=\empty, 
ytick=\empty,
axis equal image,
xmin=0, xmax=224,
ymin=0, ymax=224,
title={$t_1$},
title style={yshift=-7pt},
]
\addplot graphics [includegraphics cmd=\pgfimage, xmin=0, xmax=224, ymin=0, ymax=224] {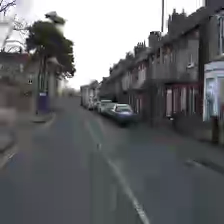};

\nextgroupplot[
width=0.23\columnwidth,
ticks=none,
axis lines=none, 
xtick=\empty, 
ytick=\empty,
axis equal image,
xmin=0, xmax=224,
ymin=0, ymax=224,
title={$t_2$},
title style={yshift=-7pt},
]
\addplot graphics [includegraphics cmd=\pgfimage, xmin=0, xmax=224, ymin=0, ymax=224] {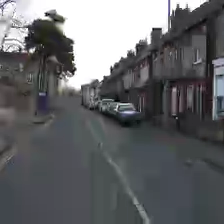};

\nextgroupplot[
width=0.23\columnwidth,
ticks=none,
axis lines=none, 
xtick=\empty, 
ytick=\empty,
axis equal image,
xmin=0, xmax=224,
ymin=0, ymax=224,
title={$t_3$},
title style={yshift=-7pt},
]
\addplot graphics [includegraphics cmd=\pgfimage, xmin=0, xmax=224, ymin=0, ymax=224] {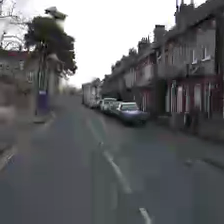};

\nextgroupplot[
width=0.23\columnwidth,
ticks=none,
axis lines=none, 
xtick=\empty, 
ytick=\empty,
axis equal image,
xmin=0, xmax=224,
ymin=0, ymax=224,
title={$t_4$},
title style={yshift=-7pt},
]
\addplot graphics [includegraphics cmd=\pgfimage, xmin=0, xmax=224, ymin=0, ymax=224] {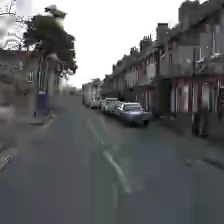};

\nextgroupplot[
width=0.23\columnwidth,
ticks=none,
axis lines=none, 
xtick=\empty, 
ytick=\empty,
axis equal image,
xmin=0, xmax=224,
ymin=0, ymax=224,
title={$t_5$},
title style={yshift=-7pt},
]
\addplot graphics [includegraphics cmd=\pgfimage, xmin=0, xmax=224, ymin=0, ymax=224] {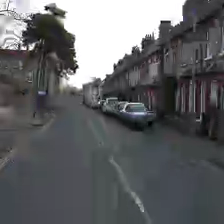};

\nextgroupplot[
width=0.23\columnwidth,
ticks=none,
axis lines=none, 
xtick=\empty, 
ytick=\empty,
axis equal image,
xmin=0, xmax=224,
ymin=0, ymax=224,
title={$t_6$},
title style={yshift=-7pt},
]
\addplot graphics [includegraphics cmd=\pgfimage, xmin=0, xmax=224, ymin=0, ymax=224] {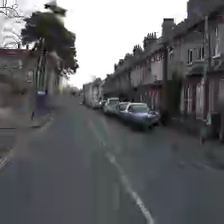};

\nextgroupplot[
width=0.23\columnwidth,
ticks=none,
axis lines=none, 
xtick=\empty, 
ytick=\empty,
axis equal image,
xmin=0, xmax=224,
ymin=0, ymax=224,
title={$t_7$},
title style={yshift=-7pt},
]
\addplot graphics [includegraphics cmd=\pgfimage, xmin=0, xmax=224, ymin=0, ymax=224] {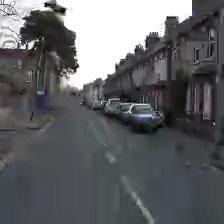};

\nextgroupplot[
width=0.23\columnwidth,
ticks=none,
axis line style={draw=none}, 
xtick=\empty, 
ytick=\empty,
axis equal image,
xmin=0, xmax=224,
ymin=0, ymax=224,
ylabel={Surrogate},
title style={yshift=-7pt},
]
\addplot graphics [includegraphics cmd=\pgfimage, xmin=0, xmax=224, ymin=0, ymax=224] {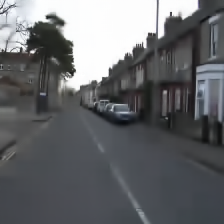};

\nextgroupplot[
width=0.23\columnwidth,
ticks=none,
axis lines=none, 
xtick=\empty, 
ytick=\empty,
axis equal image,
xmin=0, xmax=224,
ymin=0, ymax=224,
title style={yshift=-7pt},
]
\addplot graphics [includegraphics cmd=\pgfimage, xmin=0, xmax=224, ymin=0, ymax=224] {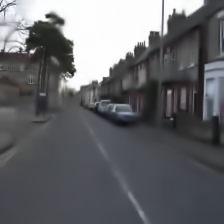};

\nextgroupplot[
width=0.23\columnwidth,
ticks=none,
axis lines=none, 
xtick=\empty, 
ytick=\empty,
axis equal image,
xmin=0, xmax=224,
ymin=0, ymax=224,
title style={yshift=-7pt},
]
\addplot graphics [includegraphics cmd=\pgfimage, xmin=0, xmax=224, ymin=0, ymax=224] {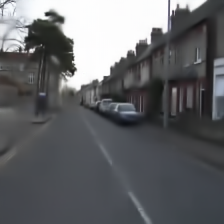};

\nextgroupplot[
width=0.23\columnwidth,
ticks=none,
axis lines=none, 
xtick=\empty, 
ytick=\empty,
axis equal image,
xmin=0, xmax=224,
ymin=0, ymax=224,
title style={yshift=-7pt},
]
\addplot graphics [includegraphics cmd=\pgfimage, xmin=0, xmax=224, ymin=0, ymax=224] {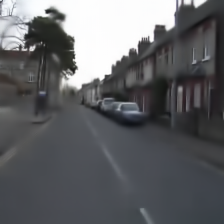};

\nextgroupplot[
width=0.23\columnwidth,
ticks=none,
axis lines=none, 
xtick=\empty, 
ytick=\empty,
axis equal image,
xmin=0, xmax=224,
ymin=0, ymax=224,
title style={yshift=-7pt},
]
\addplot graphics [includegraphics cmd=\pgfimage, xmin=0, xmax=224, ymin=0, ymax=224] {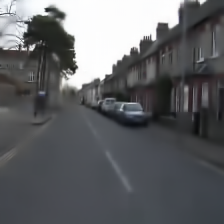};

\nextgroupplot[
width=0.23\columnwidth,
ticks=none,
axis lines=none, 
xtick=\empty, 
ytick=\empty,
axis equal image,
xmin=0, xmax=224,
ymin=0, ymax=224,
title style={yshift=-7pt},
]
\addplot graphics [includegraphics cmd=\pgfimage, xmin=0, xmax=224, ymin=0, ymax=224] {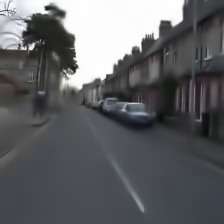};

\nextgroupplot[
width=0.23\columnwidth,
ticks=none,
axis lines=none, 
xtick=\empty, 
ytick=\empty,
axis equal image,
xmin=0, xmax=224,
ymin=0, ymax=224,
title style={yshift=-7pt},
]
\addplot graphics [includegraphics cmd=\pgfimage, xmin=0, xmax=224, ymin=0, ymax=224] {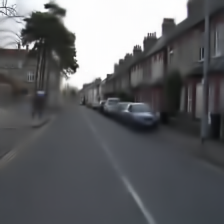};

\nextgroupplot[
width=0.23\columnwidth,
ticks=none,
axis lines=none, 
xtick=\empty, 
ytick=\empty,
axis equal image,
xmin=0, xmax=224,
ymin=0, ymax=224,
title style={yshift=-7pt},
]
\addplot graphics [includegraphics cmd=\pgfimage, xmin=0, xmax=224, ymin=0, ymax=224] {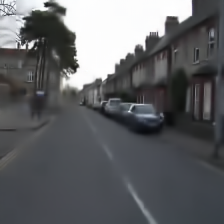};

\nextgroupplot[
width=0.23\columnwidth,
ticks=none,
axis line style={draw=none}, 
xtick=\empty, 
ytick=\empty,
axis equal image,
xmin=0, xmax=224,
ymin=0, ymax=224,
ylabel={$\qp$ map},
title style={yshift=-7pt},
]
\addplot graphics [includegraphics cmd=\pgfimage, xmin=0, xmax=224, ymin=0, ymax=224] {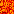};

\nextgroupplot[
width=0.23\columnwidth,
ticks=none,
axis lines=none, 
xtick=\empty, 
ytick=\empty,
axis equal image,
xmin=0, xmax=224,
ymin=0, ymax=224,
title style={yshift=-7pt},
]
\addplot graphics [includegraphics cmd=\pgfimage, xmin=0, xmax=224, ymin=0, ymax=224] {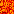};

\nextgroupplot[
width=0.23\columnwidth,
ticks=none,
axis lines=none, 
xtick=\empty, 
ytick=\empty,
axis equal image,
xmin=0, xmax=224,
ymin=0, ymax=224,
title style={yshift=-7pt},
]
\addplot graphics [includegraphics cmd=\pgfimage, xmin=0, xmax=224, ymin=0, ymax=224] {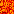};

\nextgroupplot[
width=0.23\columnwidth,
ticks=none,
axis lines=none, 
xtick=\empty, 
ytick=\empty,
axis equal image,
xmin=0, xmax=224,
ymin=0, ymax=224,
title style={yshift=-7pt},
]
\addplot graphics [includegraphics cmd=\pgfimage, xmin=0, xmax=224, ymin=0, ymax=224] {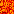};

\nextgroupplot[
width=0.23\columnwidth,
ticks=none,
axis lines=none, 
xtick=\empty, 
ytick=\empty,
axis equal image,
xmin=0, xmax=224,
ymin=0, ymax=224,
title style={yshift=-7pt},
]
\addplot graphics [includegraphics cmd=\pgfimage, xmin=0, xmax=224, ymin=0, ymax=224] {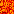};

\nextgroupplot[
width=0.23\columnwidth,
ticks=none,
axis lines=none, 
xtick=\empty, 
ytick=\empty,
axis equal image,
xmin=0, xmax=224,
ymin=0, ymax=224,
title style={yshift=-7pt},
]
\addplot graphics [includegraphics cmd=\pgfimage, xmin=0, xmax=224, ymin=0, ymax=224] {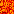};

\nextgroupplot[
width=0.23\columnwidth,
ticks=none,
axis lines=none, 
xtick=\empty, 
ytick=\empty,
axis equal image,
xmin=0, xmax=224,
ymin=0, ymax=224,
title style={yshift=-7pt},
]
\addplot graphics [includegraphics cmd=\pgfimage, xmin=0, xmax=224, ymin=0, ymax=224] {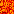};

\nextgroupplot[
width=0.23\columnwidth,
ticks=none,
axis lines=none, 
xtick=\empty, 
ytick=\empty,
axis equal image,
xmin=0, xmax=224,
ymin=0, ymax=224,
title style={yshift=-7pt},
colorbar,
colorbar style={
    ylabel={},
    ytick={5, 25, 45},
    yticklabels={5, 25, 45},
    at={(1.025,1)},
    yticklabel style={xshift=-1.5pt},
},
colormap={mymap}{[1pt]
    rgb(0pt)=(0,0,0.5);
    rgb(22pt)=(0,0,1);
    rgb(25pt)=(0,0,1);
    rgb(68pt)=(0,0.86,1);
    rgb(70pt)=(0,0.9,0.967741935483871);
    rgb(75pt)=(0.0806451612903226,1,0.887096774193548);
    rgb(128pt)=(0.935483870967742,1,0.0322580645161291);
    rgb(130pt)=(0.967741935483871,0.962962962962963,0);
    rgb(132pt)=(1,0.925925925925926,0);
    rgb(178pt)=(1,0.0740740740740741,0);
    rgb(182pt)=(0.909090909090909,0,0);
    rgb(200pt)=(0.5,0,0)
},
point meta max=51,
point meta min=0,
]
\addplot graphics [includegraphics cmd=\pgfimage, xmin=0, xmax=224, ymin=0, ymax=224] {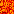};

\end{groupplot}

\end{tikzpicture}

%% file: artwork/surrogate_results/camvid/supplement/surrogate_results_camvid_high_3.tex
\begin{tikzpicture}[every node/.style={font=\small}, spy using outlines={tud6a, line width=0.80mm, dashed, dash pattern=on 1.5pt off 1.5pt, magnification=2, size=0.75cm, connect spies}]

\begin{groupplot}[group style={group size=8 by 3, horizontal sep=1.25pt, vertical sep=1.25pt}]

\nextgroupplot[
width=0.23\columnwidth,
ticks=none,
axis line style={draw=none}, 
xtick=\empty, 
ytick=\empty,
axis equal image,
xmin=0, xmax=224,
ymin=0, ymax=224,
ylabel={H.264},
title={$t_0$},
title style={yshift=-7pt},
]
\addplot graphics [includegraphics cmd=\pgfimage, xmin=0, xmax=224, ymin=0, ymax=224] {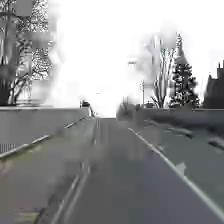};

\nextgroupplot[
width=0.23\columnwidth,
ticks=none,
axis lines=none, 
xtick=\empty, 
ytick=\empty,
axis equal image,
xmin=0, xmax=224,
ymin=0, ymax=224,
title={$t_1$},
title style={yshift=-7pt},
]
\addplot graphics [includegraphics cmd=\pgfimage, xmin=0, xmax=224, ymin=0, ymax=224] {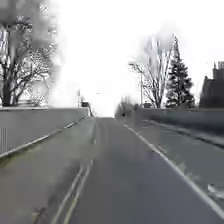};

\nextgroupplot[
width=0.23\columnwidth,
ticks=none,
axis lines=none, 
xtick=\empty, 
ytick=\empty,
axis equal image,
xmin=0, xmax=224,
ymin=0, ymax=224,
title={$t_2$},
title style={yshift=-7pt},
]
\addplot graphics [includegraphics cmd=\pgfimage, xmin=0, xmax=224, ymin=0, ymax=224] {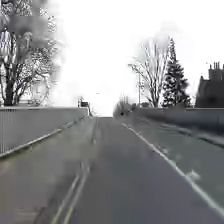};

\nextgroupplot[
width=0.23\columnwidth,
ticks=none,
axis lines=none, 
xtick=\empty, 
ytick=\empty,
axis equal image,
xmin=0, xmax=224,
ymin=0, ymax=224,
title={$t_3$},
title style={yshift=-7pt},
]
\addplot graphics [includegraphics cmd=\pgfimage, xmin=0, xmax=224, ymin=0, ymax=224] {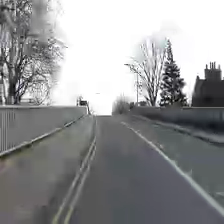};

\nextgroupplot[
width=0.23\columnwidth,
ticks=none,
axis lines=none, 
xtick=\empty, 
ytick=\empty,
axis equal image,
xmin=0, xmax=224,
ymin=0, ymax=224,
title={$t_4$},
title style={yshift=-7pt},
]
\addplot graphics [includegraphics cmd=\pgfimage, xmin=0, xmax=224, ymin=0, ymax=224] {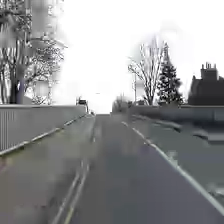};

\nextgroupplot[
width=0.23\columnwidth,
ticks=none,
axis lines=none, 
xtick=\empty, 
ytick=\empty,
axis equal image,
xmin=0, xmax=224,
ymin=0, ymax=224,
title={$t_5$},
title style={yshift=-7pt},
]
\addplot graphics [includegraphics cmd=\pgfimage, xmin=0, xmax=224, ymin=0, ymax=224] {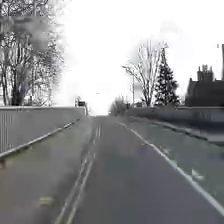};

\nextgroupplot[
width=0.23\columnwidth,
ticks=none,
axis lines=none, 
xtick=\empty, 
ytick=\empty,
axis equal image,
xmin=0, xmax=224,
ymin=0, ymax=224,
title={$t_6$},
title style={yshift=-7pt},
]
\addplot graphics [includegraphics cmd=\pgfimage, xmin=0, xmax=224, ymin=0, ymax=224] {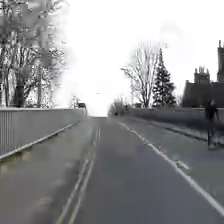};

\nextgroupplot[
width=0.23\columnwidth,
ticks=none,
axis lines=none, 
xtick=\empty, 
ytick=\empty,
axis equal image,
xmin=0, xmax=224,
ymin=0, ymax=224,
title={$t_7$},
title style={yshift=-7pt},
]
\addplot graphics [includegraphics cmd=\pgfimage, xmin=0, xmax=224, ymin=0, ymax=224] {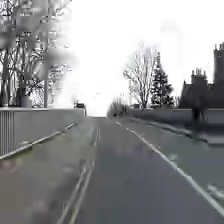};

\nextgroupplot[
width=0.23\columnwidth,
ticks=none,
axis line style={draw=none}, 
xtick=\empty, 
ytick=\empty,
axis equal image,
xmin=0, xmax=224,
ymin=0, ymax=224,
ylabel={Surrogate},
title style={yshift=-7pt},
]
\addplot graphics [includegraphics cmd=\pgfimage, xmin=0, xmax=224, ymin=0, ymax=224] {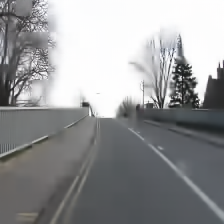};

\nextgroupplot[
width=0.23\columnwidth,
ticks=none,
axis lines=none, 
xtick=\empty, 
ytick=\empty,
axis equal image,
xmin=0, xmax=224,
ymin=0, ymax=224,
title style={yshift=-7pt},
]
\addplot graphics [includegraphics cmd=\pgfimage, xmin=0, xmax=224, ymin=0, ymax=224] {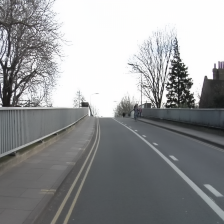};

\nextgroupplot[
width=0.23\columnwidth,
ticks=none,
axis lines=none, 
xtick=\empty, 
ytick=\empty,
axis equal image,
xmin=0, xmax=224,
ymin=0, ymax=224,
title style={yshift=-7pt},
]
\addplot graphics [includegraphics cmd=\pgfimage, xmin=0, xmax=224, ymin=0, ymax=224] {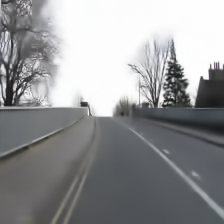};

\nextgroupplot[
width=0.23\columnwidth,
ticks=none,
axis lines=none, 
xtick=\empty, 
ytick=\empty,
axis equal image,
xmin=0, xmax=224,
ymin=0, ymax=224,
title style={yshift=-7pt},
]
\addplot graphics [includegraphics cmd=\pgfimage, xmin=0, xmax=224, ymin=0, ymax=224] {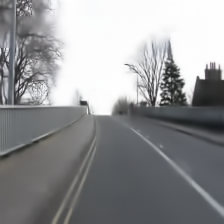};

\nextgroupplot[
width=0.23\columnwidth,
ticks=none,
axis lines=none, 
xtick=\empty, 
ytick=\empty,
axis equal image,
xmin=0, xmax=224,
ymin=0, ymax=224,
title style={yshift=-7pt},
]
\addplot graphics [includegraphics cmd=\pgfimage, xmin=0, xmax=224, ymin=0, ymax=224] {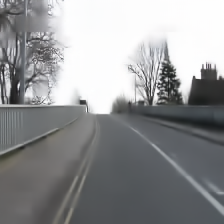};

\nextgroupplot[
width=0.23\columnwidth,
ticks=none,
axis lines=none, 
xtick=\empty, 
ytick=\empty,
axis equal image,
xmin=0, xmax=224,
ymin=0, ymax=224,
title style={yshift=-7pt},
]
\addplot graphics [includegraphics cmd=\pgfimage, xmin=0, xmax=224, ymin=0, ymax=224] {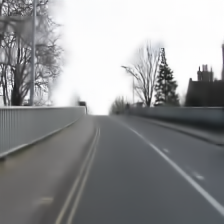};

\nextgroupplot[
width=0.23\columnwidth,
ticks=none,
axis lines=none, 
xtick=\empty, 
ytick=\empty,
axis equal image,
xmin=0, xmax=224,
ymin=0, ymax=224,
title style={yshift=-7pt},
]
\addplot graphics [includegraphics cmd=\pgfimage, xmin=0, xmax=224, ymin=0, ymax=224] {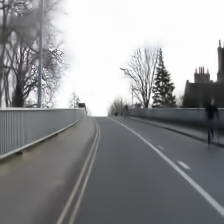};

\nextgroupplot[
width=0.23\columnwidth,
ticks=none,
axis lines=none, 
xtick=\empty, 
ytick=\empty,
axis equal image,
xmin=0, xmax=224,
ymin=0, ymax=224,
title style={yshift=-7pt},
]
\addplot graphics [includegraphics cmd=\pgfimage, xmin=0, xmax=224, ymin=0, ymax=224] {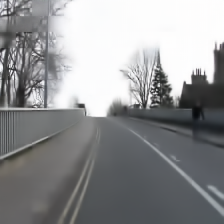};

\nextgroupplot[
width=0.23\columnwidth,
ticks=none,
axis line style={draw=none}, 
xtick=\empty, 
ytick=\empty,
axis equal image,
xmin=0, xmax=224,
ymin=0, ymax=224,
ylabel={$\qp$ map},
title style={yshift=-7pt},
]
\addplot graphics [includegraphics cmd=\pgfimage, xmin=0, xmax=224, ymin=0, ymax=224] {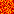};

\nextgroupplot[
width=0.23\columnwidth,
ticks=none,
axis lines=none, 
xtick=\empty, 
ytick=\empty,
axis equal image,
xmin=0, xmax=224,
ymin=0, ymax=224,
title style={yshift=-7pt},
]
\addplot graphics [includegraphics cmd=\pgfimage, xmin=0, xmax=224, ymin=0, ymax=224] {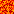};

\nextgroupplot[
width=0.23\columnwidth,
ticks=none,
axis lines=none, 
xtick=\empty, 
ytick=\empty,
axis equal image,
xmin=0, xmax=224,
ymin=0, ymax=224,
title style={yshift=-7pt},
]
\addplot graphics [includegraphics cmd=\pgfimage, xmin=0, xmax=224, ymin=0, ymax=224] {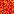};

\nextgroupplot[
width=0.23\columnwidth,
ticks=none,
axis lines=none, 
xtick=\empty, 
ytick=\empty,
axis equal image,
xmin=0, xmax=224,
ymin=0, ymax=224,
title style={yshift=-7pt},
]
\addplot graphics [includegraphics cmd=\pgfimage, xmin=0, xmax=224, ymin=0, ymax=224] {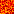};

\nextgroupplot[
width=0.23\columnwidth,
ticks=none,
axis lines=none, 
xtick=\empty, 
ytick=\empty,
axis equal image,
xmin=0, xmax=224,
ymin=0, ymax=224,
title style={yshift=-7pt},
]
\addplot graphics [includegraphics cmd=\pgfimage, xmin=0, xmax=224, ymin=0, ymax=224] {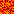};

\nextgroupplot[
width=0.23\columnwidth,
ticks=none,
axis lines=none, 
xtick=\empty, 
ytick=\empty,
axis equal image,
xmin=0, xmax=224,
ymin=0, ymax=224,
title style={yshift=-7pt},
]
\addplot graphics [includegraphics cmd=\pgfimage, xmin=0, xmax=224, ymin=0, ymax=224] {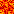};

\nextgroupplot[
width=0.23\columnwidth,
ticks=none,
axis lines=none, 
xtick=\empty, 
ytick=\empty,
axis equal image,
xmin=0, xmax=224,
ymin=0, ymax=224,
title style={yshift=-7pt},
]
\addplot graphics [includegraphics cmd=\pgfimage, xmin=0, xmax=224, ymin=0, ymax=224] {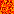};

\nextgroupplot[
width=0.23\columnwidth,
ticks=none,
axis lines=none, 
xtick=\empty, 
ytick=\empty,
axis equal image,
xmin=0, xmax=224,
ymin=0, ymax=224,
title style={yshift=-7pt},
colorbar,
colorbar style={
    ylabel={},
    ytick={5, 25, 45},
    yticklabels={5, 25, 45},
    at={(1.025,1)},
    yticklabel style={xshift=-1.5pt},
},
colormap={mymap}{[1pt]
    rgb(0pt)=(0,0,0.5);
    rgb(22pt)=(0,0,1);
    rgb(25pt)=(0,0,1);
    rgb(68pt)=(0,0.86,1);
    rgb(70pt)=(0,0.9,0.967741935483871);
    rgb(75pt)=(0.0806451612903226,1,0.887096774193548);
    rgb(128pt)=(0.935483870967742,1,0.0322580645161291);
    rgb(130pt)=(0.967741935483871,0.962962962962963,0);
    rgb(132pt)=(1,0.925925925925926,0);
    rgb(178pt)=(1,0.0740740740740741,0);
    rgb(182pt)=(0.909090909090909,0,0);
    rgb(200pt)=(0.5,0,0)
},
point meta max=51,
point meta min=0,
]
\addplot graphics [includegraphics cmd=\pgfimage, xmin=0, xmax=224, ymin=0, ymax=224] {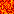};

\end{groupplot}

\end{tikzpicture}

%% file: artwork/surrogate_results/camvid/supplement/surrogate_results_camvid_low_1.tex
\begin{tikzpicture}[every node/.style={font=\small}, spy using outlines={tud6a, line width=0.80mm, dashed, dash pattern=on 1.5pt off 1.5pt, magnification=2, size=0.75cm, connect spies}]

\begin{groupplot}[group style={group size=8 by 3, horizontal sep=1.25pt, vertical sep=1.25pt}]

\nextgroupplot[
width=0.23\columnwidth,
ticks=none,
axis line style={draw=none}, 
xtick=\empty, 
ytick=\empty,
axis equal image,
xmin=0, xmax=224,
ymin=0, ymax=224,
ylabel={H.264},
title={$t_0$},
title style={yshift=-7pt},
]
\addplot graphics [includegraphics cmd=\pgfimage, xmin=0, xmax=224, ymin=0, ymax=224] {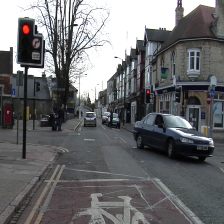};

\nextgroupplot[
width=0.23\columnwidth,
ticks=none,
axis lines=none, 
xtick=\empty, 
ytick=\empty,
axis equal image,
xmin=0, xmax=224,
ymin=0, ymax=224,
title={$t_1$},
title style={yshift=-7pt},
]
\addplot graphics [includegraphics cmd=\pgfimage, xmin=0, xmax=224, ymin=0, ymax=224] {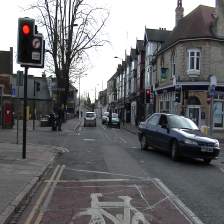};

\nextgroupplot[
width=0.23\columnwidth,
ticks=none,
axis lines=none, 
xtick=\empty, 
ytick=\empty,
axis equal image,
xmin=0, xmax=224,
ymin=0, ymax=224,
title={$t_2$},
title style={yshift=-7pt},
]
\addplot graphics [includegraphics cmd=\pgfimage, xmin=0, xmax=224, ymin=0, ymax=224] {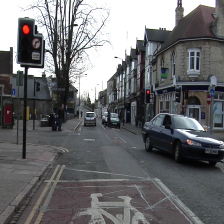};

\nextgroupplot[
width=0.23\columnwidth,
ticks=none,
axis lines=none, 
xtick=\empty, 
ytick=\empty,
axis equal image,
xmin=0, xmax=224,
ymin=0, ymax=224,
title={$t_3$},
title style={yshift=-7pt},
]
\addplot graphics [includegraphics cmd=\pgfimage, xmin=0, xmax=224, ymin=0, ymax=224] {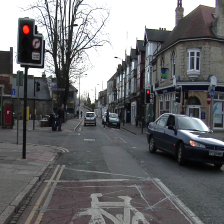};

\nextgroupplot[
width=0.23\columnwidth,
ticks=none,
axis lines=none, 
xtick=\empty, 
ytick=\empty,
axis equal image,
xmin=0, xmax=224,
ymin=0, ymax=224,
title={$t_4$},
title style={yshift=-7pt},
]
\addplot graphics [includegraphics cmd=\pgfimage, xmin=0, xmax=224, ymin=0, ymax=224] {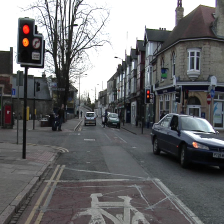};

\nextgroupplot[
width=0.23\columnwidth,
ticks=none,
axis lines=none, 
xtick=\empty, 
ytick=\empty,
axis equal image,
xmin=0, xmax=224,
ymin=0, ymax=224,
title={$t_5$},
title style={yshift=-7pt},
]
\addplot graphics [includegraphics cmd=\pgfimage, xmin=0, xmax=224, ymin=0, ymax=224] {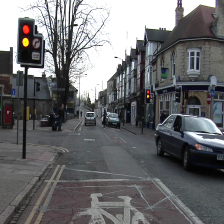};

\nextgroupplot[
width=0.23\columnwidth,
ticks=none,
axis lines=none, 
xtick=\empty, 
ytick=\empty,
axis equal image,
xmin=0, xmax=224,
ymin=0, ymax=224,
title={$t_6$},
title style={yshift=-7pt},
]
\addplot graphics [includegraphics cmd=\pgfimage, xmin=0, xmax=224, ymin=0, ymax=224] {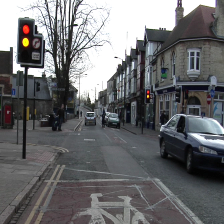};

\nextgroupplot[
width=0.23\columnwidth,
ticks=none,
axis lines=none, 
xtick=\empty, 
ytick=\empty,
axis equal image,
xmin=0, xmax=224,
ymin=0, ymax=224,
title={$t_7$},
title style={yshift=-7pt},
]
\addplot graphics [includegraphics cmd=\pgfimage, xmin=0, xmax=224, ymin=0, ymax=224] {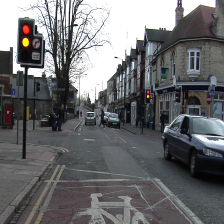};

\nextgroupplot[
width=0.23\columnwidth,
ticks=none,
axis line style={draw=none}, 
xtick=\empty, 
ytick=\empty,
axis equal image,
xmin=0, xmax=224,
ymin=0, ymax=224,
ylabel={Surrogate},
title style={yshift=-7pt},
]
\addplot graphics [includegraphics cmd=\pgfimage, xmin=0, xmax=224, ymin=0, ymax=224] {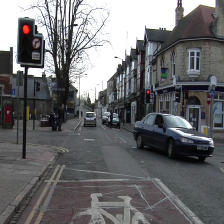};

\nextgroupplot[
width=0.23\columnwidth,
ticks=none,
axis lines=none, 
xtick=\empty, 
ytick=\empty,
axis equal image,
xmin=0, xmax=224,
ymin=0, ymax=224,
title style={yshift=-7pt},
]
\addplot graphics [includegraphics cmd=\pgfimage, xmin=0, xmax=224, ymin=0, ymax=224] {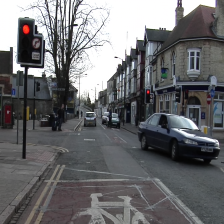};

\nextgroupplot[
width=0.23\columnwidth,
ticks=none,
axis lines=none, 
xtick=\empty, 
ytick=\empty,
axis equal image,
xmin=0, xmax=224,
ymin=0, ymax=224,
title style={yshift=-7pt},
]
\addplot graphics [includegraphics cmd=\pgfimage, xmin=0, xmax=224, ymin=0, ymax=224] {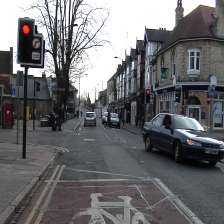};

\nextgroupplot[
width=0.23\columnwidth,
ticks=none,
axis lines=none, 
xtick=\empty, 
ytick=\empty,
axis equal image,
xmin=0, xmax=224,
ymin=0, ymax=224,
title style={yshift=-7pt},
]
\addplot graphics [includegraphics cmd=\pgfimage, xmin=0, xmax=224, ymin=0, ymax=224] {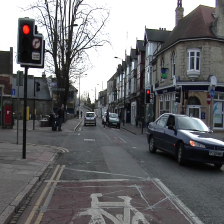};

\nextgroupplot[
width=0.23\columnwidth,
ticks=none,
axis lines=none, 
xtick=\empty, 
ytick=\empty,
axis equal image,
xmin=0, xmax=224,
ymin=0, ymax=224,
title style={yshift=-7pt},
]
\addplot graphics [includegraphics cmd=\pgfimage, xmin=0, xmax=224, ymin=0, ymax=224] {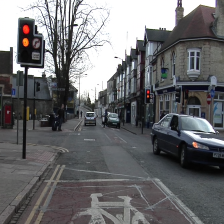};

\nextgroupplot[
width=0.23\columnwidth,
ticks=none,
axis lines=none, 
xtick=\empty, 
ytick=\empty,
axis equal image,
xmin=0, xmax=224,
ymin=0, ymax=224,
title style={yshift=-7pt},
]
\addplot graphics [includegraphics cmd=\pgfimage, xmin=0, xmax=224, ymin=0, ymax=224] {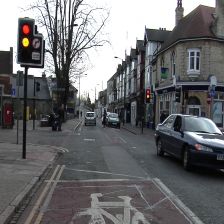};

\nextgroupplot[
width=0.23\columnwidth,
ticks=none,
axis lines=none, 
xtick=\empty, 
ytick=\empty,
axis equal image,
xmin=0, xmax=224,
ymin=0, ymax=224,
title style={yshift=-7pt},
]
\addplot graphics [includegraphics cmd=\pgfimage, xmin=0, xmax=224, ymin=0, ymax=224] {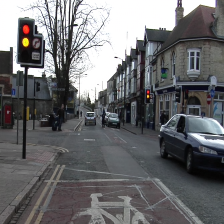};

\nextgroupplot[
width=0.23\columnwidth,
ticks=none,
axis lines=none, 
xtick=\empty, 
ytick=\empty,
axis equal image,
xmin=0, xmax=224,
ymin=0, ymax=224,
title style={yshift=-7pt},
]
\addplot graphics [includegraphics cmd=\pgfimage, xmin=0, xmax=224, ymin=0, ymax=224] {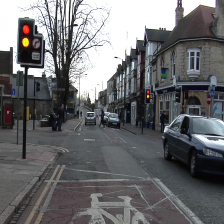};

\nextgroupplot[
width=0.23\columnwidth,
ticks=none,
axis line style={draw=none}, 
xtick=\empty, 
ytick=\empty,
axis equal image,
xmin=0, xmax=224,
ymin=0, ymax=224,
ylabel={$\qp$ map},
title style={yshift=-7pt},
]
\addplot graphics [includegraphics cmd=\pgfimage, xmin=0, xmax=224, ymin=0, ymax=224] {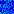};

\nextgroupplot[
width=0.23\columnwidth,
ticks=none,
axis lines=none, 
xtick=\empty, 
ytick=\empty,
axis equal image,
xmin=0, xmax=224,
ymin=0, ymax=224,
title style={yshift=-7pt},
]
\addplot graphics [includegraphics cmd=\pgfimage, xmin=0, xmax=224, ymin=0, ymax=224] {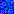};

\nextgroupplot[
width=0.23\columnwidth,
ticks=none,
axis lines=none, 
xtick=\empty, 
ytick=\empty,
axis equal image,
xmin=0, xmax=224,
ymin=0, ymax=224,
title style={yshift=-7pt},
]
\addplot graphics [includegraphics cmd=\pgfimage, xmin=0, xmax=224, ymin=0, ymax=224] {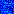};

\nextgroupplot[
width=0.23\columnwidth,
ticks=none,
axis lines=none, 
xtick=\empty, 
ytick=\empty,
axis equal image,
xmin=0, xmax=224,
ymin=0, ymax=224,
title style={yshift=-7pt},
]
\addplot graphics [includegraphics cmd=\pgfimage, xmin=0, xmax=224, ymin=0, ymax=224] {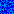};

\nextgroupplot[
width=0.23\columnwidth,
ticks=none,
axis lines=none, 
xtick=\empty, 
ytick=\empty,
axis equal image,
xmin=0, xmax=224,
ymin=0, ymax=224,
title style={yshift=-7pt},
]
\addplot graphics [includegraphics cmd=\pgfimage, xmin=0, xmax=224, ymin=0, ymax=224] {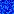};

\nextgroupplot[
width=0.23\columnwidth,
ticks=none,
axis lines=none, 
xtick=\empty, 
ytick=\empty,
axis equal image,
xmin=0, xmax=224,
ymin=0, ymax=224,
title style={yshift=-7pt},
]
\addplot graphics [includegraphics cmd=\pgfimage, xmin=0, xmax=224, ymin=0, ymax=224] {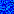};

\nextgroupplot[
width=0.23\columnwidth,
ticks=none,
axis lines=none, 
xtick=\empty, 
ytick=\empty,
axis equal image,
xmin=0, xmax=224,
ymin=0, ymax=224,
title style={yshift=-7pt},
]
\addplot graphics [includegraphics cmd=\pgfimage, xmin=0, xmax=224, ymin=0, ymax=224] {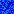};

\nextgroupplot[
width=0.23\columnwidth,
ticks=none,
axis lines=none, 
xtick=\empty, 
ytick=\empty,
axis equal image,
xmin=0, xmax=224,
ymin=0, ymax=224,
title style={yshift=-7pt},
colorbar,
colorbar style={
    ylabel={},
    ytick={5, 25, 45},
    yticklabels={5, 25, 45},
    at={(1.025,1)},
    yticklabel style={xshift=-1.5pt},
},
colormap={mymap}{[1pt]
    rgb(0pt)=(0,0,0.5);
    rgb(22pt)=(0,0,1);
    rgb(25pt)=(0,0,1);
    rgb(68pt)=(0,0.86,1);
    rgb(70pt)=(0,0.9,0.967741935483871);
    rgb(75pt)=(0.0806451612903226,1,0.887096774193548);
    rgb(128pt)=(0.935483870967742,1,0.0322580645161291);
    rgb(130pt)=(0.967741935483871,0.962962962962963,0);
    rgb(132pt)=(1,0.925925925925926,0);
    rgb(178pt)=(1,0.0740740740740741,0);
    rgb(182pt)=(0.909090909090909,0,0);
    rgb(200pt)=(0.5,0,0)
},
point meta max=51,
point meta min=0,
]
\addplot graphics [includegraphics cmd=\pgfimage, xmin=0, xmax=224, ymin=0, ymax=224] {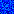};

\end{groupplot}

\end{tikzpicture}

%% file: artwork/surrogate_results/camvid/supplement/surrogate_results_camvid_low_2.tex
\begin{tikzpicture}[every node/.style={font=\small}, spy using outlines={tud6a, line width=0.80mm, dashed, dash pattern=on 1.5pt off 1.5pt, magnification=2, size=0.75cm, connect spies}]

\begin{groupplot}[group style={group size=8 by 3, horizontal sep=1.25pt, vertical sep=1.25pt}]

\nextgroupplot[
width=0.23\columnwidth,
ticks=none,
axis line style={draw=none}, 
xtick=\empty, 
ytick=\empty,
axis equal image,
xmin=0, xmax=224,
ymin=0, ymax=224,
ylabel={H.264},
title={$t_0$},
title style={yshift=-7pt},
]
\addplot graphics [includegraphics cmd=\pgfimage, xmin=0, xmax=224, ymin=0, ymax=224] {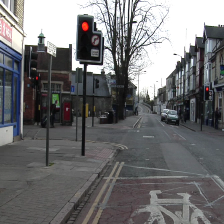};

\nextgroupplot[
width=0.23\columnwidth,
ticks=none,
axis lines=none, 
xtick=\empty, 
ytick=\empty,
axis equal image,
xmin=0, xmax=224,
ymin=0, ymax=224,
title={$t_1$},
title style={yshift=-7pt},
]
\addplot graphics [includegraphics cmd=\pgfimage, xmin=0, xmax=224, ymin=0, ymax=224] {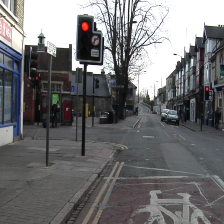};

\nextgroupplot[
width=0.23\columnwidth,
ticks=none,
axis lines=none, 
xtick=\empty, 
ytick=\empty,
axis equal image,
xmin=0, xmax=224,
ymin=0, ymax=224,
title={$t_2$},
title style={yshift=-7pt},
]
\addplot graphics [includegraphics cmd=\pgfimage, xmin=0, xmax=224, ymin=0, ymax=224] {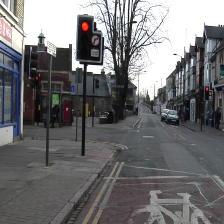};

\nextgroupplot[
width=0.23\columnwidth,
ticks=none,
axis lines=none, 
xtick=\empty, 
ytick=\empty,
axis equal image,
xmin=0, xmax=224,
ymin=0, ymax=224,
title={$t_3$},
title style={yshift=-7pt},
]
\addplot graphics [includegraphics cmd=\pgfimage, xmin=0, xmax=224, ymin=0, ymax=224] {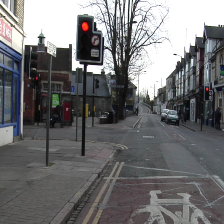};

\nextgroupplot[
width=0.23\columnwidth,
ticks=none,
axis lines=none, 
xtick=\empty, 
ytick=\empty,
axis equal image,
xmin=0, xmax=224,
ymin=0, ymax=224,
title={$t_4$},
title style={yshift=-7pt},
]
\addplot graphics [includegraphics cmd=\pgfimage, xmin=0, xmax=224, ymin=0, ymax=224] {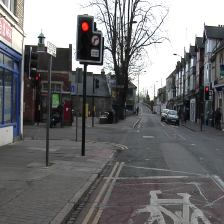};

\nextgroupplot[
width=0.23\columnwidth,
ticks=none,
axis lines=none, 
xtick=\empty, 
ytick=\empty,
axis equal image,
xmin=0, xmax=224,
ymin=0, ymax=224,
title={$t_5$},
title style={yshift=-7pt},
]
\addplot graphics [includegraphics cmd=\pgfimage, xmin=0, xmax=224, ymin=0, ymax=224] {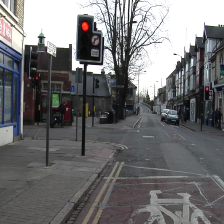};

\nextgroupplot[
width=0.23\columnwidth,
ticks=none,
axis lines=none, 
xtick=\empty, 
ytick=\empty,
axis equal image,
xmin=0, xmax=224,
ymin=0, ymax=224,
title={$t_6$},
title style={yshift=-7pt},
]
\addplot graphics [includegraphics cmd=\pgfimage, xmin=0, xmax=224, ymin=0, ymax=224] {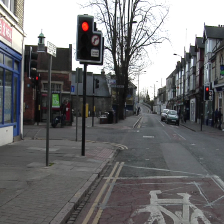};

\nextgroupplot[
width=0.23\columnwidth,
ticks=none,
axis lines=none, 
xtick=\empty, 
ytick=\empty,
axis equal image,
xmin=0, xmax=224,
ymin=0, ymax=224,
title={$t_7$},
title style={yshift=-7pt},
]
\addplot graphics [includegraphics cmd=\pgfimage, xmin=0, xmax=224, ymin=0, ymax=224] {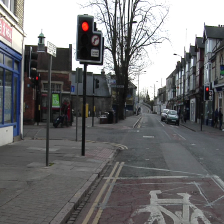};

\nextgroupplot[
width=0.23\columnwidth,
ticks=none,
axis line style={draw=none}, 
xtick=\empty, 
ytick=\empty,
axis equal image,
xmin=0, xmax=224,
ymin=0, ymax=224,
ylabel={Surrogate},
title style={yshift=-7pt},
]
\addplot graphics [includegraphics cmd=\pgfimage, xmin=0, xmax=224, ymin=0, ymax=224] {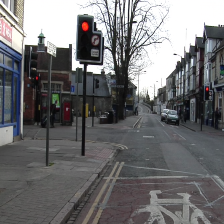};

\nextgroupplot[
width=0.23\columnwidth,
ticks=none,
axis lines=none, 
xtick=\empty, 
ytick=\empty,
axis equal image,
xmin=0, xmax=224,
ymin=0, ymax=224,
title style={yshift=-7pt},
]
\addplot graphics [includegraphics cmd=\pgfimage, xmin=0, xmax=224, ymin=0, ymax=224] {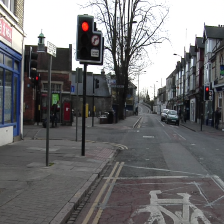};

\nextgroupplot[
width=0.23\columnwidth,
ticks=none,
axis lines=none, 
xtick=\empty, 
ytick=\empty,
axis equal image,
xmin=0, xmax=224,
ymin=0, ymax=224,
title style={yshift=-7pt},
]
\addplot graphics [includegraphics cmd=\pgfimage, xmin=0, xmax=224, ymin=0, ymax=224] {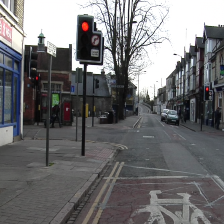};

\nextgroupplot[
width=0.23\columnwidth,
ticks=none,
axis lines=none, 
xtick=\empty, 
ytick=\empty,
axis equal image,
xmin=0, xmax=224,
ymin=0, ymax=224,
title style={yshift=-7pt},
]
\addplot graphics [includegraphics cmd=\pgfimage, xmin=0, xmax=224, ymin=0, ymax=224] {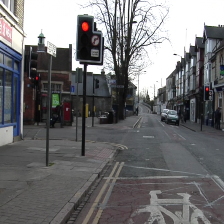};

\nextgroupplot[
width=0.23\columnwidth,
ticks=none,
axis lines=none, 
xtick=\empty, 
ytick=\empty,
axis equal image,
xmin=0, xmax=224,
ymin=0, ymax=224,
title style={yshift=-7pt},
]
\addplot graphics [includegraphics cmd=\pgfimage, xmin=0, xmax=224, ymin=0, ymax=224] {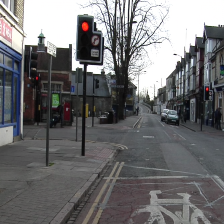};

\nextgroupplot[
width=0.23\columnwidth,
ticks=none,
axis lines=none, 
xtick=\empty, 
ytick=\empty,
axis equal image,
xmin=0, xmax=224,
ymin=0, ymax=224,
title style={yshift=-7pt},
]
\addplot graphics [includegraphics cmd=\pgfimage, xmin=0, xmax=224, ymin=0, ymax=224] {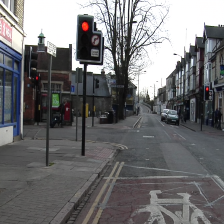};

\nextgroupplot[
width=0.23\columnwidth,
ticks=none,
axis lines=none, 
xtick=\empty, 
ytick=\empty,
axis equal image,
xmin=0, xmax=224,
ymin=0, ymax=224,
title style={yshift=-7pt},
]
\addplot graphics [includegraphics cmd=\pgfimage, xmin=0, xmax=224, ymin=0, ymax=224] {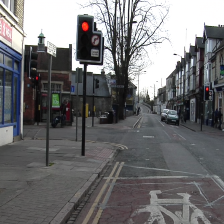};

\nextgroupplot[
width=0.23\columnwidth,
ticks=none,
axis lines=none, 
xtick=\empty, 
ytick=\empty,
axis equal image,
xmin=0, xmax=224,
ymin=0, ymax=224,
title style={yshift=-7pt},
]
\addplot graphics [includegraphics cmd=\pgfimage, xmin=0, xmax=224, ymin=0, ymax=224] {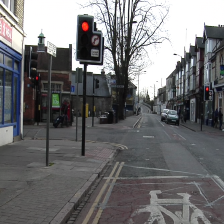};

\nextgroupplot[
width=0.23\columnwidth,
ticks=none,
axis line style={draw=none}, 
xtick=\empty, 
ytick=\empty,
axis equal image,
xmin=0, xmax=224,
ymin=0, ymax=224,
ylabel={$\qp$ map},
title style={yshift=-7pt},
]
\addplot graphics [includegraphics cmd=\pgfimage, xmin=0, xmax=224, ymin=0, ymax=224] {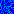};

\nextgroupplot[
width=0.23\columnwidth,
ticks=none,
axis lines=none, 
xtick=\empty, 
ytick=\empty,
axis equal image,
xmin=0, xmax=224,
ymin=0, ymax=224,
title style={yshift=-7pt},
]
\addplot graphics [includegraphics cmd=\pgfimage, xmin=0, xmax=224, ymin=0, ymax=224] {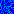};

\nextgroupplot[
width=0.23\columnwidth,
ticks=none,
axis lines=none, 
xtick=\empty, 
ytick=\empty,
axis equal image,
xmin=0, xmax=224,
ymin=0, ymax=224,
title style={yshift=-7pt},
]
\addplot graphics [includegraphics cmd=\pgfimage, xmin=0, xmax=224, ymin=0, ymax=224] {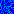};

\nextgroupplot[
width=0.23\columnwidth,
ticks=none,
axis lines=none, 
xtick=\empty, 
ytick=\empty,
axis equal image,
xmin=0, xmax=224,
ymin=0, ymax=224,
title style={yshift=-7pt},
]
\addplot graphics [includegraphics cmd=\pgfimage, xmin=0, xmax=224, ymin=0, ymax=224] {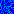};

\nextgroupplot[
width=0.23\columnwidth,
ticks=none,
axis lines=none, 
xtick=\empty, 
ytick=\empty,
axis equal image,
xmin=0, xmax=224,
ymin=0, ymax=224,
title style={yshift=-7pt},
]
\addplot graphics [includegraphics cmd=\pgfimage, xmin=0, xmax=224, ymin=0, ymax=224] {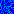};

\nextgroupplot[
width=0.23\columnwidth,
ticks=none,
axis lines=none, 
xtick=\empty, 
ytick=\empty,
axis equal image,
xmin=0, xmax=224,
ymin=0, ymax=224,
title style={yshift=-7pt},
]
\addplot graphics [includegraphics cmd=\pgfimage, xmin=0, xmax=224, ymin=0, ymax=224] {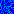};

\nextgroupplot[
width=0.23\columnwidth,
ticks=none,
axis lines=none, 
xtick=\empty, 
ytick=\empty,
axis equal image,
xmin=0, xmax=224,
ymin=0, ymax=224,
title style={yshift=-7pt},
]
\addplot graphics [includegraphics cmd=\pgfimage, xmin=0, xmax=224, ymin=0, ymax=224] {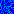};

\nextgroupplot[
width=0.23\columnwidth,
ticks=none,
axis lines=none, 
xtick=\empty, 
ytick=\empty,
axis equal image,
xmin=0, xmax=224,
ymin=0, ymax=224,
title style={yshift=-7pt},
colorbar,
colorbar style={
    ylabel={},
    ytick={5, 25, 45},
    yticklabels={5, 25, 45},
    at={(1.025,1)},
    yticklabel style={xshift=-1.5pt},
},
colormap={mymap}{[1pt]
    rgb(0pt)=(0,0,0.5);
    rgb(22pt)=(0,0,1);
    rgb(25pt)=(0,0,1);
    rgb(68pt)=(0,0.86,1);
    rgb(70pt)=(0,0.9,0.967741935483871);
    rgb(75pt)=(0.0806451612903226,1,0.887096774193548);
    rgb(128pt)=(0.935483870967742,1,0.0322580645161291);
    rgb(130pt)=(0.967741935483871,0.962962962962963,0);
    rgb(132pt)=(1,0.925925925925926,0);
    rgb(178pt)=(1,0.0740740740740741,0);
    rgb(182pt)=(0.909090909090909,0,0);
    rgb(200pt)=(0.5,0,0)
},
point meta max=51,
point meta min=0,
]
\addplot graphics [includegraphics cmd=\pgfimage, xmin=0, xmax=224, ymin=0, ymax=224] {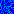};

\end{groupplot}

\end{tikzpicture}

%% file: artwork/surrogate_results/camvid/supplement/surrogate_results_camvid_low_3.tex
\begin{tikzpicture}[every node/.style={font=\small}, spy using outlines={tud6a, line width=0.80mm, dashed, dash pattern=on 1.5pt off 1.5pt, magnification=2, size=0.75cm, connect spies}]

\begin{groupplot}[group style={group size=8 by 3, horizontal sep=1.25pt, vertical sep=1.25pt}]

\nextgroupplot[
width=0.23\columnwidth,
ticks=none,
axis line style={draw=none}, 
xtick=\empty, 
ytick=\empty,
axis equal image,
xmin=0, xmax=224,
ymin=0, ymax=224,
ylabel={H.264},
title={$t_0$},
title style={yshift=-7pt},
]
\addplot graphics [includegraphics cmd=\pgfimage, xmin=0, xmax=224, ymin=0, ymax=224] {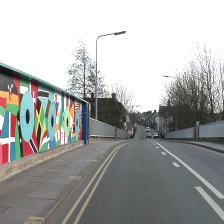};

\nextgroupplot[
width=0.23\columnwidth,
ticks=none,
axis lines=none, 
xtick=\empty, 
ytick=\empty,
axis equal image,
xmin=0, xmax=224,
ymin=0, ymax=224,
title={$t_1$},
title style={yshift=-7pt},
]
\addplot graphics [includegraphics cmd=\pgfimage, xmin=0, xmax=224, ymin=0, ymax=224] {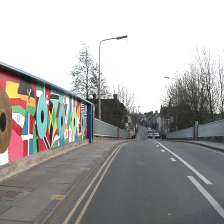};

\nextgroupplot[
width=0.23\columnwidth,
ticks=none,
axis lines=none, 
xtick=\empty, 
ytick=\empty,
axis equal image,
xmin=0, xmax=224,
ymin=0, ymax=224,
title={$t_2$},
title style={yshift=-7pt},
]
\addplot graphics [includegraphics cmd=\pgfimage, xmin=0, xmax=224, ymin=0, ymax=224] {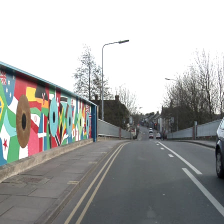};

\nextgroupplot[
width=0.23\columnwidth,
ticks=none,
axis lines=none, 
xtick=\empty, 
ytick=\empty,
axis equal image,
xmin=0, xmax=224,
ymin=0, ymax=224,
title={$t_3$},
title style={yshift=-7pt},
]
\addplot graphics [includegraphics cmd=\pgfimage, xmin=0, xmax=224, ymin=0, ymax=224] {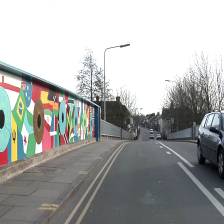};

\nextgroupplot[
width=0.23\columnwidth,
ticks=none,
axis lines=none, 
xtick=\empty, 
ytick=\empty,
axis equal image,
xmin=0, xmax=224,
ymin=0, ymax=224,
title={$t_4$},
title style={yshift=-7pt},
]
\addplot graphics [includegraphics cmd=\pgfimage, xmin=0, xmax=224, ymin=0, ymax=224] {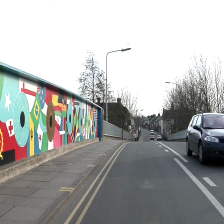};

\nextgroupplot[
width=0.23\columnwidth,
ticks=none,
axis lines=none, 
xtick=\empty, 
ytick=\empty,
axis equal image,
xmin=0, xmax=224,
ymin=0, ymax=224,
title={$t_5$},
title style={yshift=-7pt},
]
\addplot graphics [includegraphics cmd=\pgfimage, xmin=0, xmax=224, ymin=0, ymax=224] {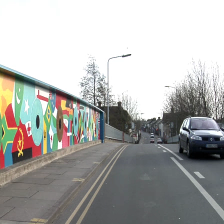};

\nextgroupplot[
width=0.23\columnwidth,
ticks=none,
axis lines=none, 
xtick=\empty, 
ytick=\empty,
axis equal image,
xmin=0, xmax=224,
ymin=0, ymax=224,
title={$t_6$},
title style={yshift=-7pt},
]
\addplot graphics [includegraphics cmd=\pgfimage, xmin=0, xmax=224, ymin=0, ymax=224] {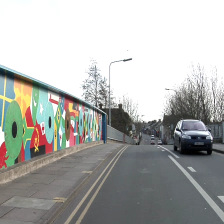};

\nextgroupplot[
width=0.23\columnwidth,
ticks=none,
axis lines=none, 
xtick=\empty, 
ytick=\empty,
axis equal image,
xmin=0, xmax=224,
ymin=0, ymax=224,
title={$t_7$},
title style={yshift=-7pt},
]
\addplot graphics [includegraphics cmd=\pgfimage, xmin=0, xmax=224, ymin=0, ymax=224] {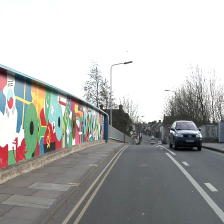};

\nextgroupplot[
width=0.23\columnwidth,
ticks=none,
axis line style={draw=none}, 
xtick=\empty, 
ytick=\empty,
axis equal image,
xmin=0, xmax=224,
ymin=0, ymax=224,
ylabel={Surrogate},
title style={yshift=-7pt},
]
\addplot graphics [includegraphics cmd=\pgfimage, xmin=0, xmax=224, ymin=0, ymax=224] {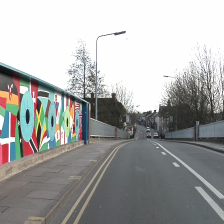};

\nextgroupplot[
width=0.23\columnwidth,
ticks=none,
axis lines=none, 
xtick=\empty, 
ytick=\empty,
axis equal image,
xmin=0, xmax=224,
ymin=0, ymax=224,
title style={yshift=-7pt},
]
\addplot graphics [includegraphics cmd=\pgfimage, xmin=0, xmax=224, ymin=0, ymax=224] {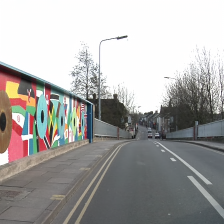};

\nextgroupplot[
width=0.23\columnwidth,
ticks=none,
axis lines=none, 
xtick=\empty, 
ytick=\empty,
axis equal image,
xmin=0, xmax=224,
ymin=0, ymax=224,
title style={yshift=-7pt},
]
\addplot graphics [includegraphics cmd=\pgfimage, xmin=0, xmax=224, ymin=0, ymax=224] {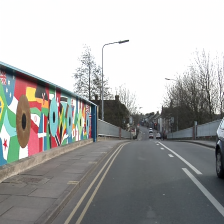};

\nextgroupplot[
width=0.23\columnwidth,
ticks=none,
axis lines=none, 
xtick=\empty, 
ytick=\empty,
axis equal image,
xmin=0, xmax=224,
ymin=0, ymax=224,
title style={yshift=-7pt},
]
\addplot graphics [includegraphics cmd=\pgfimage, xmin=0, xmax=224, ymin=0, ymax=224] {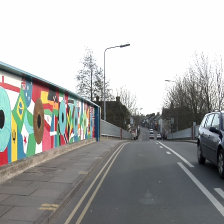};

\nextgroupplot[
width=0.23\columnwidth,
ticks=none,
axis lines=none, 
xtick=\empty, 
ytick=\empty,
axis equal image,
xmin=0, xmax=224,
ymin=0, ymax=224,
title style={yshift=-7pt},
]
\addplot graphics [includegraphics cmd=\pgfimage, xmin=0, xmax=224, ymin=0, ymax=224] {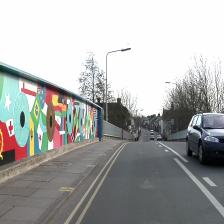};

\nextgroupplot[
width=0.23\columnwidth,
ticks=none,
axis lines=none, 
xtick=\empty, 
ytick=\empty,
axis equal image,
xmin=0, xmax=224,
ymin=0, ymax=224,
title style={yshift=-7pt},
]
\addplot graphics [includegraphics cmd=\pgfimage, xmin=0, xmax=224, ymin=0, ymax=224] {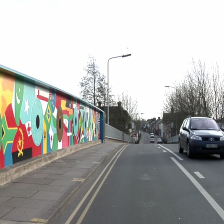};

\nextgroupplot[
width=0.23\columnwidth,
ticks=none,
axis lines=none, 
xtick=\empty, 
ytick=\empty,
axis equal image,
xmin=0, xmax=224,
ymin=0, ymax=224,
title style={yshift=-7pt},
]
\addplot graphics [includegraphics cmd=\pgfimage, xmin=0, xmax=224, ymin=0, ymax=224] {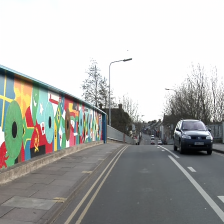};

\nextgroupplot[
width=0.23\columnwidth,
ticks=none,
axis lines=none, 
xtick=\empty, 
ytick=\empty,
axis equal image,
xmin=0, xmax=224,
ymin=0, ymax=224,
title style={yshift=-7pt},
]
\addplot graphics [includegraphics cmd=\pgfimage, xmin=0, xmax=224, ymin=0, ymax=224] {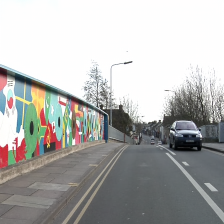};

\nextgroupplot[
width=0.23\columnwidth,
ticks=none,
axis line style={draw=none}, 
xtick=\empty, 
ytick=\empty,
axis equal image,
xmin=0, xmax=224,
ymin=0, ymax=224,
ylabel={$\qp$ map},
title style={yshift=-7pt},
]
\addplot graphics [includegraphics cmd=\pgfimage, xmin=0, xmax=224, ymin=0, ymax=224] {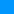};

\nextgroupplot[
width=0.23\columnwidth,
ticks=none,
axis lines=none, 
xtick=\empty, 
ytick=\empty,
axis equal image,
xmin=0, xmax=224,
ymin=0, ymax=224,
title style={yshift=-7pt},
]
\addplot graphics [includegraphics cmd=\pgfimage, xmin=0, xmax=224, ymin=0, ymax=224] {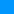};

\nextgroupplot[
width=0.23\columnwidth,
ticks=none,
axis lines=none, 
xtick=\empty, 
ytick=\empty,
axis equal image,
xmin=0, xmax=224,
ymin=0, ymax=224,
title style={yshift=-7pt},
]
\addplot graphics [includegraphics cmd=\pgfimage, xmin=0, xmax=224, ymin=0, ymax=224] {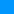};

\nextgroupplot[
width=0.23\columnwidth,
ticks=none,
axis lines=none, 
xtick=\empty, 
ytick=\empty,
axis equal image,
xmin=0, xmax=224,
ymin=0, ymax=224,
title style={yshift=-7pt},
]
\addplot graphics [includegraphics cmd=\pgfimage, xmin=0, xmax=224, ymin=0, ymax=224] {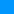};

\nextgroupplot[
width=0.23\columnwidth,
ticks=none,
axis lines=none, 
xtick=\empty, 
ytick=\empty,
axis equal image,
xmin=0, xmax=224,
ymin=0, ymax=224,
title style={yshift=-7pt},
]
\addplot graphics [includegraphics cmd=\pgfimage, xmin=0, xmax=224, ymin=0, ymax=224] {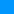};

\nextgroupplot[
width=0.23\columnwidth,
ticks=none,
axis lines=none, 
xtick=\empty, 
ytick=\empty,
axis equal image,
xmin=0, xmax=224,
ymin=0, ymax=224,
title style={yshift=-7pt},
]
\addplot graphics [includegraphics cmd=\pgfimage, xmin=0, xmax=224, ymin=0, ymax=224] {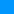};

\nextgroupplot[
width=0.23\columnwidth,
ticks=none,
axis lines=none, 
xtick=\empty, 
ytick=\empty,
axis equal image,
xmin=0, xmax=224,
ymin=0, ymax=224,
title style={yshift=-7pt},
]
\addplot graphics [includegraphics cmd=\pgfimage, xmin=0, xmax=224, ymin=0, ymax=224] {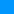};

\nextgroupplot[
width=0.23\columnwidth,
ticks=none,
axis lines=none, 
xtick=\empty, 
ytick=\empty,
axis equal image,
xmin=0, xmax=224,
ymin=0, ymax=224,
title style={yshift=-7pt},
colorbar,
colorbar style={
    ylabel={},
    ytick={5, 25, 45},
    yticklabels={5, 25, 45},
    at={(1.025,1)},
    yticklabel style={xshift=-1.5pt},
},
colormap={mymap}{[1pt]
    rgb(0pt)=(0,0,0.5);
    rgb(22pt)=(0,0,1);
    rgb(25pt)=(0,0,1);
    rgb(68pt)=(0,0.86,1);
    rgb(70pt)=(0,0.9,0.967741935483871);
    rgb(75pt)=(0.0806451612903226,1,0.887096774193548);
    rgb(128pt)=(0.935483870967742,1,0.0322580645161291);
    rgb(130pt)=(0.967741935483871,0.962962962962963,0);
    rgb(132pt)=(1,0.925925925925926,0);
    rgb(178pt)=(1,0.0740740740740741,0);
    rgb(182pt)=(0.909090909090909,0,0);
    rgb(200pt)=(0.5,0,0)
},
point meta max=51,
point meta min=0,
]
\addplot graphics [includegraphics cmd=\pgfimage, xmin=0, xmax=224, ymin=0, ymax=224] {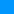};

\end{groupplot}

\end{tikzpicture}

%% file: artwork/surrogate_results/cityscapes/supplement/surrogate_results_cityscapes_full_1.tex
\begin{tikzpicture}[every node/.style={font=\small}, spy using outlines={tud6a, line width=0.80mm, dashed, dash pattern=on 1.5pt off 1.5pt, magnification=2, size=0.75cm, connect spies}]

\begin{groupplot}[group style={group size=8 by 3, horizontal sep=1.25pt, vertical sep=1.25pt}]

\nextgroupplot[
width=0.23\columnwidth,
ticks=none,
axis line style={draw=none}, 
xtick=\empty, 
ytick=\empty,
axis equal image,
xmin=0, xmax=224,
ymin=0, ymax=224,
ylabel={H.264},
title={$t_0$},
title style={yshift=-7pt},
]
\addplot graphics [includegraphics cmd=\pgfimage, xmin=0, xmax=224, ymin=0, ymax=224] {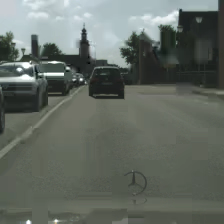};

\nextgroupplot[
width=0.23\columnwidth,
ticks=none,
axis lines=none, 
xtick=\empty, 
ytick=\empty,
axis equal image,
xmin=0, xmax=224,
ymin=0, ymax=224,
title={$t_1$},
title style={yshift=-7pt},
]
\addplot graphics [includegraphics cmd=\pgfimage, xmin=0, xmax=224, ymin=0, ymax=224] {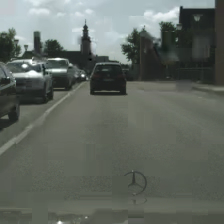};

\nextgroupplot[
width=0.23\columnwidth,
ticks=none,
axis lines=none, 
xtick=\empty, 
ytick=\empty,
axis equal image,
xmin=0, xmax=224,
ymin=0, ymax=224,
title={$t_2$},
title style={yshift=-7pt},
]
\addplot graphics [includegraphics cmd=\pgfimage, xmin=0, xmax=224, ymin=0, ymax=224] {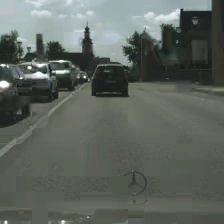};

\nextgroupplot[
width=0.23\columnwidth,
ticks=none,
axis lines=none, 
xtick=\empty, 
ytick=\empty,
axis equal image,
xmin=0, xmax=224,
ymin=0, ymax=224,
title={$t_3$},
title style={yshift=-7pt},
]
\addplot graphics [includegraphics cmd=\pgfimage, xmin=0, xmax=224, ymin=0, ymax=224] {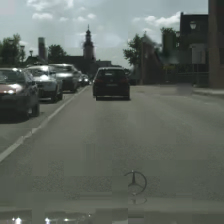};

\nextgroupplot[
width=0.23\columnwidth,
ticks=none,
axis lines=none, 
xtick=\empty, 
ytick=\empty,
axis equal image,
xmin=0, xmax=224,
ymin=0, ymax=224,
title={$t_4$},
title style={yshift=-7pt},
]
\addplot graphics [includegraphics cmd=\pgfimage, xmin=0, xmax=224, ymin=0, ymax=224] {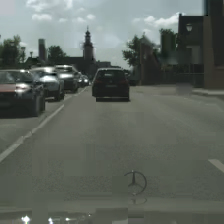};

\nextgroupplot[
width=0.23\columnwidth,
ticks=none,
axis lines=none, 
xtick=\empty, 
ytick=\empty,
axis equal image,
xmin=0, xmax=224,
ymin=0, ymax=224,
title={$t_5$},
title style={yshift=-7pt},
]
\addplot graphics [includegraphics cmd=\pgfimage, xmin=0, xmax=224, ymin=0, ymax=224] {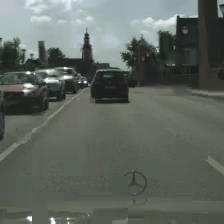};

\nextgroupplot[
width=0.23\columnwidth,
ticks=none,
axis lines=none, 
xtick=\empty, 
ytick=\empty,
axis equal image,
xmin=0, xmax=224,
ymin=0, ymax=224,
title={$t_6$},
title style={yshift=-7pt},
]
\addplot graphics [includegraphics cmd=\pgfimage, xmin=0, xmax=224, ymin=0, ymax=224] {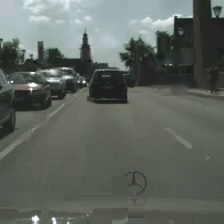};

\nextgroupplot[
width=0.23\columnwidth,
ticks=none,
axis lines=none, 
xtick=\empty, 
ytick=\empty,
axis equal image,
xmin=0, xmax=224,
ymin=0, ymax=224,
title={$t_7$},
title style={yshift=-7pt},
]
\addplot graphics [includegraphics cmd=\pgfimage, xmin=0, xmax=224, ymin=0, ymax=224] {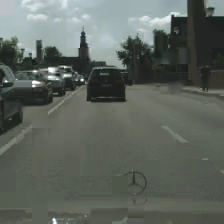};

\nextgroupplot[
width=0.23\columnwidth,
ticks=none,
axis line style={draw=none}, 
xtick=\empty, 
ytick=\empty,
axis equal image,
xmin=0, xmax=224,
ymin=0, ymax=224,
ylabel={Surrogate},
title style={yshift=-7pt},
]
\addplot graphics [includegraphics cmd=\pgfimage, xmin=0, xmax=224, ymin=0, ymax=224] {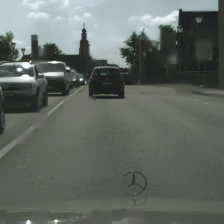};

\nextgroupplot[
width=0.23\columnwidth,
ticks=none,
axis lines=none, 
xtick=\empty, 
ytick=\empty,
axis equal image,
xmin=0, xmax=224,
ymin=0, ymax=224,
title style={yshift=-7pt},
]
\addplot graphics [includegraphics cmd=\pgfimage, xmin=0, xmax=224, ymin=0, ymax=224] {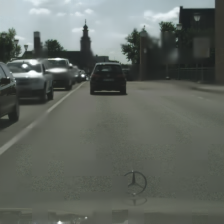};

\nextgroupplot[
width=0.23\columnwidth,
ticks=none,
axis lines=none, 
xtick=\empty, 
ytick=\empty,
axis equal image,
xmin=0, xmax=224,
ymin=0, ymax=224,
title style={yshift=-7pt},
]
\addplot graphics [includegraphics cmd=\pgfimage, xmin=0, xmax=224, ymin=0, ymax=224] {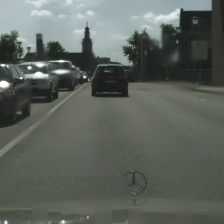};

\nextgroupplot[
width=0.23\columnwidth,
ticks=none,
axis lines=none, 
xtick=\empty, 
ytick=\empty,
axis equal image,
xmin=0, xmax=224,
ymin=0, ymax=224,
title style={yshift=-7pt},
]
\addplot graphics [includegraphics cmd=\pgfimage, xmin=0, xmax=224, ymin=0, ymax=224] {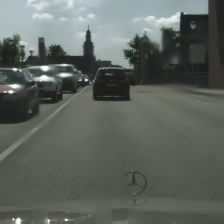};

\nextgroupplot[
width=0.23\columnwidth,
ticks=none,
axis lines=none, 
xtick=\empty, 
ytick=\empty,
axis equal image,
xmin=0, xmax=224,
ymin=0, ymax=224,
title style={yshift=-7pt},
]
\addplot graphics [includegraphics cmd=\pgfimage, xmin=0, xmax=224, ymin=0, ymax=224] {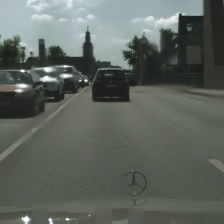};

\nextgroupplot[
width=0.23\columnwidth,
ticks=none,
axis lines=none, 
xtick=\empty, 
ytick=\empty,
axis equal image,
xmin=0, xmax=224,
ymin=0, ymax=224,
title style={yshift=-7pt},
]
\addplot graphics [includegraphics cmd=\pgfimage, xmin=0, xmax=224, ymin=0, ymax=224] {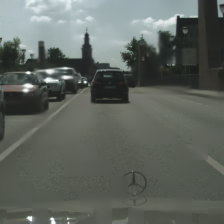};

\nextgroupplot[
width=0.23\columnwidth,
ticks=none,
axis lines=none, 
xtick=\empty, 
ytick=\empty,
axis equal image,
xmin=0, xmax=224,
ymin=0, ymax=224,
title style={yshift=-7pt},
]
\addplot graphics [includegraphics cmd=\pgfimage, xmin=0, xmax=224, ymin=0, ymax=224] {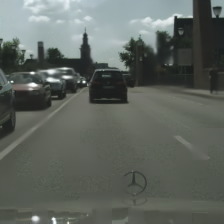};

\nextgroupplot[
width=0.23\columnwidth,
ticks=none,
axis lines=none, 
xtick=\empty, 
ytick=\empty,
axis equal image,
xmin=0, xmax=224,
ymin=0, ymax=224,
title style={yshift=-7pt},
]
\addplot graphics [includegraphics cmd=\pgfimage, xmin=0, xmax=224, ymin=0, ymax=224] {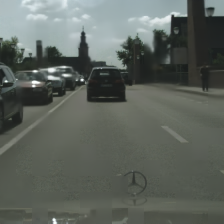};

\nextgroupplot[
width=0.23\columnwidth,
ticks=none,
axis line style={draw=none}, 
xtick=\empty, 
ytick=\empty,
axis equal image,
xmin=0, xmax=224,
ymin=0, ymax=224,
ylabel={$\qp$ map},
title style={yshift=-7pt},
]
\addplot graphics [includegraphics cmd=\pgfimage, xmin=0, xmax=224, ymin=0, ymax=224] {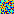};

\nextgroupplot[
width=0.23\columnwidth,
ticks=none,
axis lines=none, 
xtick=\empty, 
ytick=\empty,
axis equal image,
xmin=0, xmax=224,
ymin=0, ymax=224,
title style={yshift=-7pt},
]
\addplot graphics [includegraphics cmd=\pgfimage, xmin=0, xmax=224, ymin=0, ymax=224] {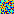};

\nextgroupplot[
width=0.23\columnwidth,
ticks=none,
axis lines=none, 
xtick=\empty, 
ytick=\empty,
axis equal image,
xmin=0, xmax=224,
ymin=0, ymax=224,
title style={yshift=-7pt},
]
\addplot graphics [includegraphics cmd=\pgfimage, xmin=0, xmax=224, ymin=0, ymax=224] {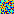};

\nextgroupplot[
width=0.23\columnwidth,
ticks=none,
axis lines=none, 
xtick=\empty, 
ytick=\empty,
axis equal image,
xmin=0, xmax=224,
ymin=0, ymax=224,
title style={yshift=-7pt},
]
\addplot graphics [includegraphics cmd=\pgfimage, xmin=0, xmax=224, ymin=0, ymax=224] {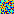};

\nextgroupplot[
width=0.23\columnwidth,
ticks=none,
axis lines=none, 
xtick=\empty, 
ytick=\empty,
axis equal image,
xmin=0, xmax=224,
ymin=0, ymax=224,
title style={yshift=-7pt},
]
\addplot graphics [includegraphics cmd=\pgfimage, xmin=0, xmax=224, ymin=0, ymax=224] {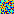};

\nextgroupplot[
width=0.23\columnwidth,
ticks=none,
axis lines=none, 
xtick=\empty, 
ytick=\empty,
axis equal image,
xmin=0, xmax=224,
ymin=0, ymax=224,
title style={yshift=-7pt},
]
\addplot graphics [includegraphics cmd=\pgfimage, xmin=0, xmax=224, ymin=0, ymax=224] {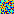};

\nextgroupplot[
width=0.23\columnwidth,
ticks=none,
axis lines=none, 
xtick=\empty, 
ytick=\empty,
axis equal image,
xmin=0, xmax=224,
ymin=0, ymax=224,
title style={yshift=-7pt},
]
\addplot graphics [includegraphics cmd=\pgfimage, xmin=0, xmax=224, ymin=0, ymax=224] {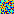};

\nextgroupplot[
width=0.23\columnwidth,
ticks=none,
axis lines=none, 
xtick=\empty, 
ytick=\empty,
axis equal image,
xmin=0, xmax=224,
ymin=0, ymax=224,
title style={yshift=-7pt},
colorbar,
colorbar style={
    ylabel={},
    ytick={5, 25, 45},
    yticklabels={5, 25, 45},
    at={(1.025,1)},
    yticklabel style={xshift=-1.5pt},
},
colormap={mymap}{[1pt]
    rgb(0pt)=(0,0,0.5);
    rgb(22pt)=(0,0,1);
    rgb(25pt)=(0,0,1);
    rgb(68pt)=(0,0.86,1);
    rgb(70pt)=(0,0.9,0.967741935483871);
    rgb(75pt)=(0.0806451612903226,1,0.887096774193548);
    rgb(128pt)=(0.935483870967742,1,0.0322580645161291);
    rgb(130pt)=(0.967741935483871,0.962962962962963,0);
    rgb(132pt)=(1,0.925925925925926,0);
    rgb(178pt)=(1,0.0740740740740741,0);
    rgb(182pt)=(0.909090909090909,0,0);
    rgb(200pt)=(0.5,0,0)
},
point meta max=51,
point meta min=0,
]
\addplot graphics [includegraphics cmd=\pgfimage, xmin=0, xmax=224, ymin=0, ymax=224] {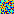};

\end{groupplot}

\end{tikzpicture}

%% file: artwork/surrogate_results/cityscapes/supplement/surrogate_results_cityscapes_full_2.tex
\begin{tikzpicture}[every node/.style={font=\small}, spy using outlines={tud6a, line width=0.80mm, dashed, dash pattern=on 1.5pt off 1.5pt, magnification=2, size=0.75cm, connect spies}]

\begin{groupplot}[group style={group size=8 by 3, horizontal sep=1.25pt, vertical sep=1.25pt}]

\nextgroupplot[
width=0.23\columnwidth,
ticks=none,
axis line style={draw=none}, 
xtick=\empty, 
ytick=\empty,
axis equal image,
xmin=0, xmax=224,
ymin=0, ymax=224,
ylabel={H.264},
title={$t_0$},
title style={yshift=-7pt},
]
\addplot graphics [includegraphics cmd=\pgfimage, xmin=0, xmax=224, ymin=0, ymax=224] {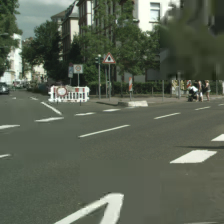};

\nextgroupplot[
width=0.23\columnwidth,
ticks=none,
axis lines=none, 
xtick=\empty, 
ytick=\empty,
axis equal image,
xmin=0, xmax=224,
ymin=0, ymax=224,
title={$t_1$},
title style={yshift=-7pt},
]
\addplot graphics [includegraphics cmd=\pgfimage, xmin=0, xmax=224, ymin=0, ymax=224] {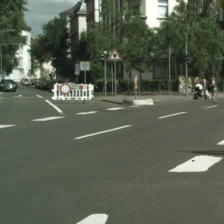};

\nextgroupplot[
width=0.23\columnwidth,
ticks=none,
axis lines=none, 
xtick=\empty, 
ytick=\empty,
axis equal image,
xmin=0, xmax=224,
ymin=0, ymax=224,
title={$t_2$},
title style={yshift=-7pt},
]
\addplot graphics [includegraphics cmd=\pgfimage, xmin=0, xmax=224, ymin=0, ymax=224] {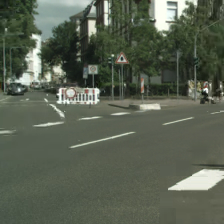};

\nextgroupplot[
width=0.23\columnwidth,
ticks=none,
axis lines=none, 
xtick=\empty, 
ytick=\empty,
axis equal image,
xmin=0, xmax=224,
ymin=0, ymax=224,
title={$t_3$},
title style={yshift=-7pt},
]
\addplot graphics [includegraphics cmd=\pgfimage, xmin=0, xmax=224, ymin=0, ymax=224] {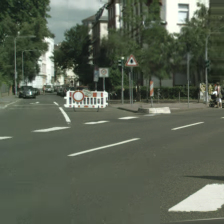};

\nextgroupplot[
width=0.23\columnwidth,
ticks=none,
axis lines=none, 
xtick=\empty, 
ytick=\empty,
axis equal image,
xmin=0, xmax=224,
ymin=0, ymax=224,
title={$t_4$},
title style={yshift=-7pt},
]
\addplot graphics [includegraphics cmd=\pgfimage, xmin=0, xmax=224, ymin=0, ymax=224] {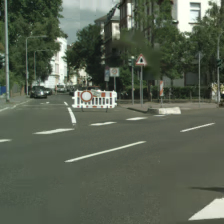};

\nextgroupplot[
width=0.23\columnwidth,
ticks=none,
axis lines=none, 
xtick=\empty, 
ytick=\empty,
axis equal image,
xmin=0, xmax=224,
ymin=0, ymax=224,
title={$t_5$},
title style={yshift=-7pt},
]
\addplot graphics [includegraphics cmd=\pgfimage, xmin=0, xmax=224, ymin=0, ymax=224] {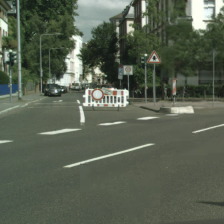};

\nextgroupplot[
width=0.23\columnwidth,
ticks=none,
axis lines=none, 
xtick=\empty, 
ytick=\empty,
axis equal image,
xmin=0, xmax=224,
ymin=0, ymax=224,
title={$t_6$},
title style={yshift=-7pt},
]
\addplot graphics [includegraphics cmd=\pgfimage, xmin=0, xmax=224, ymin=0, ymax=224] {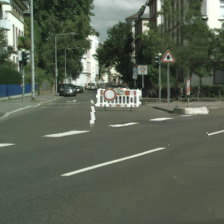};

\nextgroupplot[
width=0.23\columnwidth,
ticks=none,
axis lines=none, 
xtick=\empty, 
ytick=\empty,
axis equal image,
xmin=0, xmax=224,
ymin=0, ymax=224,
title={$t_7$},
title style={yshift=-7pt},
]
\addplot graphics [includegraphics cmd=\pgfimage, xmin=0, xmax=224, ymin=0, ymax=224] {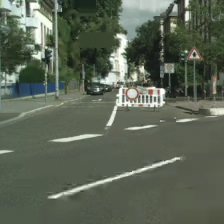};

\nextgroupplot[
width=0.23\columnwidth,
ticks=none,
axis line style={draw=none}, 
xtick=\empty, 
ytick=\empty,
axis equal image,
xmin=0, xmax=224,
ymin=0, ymax=224,
ylabel={Surrogate},
title style={yshift=-7pt},
]
\addplot graphics [includegraphics cmd=\pgfimage, xmin=0, xmax=224, ymin=0, ymax=224] {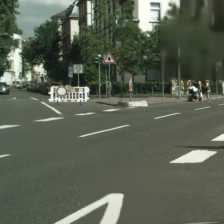};

\nextgroupplot[
width=0.23\columnwidth,
ticks=none,
axis lines=none, 
xtick=\empty, 
ytick=\empty,
axis equal image,
xmin=0, xmax=224,
ymin=0, ymax=224,
title style={yshift=-7pt},
]
\addplot graphics [includegraphics cmd=\pgfimage, xmin=0, xmax=224, ymin=0, ymax=224] {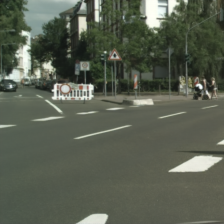};

\nextgroupplot[
width=0.23\columnwidth,
ticks=none,
axis lines=none, 
xtick=\empty, 
ytick=\empty,
axis equal image,
xmin=0, xmax=224,
ymin=0, ymax=224,
title style={yshift=-7pt},
]
\addplot graphics [includegraphics cmd=\pgfimage, xmin=0, xmax=224, ymin=0, ymax=224] {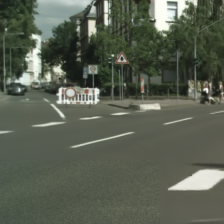};

\nextgroupplot[
width=0.23\columnwidth,
ticks=none,
axis lines=none, 
xtick=\empty, 
ytick=\empty,
axis equal image,
xmin=0, xmax=224,
ymin=0, ymax=224,
title style={yshift=-7pt},
]
\addplot graphics [includegraphics cmd=\pgfimage, xmin=0, xmax=224, ymin=0, ymax=224] {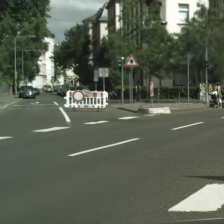};

\nextgroupplot[
width=0.23\columnwidth,
ticks=none,
axis lines=none, 
xtick=\empty, 
ytick=\empty,
axis equal image,
xmin=0, xmax=224,
ymin=0, ymax=224,
title style={yshift=-7pt},
]
\addplot graphics [includegraphics cmd=\pgfimage, xmin=0, xmax=224, ymin=0, ymax=224] {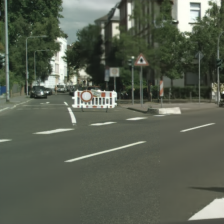};

\nextgroupplot[
width=0.23\columnwidth,
ticks=none,
axis lines=none, 
xtick=\empty, 
ytick=\empty,
axis equal image,
xmin=0, xmax=224,
ymin=0, ymax=224,
title style={yshift=-7pt},
]
\addplot graphics [includegraphics cmd=\pgfimage, xmin=0, xmax=224, ymin=0, ymax=224] {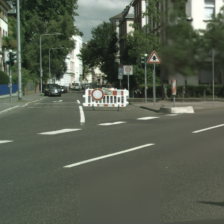};

\nextgroupplot[
width=0.23\columnwidth,
ticks=none,
axis lines=none, 
xtick=\empty, 
ytick=\empty,
axis equal image,
xmin=0, xmax=224,
ymin=0, ymax=224,
title style={yshift=-7pt},
]
\addplot graphics [includegraphics cmd=\pgfimage, xmin=0, xmax=224, ymin=0, ymax=224] {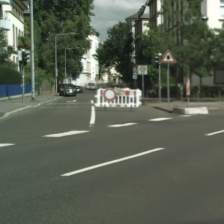};

\nextgroupplot[
width=0.23\columnwidth,
ticks=none,
axis lines=none, 
xtick=\empty, 
ytick=\empty,
axis equal image,
xmin=0, xmax=224,
ymin=0, ymax=224,
title style={yshift=-7pt},
]
\addplot graphics [includegraphics cmd=\pgfimage, xmin=0, xmax=224, ymin=0, ymax=224] {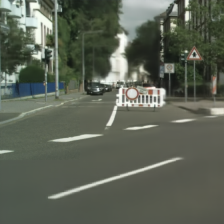};

\nextgroupplot[
width=0.23\columnwidth,
ticks=none,
axis line style={draw=none}, 
xtick=\empty, 
ytick=\empty,
axis equal image,
xmin=0, xmax=224,
ymin=0, ymax=224,
ylabel={$\qp$ map},
title style={yshift=-7pt},
]
\addplot graphics [includegraphics cmd=\pgfimage, xmin=0, xmax=224, ymin=0, ymax=224] {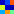};

\nextgroupplot[
width=0.23\columnwidth,
ticks=none,
axis lines=none, 
xtick=\empty, 
ytick=\empty,
axis equal image,
xmin=0, xmax=224,
ymin=0, ymax=224,
title style={yshift=-7pt},
]
\addplot graphics [includegraphics cmd=\pgfimage, xmin=0, xmax=224, ymin=0, ymax=224] {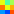};

\nextgroupplot[
width=0.23\columnwidth,
ticks=none,
axis lines=none, 
xtick=\empty, 
ytick=\empty,
axis equal image,
xmin=0, xmax=224,
ymin=0, ymax=224,
title style={yshift=-7pt},
]
\addplot graphics [includegraphics cmd=\pgfimage, xmin=0, xmax=224, ymin=0, ymax=224] {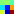};

\nextgroupplot[
width=0.23\columnwidth,
ticks=none,
axis lines=none, 
xtick=\empty, 
ytick=\empty,
axis equal image,
xmin=0, xmax=224,
ymin=0, ymax=224,
title style={yshift=-7pt},
]
\addplot graphics [includegraphics cmd=\pgfimage, xmin=0, xmax=224, ymin=0, ymax=224] {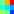};

\nextgroupplot[
width=0.23\columnwidth,
ticks=none,
axis lines=none, 
xtick=\empty, 
ytick=\empty,
axis equal image,
xmin=0, xmax=224,
ymin=0, ymax=224,
title style={yshift=-7pt},
]
\addplot graphics [includegraphics cmd=\pgfimage, xmin=0, xmax=224, ymin=0, ymax=224] {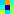};

\nextgroupplot[
width=0.23\columnwidth,
ticks=none,
axis lines=none, 
xtick=\empty, 
ytick=\empty,
axis equal image,
xmin=0, xmax=224,
ymin=0, ymax=224,
title style={yshift=-7pt},
]
\addplot graphics [includegraphics cmd=\pgfimage, xmin=0, xmax=224, ymin=0, ymax=224] {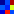};

\nextgroupplot[
width=0.23\columnwidth,
ticks=none,
axis lines=none, 
xtick=\empty, 
ytick=\empty,
axis equal image,
xmin=0, xmax=224,
ymin=0, ymax=224,
title style={yshift=-7pt},
]
\addplot graphics [includegraphics cmd=\pgfimage, xmin=0, xmax=224, ymin=0, ymax=224] {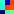};

\nextgroupplot[
width=0.23\columnwidth,
ticks=none,
axis lines=none, 
xtick=\empty, 
ytick=\empty,
axis equal image,
xmin=0, xmax=224,
ymin=0, ymax=224,
title style={yshift=-7pt},
colorbar,
colorbar style={
    ylabel={},
    ytick={5, 25, 45},
    yticklabels={5, 25, 45},
    at={(1.025,1)},
    yticklabel style={xshift=-1.5pt},
},
colormap={mymap}{[1pt]
    rgb(0pt)=(0,0,0.5);
    rgb(22pt)=(0,0,1);
    rgb(25pt)=(0,0,1);
    rgb(68pt)=(0,0.86,1);
    rgb(70pt)=(0,0.9,0.967741935483871);
    rgb(75pt)=(0.0806451612903226,1,0.887096774193548);
    rgb(128pt)=(0.935483870967742,1,0.0322580645161291);
    rgb(130pt)=(0.967741935483871,0.962962962962963,0);
    rgb(132pt)=(1,0.925925925925926,0);
    rgb(178pt)=(1,0.0740740740740741,0);
    rgb(182pt)=(0.909090909090909,0,0);
    rgb(200pt)=(0.5,0,0)
},
point meta max=51,
point meta min=0,
]
\addplot graphics [includegraphics cmd=\pgfimage, xmin=0, xmax=224, ymin=0, ymax=224] {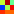};

\end{groupplot}

\end{tikzpicture}

%% file: artwork/surrogate_results/cityscapes/supplement/surrogate_results_cityscapes_full_3.tex
\begin{tikzpicture}[every node/.style={font=\small}, spy using outlines={tud6a, line width=0.80mm, dashed, dash pattern=on 1.5pt off 1.5pt, magnification=2, size=0.75cm, connect spies}]

\begin{groupplot}[group style={group size=8 by 3, horizontal sep=1.25pt, vertical sep=1.25pt}]

\nextgroupplot[
width=0.23\columnwidth,
ticks=none,
axis line style={draw=none}, 
xtick=\empty, 
ytick=\empty,
axis equal image,
xmin=0, xmax=224,
ymin=0, ymax=224,
ylabel={H.264},
title={$t_0$},
title style={yshift=-7pt},
]
\addplot graphics [includegraphics cmd=\pgfimage, xmin=0, xmax=224, ymin=0, ymax=224] {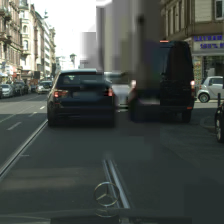};

\nextgroupplot[
width=0.23\columnwidth,
ticks=none,
axis lines=none, 
xtick=\empty, 
ytick=\empty,
axis equal image,
xmin=0, xmax=224,
ymin=0, ymax=224,
title={$t_1$},
title style={yshift=-7pt},
]
\addplot graphics [includegraphics cmd=\pgfimage, xmin=0, xmax=224, ymin=0, ymax=224] {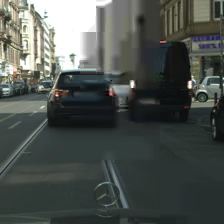};

\nextgroupplot[
width=0.23\columnwidth,
ticks=none,
axis lines=none, 
xtick=\empty, 
ytick=\empty,
axis equal image,
xmin=0, xmax=224,
ymin=0, ymax=224,
title={$t_2$},
title style={yshift=-7pt},
]
\addplot graphics [includegraphics cmd=\pgfimage, xmin=0, xmax=224, ymin=0, ymax=224] {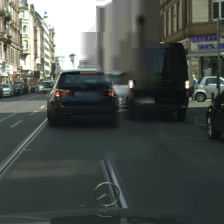};

\nextgroupplot[
width=0.23\columnwidth,
ticks=none,
axis lines=none, 
xtick=\empty, 
ytick=\empty,
axis equal image,
xmin=0, xmax=224,
ymin=0, ymax=224,
title={$t_3$},
title style={yshift=-7pt},
]
\addplot graphics [includegraphics cmd=\pgfimage, xmin=0, xmax=224, ymin=0, ymax=224] {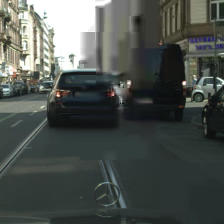};

\nextgroupplot[
width=0.23\columnwidth,
ticks=none,
axis lines=none, 
xtick=\empty, 
ytick=\empty,
axis equal image,
xmin=0, xmax=224,
ymin=0, ymax=224,
title={$t_4$},
title style={yshift=-7pt},
]
\addplot graphics [includegraphics cmd=\pgfimage, xmin=0, xmax=224, ymin=0, ymax=224] {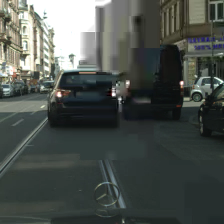};

\nextgroupplot[
width=0.23\columnwidth,
ticks=none,
axis lines=none, 
xtick=\empty, 
ytick=\empty,
axis equal image,
xmin=0, xmax=224,
ymin=0, ymax=224,
title={$t_5$},
title style={yshift=-7pt},
]
\addplot graphics [includegraphics cmd=\pgfimage, xmin=0, xmax=224, ymin=0, ymax=224] {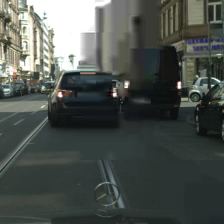};

\nextgroupplot[
width=0.23\columnwidth,
ticks=none,
axis lines=none, 
xtick=\empty, 
ytick=\empty,
axis equal image,
xmin=0, xmax=224,
ymin=0, ymax=224,
title={$t_6$},
title style={yshift=-7pt},
]
\addplot graphics [includegraphics cmd=\pgfimage, xmin=0, xmax=224, ymin=0, ymax=224] {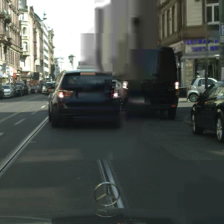};

\nextgroupplot[
width=0.23\columnwidth,
ticks=none,
axis lines=none, 
xtick=\empty, 
ytick=\empty,
axis equal image,
xmin=0, xmax=224,
ymin=0, ymax=224,
title={$t_7$},
title style={yshift=-7pt},
]
\addplot graphics [includegraphics cmd=\pgfimage, xmin=0, xmax=224, ymin=0, ymax=224] {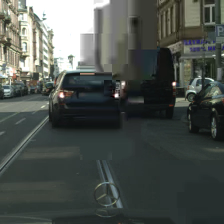};

\nextgroupplot[
width=0.23\columnwidth,
ticks=none,
axis line style={draw=none}, 
xtick=\empty, 
ytick=\empty,
axis equal image,
xmin=0, xmax=224,
ymin=0, ymax=224,
ylabel={Surrogate},
title style={yshift=-7pt},
]
\addplot graphics [includegraphics cmd=\pgfimage, xmin=0, xmax=224, ymin=0, ymax=224] {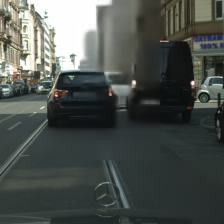};

\nextgroupplot[
width=0.23\columnwidth,
ticks=none,
axis lines=none, 
xtick=\empty, 
ytick=\empty,
axis equal image,
xmin=0, xmax=224,
ymin=0, ymax=224,
title style={yshift=-7pt},
]
\addplot graphics [includegraphics cmd=\pgfimage, xmin=0, xmax=224, ymin=0, ymax=224] {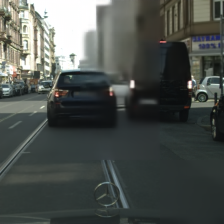};

\nextgroupplot[
width=0.23\columnwidth,
ticks=none,
axis lines=none, 
xtick=\empty, 
ytick=\empty,
axis equal image,
xmin=0, xmax=224,
ymin=0, ymax=224,
title style={yshift=-7pt},
]
\addplot graphics [includegraphics cmd=\pgfimage, xmin=0, xmax=224, ymin=0, ymax=224] {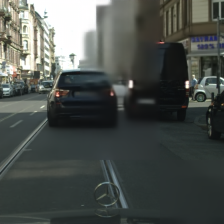};

\nextgroupplot[
width=0.23\columnwidth,
ticks=none,
axis lines=none, 
xtick=\empty, 
ytick=\empty,
axis equal image,
xmin=0, xmax=224,
ymin=0, ymax=224,
title style={yshift=-7pt},
]
\addplot graphics [includegraphics cmd=\pgfimage, xmin=0, xmax=224, ymin=0, ymax=224] {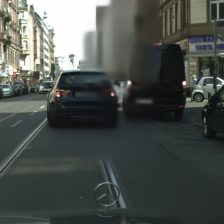};

\nextgroupplot[
width=0.23\columnwidth,
ticks=none,
axis lines=none, 
xtick=\empty, 
ytick=\empty,
axis equal image,
xmin=0, xmax=224,
ymin=0, ymax=224,
title style={yshift=-7pt},
]
\addplot graphics [includegraphics cmd=\pgfimage, xmin=0, xmax=224, ymin=0, ymax=224] {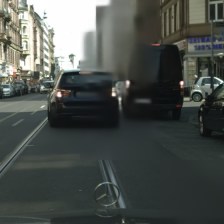};

\nextgroupplot[
width=0.23\columnwidth,
ticks=none,
axis lines=none, 
xtick=\empty, 
ytick=\empty,
axis equal image,
xmin=0, xmax=224,
ymin=0, ymax=224,
title style={yshift=-7pt},
]
\addplot graphics [includegraphics cmd=\pgfimage, xmin=0, xmax=224, ymin=0, ymax=224] {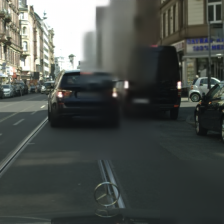};

\nextgroupplot[
width=0.23\columnwidth,
ticks=none,
axis lines=none, 
xtick=\empty, 
ytick=\empty,
axis equal image,
xmin=0, xmax=224,
ymin=0, ymax=224,
title style={yshift=-7pt},
]
\addplot graphics [includegraphics cmd=\pgfimage, xmin=0, xmax=224, ymin=0, ymax=224] {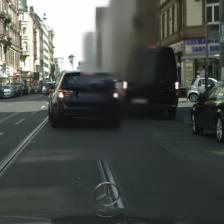};

\nextgroupplot[
width=0.23\columnwidth,
ticks=none,
axis lines=none, 
xtick=\empty, 
ytick=\empty,
axis equal image,
xmin=0, xmax=224,
ymin=0, ymax=224,
title style={yshift=-7pt},
]
\addplot graphics [includegraphics cmd=\pgfimage, xmin=0, xmax=224, ymin=0, ymax=224] {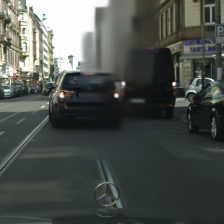};

\nextgroupplot[
width=0.23\columnwidth,
ticks=none,
axis line style={draw=none}, 
xtick=\empty, 
ytick=\empty,
axis equal image,
xmin=0, xmax=224,
ymin=0, ymax=224,
ylabel={$\qp$ map},
title style={yshift=-7pt},
]
\addplot graphics [includegraphics cmd=\pgfimage, xmin=0, xmax=224, ymin=0, ymax=224] {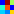};

\nextgroupplot[
width=0.23\columnwidth,
ticks=none,
axis lines=none, 
xtick=\empty, 
ytick=\empty,
axis equal image,
xmin=0, xmax=224,
ymin=0, ymax=224,
title style={yshift=-7pt},
]
\addplot graphics [includegraphics cmd=\pgfimage, xmin=0, xmax=224, ymin=0, ymax=224] {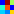};

\nextgroupplot[
width=0.23\columnwidth,
ticks=none,
axis lines=none, 
xtick=\empty, 
ytick=\empty,
axis equal image,
xmin=0, xmax=224,
ymin=0, ymax=224,
title style={yshift=-7pt},
]
\addplot graphics [includegraphics cmd=\pgfimage, xmin=0, xmax=224, ymin=0, ymax=224] {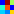};

\nextgroupplot[
width=0.23\columnwidth,
ticks=none,
axis lines=none, 
xtick=\empty, 
ytick=\empty,
axis equal image,
xmin=0, xmax=224,
ymin=0, ymax=224,
title style={yshift=-7pt},
]
\addplot graphics [includegraphics cmd=\pgfimage, xmin=0, xmax=224, ymin=0, ymax=224] {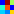};

\nextgroupplot[
width=0.23\columnwidth,
ticks=none,
axis lines=none, 
xtick=\empty, 
ytick=\empty,
axis equal image,
xmin=0, xmax=224,
ymin=0, ymax=224,
title style={yshift=-7pt},
]
\addplot graphics [includegraphics cmd=\pgfimage, xmin=0, xmax=224, ymin=0, ymax=224] {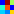};

\nextgroupplot[
width=0.23\columnwidth,
ticks=none,
axis lines=none, 
xtick=\empty, 
ytick=\empty,
axis equal image,
xmin=0, xmax=224,
ymin=0, ymax=224,
title style={yshift=-7pt},
]
\addplot graphics [includegraphics cmd=\pgfimage, xmin=0, xmax=224, ymin=0, ymax=224] {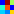};

\nextgroupplot[
width=0.23\columnwidth,
ticks=none,
axis lines=none, 
xtick=\empty, 
ytick=\empty,
axis equal image,
xmin=0, xmax=224,
ymin=0, ymax=224,
title style={yshift=-7pt},
]
\addplot graphics [includegraphics cmd=\pgfimage, xmin=0, xmax=224, ymin=0, ymax=224] {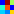};

\nextgroupplot[
width=0.23\columnwidth,
ticks=none,
axis lines=none, 
xtick=\empty, 
ytick=\empty,
axis equal image,
xmin=0, xmax=224,
ymin=0, ymax=224,
title style={yshift=-7pt},
colorbar,
colorbar style={
    ylabel={},
    ytick={5, 25, 45},
    yticklabels={5, 25, 45},
    at={(1.025,1)},
    yticklabel style={xshift=-1.5pt},
},
colormap={mymap}{[1pt]
    rgb(0pt)=(0,0,0.5);
    rgb(22pt)=(0,0,1);
    rgb(25pt)=(0,0,1);
    rgb(68pt)=(0,0.86,1);
    rgb(70pt)=(0,0.9,0.967741935483871);
    rgb(75pt)=(0.0806451612903226,1,0.887096774193548);
    rgb(128pt)=(0.935483870967742,1,0.0322580645161291);
    rgb(130pt)=(0.967741935483871,0.962962962962963,0);
    rgb(132pt)=(1,0.925925925925926,0);
    rgb(178pt)=(1,0.0740740740740741,0);
    rgb(182pt)=(0.909090909090909,0,0);
    rgb(200pt)=(0.5,0,0)
},
point meta max=51,
point meta min=0,
]
\addplot graphics [includegraphics cmd=\pgfimage, xmin=0, xmax=224, ymin=0, ymax=224] {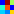};

\end{groupplot}

\end{tikzpicture}

%% file: artwork/surrogate_results/cityscapes/supplement/surrogate_results_cityscapes_high_1.tex
\begin{tikzpicture}[every node/.style={font=\small}, spy using outlines={tud6a, line width=0.80mm, dashed, dash pattern=on 1.5pt off 1.5pt, magnification=2, size=0.75cm, connect spies}]

\begin{groupplot}[group style={group size=8 by 3, horizontal sep=1.25pt, vertical sep=1.25pt}]

\nextgroupplot[
width=0.23\columnwidth,
ticks=none,
axis line style={draw=none}, 
xtick=\empty, 
ytick=\empty,
axis equal image,
xmin=0, xmax=224,
ymin=0, ymax=224,
ylabel={H.264},
title={$t_0$},
title style={yshift=-7pt},
]
\addplot graphics [includegraphics cmd=\pgfimage, xmin=0, xmax=224, ymin=0, ymax=224] {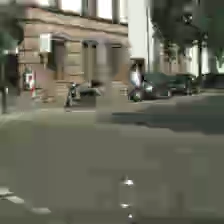};

\nextgroupplot[
width=0.23\columnwidth,
ticks=none,
axis lines=none, 
xtick=\empty, 
ytick=\empty,
axis equal image,
xmin=0, xmax=224,
ymin=0, ymax=224,
title={$t_1$},
title style={yshift=-7pt},
]
\addplot graphics [includegraphics cmd=\pgfimage, xmin=0, xmax=224, ymin=0, ymax=224] {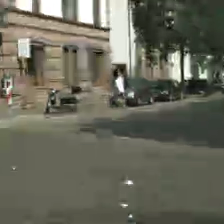};

\nextgroupplot[
width=0.23\columnwidth,
ticks=none,
axis lines=none, 
xtick=\empty, 
ytick=\empty,
axis equal image,
xmin=0, xmax=224,
ymin=0, ymax=224,
title={$t_2$},
title style={yshift=-7pt},
]
\addplot graphics [includegraphics cmd=\pgfimage, xmin=0, xmax=224, ymin=0, ymax=224] {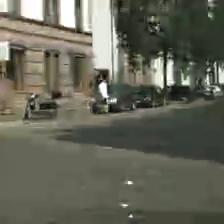};

\nextgroupplot[
width=0.23\columnwidth,
ticks=none,
axis lines=none, 
xtick=\empty, 
ytick=\empty,
axis equal image,
xmin=0, xmax=224,
ymin=0, ymax=224,
title={$t_3$},
title style={yshift=-7pt},
]
\addplot graphics [includegraphics cmd=\pgfimage, xmin=0, xmax=224, ymin=0, ymax=224] {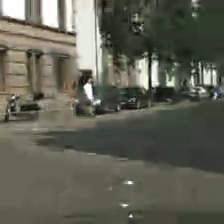};

\nextgroupplot[
width=0.23\columnwidth,
ticks=none,
axis lines=none, 
xtick=\empty, 
ytick=\empty,
axis equal image,
xmin=0, xmax=224,
ymin=0, ymax=224,
title={$t_4$},
title style={yshift=-7pt},
]
\addplot graphics [includegraphics cmd=\pgfimage, xmin=0, xmax=224, ymin=0, ymax=224] {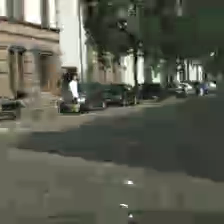};

\nextgroupplot[
width=0.23\columnwidth,
ticks=none,
axis lines=none, 
xtick=\empty, 
ytick=\empty,
axis equal image,
xmin=0, xmax=224,
ymin=0, ymax=224,
title={$t_5$},
title style={yshift=-7pt},
]
\addplot graphics [includegraphics cmd=\pgfimage, xmin=0, xmax=224, ymin=0, ymax=224] {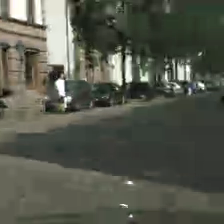};

\nextgroupplot[
width=0.23\columnwidth,
ticks=none,
axis lines=none, 
xtick=\empty, 
ytick=\empty,
axis equal image,
xmin=0, xmax=224,
ymin=0, ymax=224,
title={$t_6$},
title style={yshift=-7pt},
]
\addplot graphics [includegraphics cmd=\pgfimage, xmin=0, xmax=224, ymin=0, ymax=224] {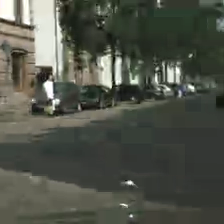};

\nextgroupplot[
width=0.23\columnwidth,
ticks=none,
axis lines=none, 
xtick=\empty, 
ytick=\empty,
axis equal image,
xmin=0, xmax=224,
ymin=0, ymax=224,
title={$t_7$},
title style={yshift=-7pt},
]
\addplot graphics [includegraphics cmd=\pgfimage, xmin=0, xmax=224, ymin=0, ymax=224] {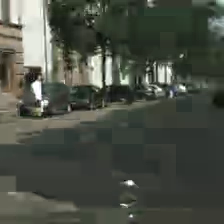};

\nextgroupplot[
width=0.23\columnwidth,
ticks=none,
axis line style={draw=none}, 
xtick=\empty, 
ytick=\empty,
axis equal image,
xmin=0, xmax=224,
ymin=0, ymax=224,
ylabel={Surrogate},
title style={yshift=-7pt},
]
\addplot graphics [includegraphics cmd=\pgfimage, xmin=0, xmax=224, ymin=0, ymax=224] {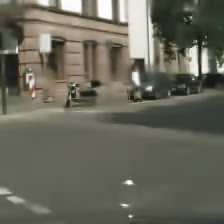};

\nextgroupplot[
width=0.23\columnwidth,
ticks=none,
axis lines=none, 
xtick=\empty, 
ytick=\empty,
axis equal image,
xmin=0, xmax=224,
ymin=0, ymax=224,
title style={yshift=-7pt},
]
\addplot graphics [includegraphics cmd=\pgfimage, xmin=0, xmax=224, ymin=0, ymax=224] {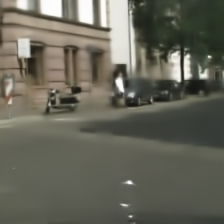};

\nextgroupplot[
width=0.23\columnwidth,
ticks=none,
axis lines=none, 
xtick=\empty, 
ytick=\empty,
axis equal image,
xmin=0, xmax=224,
ymin=0, ymax=224,
title style={yshift=-7pt},
]
\addplot graphics [includegraphics cmd=\pgfimage, xmin=0, xmax=224, ymin=0, ymax=224] {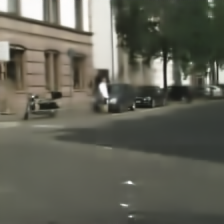};

\nextgroupplot[
width=0.23\columnwidth,
ticks=none,
axis lines=none, 
xtick=\empty, 
ytick=\empty,
axis equal image,
xmin=0, xmax=224,
ymin=0, ymax=224,
title style={yshift=-7pt},
]
\addplot graphics [includegraphics cmd=\pgfimage, xmin=0, xmax=224, ymin=0, ymax=224] {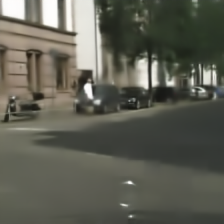};

\nextgroupplot[
width=0.23\columnwidth,
ticks=none,
axis lines=none, 
xtick=\empty, 
ytick=\empty,
axis equal image,
xmin=0, xmax=224,
ymin=0, ymax=224,
title style={yshift=-7pt},
]
\addplot graphics [includegraphics cmd=\pgfimage, xmin=0, xmax=224, ymin=0, ymax=224] {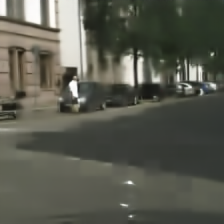};

\nextgroupplot[
width=0.23\columnwidth,
ticks=none,
axis lines=none, 
xtick=\empty, 
ytick=\empty,
axis equal image,
xmin=0, xmax=224,
ymin=0, ymax=224,
title style={yshift=-7pt},
]
\addplot graphics [includegraphics cmd=\pgfimage, xmin=0, xmax=224, ymin=0, ymax=224] {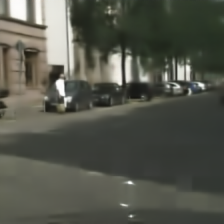};

\nextgroupplot[
width=0.23\columnwidth,
ticks=none,
axis lines=none, 
xtick=\empty, 
ytick=\empty,
axis equal image,
xmin=0, xmax=224,
ymin=0, ymax=224,
title style={yshift=-7pt},
]
\addplot graphics [includegraphics cmd=\pgfimage, xmin=0, xmax=224, ymin=0, ymax=224] {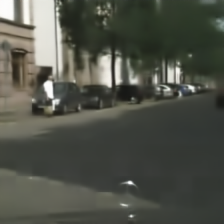};

\nextgroupplot[
width=0.23\columnwidth,
ticks=none,
axis lines=none, 
xtick=\empty, 
ytick=\empty,
axis equal image,
xmin=0, xmax=224,
ymin=0, ymax=224,
title style={yshift=-7pt},
]
\addplot graphics [includegraphics cmd=\pgfimage, xmin=0, xmax=224, ymin=0, ymax=224] {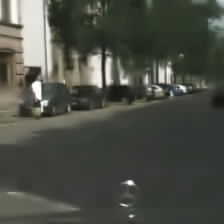};

\nextgroupplot[
width=0.23\columnwidth,
ticks=none,
axis line style={draw=none}, 
xtick=\empty, 
ytick=\empty,
axis equal image,
xmin=0, xmax=224,
ymin=0, ymax=224,
ylabel={$\qp$ map},
title style={yshift=-7pt},
]
\addplot graphics [includegraphics cmd=\pgfimage, xmin=0, xmax=224, ymin=0, ymax=224] {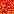};

\nextgroupplot[
width=0.23\columnwidth,
ticks=none,
axis lines=none, 
xtick=\empty, 
ytick=\empty,
axis equal image,
xmin=0, xmax=224,
ymin=0, ymax=224,
title style={yshift=-7pt},
]
\addplot graphics [includegraphics cmd=\pgfimage, xmin=0, xmax=224, ymin=0, ymax=224] {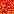};

\nextgroupplot[
width=0.23\columnwidth,
ticks=none,
axis lines=none, 
xtick=\empty, 
ytick=\empty,
axis equal image,
xmin=0, xmax=224,
ymin=0, ymax=224,
title style={yshift=-7pt},
]
\addplot graphics [includegraphics cmd=\pgfimage, xmin=0, xmax=224, ymin=0, ymax=224] {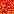};

\nextgroupplot[
width=0.23\columnwidth,
ticks=none,
axis lines=none, 
xtick=\empty, 
ytick=\empty,
axis equal image,
xmin=0, xmax=224,
ymin=0, ymax=224,
title style={yshift=-7pt},
]
\addplot graphics [includegraphics cmd=\pgfimage, xmin=0, xmax=224, ymin=0, ymax=224] {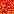};

\nextgroupplot[
width=0.23\columnwidth,
ticks=none,
axis lines=none, 
xtick=\empty, 
ytick=\empty,
axis equal image,
xmin=0, xmax=224,
ymin=0, ymax=224,
title style={yshift=-7pt},
]
\addplot graphics [includegraphics cmd=\pgfimage, xmin=0, xmax=224, ymin=0, ymax=224] {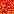};

\nextgroupplot[
width=0.23\columnwidth,
ticks=none,
axis lines=none, 
xtick=\empty, 
ytick=\empty,
axis equal image,
xmin=0, xmax=224,
ymin=0, ymax=224,
title style={yshift=-7pt},
]
\addplot graphics [includegraphics cmd=\pgfimage, xmin=0, xmax=224, ymin=0, ymax=224] {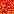};

\nextgroupplot[
width=0.23\columnwidth,
ticks=none,
axis lines=none, 
xtick=\empty, 
ytick=\empty,
axis equal image,
xmin=0, xmax=224,
ymin=0, ymax=224,
title style={yshift=-7pt},
]
\addplot graphics [includegraphics cmd=\pgfimage, xmin=0, xmax=224, ymin=0, ymax=224] {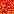};

\nextgroupplot[
width=0.23\columnwidth,
ticks=none,
axis lines=none, 
xtick=\empty, 
ytick=\empty,
axis equal image,
xmin=0, xmax=224,
ymin=0, ymax=224,
title style={yshift=-7pt},
colorbar,
colorbar style={
    ylabel={},
    ytick={5, 25, 45},
    yticklabels={5, 25, 45},
    at={(1.025,1)},
    yticklabel style={xshift=-1.5pt},
},
colormap={mymap}{[1pt]
    rgb(0pt)=(0,0,0.5);
    rgb(22pt)=(0,0,1);
    rgb(25pt)=(0,0,1);
    rgb(68pt)=(0,0.86,1);
    rgb(70pt)=(0,0.9,0.967741935483871);
    rgb(75pt)=(0.0806451612903226,1,0.887096774193548);
    rgb(128pt)=(0.935483870967742,1,0.0322580645161291);
    rgb(130pt)=(0.967741935483871,0.962962962962963,0);
    rgb(132pt)=(1,0.925925925925926,0);
    rgb(178pt)=(1,0.0740740740740741,0);
    rgb(182pt)=(0.909090909090909,0,0);
    rgb(200pt)=(0.5,0,0)
},
point meta max=51,
point meta min=0,
]
\addplot graphics [includegraphics cmd=\pgfimage, xmin=0, xmax=224, ymin=0, ymax=224] {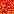};

\end{groupplot}

\end{tikzpicture}

%% file: artwork/surrogate_results/cityscapes/supplement/surrogate_results_cityscapes_high_2.tex
\begin{tikzpicture}[every node/.style={font=\small}, spy using outlines={tud6a, line width=0.80mm, dashed, dash pattern=on 1.5pt off 1.5pt, magnification=2, size=0.75cm, connect spies}]

\begin{groupplot}[group style={group size=8 by 3, horizontal sep=1.25pt, vertical sep=1.25pt}]

\nextgroupplot[
width=0.23\columnwidth,
ticks=none,
axis line style={draw=none}, 
xtick=\empty, 
ytick=\empty,
axis equal image,
xmin=0, xmax=224,
ymin=0, ymax=224,
ylabel={H.264},
title={$t_0$},
title style={yshift=-7pt},
]
\addplot graphics [includegraphics cmd=\pgfimage, xmin=0, xmax=224, ymin=0, ymax=224] {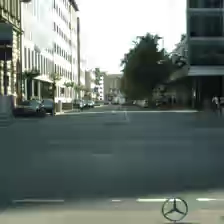};

\nextgroupplot[
width=0.23\columnwidth,
ticks=none,
axis lines=none, 
xtick=\empty, 
ytick=\empty,
axis equal image,
xmin=0, xmax=224,
ymin=0, ymax=224,
title={$t_1$},
title style={yshift=-7pt},
]
\addplot graphics [includegraphics cmd=\pgfimage, xmin=0, xmax=224, ymin=0, ymax=224] {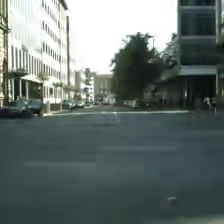};

\nextgroupplot[
width=0.23\columnwidth,
ticks=none,
axis lines=none, 
xtick=\empty, 
ytick=\empty,
axis equal image,
xmin=0, xmax=224,
ymin=0, ymax=224,
title={$t_2$},
title style={yshift=-7pt},
]
\addplot graphics [includegraphics cmd=\pgfimage, xmin=0, xmax=224, ymin=0, ymax=224] {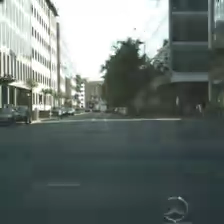};

\nextgroupplot[
width=0.23\columnwidth,
ticks=none,
axis lines=none, 
xtick=\empty, 
ytick=\empty,
axis equal image,
xmin=0, xmax=224,
ymin=0, ymax=224,
title={$t_3$},
title style={yshift=-7pt},
]
\addplot graphics [includegraphics cmd=\pgfimage, xmin=0, xmax=224, ymin=0, ymax=224] {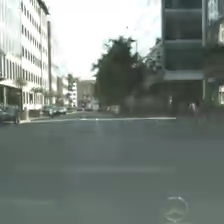};

\nextgroupplot[
width=0.23\columnwidth,
ticks=none,
axis lines=none, 
xtick=\empty, 
ytick=\empty,
axis equal image,
xmin=0, xmax=224,
ymin=0, ymax=224,
title={$t_4$},
title style={yshift=-7pt},
]
\addplot graphics [includegraphics cmd=\pgfimage, xmin=0, xmax=224, ymin=0, ymax=224] {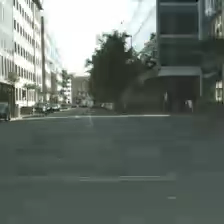};

\nextgroupplot[
width=0.23\columnwidth,
ticks=none,
axis lines=none, 
xtick=\empty, 
ytick=\empty,
axis equal image,
xmin=0, xmax=224,
ymin=0, ymax=224,
title={$t_5$},
title style={yshift=-7pt},
]
\addplot graphics [includegraphics cmd=\pgfimage, xmin=0, xmax=224, ymin=0, ymax=224] {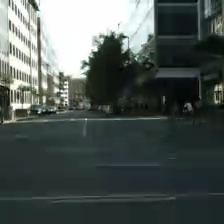};

\nextgroupplot[
width=0.23\columnwidth,
ticks=none,
axis lines=none, 
xtick=\empty, 
ytick=\empty,
axis equal image,
xmin=0, xmax=224,
ymin=0, ymax=224,
title={$t_6$},
title style={yshift=-7pt},
]
\addplot graphics [includegraphics cmd=\pgfimage, xmin=0, xmax=224, ymin=0, ymax=224] {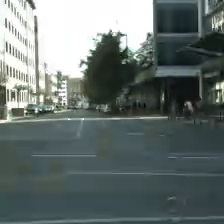};

\nextgroupplot[
width=0.23\columnwidth,
ticks=none,
axis lines=none, 
xtick=\empty, 
ytick=\empty,
axis equal image,
xmin=0, xmax=224,
ymin=0, ymax=224,
title={$t_7$},
title style={yshift=-7pt},
]
\addplot graphics [includegraphics cmd=\pgfimage, xmin=0, xmax=224, ymin=0, ymax=224] {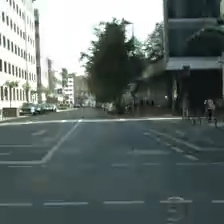};

\nextgroupplot[
width=0.23\columnwidth,
ticks=none,
axis line style={draw=none}, 
xtick=\empty, 
ytick=\empty,
axis equal image,
xmin=0, xmax=224,
ymin=0, ymax=224,
ylabel={Surrogate},
title style={yshift=-7pt},
]
\addplot graphics [includegraphics cmd=\pgfimage, xmin=0, xmax=224, ymin=0, ymax=224] {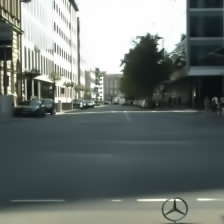};

\nextgroupplot[
width=0.23\columnwidth,
ticks=none,
axis lines=none, 
xtick=\empty, 
ytick=\empty,
axis equal image,
xmin=0, xmax=224,
ymin=0, ymax=224,
title style={yshift=-7pt},
]
\addplot graphics [includegraphics cmd=\pgfimage, xmin=0, xmax=224, ymin=0, ymax=224] {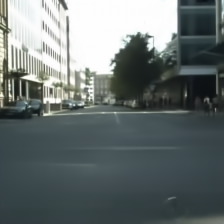};

\nextgroupplot[
width=0.23\columnwidth,
ticks=none,
axis lines=none, 
xtick=\empty, 
ytick=\empty,
axis equal image,
xmin=0, xmax=224,
ymin=0, ymax=224,
title style={yshift=-7pt},
]
\addplot graphics [includegraphics cmd=\pgfimage, xmin=0, xmax=224, ymin=0, ymax=224] {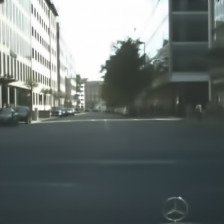};

\nextgroupplot[
width=0.23\columnwidth,
ticks=none,
axis lines=none, 
xtick=\empty, 
ytick=\empty,
axis equal image,
xmin=0, xmax=224,
ymin=0, ymax=224,
title style={yshift=-7pt},
]
\addplot graphics [includegraphics cmd=\pgfimage, xmin=0, xmax=224, ymin=0, ymax=224] {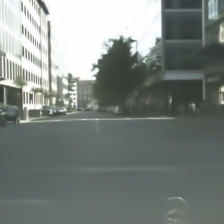};

\nextgroupplot[
width=0.23\columnwidth,
ticks=none,
axis lines=none, 
xtick=\empty, 
ytick=\empty,
axis equal image,
xmin=0, xmax=224,
ymin=0, ymax=224,
title style={yshift=-7pt},
]
\addplot graphics [includegraphics cmd=\pgfimage, xmin=0, xmax=224, ymin=0, ymax=224] {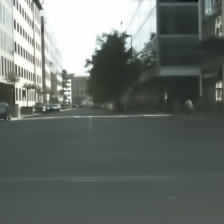};

\nextgroupplot[
width=0.23\columnwidth,
ticks=none,
axis lines=none, 
xtick=\empty, 
ytick=\empty,
axis equal image,
xmin=0, xmax=224,
ymin=0, ymax=224,
title style={yshift=-7pt},
]
\addplot graphics [includegraphics cmd=\pgfimage, xmin=0, xmax=224, ymin=0, ymax=224] {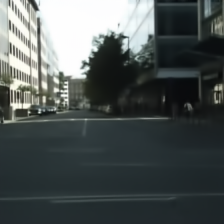};

\nextgroupplot[
width=0.23\columnwidth,
ticks=none,
axis lines=none, 
xtick=\empty, 
ytick=\empty,
axis equal image,
xmin=0, xmax=224,
ymin=0, ymax=224,
title style={yshift=-7pt},
]
\addplot graphics [includegraphics cmd=\pgfimage, xmin=0, xmax=224, ymin=0, ymax=224] {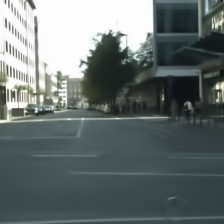};

\nextgroupplot[
width=0.23\columnwidth,
ticks=none,
axis lines=none, 
xtick=\empty, 
ytick=\empty,
axis equal image,
xmin=0, xmax=224,
ymin=0, ymax=224,
title style={yshift=-7pt},
]
\addplot graphics [includegraphics cmd=\pgfimage, xmin=0, xmax=224, ymin=0, ymax=224] {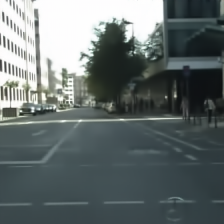};

\nextgroupplot[
width=0.23\columnwidth,
ticks=none,
axis line style={draw=none}, 
xtick=\empty, 
ytick=\empty,
axis equal image,
xmin=0, xmax=224,
ymin=0, ymax=224,
ylabel={$\qp$ map},
title style={yshift=-7pt},
]
\addplot graphics [includegraphics cmd=\pgfimage, xmin=0, xmax=224, ymin=0, ymax=224] {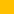};

\nextgroupplot[
width=0.23\columnwidth,
ticks=none,
axis lines=none, 
xtick=\empty, 
ytick=\empty,
axis equal image,
xmin=0, xmax=224,
ymin=0, ymax=224,
title style={yshift=-7pt},
]
\addplot graphics [includegraphics cmd=\pgfimage, xmin=0, xmax=224, ymin=0, ymax=224] {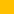};

\nextgroupplot[
width=0.23\columnwidth,
ticks=none,
axis lines=none, 
xtick=\empty, 
ytick=\empty,
axis equal image,
xmin=0, xmax=224,
ymin=0, ymax=224,
title style={yshift=-7pt},
]
\addplot graphics [includegraphics cmd=\pgfimage, xmin=0, xmax=224, ymin=0, ymax=224] {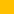};

\nextgroupplot[
width=0.23\columnwidth,
ticks=none,
axis lines=none, 
xtick=\empty, 
ytick=\empty,
axis equal image,
xmin=0, xmax=224,
ymin=0, ymax=224,
title style={yshift=-7pt},
]
\addplot graphics [includegraphics cmd=\pgfimage, xmin=0, xmax=224, ymin=0, ymax=224] {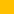};

\nextgroupplot[
width=0.23\columnwidth,
ticks=none,
axis lines=none, 
xtick=\empty, 
ytick=\empty,
axis equal image,
xmin=0, xmax=224,
ymin=0, ymax=224,
title style={yshift=-7pt},
]
\addplot graphics [includegraphics cmd=\pgfimage, xmin=0, xmax=224, ymin=0, ymax=224] {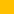};

\nextgroupplot[
width=0.23\columnwidth,
ticks=none,
axis lines=none, 
xtick=\empty, 
ytick=\empty,
axis equal image,
xmin=0, xmax=224,
ymin=0, ymax=224,
title style={yshift=-7pt},
]
\addplot graphics [includegraphics cmd=\pgfimage, xmin=0, xmax=224, ymin=0, ymax=224] {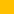};

\nextgroupplot[
width=0.23\columnwidth,
ticks=none,
axis lines=none, 
xtick=\empty, 
ytick=\empty,
axis equal image,
xmin=0, xmax=224,
ymin=0, ymax=224,
title style={yshift=-7pt},
]
\addplot graphics [includegraphics cmd=\pgfimage, xmin=0, xmax=224, ymin=0, ymax=224] {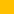};

\nextgroupplot[
width=0.23\columnwidth,
ticks=none,
axis lines=none, 
xtick=\empty, 
ytick=\empty,
axis equal image,
xmin=0, xmax=224,
ymin=0, ymax=224,
title style={yshift=-7pt},
colorbar,
colorbar style={
    ylabel={},
    ytick={5, 25, 45},
    yticklabels={5, 25, 45},
    at={(1.025,1)},
    yticklabel style={xshift=-1.5pt},
},
colormap={mymap}{[1pt]
    rgb(0pt)=(0,0,0.5);
    rgb(22pt)=(0,0,1);
    rgb(25pt)=(0,0,1);
    rgb(68pt)=(0,0.86,1);
    rgb(70pt)=(0,0.9,0.967741935483871);
    rgb(75pt)=(0.0806451612903226,1,0.887096774193548);
    rgb(128pt)=(0.935483870967742,1,0.0322580645161291);
    rgb(130pt)=(0.967741935483871,0.962962962962963,0);
    rgb(132pt)=(1,0.925925925925926,0);
    rgb(178pt)=(1,0.0740740740740741,0);
    rgb(182pt)=(0.909090909090909,0,0);
    rgb(200pt)=(0.5,0,0)
},
point meta max=51,
point meta min=0,
]
\addplot graphics [includegraphics cmd=\pgfimage, xmin=0, xmax=224, ymin=0, ymax=224] {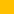};

\end{groupplot}

\end{tikzpicture}

%% file: artwork/surrogate_results/cityscapes/supplement/surrogate_results_cityscapes_high_3.tex
\begin{tikzpicture}[every node/.style={font=\small}, spy using outlines={tud6a, line width=0.80mm, dashed, dash pattern=on 1.5pt off 1.5pt, magnification=2, size=0.75cm, connect spies}]

\begin{groupplot}[group style={group size=8 by 3, horizontal sep=1.25pt, vertical sep=1.25pt}]

\nextgroupplot[
width=0.23\columnwidth,
ticks=none,
axis line style={draw=none}, 
xtick=\empty, 
ytick=\empty,
axis equal image,
xmin=0, xmax=224,
ymin=0, ymax=224,
ylabel={H.264},
title={$t_0$},
title style={yshift=-7pt},
]
\addplot graphics [includegraphics cmd=\pgfimage, xmin=0, xmax=224, ymin=0, ymax=224] {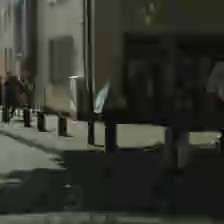};

\nextgroupplot[
width=0.23\columnwidth,
ticks=none,
axis lines=none, 
xtick=\empty, 
ytick=\empty,
axis equal image,
xmin=0, xmax=224,
ymin=0, ymax=224,
title={$t_1$},
title style={yshift=-7pt},
]
\addplot graphics [includegraphics cmd=\pgfimage, xmin=0, xmax=224, ymin=0, ymax=224] {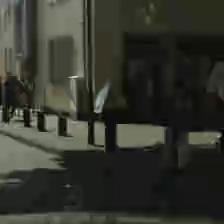};

\nextgroupplot[
width=0.23\columnwidth,
ticks=none,
axis lines=none, 
xtick=\empty, 
ytick=\empty,
axis equal image,
xmin=0, xmax=224,
ymin=0, ymax=224,
title={$t_2$},
title style={yshift=-7pt},
]
\addplot graphics [includegraphics cmd=\pgfimage, xmin=0, xmax=224, ymin=0, ymax=224] {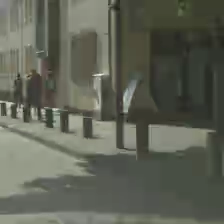};

\nextgroupplot[
width=0.23\columnwidth,
ticks=none,
axis lines=none, 
xtick=\empty, 
ytick=\empty,
axis equal image,
xmin=0, xmax=224,
ymin=0, ymax=224,
title={$t_3$},
title style={yshift=-7pt},
]
\addplot graphics [includegraphics cmd=\pgfimage, xmin=0, xmax=224, ymin=0, ymax=224] {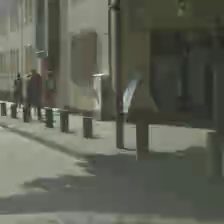};

\nextgroupplot[
width=0.23\columnwidth,
ticks=none,
axis lines=none, 
xtick=\empty, 
ytick=\empty,
axis equal image,
xmin=0, xmax=224,
ymin=0, ymax=224,
title={$t_4$},
title style={yshift=-7pt},
]
\addplot graphics [includegraphics cmd=\pgfimage, xmin=0, xmax=224, ymin=0, ymax=224] {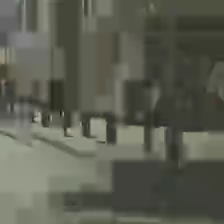};

\nextgroupplot[
width=0.23\columnwidth,
ticks=none,
axis lines=none, 
xtick=\empty, 
ytick=\empty,
axis equal image,
xmin=0, xmax=224,
ymin=0, ymax=224,
title={$t_5$},
title style={yshift=-7pt},
]
\addplot graphics [includegraphics cmd=\pgfimage, xmin=0, xmax=224, ymin=0, ymax=224] {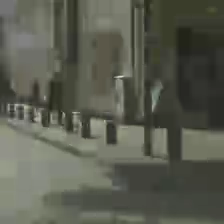};

\nextgroupplot[
width=0.23\columnwidth,
ticks=none,
axis lines=none, 
xtick=\empty, 
ytick=\empty,
axis equal image,
xmin=0, xmax=224,
ymin=0, ymax=224,
title={$t_6$},
title style={yshift=-7pt},
]
\addplot graphics [includegraphics cmd=\pgfimage, xmin=0, xmax=224, ymin=0, ymax=224] {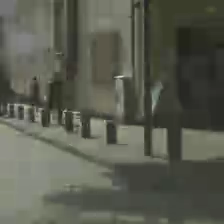};

\nextgroupplot[
width=0.23\columnwidth,
ticks=none,
axis lines=none, 
xtick=\empty, 
ytick=\empty,
axis equal image,
xmin=0, xmax=224,
ymin=0, ymax=224,
title={$t_7$},
title style={yshift=-7pt},
]
\addplot graphics [includegraphics cmd=\pgfimage, xmin=0, xmax=224, ymin=0, ymax=224] {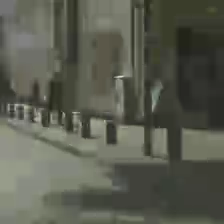};

\nextgroupplot[
width=0.23\columnwidth,
ticks=none,
axis line style={draw=none}, 
xtick=\empty, 
ytick=\empty,
axis equal image,
xmin=0, xmax=224,
ymin=0, ymax=224,
ylabel={Surrogate},
title style={yshift=-7pt},
]
\addplot graphics [includegraphics cmd=\pgfimage, xmin=0, xmax=224, ymin=0, ymax=224] {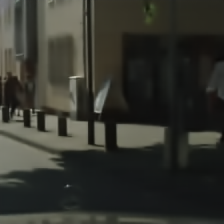};

\nextgroupplot[
width=0.23\columnwidth,
ticks=none,
axis lines=none, 
xtick=\empty, 
ytick=\empty,
axis equal image,
xmin=0, xmax=224,
ymin=0, ymax=224,
title style={yshift=-7pt},
]
\addplot graphics [includegraphics cmd=\pgfimage, xmin=0, xmax=224, ymin=0, ymax=224] {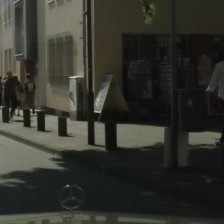};

\nextgroupplot[
width=0.23\columnwidth,
ticks=none,
axis lines=none, 
xtick=\empty, 
ytick=\empty,
axis equal image,
xmin=0, xmax=224,
ymin=0, ymax=224,
title style={yshift=-7pt},
]
\addplot graphics [includegraphics cmd=\pgfimage, xmin=0, xmax=224, ymin=0, ymax=224] {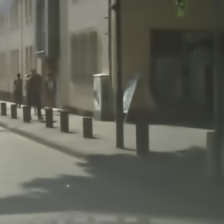};

\nextgroupplot[
width=0.23\columnwidth,
ticks=none,
axis lines=none, 
xtick=\empty, 
ytick=\empty,
axis equal image,
xmin=0, xmax=224,
ymin=0, ymax=224,
title style={yshift=-7pt},
]
\addplot graphics [includegraphics cmd=\pgfimage, xmin=0, xmax=224, ymin=0, ymax=224] {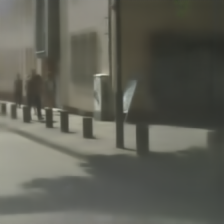};

\nextgroupplot[
width=0.23\columnwidth,
ticks=none,
axis lines=none, 
xtick=\empty, 
ytick=\empty,
axis equal image,
xmin=0, xmax=224,
ymin=0, ymax=224,
title style={yshift=-7pt},
]
\addplot graphics [includegraphics cmd=\pgfimage, xmin=0, xmax=224, ymin=0, ymax=224] {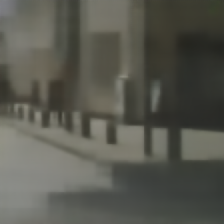};

\nextgroupplot[
width=0.23\columnwidth,
ticks=none,
axis lines=none, 
xtick=\empty, 
ytick=\empty,
axis equal image,
xmin=0, xmax=224,
ymin=0, ymax=224,
title style={yshift=-7pt},
]
\addplot graphics [includegraphics cmd=\pgfimage, xmin=0, xmax=224, ymin=0, ymax=224] {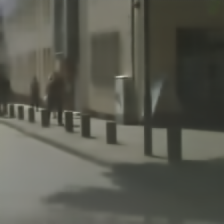};

\nextgroupplot[
width=0.23\columnwidth,
ticks=none,
axis lines=none, 
xtick=\empty, 
ytick=\empty,
axis equal image,
xmin=0, xmax=224,
ymin=0, ymax=224,
title style={yshift=-7pt},
]
\addplot graphics [includegraphics cmd=\pgfimage, xmin=0, xmax=224, ymin=0, ymax=224] {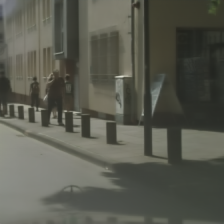};

\nextgroupplot[
width=0.23\columnwidth,
ticks=none,
axis lines=none, 
xtick=\empty, 
ytick=\empty,
axis equal image,
xmin=0, xmax=224,
ymin=0, ymax=224,
title style={yshift=-7pt},
]
\addplot graphics [includegraphics cmd=\pgfimage, xmin=0, xmax=224, ymin=0, ymax=224] {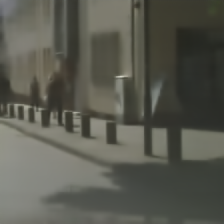};

\nextgroupplot[
width=0.23\columnwidth,
ticks=none,
axis line style={draw=none}, 
xtick=\empty, 
ytick=\empty,
axis equal image,
xmin=0, xmax=224,
ymin=0, ymax=224,
ylabel={$\qp$ map},
title style={yshift=-7pt},
]
\addplot graphics [includegraphics cmd=\pgfimage, xmin=0, xmax=224, ymin=0, ymax=224] {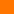};

\nextgroupplot[
width=0.23\columnwidth,
ticks=none,
axis lines=none, 
xtick=\empty, 
ytick=\empty,
axis equal image,
xmin=0, xmax=224,
ymin=0, ymax=224,
title style={yshift=-7pt},
]
\addplot graphics [includegraphics cmd=\pgfimage, xmin=0, xmax=224, ymin=0, ymax=224] {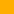};

\nextgroupplot[
width=0.23\columnwidth,
ticks=none,
axis lines=none, 
xtick=\empty, 
ytick=\empty,
axis equal image,
xmin=0, xmax=224,
ymin=0, ymax=224,
title style={yshift=-7pt},
]
\addplot graphics [includegraphics cmd=\pgfimage, xmin=0, xmax=224, ymin=0, ymax=224] {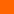};

\nextgroupplot[
width=0.23\columnwidth,
ticks=none,
axis lines=none, 
xtick=\empty, 
ytick=\empty,
axis equal image,
xmin=0, xmax=224,
ymin=0, ymax=224,
title style={yshift=-7pt},
]
\addplot graphics [includegraphics cmd=\pgfimage, xmin=0, xmax=224, ymin=0, ymax=224] {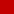};

\nextgroupplot[
width=0.23\columnwidth,
ticks=none,
axis lines=none, 
xtick=\empty, 
ytick=\empty,
axis equal image,
xmin=0, xmax=224,
ymin=0, ymax=224,
title style={yshift=-7pt},
]
\addplot graphics [includegraphics cmd=\pgfimage, xmin=0, xmax=224, ymin=0, ymax=224] {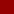};

\nextgroupplot[
width=0.23\columnwidth,
ticks=none,
axis lines=none, 
xtick=\empty, 
ytick=\empty,
axis equal image,
xmin=0, xmax=224,
ymin=0, ymax=224,
title style={yshift=-7pt},
]
\addplot graphics [includegraphics cmd=\pgfimage, xmin=0, xmax=224, ymin=0, ymax=224] {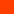};

\nextgroupplot[
width=0.23\columnwidth,
ticks=none,
axis lines=none, 
xtick=\empty, 
ytick=\empty,
axis equal image,
xmin=0, xmax=224,
ymin=0, ymax=224,
title style={yshift=-7pt},
]
\addplot graphics [includegraphics cmd=\pgfimage, xmin=0, xmax=224, ymin=0, ymax=224] {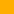};

\nextgroupplot[
width=0.23\columnwidth,
ticks=none,
axis lines=none, 
xtick=\empty, 
ytick=\empty,
axis equal image,
xmin=0, xmax=224,
ymin=0, ymax=224,
title style={yshift=-7pt},
colorbar,
colorbar style={
    ylabel={},
    ytick={5, 25, 45},
    yticklabels={5, 25, 45},
    at={(1.025,1)},
    yticklabel style={xshift=-1.5pt},
},
colormap={mymap}{[1pt]
    rgb(0pt)=(0,0,0.5);
    rgb(22pt)=(0,0,1);
    rgb(25pt)=(0,0,1);
    rgb(68pt)=(0,0.86,1);
    rgb(70pt)=(0,0.9,0.967741935483871);
    rgb(75pt)=(0.0806451612903226,1,0.887096774193548);
    rgb(128pt)=(0.935483870967742,1,0.0322580645161291);
    rgb(130pt)=(0.967741935483871,0.962962962962963,0);
    rgb(132pt)=(1,0.925925925925926,0);
    rgb(178pt)=(1,0.0740740740740741,0);
    rgb(182pt)=(0.909090909090909,0,0);
    rgb(200pt)=(0.5,0,0)
},
point meta max=51,
point meta min=0,
]
\addplot graphics [includegraphics cmd=\pgfimage, xmin=0, xmax=224, ymin=0, ymax=224] {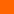};

\end{groupplot}

\end{tikzpicture}

%% file: artwork/surrogate_results/cityscapes/supplement/surrogate_results_cityscapes_low_1.tex
\begin{tikzpicture}[every node/.style={font=\small}, spy using outlines={tud6a, line width=0.80mm, dashed, dash pattern=on 1.5pt off 1.5pt, magnification=2, size=0.75cm, connect spies}]

\begin{groupplot}[group style={group size=8 by 3, horizontal sep=1.25pt, vertical sep=1.25pt}]

\nextgroupplot[
width=0.23\columnwidth,
ticks=none,
axis line style={draw=none}, 
xtick=\empty, 
ytick=\empty,
axis equal image,
xmin=0, xmax=224,
ymin=0, ymax=224,
ylabel={H.264},
title={$t_0$},
title style={yshift=-7pt},
]
\addplot graphics [includegraphics cmd=\pgfimage, xmin=0, xmax=224, ymin=0, ymax=224] {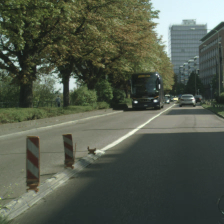};

\nextgroupplot[
width=0.23\columnwidth,
ticks=none,
axis lines=none, 
xtick=\empty, 
ytick=\empty,
axis equal image,
xmin=0, xmax=224,
ymin=0, ymax=224,
title={$t_1$},
title style={yshift=-7pt},
]
\addplot graphics [includegraphics cmd=\pgfimage, xmin=0, xmax=224, ymin=0, ymax=224] {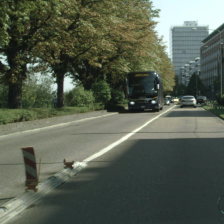};

\nextgroupplot[
width=0.23\columnwidth,
ticks=none,
axis lines=none, 
xtick=\empty, 
ytick=\empty,
axis equal image,
xmin=0, xmax=224,
ymin=0, ymax=224,
title={$t_2$},
title style={yshift=-7pt},
]
\addplot graphics [includegraphics cmd=\pgfimage, xmin=0, xmax=224, ymin=0, ymax=224] {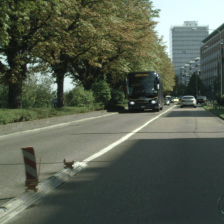};

\nextgroupplot[
width=0.23\columnwidth,
ticks=none,
axis lines=none, 
xtick=\empty, 
ytick=\empty,
axis equal image,
xmin=0, xmax=224,
ymin=0, ymax=224,
title={$t_3$},
title style={yshift=-7pt},
]
\addplot graphics [includegraphics cmd=\pgfimage, xmin=0, xmax=224, ymin=0, ymax=224] {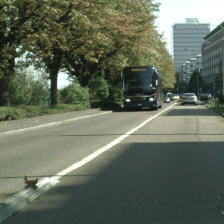};

\nextgroupplot[
width=0.23\columnwidth,
ticks=none,
axis lines=none, 
xtick=\empty, 
ytick=\empty,
axis equal image,
xmin=0, xmax=224,
ymin=0, ymax=224,
title={$t_4$},
title style={yshift=-7pt},
]
\addplot graphics [includegraphics cmd=\pgfimage, xmin=0, xmax=224, ymin=0, ymax=224] {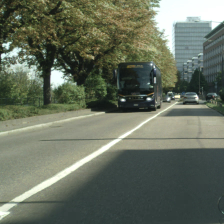};

\nextgroupplot[
width=0.23\columnwidth,
ticks=none,
axis lines=none, 
xtick=\empty, 
ytick=\empty,
axis equal image,
xmin=0, xmax=224,
ymin=0, ymax=224,
title={$t_5$},
title style={yshift=-7pt},
]
\addplot graphics [includegraphics cmd=\pgfimage, xmin=0, xmax=224, ymin=0, ymax=224] {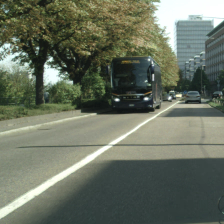};

\nextgroupplot[
width=0.23\columnwidth,
ticks=none,
axis lines=none, 
xtick=\empty, 
ytick=\empty,
axis equal image,
xmin=0, xmax=224,
ymin=0, ymax=224,
title={$t_6$},
title style={yshift=-7pt},
]
\addplot graphics [includegraphics cmd=\pgfimage, xmin=0, xmax=224, ymin=0, ymax=224] {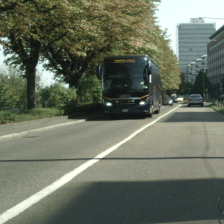};

\nextgroupplot[
width=0.23\columnwidth,
ticks=none,
axis lines=none, 
xtick=\empty, 
ytick=\empty,
axis equal image,
xmin=0, xmax=224,
ymin=0, ymax=224,
title={$t_7$},
title style={yshift=-7pt},
]
\addplot graphics [includegraphics cmd=\pgfimage, xmin=0, xmax=224, ymin=0, ymax=224] {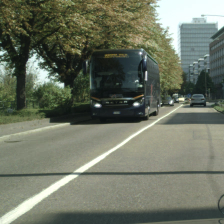};

\nextgroupplot[
width=0.23\columnwidth,
ticks=none,
axis line style={draw=none}, 
xtick=\empty, 
ytick=\empty,
axis equal image,
xmin=0, xmax=224,
ymin=0, ymax=224,
ylabel={Surrogate},
title style={yshift=-7pt},
]
\addplot graphics [includegraphics cmd=\pgfimage, xmin=0, xmax=224, ymin=0, ymax=224] {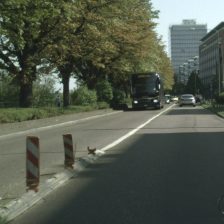};

\nextgroupplot[
width=0.23\columnwidth,
ticks=none,
axis lines=none, 
xtick=\empty, 
ytick=\empty,
axis equal image,
xmin=0, xmax=224,
ymin=0, ymax=224,
title style={yshift=-7pt},
]
\addplot graphics [includegraphics cmd=\pgfimage, xmin=0, xmax=224, ymin=0, ymax=224] {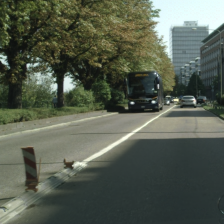};

\nextgroupplot[
width=0.23\columnwidth,
ticks=none,
axis lines=none, 
xtick=\empty, 
ytick=\empty,
axis equal image,
xmin=0, xmax=224,
ymin=0, ymax=224,
title style={yshift=-7pt},
]
\addplot graphics [includegraphics cmd=\pgfimage, xmin=0, xmax=224, ymin=0, ymax=224] {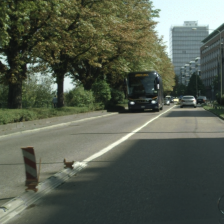};

\nextgroupplot[
width=0.23\columnwidth,
ticks=none,
axis lines=none, 
xtick=\empty, 
ytick=\empty,
axis equal image,
xmin=0, xmax=224,
ymin=0, ymax=224,
title style={yshift=-7pt},
]
\addplot graphics [includegraphics cmd=\pgfimage, xmin=0, xmax=224, ymin=0, ymax=224] {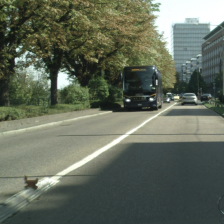};

\nextgroupplot[
width=0.23\columnwidth,
ticks=none,
axis lines=none, 
xtick=\empty, 
ytick=\empty,
axis equal image,
xmin=0, xmax=224,
ymin=0, ymax=224,
title style={yshift=-7pt},
]
\addplot graphics [includegraphics cmd=\pgfimage, xmin=0, xmax=224, ymin=0, ymax=224] {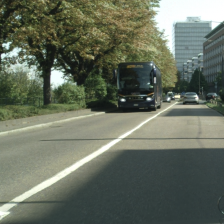};

\nextgroupplot[
width=0.23\columnwidth,
ticks=none,
axis lines=none, 
xtick=\empty, 
ytick=\empty,
axis equal image,
xmin=0, xmax=224,
ymin=0, ymax=224,
title style={yshift=-7pt},
]
\addplot graphics [includegraphics cmd=\pgfimage, xmin=0, xmax=224, ymin=0, ymax=224] {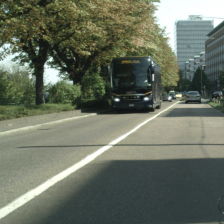};

\nextgroupplot[
width=0.23\columnwidth,
ticks=none,
axis lines=none, 
xtick=\empty, 
ytick=\empty,
axis equal image,
xmin=0, xmax=224,
ymin=0, ymax=224,
title style={yshift=-7pt},
]
\addplot graphics [includegraphics cmd=\pgfimage, xmin=0, xmax=224, ymin=0, ymax=224] {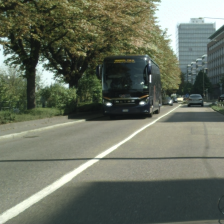};

\nextgroupplot[
width=0.23\columnwidth,
ticks=none,
axis lines=none, 
xtick=\empty, 
ytick=\empty,
axis equal image,
xmin=0, xmax=224,
ymin=0, ymax=224,
title style={yshift=-7pt},
]
\addplot graphics [includegraphics cmd=\pgfimage, xmin=0, xmax=224, ymin=0, ymax=224] {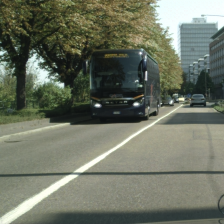};

\nextgroupplot[
width=0.23\columnwidth,
ticks=none,
axis line style={draw=none}, 
xtick=\empty, 
ytick=\empty,
axis equal image,
xmin=0, xmax=224,
ymin=0, ymax=224,
ylabel={$\qp$ map},
title style={yshift=-7pt},
]
\addplot graphics [includegraphics cmd=\pgfimage, xmin=0, xmax=224, ymin=0, ymax=224] {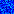};

\nextgroupplot[
width=0.23\columnwidth,
ticks=none,
axis lines=none, 
xtick=\empty, 
ytick=\empty,
axis equal image,
xmin=0, xmax=224,
ymin=0, ymax=224,
title style={yshift=-7pt},
]
\addplot graphics [includegraphics cmd=\pgfimage, xmin=0, xmax=224, ymin=0, ymax=224] {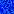};

\nextgroupplot[
width=0.23\columnwidth,
ticks=none,
axis lines=none, 
xtick=\empty, 
ytick=\empty,
axis equal image,
xmin=0, xmax=224,
ymin=0, ymax=224,
title style={yshift=-7pt},
]
\addplot graphics [includegraphics cmd=\pgfimage, xmin=0, xmax=224, ymin=0, ymax=224] {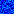};

\nextgroupplot[
width=0.23\columnwidth,
ticks=none,
axis lines=none, 
xtick=\empty, 
ytick=\empty,
axis equal image,
xmin=0, xmax=224,
ymin=0, ymax=224,
title style={yshift=-7pt},
]
\addplot graphics [includegraphics cmd=\pgfimage, xmin=0, xmax=224, ymin=0, ymax=224] {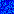};

\nextgroupplot[
width=0.23\columnwidth,
ticks=none,
axis lines=none, 
xtick=\empty, 
ytick=\empty,
axis equal image,
xmin=0, xmax=224,
ymin=0, ymax=224,
title style={yshift=-7pt},
]
\addplot graphics [includegraphics cmd=\pgfimage, xmin=0, xmax=224, ymin=0, ymax=224] {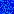};

\nextgroupplot[
width=0.23\columnwidth,
ticks=none,
axis lines=none, 
xtick=\empty, 
ytick=\empty,
axis equal image,
xmin=0, xmax=224,
ymin=0, ymax=224,
title style={yshift=-7pt},
]
\addplot graphics [includegraphics cmd=\pgfimage, xmin=0, xmax=224, ymin=0, ymax=224] {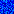};

\nextgroupplot[
width=0.23\columnwidth,
ticks=none,
axis lines=none, 
xtick=\empty, 
ytick=\empty,
axis equal image,
xmin=0, xmax=224,
ymin=0, ymax=224,
title style={yshift=-7pt},
]
\addplot graphics [includegraphics cmd=\pgfimage, xmin=0, xmax=224, ymin=0, ymax=224] {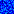};

\nextgroupplot[
width=0.23\columnwidth,
ticks=none,
axis lines=none, 
xtick=\empty, 
ytick=\empty,
axis equal image,
xmin=0, xmax=224,
ymin=0, ymax=224,
title style={yshift=-7pt},
colorbar,
colorbar style={
    ylabel={},
    ytick={5, 25, 45},
    yticklabels={5, 25, 45},
    at={(1.025,1)},
    yticklabel style={xshift=-1.5pt},
},
colormap={mymap}{[1pt]
    rgb(0pt)=(0,0,0.5);
    rgb(22pt)=(0,0,1);
    rgb(25pt)=(0,0,1);
    rgb(68pt)=(0,0.86,1);
    rgb(70pt)=(0,0.9,0.967741935483871);
    rgb(75pt)=(0.0806451612903226,1,0.887096774193548);
    rgb(128pt)=(0.935483870967742,1,0.0322580645161291);
    rgb(130pt)=(0.967741935483871,0.962962962962963,0);
    rgb(132pt)=(1,0.925925925925926,0);
    rgb(178pt)=(1,0.0740740740740741,0);
    rgb(182pt)=(0.909090909090909,0,0);
    rgb(200pt)=(0.5,0,0)
},
point meta max=51,
point meta min=0,
]
\addplot graphics [includegraphics cmd=\pgfimage, xmin=0, xmax=224, ymin=0, ymax=224] {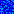};

\end{groupplot}

\end{tikzpicture}

%% file: artwork/surrogate_results/cityscapes/supplement/surrogate_results_cityscapes_low_2.tex
\begin{tikzpicture}[every node/.style={font=\small}, spy using outlines={tud6a, line width=0.80mm, dashed, dash pattern=on 1.5pt off 1.5pt, magnification=2, size=0.75cm, connect spies}]

\begin{groupplot}[group style={group size=8 by 3, horizontal sep=1.25pt, vertical sep=1.25pt}]

\nextgroupplot[
width=0.23\columnwidth,
ticks=none,
axis line style={draw=none}, 
xtick=\empty, 
ytick=\empty,
axis equal image,
xmin=0, xmax=224,
ymin=0, ymax=224,
ylabel={H.264},
title={$t_0$},
title style={yshift=-7pt},
]
\addplot graphics [includegraphics cmd=\pgfimage, xmin=0, xmax=224, ymin=0, ymax=224] {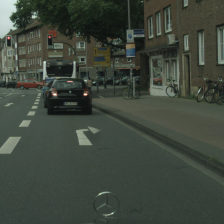};

\nextgroupplot[
width=0.23\columnwidth,
ticks=none,
axis lines=none, 
xtick=\empty, 
ytick=\empty,
axis equal image,
xmin=0, xmax=224,
ymin=0, ymax=224,
title={$t_1$},
title style={yshift=-7pt},
]
\addplot graphics [includegraphics cmd=\pgfimage, xmin=0, xmax=224, ymin=0, ymax=224] {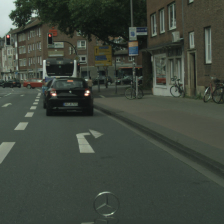};

\nextgroupplot[
width=0.23\columnwidth,
ticks=none,
axis lines=none, 
xtick=\empty, 
ytick=\empty,
axis equal image,
xmin=0, xmax=224,
ymin=0, ymax=224,
title={$t_2$},
title style={yshift=-7pt},
]
\addplot graphics [includegraphics cmd=\pgfimage, xmin=0, xmax=224, ymin=0, ymax=224] {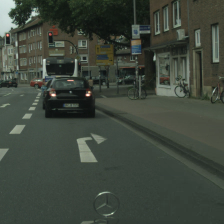};

\nextgroupplot[
width=0.23\columnwidth,
ticks=none,
axis lines=none, 
xtick=\empty, 
ytick=\empty,
axis equal image,
xmin=0, xmax=224,
ymin=0, ymax=224,
title={$t_3$},
title style={yshift=-7pt},
]
\addplot graphics [includegraphics cmd=\pgfimage, xmin=0, xmax=224, ymin=0, ymax=224] {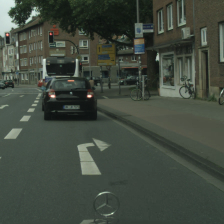};

\nextgroupplot[
width=0.23\columnwidth,
ticks=none,
axis lines=none, 
xtick=\empty, 
ytick=\empty,
axis equal image,
xmin=0, xmax=224,
ymin=0, ymax=224,
title={$t_4$},
title style={yshift=-7pt},
]
\addplot graphics [includegraphics cmd=\pgfimage, xmin=0, xmax=224, ymin=0, ymax=224] {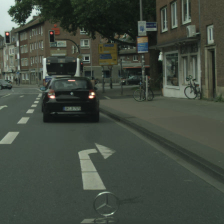};

\nextgroupplot[
width=0.23\columnwidth,
ticks=none,
axis lines=none, 
xtick=\empty, 
ytick=\empty,
axis equal image,
xmin=0, xmax=224,
ymin=0, ymax=224,
title={$t_5$},
title style={yshift=-7pt},
]
\addplot graphics [includegraphics cmd=\pgfimage, xmin=0, xmax=224, ymin=0, ymax=224] {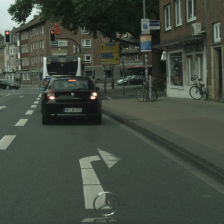};

\nextgroupplot[
width=0.23\columnwidth,
ticks=none,
axis lines=none, 
xtick=\empty, 
ytick=\empty,
axis equal image,
xmin=0, xmax=224,
ymin=0, ymax=224,
title={$t_6$},
title style={yshift=-7pt},
]
\addplot graphics [includegraphics cmd=\pgfimage, xmin=0, xmax=224, ymin=0, ymax=224] {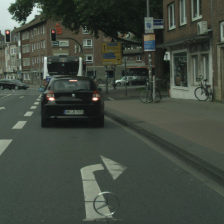};

\nextgroupplot[
width=0.23\columnwidth,
ticks=none,
axis lines=none, 
xtick=\empty, 
ytick=\empty,
axis equal image,
xmin=0, xmax=224,
ymin=0, ymax=224,
title={$t_7$},
title style={yshift=-7pt},
]
\addplot graphics [includegraphics cmd=\pgfimage, xmin=0, xmax=224, ymin=0, ymax=224] {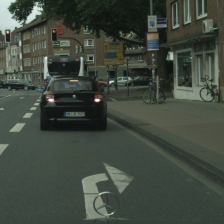};

\nextgroupplot[
width=0.23\columnwidth,
ticks=none,
axis line style={draw=none}, 
xtick=\empty, 
ytick=\empty,
axis equal image,
xmin=0, xmax=224,
ymin=0, ymax=224,
ylabel={Surrogate},
title style={yshift=-7pt},
]
\addplot graphics [includegraphics cmd=\pgfimage, xmin=0, xmax=224, ymin=0, ymax=224] {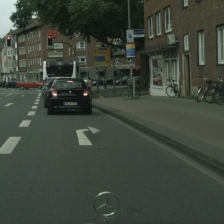};

\nextgroupplot[
width=0.23\columnwidth,
ticks=none,
axis lines=none, 
xtick=\empty, 
ytick=\empty,
axis equal image,
xmin=0, xmax=224,
ymin=0, ymax=224,
title style={yshift=-7pt},
]
\addplot graphics [includegraphics cmd=\pgfimage, xmin=0, xmax=224, ymin=0, ymax=224] {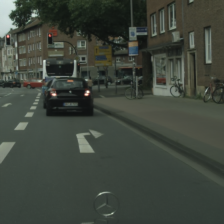};

\nextgroupplot[
width=0.23\columnwidth,
ticks=none,
axis lines=none, 
xtick=\empty, 
ytick=\empty,
axis equal image,
xmin=0, xmax=224,
ymin=0, ymax=224,
title style={yshift=-7pt},
]
\addplot graphics [includegraphics cmd=\pgfimage, xmin=0, xmax=224, ymin=0, ymax=224] {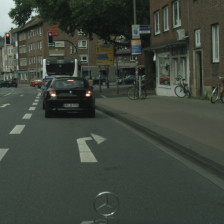};

\nextgroupplot[
width=0.23\columnwidth,
ticks=none,
axis lines=none, 
xtick=\empty, 
ytick=\empty,
axis equal image,
xmin=0, xmax=224,
ymin=0, ymax=224,
title style={yshift=-7pt},
]
\addplot graphics [includegraphics cmd=\pgfimage, xmin=0, xmax=224, ymin=0, ymax=224] {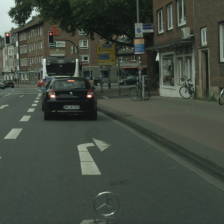};

\nextgroupplot[
width=0.23\columnwidth,
ticks=none,
axis lines=none, 
xtick=\empty, 
ytick=\empty,
axis equal image,
xmin=0, xmax=224,
ymin=0, ymax=224,
title style={yshift=-7pt},
]
\addplot graphics [includegraphics cmd=\pgfimage, xmin=0, xmax=224, ymin=0, ymax=224] {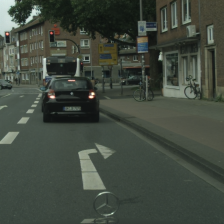};

\nextgroupplot[
width=0.23\columnwidth,
ticks=none,
axis lines=none, 
xtick=\empty, 
ytick=\empty,
axis equal image,
xmin=0, xmax=224,
ymin=0, ymax=224,
title style={yshift=-7pt},
]
\addplot graphics [includegraphics cmd=\pgfimage, xmin=0, xmax=224, ymin=0, ymax=224] {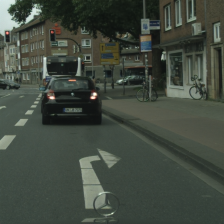};

\nextgroupplot[
width=0.23\columnwidth,
ticks=none,
axis lines=none, 
xtick=\empty, 
ytick=\empty,
axis equal image,
xmin=0, xmax=224,
ymin=0, ymax=224,
title style={yshift=-7pt},
]
\addplot graphics [includegraphics cmd=\pgfimage, xmin=0, xmax=224, ymin=0, ymax=224] {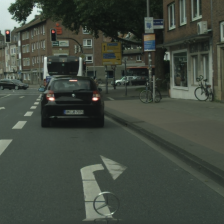};

\nextgroupplot[
width=0.23\columnwidth,
ticks=none,
axis lines=none, 
xtick=\empty, 
ytick=\empty,
axis equal image,
xmin=0, xmax=224,
ymin=0, ymax=224,
title style={yshift=-7pt},
]
\addplot graphics [includegraphics cmd=\pgfimage, xmin=0, xmax=224, ymin=0, ymax=224] {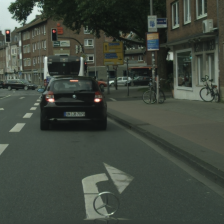};

\nextgroupplot[
width=0.23\columnwidth,
ticks=none,
axis line style={draw=none}, 
xtick=\empty, 
ytick=\empty,
axis equal image,
xmin=0, xmax=224,
ymin=0, ymax=224,
ylabel={$\qp$ map},
title style={yshift=-7pt},
]
\addplot graphics [includegraphics cmd=\pgfimage, xmin=0, xmax=224, ymin=0, ymax=224] {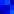};

\nextgroupplot[
width=0.23\columnwidth,
ticks=none,
axis lines=none, 
xtick=\empty, 
ytick=\empty,
axis equal image,
xmin=0, xmax=224,
ymin=0, ymax=224,
title style={yshift=-7pt},
]
\addplot graphics [includegraphics cmd=\pgfimage, xmin=0, xmax=224, ymin=0, ymax=224] {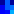};

\nextgroupplot[
width=0.23\columnwidth,
ticks=none,
axis lines=none, 
xtick=\empty, 
ytick=\empty,
axis equal image,
xmin=0, xmax=224,
ymin=0, ymax=224,
title style={yshift=-7pt},
]
\addplot graphics [includegraphics cmd=\pgfimage, xmin=0, xmax=224, ymin=0, ymax=224] {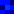};

\nextgroupplot[
width=0.23\columnwidth,
ticks=none,
axis lines=none, 
xtick=\empty, 
ytick=\empty,
axis equal image,
xmin=0, xmax=224,
ymin=0, ymax=224,
title style={yshift=-7pt},
]
\addplot graphics [includegraphics cmd=\pgfimage, xmin=0, xmax=224, ymin=0, ymax=224] {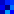};

\nextgroupplot[
width=0.23\columnwidth,
ticks=none,
axis lines=none, 
xtick=\empty, 
ytick=\empty,
axis equal image,
xmin=0, xmax=224,
ymin=0, ymax=224,
title style={yshift=-7pt},
]
\addplot graphics [includegraphics cmd=\pgfimage, xmin=0, xmax=224, ymin=0, ymax=224] {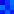};

\nextgroupplot[
width=0.23\columnwidth,
ticks=none,
axis lines=none, 
xtick=\empty, 
ytick=\empty,
axis equal image,
xmin=0, xmax=224,
ymin=0, ymax=224,
title style={yshift=-7pt},
]
\addplot graphics [includegraphics cmd=\pgfimage, xmin=0, xmax=224, ymin=0, ymax=224] {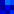};

\nextgroupplot[
width=0.23\columnwidth,
ticks=none,
axis lines=none, 
xtick=\empty, 
ytick=\empty,
axis equal image,
xmin=0, xmax=224,
ymin=0, ymax=224,
title style={yshift=-7pt},
]
\addplot graphics [includegraphics cmd=\pgfimage, xmin=0, xmax=224, ymin=0, ymax=224] {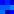};

\nextgroupplot[
width=0.23\columnwidth,
ticks=none,
axis lines=none, 
xtick=\empty, 
ytick=\empty,
axis equal image,
xmin=0, xmax=224,
ymin=0, ymax=224,
title style={yshift=-7pt},
colorbar,
colorbar style={
    ylabel={},
    ytick={5, 25, 45},
    yticklabels={5, 25, 45},
    at={(1.025,1)},
    yticklabel style={xshift=-1.5pt},
},
colormap={mymap}{[1pt]
    rgb(0pt)=(0,0,0.5);
    rgb(22pt)=(0,0,1);
    rgb(25pt)=(0,0,1);
    rgb(68pt)=(0,0.86,1);
    rgb(70pt)=(0,0.9,0.967741935483871);
    rgb(75pt)=(0.0806451612903226,1,0.887096774193548);
    rgb(128pt)=(0.935483870967742,1,0.0322580645161291);
    rgb(130pt)=(0.967741935483871,0.962962962962963,0);
    rgb(132pt)=(1,0.925925925925926,0);
    rgb(178pt)=(1,0.0740740740740741,0);
    rgb(182pt)=(0.909090909090909,0,0);
    rgb(200pt)=(0.5,0,0)
},
point meta max=51,
point meta min=0,
]
\addplot graphics [includegraphics cmd=\pgfimage, xmin=0, xmax=224, ymin=0, ymax=224] {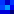};

\end{groupplot}

\end{tikzpicture}

%% file: artwork/surrogate_results/cityscapes/supplement/surrogate_results_cityscapes_low_3.tex
\begin{tikzpicture}[every node/.style={font=\small}, spy using outlines={tud6a, line width=0.80mm, dashed, dash pattern=on 1.5pt off 1.5pt, magnification=2, size=0.75cm, connect spies}]

\begin{groupplot}[group style={group size=8 by 3, horizontal sep=1.25pt, vertical sep=1.25pt}]

\nextgroupplot[
width=0.23\columnwidth,
ticks=none,
axis line style={draw=none}, 
xtick=\empty, 
ytick=\empty,
axis equal image,
xmin=0, xmax=224,
ymin=0, ymax=224,
ylabel={H.264},
title={$t_0$},
title style={yshift=-7pt},
]
\addplot graphics [includegraphics cmd=\pgfimage, xmin=0, xmax=224, ymin=0, ymax=224] {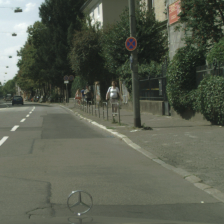};

\nextgroupplot[
width=0.23\columnwidth,
ticks=none,
axis lines=none, 
xtick=\empty, 
ytick=\empty,
axis equal image,
xmin=0, xmax=224,
ymin=0, ymax=224,
title={$t_1$},
title style={yshift=-7pt},
]
\addplot graphics [includegraphics cmd=\pgfimage, xmin=0, xmax=224, ymin=0, ymax=224] {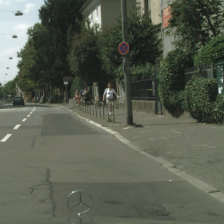};

\nextgroupplot[
width=0.23\columnwidth,
ticks=none,
axis lines=none, 
xtick=\empty, 
ytick=\empty,
axis equal image,
xmin=0, xmax=224,
ymin=0, ymax=224,
title={$t_2$},
title style={yshift=-7pt},
]
\addplot graphics [includegraphics cmd=\pgfimage, xmin=0, xmax=224, ymin=0, ymax=224] {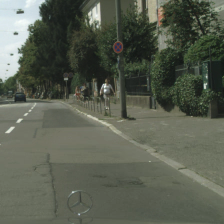};

\nextgroupplot[
width=0.23\columnwidth,
ticks=none,
axis lines=none, 
xtick=\empty, 
ytick=\empty,
axis equal image,
xmin=0, xmax=224,
ymin=0, ymax=224,
title={$t_3$},
title style={yshift=-7pt},
]
\addplot graphics [includegraphics cmd=\pgfimage, xmin=0, xmax=224, ymin=0, ymax=224] {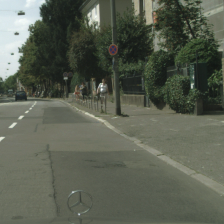};

\nextgroupplot[
width=0.23\columnwidth,
ticks=none,
axis lines=none, 
xtick=\empty, 
ytick=\empty,
axis equal image,
xmin=0, xmax=224,
ymin=0, ymax=224,
title={$t_4$},
title style={yshift=-7pt},
]
\addplot graphics [includegraphics cmd=\pgfimage, xmin=0, xmax=224, ymin=0, ymax=224] {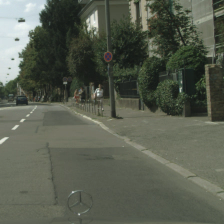};

\nextgroupplot[
width=0.23\columnwidth,
ticks=none,
axis lines=none, 
xtick=\empty, 
ytick=\empty,
axis equal image,
xmin=0, xmax=224,
ymin=0, ymax=224,
title={$t_5$},
title style={yshift=-7pt},
]
\addplot graphics [includegraphics cmd=\pgfimage, xmin=0, xmax=224, ymin=0, ymax=224] {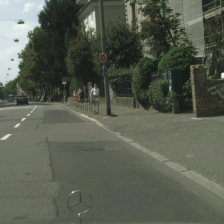};

\nextgroupplot[
width=0.23\columnwidth,
ticks=none,
axis lines=none, 
xtick=\empty, 
ytick=\empty,
axis equal image,
xmin=0, xmax=224,
ymin=0, ymax=224,
title={$t_6$},
title style={yshift=-7pt},
]
\addplot graphics [includegraphics cmd=\pgfimage, xmin=0, xmax=224, ymin=0, ymax=224] {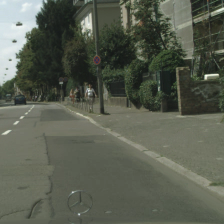};

\nextgroupplot[
width=0.23\columnwidth,
ticks=none,
axis lines=none, 
xtick=\empty, 
ytick=\empty,
axis equal image,
xmin=0, xmax=224,
ymin=0, ymax=224,
title={$t_7$},
title style={yshift=-7pt},
]
\addplot graphics [includegraphics cmd=\pgfimage, xmin=0, xmax=224, ymin=0, ymax=224] {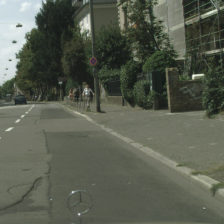};

\nextgroupplot[
width=0.23\columnwidth,
ticks=none,
axis line style={draw=none}, 
xtick=\empty, 
ytick=\empty,
axis equal image,
xmin=0, xmax=224,
ymin=0, ymax=224,
ylabel={Surrogate},
title style={yshift=-7pt},
]
\addplot graphics [includegraphics cmd=\pgfimage, xmin=0, xmax=224, ymin=0, ymax=224] {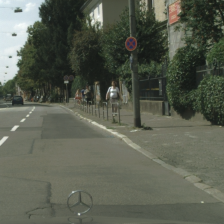};

\nextgroupplot[
width=0.23\columnwidth,
ticks=none,
axis lines=none, 
xtick=\empty, 
ytick=\empty,
axis equal image,
xmin=0, xmax=224,
ymin=0, ymax=224,
title style={yshift=-7pt},
]
\addplot graphics [includegraphics cmd=\pgfimage, xmin=0, xmax=224, ymin=0, ymax=224] {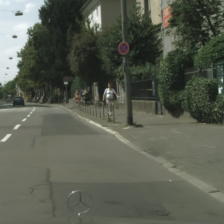};

\nextgroupplot[
width=0.23\columnwidth,
ticks=none,
axis lines=none, 
xtick=\empty, 
ytick=\empty,
axis equal image,
xmin=0, xmax=224,
ymin=0, ymax=224,
title style={yshift=-7pt},
]
\addplot graphics [includegraphics cmd=\pgfimage, xmin=0, xmax=224, ymin=0, ymax=224] {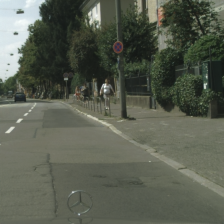};

\nextgroupplot[
width=0.23\columnwidth,
ticks=none,
axis lines=none, 
xtick=\empty, 
ytick=\empty,
axis equal image,
xmin=0, xmax=224,
ymin=0, ymax=224,
title style={yshift=-7pt},
]
\addplot graphics [includegraphics cmd=\pgfimage, xmin=0, xmax=224, ymin=0, ymax=224] {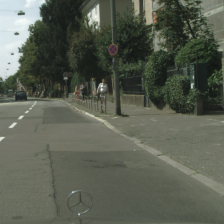};

\nextgroupplot[
width=0.23\columnwidth,
ticks=none,
axis lines=none, 
xtick=\empty, 
ytick=\empty,
axis equal image,
xmin=0, xmax=224,
ymin=0, ymax=224,
title style={yshift=-7pt},
]
\addplot graphics [includegraphics cmd=\pgfimage, xmin=0, xmax=224, ymin=0, ymax=224] {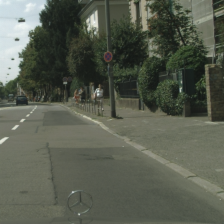};

\nextgroupplot[
width=0.23\columnwidth,
ticks=none,
axis lines=none, 
xtick=\empty, 
ytick=\empty,
axis equal image,
xmin=0, xmax=224,
ymin=0, ymax=224,
title style={yshift=-7pt},
]
\addplot graphics [includegraphics cmd=\pgfimage, xmin=0, xmax=224, ymin=0, ymax=224] {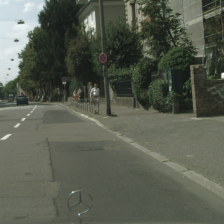};

\nextgroupplot[
width=0.23\columnwidth,
ticks=none,
axis lines=none, 
xtick=\empty, 
ytick=\empty,
axis equal image,
xmin=0, xmax=224,
ymin=0, ymax=224,
title style={yshift=-7pt},
]
\addplot graphics [includegraphics cmd=\pgfimage, xmin=0, xmax=224, ymin=0, ymax=224] {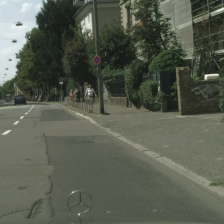};

\nextgroupplot[
width=0.23\columnwidth,
ticks=none,
axis lines=none, 
xtick=\empty, 
ytick=\empty,
axis equal image,
xmin=0, xmax=224,
ymin=0, ymax=224,
title style={yshift=-7pt},
]
\addplot graphics [includegraphics cmd=\pgfimage, xmin=0, xmax=224, ymin=0, ymax=224] {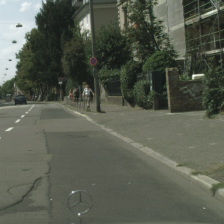};

\nextgroupplot[
width=0.23\columnwidth,
ticks=none,
axis line style={draw=none}, 
xtick=\empty, 
ytick=\empty,
axis equal image,
xmin=0, xmax=224,
ymin=0, ymax=224,
ylabel={$\qp$ map},
title style={yshift=-7pt},
]
\addplot graphics [includegraphics cmd=\pgfimage, xmin=0, xmax=224, ymin=0, ymax=224] {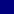};

\nextgroupplot[
width=0.23\columnwidth,
ticks=none,
axis lines=none, 
xtick=\empty, 
ytick=\empty,
axis equal image,
xmin=0, xmax=224,
ymin=0, ymax=224,
title style={yshift=-7pt},
]
\addplot graphics [includegraphics cmd=\pgfimage, xmin=0, xmax=224, ymin=0, ymax=224] {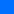};

\nextgroupplot[
width=0.23\columnwidth,
ticks=none,
axis lines=none, 
xtick=\empty, 
ytick=\empty,
axis equal image,
xmin=0, xmax=224,
ymin=0, ymax=224,
title style={yshift=-7pt},
]
\addplot graphics [includegraphics cmd=\pgfimage, xmin=0, xmax=224, ymin=0, ymax=224] {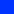};

\nextgroupplot[
width=0.23\columnwidth,
ticks=none,
axis lines=none, 
xtick=\empty, 
ytick=\empty,
axis equal image,
xmin=0, xmax=224,
ymin=0, ymax=224,
title style={yshift=-7pt},
]
\addplot graphics [includegraphics cmd=\pgfimage, xmin=0, xmax=224, ymin=0, ymax=224] {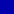};

\nextgroupplot[
width=0.23\columnwidth,
ticks=none,
axis lines=none, 
xtick=\empty, 
ytick=\empty,
axis equal image,
xmin=0, xmax=224,
ymin=0, ymax=224,
title style={yshift=-7pt},
]
\addplot graphics [includegraphics cmd=\pgfimage, xmin=0, xmax=224, ymin=0, ymax=224] {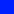};

\nextgroupplot[
width=0.23\columnwidth,
ticks=none,
axis lines=none, 
xtick=\empty, 
ytick=\empty,
axis equal image,
xmin=0, xmax=224,
ymin=0, ymax=224,
title style={yshift=-7pt},
]
\addplot graphics [includegraphics cmd=\pgfimage, xmin=0, xmax=224, ymin=0, ymax=224] {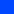};

\nextgroupplot[
width=0.23\columnwidth,
ticks=none,
axis lines=none, 
xtick=\empty, 
ytick=\empty,
axis equal image,
xmin=0, xmax=224,
ymin=0, ymax=224,
title style={yshift=-7pt},
]
\addplot graphics [includegraphics cmd=\pgfimage, xmin=0, xmax=224, ymin=0, ymax=224] {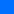};

\nextgroupplot[
width=0.23\columnwidth,
ticks=none,
axis lines=none, 
xtick=\empty, 
ytick=\empty,
axis equal image,
xmin=0, xmax=224,
ymin=0, ymax=224,
title style={yshift=-7pt},
colorbar,
colorbar style={
    ylabel={},
    ytick={5, 25, 45},
    yticklabels={5, 25, 45},
    at={(1.025,1)},
    yticklabel style={xshift=-1.5pt},
},
colormap={mymap}{[1pt]
    rgb(0pt)=(0,0,0.5);
    rgb(22pt)=(0,0,1);
    rgb(25pt)=(0,0,1);
    rgb(68pt)=(0,0.86,1);
    rgb(70pt)=(0,0.9,0.967741935483871);
    rgb(75pt)=(0.0806451612903226,1,0.887096774193548);
    rgb(128pt)=(0.935483870967742,1,0.0322580645161291);
    rgb(130pt)=(0.967741935483871,0.962962962962963,0);
    rgb(132pt)=(1,0.925925925925926,0);
    rgb(178pt)=(1,0.0740740740740741,0);
    rgb(182pt)=(0.909090909090909,0,0);
    rgb(200pt)=(0.5,0,0)
},
point meta max=51,
point meta min=0,
]
\addplot graphics [includegraphics cmd=\pgfimage, xmin=0, xmax=224, ymin=0, ymax=224] {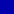};

\end{groupplot}

\end{tikzpicture}

%% file: main.bbl
\begin{thebibliography}{73}
\providecommand{\natexlab}[1]{#1}
\providecommand{\url}[1]{\texttt{#1}}
\expandafter\ifx\csname urlstyle\endcsname\relax
  \providecommand{\doi}[1]{doi: #1}\else
  \providecommand{\doi}{doi: \begingroup \urlstyle{rm}\Url}\fi

\bibitem[Araslanov and Roth(2021)]{Araslanov2021}
Nikita Araslanov and Stefan Roth.
\newblock Self-supervised augmentation consistency for adapting semantic
  segmentation.
\newblock In \emph{{CVPR}}, pages 15384--15394, 2021.

\bibitem[Ballas et~al.(2016)Ballas, Yao, Pal, and Courville]{Ballas2016}
Nicolas Ballas, Li Yao, Chris Pal, and Aaron~C. Courville.
\newblock Delving deeper into convolutional networks for learning video
  representations.
\newblock In \emph{{ICLR}}, 2016.

\bibitem[Barnett et~al.(2018)Barnett, Jain, Andra, and Khurana]{Barnett2018}
Thomas Barnett, Shruti Jain, Usha Andra, and Taru Khurana.
\newblock Cisco visual networking index {(VNI)} complete forecast update.
\newblock \emph{{Americas/EMEAR CKN Presentation}}, pages 1--30, 2018.

\bibitem[Beye et~al.(2022)Beye, Itsumi, Vitthal, and Nihei]{Beye2022}
Florian Beye, Hayato Itsumi, Charvi Vitthal, and Koichi Nihei.
\newblock Recognition-aware deep video compression for remote surveillance.
\newblock In \emph{ICIP}, pages 1986--1990, 2022.

\bibitem[Brostow et~al.(2009)Brostow, Fauqueur, and Cipolla]{Brostow2009}
Gabriel~J. Brostow, Julien Fauqueur, and Roberto Cipolla.
\newblock {Semantic object classes in video: A high-definition ground truth
  database}.
\newblock \emph{{Pattern Recognit. Lett.}}, 30\penalty0 (2):\penalty0 88--97,
  2009.

\bibitem[Bu{\c{s}}oniu et~al.(2018)Bu{\c{s}}oniu, de~Bruin, Toli{\'c}, Kober,
  and Palunko]{Bucsoniu2018}
Lucian Bu{\c{s}}oniu, Tim de Bruin, Domagoj Toli{\'c}, Jens Kober, and Ivana
  Palunko.
\newblock Reinforcement learning for control: {P}erformance, stability, and
  deep approximators.
\newblock \emph{{Annu. Rev. Control.}}, 46:\penalty0 8--28, 2018.

\bibitem[Chen et~al.(2017)Chen, Papandreou, Schroff, and Adam]{Chen2017}
Liang-Chieh Chen, George Papandreou, Florian Schroff, and Hartwig Adam.
\newblock Rethinking atrous convolution for semantic image segmentation.
\newblock \emph{{arXiv:1706.05587 [cs.CV]}}, 2017.

\bibitem[Choi et~al.(2020)Choi, Chen, Rusanovskyy, Choi, and Jang]{Choi2020}
Kiho Choi, Jianle Chen, Dmytro Rusanovskyy, Kwang-Pyo Choi, and Euee~S Jang.
\newblock An overview of the {MPEG-5} essential video coding standard.
\newblock \emph{{IEEE Signal Process. Mag.}}, 37\penalty0 (3):\penalty0
  160--167, 2020.

\bibitem[Cordts et~al.(2016)Cordts, Omran, Ramos, Rehfeld, Enzweiler, Benenson,
  Franke, Roth, and Schiele]{Cordts2016}
Marius Cordts, Mohamed Omran, Sebastian Ramos, Timo Rehfeld, Markus Enzweiler,
  Rodrigo Benenson, Uwe Franke, Stefan Roth, and Bernt Schiele.
\newblock The {Cityscapes} dataset for semantic urban scene understanding.
\newblock In \emph{CVPR}, pages 3213--3223, 2016.

\bibitem[de~Vries et~al.(2017)de~Vries, Strub, Mary, Larochelle, Pietquin, and
  Courville]{Devries2017}
Harm de Vries, Florian Strub, Jeremie Mary, Hugo Larochelle, Olivier Pietquin,
  and Aaron~C Courville.
\newblock Modulating early visual processing by language.
\newblock In \emph{NIPS}, 2017.

\bibitem[Du et~al.(2020)Du, Pervaiz, Yuan, Chowdhery, Zhang, Hoffmann, and
  Jiang]{Du2020}
Kuntai Du, Ahsan Pervaiz, Xin Yuan, Aakanksha Chowdhery, Qizheng Zhang, Henry
  Hoffmann, and Junchen Jiang.
\newblock Server-driven video streaming for deep learning inference.
\newblock In \emph{{SIGCOMM}}, pages 557--570, 2020.

\bibitem[Du et~al.(2022)Du, Zhang, Arapin, Wang, Xia, and Jiang]{Du2022}
Kuntai Du, Qizheng Zhang, Anton Arapin, Haodong Wang, Zhengxu Xia, and Junchen
  Jiang.
\newblock {AccMPEG}: {O}ptimizing video encoding for accurate video analytics.
\newblock In \emph{{MLSys}}, pages 450--466, 2022.

\bibitem[Duan et~al.(2020)Duan, Liu, Yang, Huang, and Gao]{Duan2020}
Lingyu Duan, Jiaying Liu, Wenhan Yang, Tiejun Huang, and Wen Gao.
\newblock Video coding for machines: {A} paradigm of collaborative compression
  and intelligent analytics.
\newblock \emph{{IEEE Trans. Image Process.}}, 29:\penalty0 8680--8695, 2020.

\bibitem[Emmons et~al.(2019)Emmons, Fouladi, Ananthanarayanan, Venkataraman,
  Savarese, and Winstein]{Emmons2019}
John Emmons, Sadjad Fouladi, Ganesh Ananthanarayanan, Shivaram Venkataraman,
  Silvio Savarese, and Keith Winstein.
\newblock Cracking open the {DNN} black-box: {V}ideo analytics with dnns across
  the camera-cloud boundary.
\newblock In \emph{{MobiCom HotEdgeVideo Workshop}}, pages 27--32, 2019.

\bibitem[Falcon and {The PyTorch Lightning Team}(2019)]{Falcon2019}
William Falcon and {The PyTorch Lightning Team}.
\newblock {PyTorch Lightning}.
\newblock {https://github.com/Lightning-AI/lightning}, 2019.

\bibitem[Fan et~al.(2021)Fan, Murrell, Wang, Alwala, Li, Li, Xiong, Ravi, Li,
  Yang, et~al.]{Fan2021b}
Haoqi Fan, Tullie Murrell, Heng Wang, Kalyan~Vasudev Alwala, Yanghao Li, Yilei
  Li, Bo Xiong, Nikhila Ravi, Meng Li, Haichuan Yang, et~al.
\newblock {PyTorchVideo}: {A} deep learning library for video understanding.
\newblock In \emph{ACMMM}, pages 3783--3786, 2021.

\bibitem[Feichtenhofer(2020)]{Feichtenhofer2020}
Christoph Feichtenhofer.
\newblock {X3D}: {E}xpanding architectures for efficient video recognition.
\newblock In \emph{CVPR}, pages 203--213, 2020.

\bibitem[Galteri et~al.(2018)Galteri, Bertini, Seidenari, and
  Del~Bimbo]{Galteri2018}
Leonardo Galteri, Marco Bertini, Lorenzo Seidenari, and Alberto Del~Bimbo.
\newblock Video compression for object detection algorithms.
\newblock In \emph{ICPR}, pages 3007--3012, 2018.

\bibitem[Gao et~al.(2021)Gao, Liu, Xu, Rafie, Zhang, and Curcio]{Gao2021}
Wen Gao, Shan Liu, Xiaozhong Xu, Manouchehr Rafie, Yuan Zhang, and Igor Curcio.
\newblock Recent standard development activities on video coding for machines.
\newblock \emph{{arXiv:2105.12653 [cs.CV]}}, 2021.

\bibitem[Geiger et~al.(2013)Geiger, Lenz, Stiller, and Urtasun]{Geiger2013}
Andreas Geiger, Philip Lenz, Christoph Stiller, and Raquel Urtasun.
\newblock Vision meets robotics: {T}he {KITTI} dataset.
\newblock \emph{{Int. J. Robot. Res.}}, 32\penalty0 (11):\penalty0 1231--1237,
  2013.

\bibitem[Glynn and Szechtman(2002)]{Glynn2002}
Peter~W Glynn and Roberto Szechtman.
\newblock Some new perspectives on the method of control variates.
\newblock In \emph{{Monte Carlo and Quasi-Monte Carlo Methods 2000}}, pages
  27--49. Springer, 2002.

\bibitem[Grathwohl et~al.(2018)Grathwohl, Choi, Wu, Roeder, and
  Duvenaud]{Grathwohl2018}
Will Grathwohl, Dami Choi, Yuhuai Wu, Geoff Roeder, and David Duvenaud.
\newblock Backpropagation through the void: {O}ptimizing control variates for
  black-box gradient estimation.
\newblock In \emph{ICLR}, 2018.

\bibitem[He et~al.(2016)He, Zhang, Ren, and Sun]{He2016}
Kaiming He, Xiangyu Zhang, Shaoqing Ren, and Jian Sun.
\newblock Deep residual learning for image recognition.
\newblock In \emph{CVPR}, pages 770--778, 2016.

\bibitem[Hu et~al.(2023)Hu, Luo, Pasdar, Lee, Zhou, and Wu]{Hu2023}
Miao Hu, Zhenxiao Luo, Amirmohammad Pasdar, Young~Choon Lee, Yipeng Zhou, and
  Di Wu.
\newblock Edge-based video analytics: {A} survey.
\newblock \emph{{arXiv:2303.14329 [cs.DC]}}, 2023.

\bibitem[Isik et~al.(2023)Isik, Guleryuz, Tang, Taylor, and Chou]{Isik2023}
Berivan Isik, Onur~G Guleryuz, Danhang Tang, Jonathan Taylor, and Philip~A
  Chou.
\newblock Sandwiched video compression: {E}fficiently extending the reach of
  standard codecs with neural wrappers.
\newblock \emph{{arXiv:2303.11473 [eess.IV]}}, 2023.

\bibitem[Itsumi et~al.(2022)Itsumi, Beye, Charvi, and Nihei]{Itsumi2022}
Hayato Itsumi, Florian Beye, Vitthal Charvi, and Koichi Nihei.
\newblock Learning important regions via attention for video streaming on cloud
  robotics.
\newblock In \emph{{IROS}}, 2022.

\bibitem[Jang et~al.(2017)Jang, Gu, and Poole]{Jang2017}
Eric Jang, Shixiang Gu, and Ben Poole.
\newblock Categorical reparameterization with {Gumbel-Softmax}.
\newblock In \emph{ICLR}, 2017.

\bibitem[Jeong et~al.(2023)Jeong, Cai, Garrepalli, and Porikli]{Jeong2023}
Jisoo Jeong, Hong Cai, Risheek Garrepalli, and Fatih Porikli.
\newblock {DistractFlow}: {I}mproving optical flow estimation via realistic
  distractions and pseudo-labeling.
\newblock In \emph{{CVPR}}, pages 13691--13700, 2023.

\bibitem[Jiang et~al.(2021)Jiang, Dai, Wu, and Loy]{Jiang2021}
Liming Jiang, Bo Dai, Wayne Wu, and Chen~Change Loy.
\newblock Focal frequency loss for image reconstruction and synthesis.
\newblock In \emph{ICCV}, pages 13919--13929, 2021.

\bibitem[Klopp et~al.(2021)Klopp, Liu, Chen, and Chien]{Klopp2021}
Jan~P Klopp, Keng-Chi Liu, Liang-Gee Chen, and Shao-Yi Chien.
\newblock How to exploit the transferability of learned image compression to
  conventional codecs.
\newblock In \emph{{CVPR}}, pages 16165--16174, 2021.

\bibitem[Lampert(2006)]{Lampert2006}
Christoph~H. Lampert.
\newblock Machine learning for video compression: {M}acroblock mode decision.
\newblock In \emph{{ICPR}}, pages 936--940, 2006.

\bibitem[Lederer(2019)]{Lederer2019}
Stefan Lederer.
\newblock 2019 video developer report – {T}he future of video: {AV1} codec,
  {AI} \& machine learning, and low latency.
\newblock
  {https://bitmovin.com/bitmovin-2019-video-developer-report-av1-codec-ai-machine-learning-low-latency/},
  2019.

\bibitem[Li et~al.(2018)Li, Xu, Taylor, Studer, and Goldstein]{Li2018}
Hao Li, Zheng Xu, Gavin Taylor, Christoph Studer, and Tom Goldstein.
\newblock Visualizing the loss landscape of neural nets.
\newblock In \emph{{NeurIPS}}, 2018.

\bibitem[Li et~al.(2020)Li, Padmanabhan, Zhao, Wang, Xu, and
  Netravali]{Li2020b}
Yuanqi Li, Arthi Padmanabhan, Pengzhan Zhao, Yufei Wang, Guoqing~Harry Xu, and
  Ravi Netravali.
\newblock {Reducto}: {O}n-camera filtering for resource-efficient real-time
  video analytics.
\newblock In \emph{{SIGCOMM}}, pages 359--376, 2020.

\bibitem[Liu et~al.(2020)Liu, Li, Lin, Li, and Wu]{Liu2020}
Dong Liu, Yue Li, Jianping Lin, Houqiang Li, and Feng Wu.
\newblock Deep learning-based video coding: {A} review and a case study.
\newblock \emph{{ACM Comput. Surv.}}, 53\penalty0 (1):\penalty0 1--35, 2020.

\bibitem[Liu et~al.(2019)Liu, Li, and Gruteser]{Liu2019}
Luyang Liu, Hongyu Li, and Marco Gruteser.
\newblock Edge assisted real-time object detection for mobile augmented
  reality.
\newblock In \emph{{ACM MobiCom}}, pages 1--16, 2019.

\bibitem[Loshchilov and Hutter(2017)]{Loshchilov2017}
Ilya Loshchilov and Frank Hutter.
\newblock {SGDR}: {S}tochastic gradient descent with warm restarts.
\newblock In \emph{ICLR}, 2017.

\bibitem[Luo et~al.(2021)Luo, Talebi, Yang, Elad, and Milanfar]{Luo2021}
Xiyang Luo, Hossein Talebi, Feng Yang, Michael Elad, and Peyman Milanfar.
\newblock The rate-distortion-accuracy tradeoff: {JPEG} case study.
\newblock In \emph{{DCC}}, pages 354--354, 2021.

\bibitem[Maas et~al.(2013)Maas, Hannun, Ng, et~al.]{Maas2013}
Andrew~L Maas, Awni~Y Hannun, Andrew~Y Ng, et~al.
\newblock {Empirical Evaluation of Rectified Activations in Convolutional
  Network}.
\newblock In \emph{{ICML}}, page~3, 2013.

\bibitem[Mandhane et~al.(2022)Mandhane, Zhernov, Rauh, Gu, Wang, Xue, Shang,
  Pang, Claus, Chiang, et~al.]{Mandhane2022}
Amol Mandhane, Anton Zhernov, Maribeth Rauh, Chenjie Gu, Miaosen Wang, Flora
  Xue, Wendy Shang, Derek Pang, Rene Claus, Ching-Han Chiang, et~al.
\newblock {MuZero} with self-competition for rate control in vp9 video
  compression.
\newblock \emph{{arXiv:2202.06626 [eess.IV]}}, 2022.

\bibitem[Matsubara et~al.(2019)Matsubara, Baidya, Callegaro, Levorato, and
  Singh]{Matsubara2019}
Yoshitomo Matsubara, Sabur Baidya, Davide Callegaro, Marco Levorato, and Sameer
  Singh.
\newblock Distilled split deep neural networks for edge-assisted real-time
  systems.
\newblock In \emph{{MobiCom HotEdgeVideo Workshop}}, pages 21--26, 2019.

\bibitem[{MMSegmentation Contributors}(2020)]{Mmseg2020}
{MMSegmentation Contributors}.
\newblock {MMSegmentation}: {O}penmmlab semantic segmentation toolbox and
  benchmark.
\newblock {https://github.com/open-mmlab/mmsegmentation}, 2020.

\bibitem[Nah et~al.(2019)Nah, Baik, Hong, Moon, Son, Timofte, and
  Mu~Lee]{Nah2019}
Seungjun Nah, Sungyong Baik, Seokil Hong, Gyeongsik Moon, Sanghyun Son, Radu
  Timofte, and Kyoung Mu~Lee.
\newblock {NTIRE} 2019 challenge on video deblurring and super-resolution:
  Dataset and study.
\newblock In \emph{CVPRW}, 2019.

\bibitem[Nair and Hinton(2010)]{Nair2010}
Vinod Nair and Geoffrey~E Hinton.
\newblock Rectified linear units improve restricted boltzmann machines.
\newblock In \emph{ICML}, pages 807--814, 2010.

\bibitem[Otani et~al.(2022)Otani, Hashiguchi, Omi, Fukushima, and
  Tamaki]{Otani2022}
Aoi Otani, Ryota Hashiguchi, Kazuki Omi, Norishige Fukushima, and Toru Tamaki.
\newblock Performance evaluation of action recognition models on low quality
  videos.
\newblock \emph{{IEEE Access}}, 10:\penalty0 94898--94907, 2022.

\bibitem[Paszke et~al.(2019)Paszke, Gross, Massa, Lerer, Bradbury, Chanan,
  Killeen, Lin, Gimelshein, Antiga, et~al.]{Paszke2019}
Adam Paszke, Sam Gross, Francisco Massa, Adam Lerer, James Bradbury, Gregory
  Chanan, Trevor Killeen, Zeming Lin, Natalia Gimelshein, Luca Antiga, et~al.
\newblock {PyTorch}: {A}n imperative style, high-performance deep learning
  library.
\newblock \emph{NeurIPS}, 32, 2019.

\bibitem[Qiu et~al.(2021)Qiu, Yu, and Li]{Qiu2021}
Kaitian Qiu, Lu Yu, and Daowen Li.
\newblock Codec-simulation network for joint optimization of video coding with
  pre- and post-processing.
\newblock \emph{{IEEE Open J. Circuits Syst.}}, 2:\penalty0 648--659, 2021.

\bibitem[Riba et~al.(2020)Riba, Mishkin, Ponsa, Rublee, and Bradski]{Riba2020}
Edgar Riba, Dmytro Mishkin, Daniel Ponsa, Ethan Rublee, and Gary Bradski.
\newblock {Kornia}: {A}n open source differentiable computer vision library for
  {PyTorch}.
\newblock In \emph{{WACV}}, pages 3674--3683, 2020.

\bibitem[Richardson(2004)]{Richardson2004}
Iain~E Richardson.
\newblock \emph{{H.264} and {MPEG-4} video compression: {Video} coding for
  next-generation multimedia}.
\newblock John Wiley \& Sons, 2004.

\bibitem[Richardson(2011)]{Richardson2011}
Iain~E Richardson.
\newblock \emph{The {H. 264} Advanced Video Compression Standard}.
\newblock John Wiley \& Sons, 2011.

\bibitem[Ronneberger et~al.(2015)Ronneberger, Fischer, and
  Brox]{Ronneberger2015}
Olaf Ronneberger, Philipp Fischer, and Thomas Brox.
\newblock {U-Net}: {C}onvolutional networks for biomedical image segmentation.
\newblock In \emph{MICCAI}, pages 234--241, 2015.

\bibitem[Simonyan et~al.(2014)Simonyan, Vedaldi, and Zisserman]{Simonyan2013}
Karen Simonyan, Andrea Vedaldi, and Andrew Zisserman.
\newblock Deep inside convolutional networks: {V}isualising image
  classification models and saliency maps.
\newblock In \emph{{ICLR}}, 2014.

\bibitem[Sullivan et~al.(2012)Sullivan, Ohm, Han, and Wiegand]{Sullivan2012}
Gary~J Sullivan, Jens-Rainer Ohm, Woo-Jin Han, and Thomas Wiegand.
\newblock Overview of the high efficiency video coding {(HEVC)} standard.
\newblock \emph{{IEEE Trans. Circuits Syst. Video Technol.}}, 22\penalty0
  (12):\penalty0 1649--1668, 2012.

\bibitem[Sundararajan et~al.(2017)Sundararajan, Taly, and
  Yan]{Sundararajan2017}
Mukund Sundararajan, Ankur Taly, and Qiqi Yan.
\newblock Axiomatic attribution for deep networks.
\newblock In \emph{ICML}, pages 3319--3328, 2017.

\bibitem[Teed and Deng(2020)]{Teed2020}
Zachary Teed and Jia Deng.
\newblock {RAFT}: {R}ecurrent all-pairs field transforms for optical flow.
\newblock In \emph{ECCV}, pages 402--419, 2020.

\bibitem[Tian et~al.(2021)Tian, Lu, Min, Che, Zhai, Guo, and Gao]{Tian2021}
Yuan Tian, Guo Lu, Xiongkuo Min, Zhaohui Che, Guangtao Zhai, Guodong Guo, and
  Zhiyong Gao.
\newblock Self-conditioned probabilistic learning of video rescaling.
\newblock In \emph{ICCV}, pages 4490--4499, 2021.

\bibitem[Tomar(2006)]{Tomar2006}
Suramya Tomar.
\newblock Converting video formats with {FFmpeg}.
\newblock \emph{{Linux J.}}, 2006\penalty0 (146):\penalty0 10, 2006.

\bibitem[{TorchVision Maintainers} and Contributors(2016)]{Torchvision2016}
{TorchVision Maintainers} and Contributors.
\newblock {TorchVision}: {PyTorch's} computer vision library.
\newblock {https://github.com/pytorch/vision}, 2016.

\bibitem[Vaswani et~al.(2017)Vaswani, Shazeer, Parmar, Uszkoreit, Jones, Gomez,
  Kaiser, and Polosukhin]{Vaswani2017}
Ashish Vaswani, Noam Shazeer, Niki Parmar, Jakob Uszkoreit, Llion Jones,
  Aidan~N Gomez, \L{}ukasz Kaiser, and Illia Polosukhin.
\newblock Attention is all you need.
\newblock In \emph{NIPS}, 2017.

\bibitem[Wallace(1992)]{Wallace1992}
Gregory~K Wallace.
\newblock The {JPEG} still picture compression standard.
\newblock \emph{{IEEE Trans. Consum. Electron.}}, 38\penalty0 (1):\penalty0
  xviii--xxxiv, 1992.

\bibitem[Wang et~al.(2018)Wang, Yu, Dong, and Loy]{Wang2018}
Xintao Wang, Ke Yu, Chao Dong, and Chen~Change Loy.
\newblock Recovering realistic texture in image super-resolution by deep
  spatial feature transform.
\newblock In \emph{CVPR}, pages 606--615, 2018.

\bibitem[Wang et~al.(2004)Wang, Bovik, Sheikh, and Simoncelli]{Wang2004}
Zhou Wang, Alan~C Bovik, Hamid~R Sheikh, and Eero~P Simoncelli.
\newblock Image quality assessment: {F}rom error visibility to structural
  similarity.
\newblock \emph{{IEEE Trans. Image Process.}}, 13\penalty0 (4):\penalty0
  600--612, 2004.

\bibitem[Wiegand et~al.(2003)Wiegand, Sullivan, Bjontegaard, and
  Luthra]{Wiegand2003}
Thomas Wiegand, Gary~J Sullivan, Gisle Bjontegaard, and Ajay Luthra.
\newblock Overview of the {H.264/AVC} video coding standard.
\newblock \emph{{IEEE Trans. Circuits Syst. Video Technol.}}, 13\penalty0
  (7):\penalty0 560--576, 2003.

\bibitem[Wood(2022)]{Wood2022}
Daniel Wood.
\newblock Task oriented video coding: {A} survey.
\newblock \emph{{arXiv:2208.07313 [eess.IV]}}, 2022.

\bibitem[Wu et~al.(2001)Wu, Hou, Zhu, Zhang, and Peha]{Wu2001}
Dapeng Wu, Yiwei~Thomas Hou, Wenwu Zhu, Ya-Qin Zhang, and Jon~M Peha.
\newblock Streaming video over the internet: {A}pproaches and directions.
\newblock \emph{{IEEE Trans. Circuits Syst. Video Technol.}}, 11\penalty0
  (3):\penalty0 282--300, 2001.

\bibitem[Wu and He(2018)]{Wu2018}
Yuxin Wu and Kaiming He.
\newblock Group normalization.
\newblock In \emph{ECCV}, pages 3--19, 2018.

\bibitem[Xia et~al.(2020)Xia, Liang, Yang, Duan, and Liu]{Xia2020}
Sifeng Xia, Kunchangtai Liang, Wenhan Yang, Ling-Yu Duan, and Jiaying Liu.
\newblock An emerging coding paradigm vcm: {A} scalable coding approach beyond
  feature and signal.
\newblock In \emph{{ICME}}, pages 1--6, 2020.

\bibitem[Xie et~al.(2022)Xie, Zhou, Zhu, and Liu]{Xie2022}
Xiufeng Xie, Ning Zhou, Wentao Zhu, and Ji Liu.
\newblock Bandwidth-aware adaptive codec for dnn inference offloading in iot.
\newblock In \emph{{ECCV}}, pages 88--104, 2022.

\bibitem[Zhang et~al.(2018)Zhang, Jin, Ratnasamy, Wawrzynek, and
  Lee]{Zhang2018b}
Ben Zhang, Xin Jin, Sylvia Ratnasamy, John Wawrzynek, and Edward~A. Lee.
\newblock {AWStream}: {A}daptive wide-area streaming analytics.
\newblock In \emph{{SIGCOMM}}, page 236–252, 2018.

\bibitem[Zhang et~al.(2021)Zhang, He, Liu, Jia, Liu, Gruteser, Raychaudhuri,
  and Zhang]{Zhang2021}
Wuyang Zhang, Zhezhi He, Luyang Liu, Zhenhua Jia, Yunxin Liu, Marco Gruteser,
  Dipankar Raychaudhuri, and Yanyong Zhang.
\newblock {ELF}: {A}ccelerate high-resolution mobile deep vision with
  content-aware parallel offloading.
\newblock In \emph{{ACM MobiCom}}, pages 201--214, 2021.

\bibitem[Zhang et~al.(2023)Zhang, Zhu, Jiang, Kwong, and Kuo]{Zhang2023}
Yun Zhang, Linwei Zhu, Gangyi Jiang, Sam Kwong, and C-C~Jay Kuo.
\newblock A survey on perceptually optimized video coding.
\newblock \emph{{ACM Comput. Surv.}}, 55\penalty0 (12):\penalty0 1--37, 2023.

\bibitem[Zhao et~al.(2016)Zhao, Gallo, Frosio, and Kautz]{Zhao2016}
Hang Zhao, Orazio Gallo, Iuri Frosio, and Jan Kautz.
\newblock Loss functions for image restoration with neural networks.
\newblock \emph{{IEEE Trans. Comput. Imaging.}}, 3\penalty0 (1):\penalty0
  47--57, 2016.

\bibitem[Zhao et~al.(2019)Zhao, Bai, Wang, and Zhao]{Zhao2019}
Lijun Zhao, Huihui Bai, Anhong Wang, and Yao Zhao.
\newblock Learning a virtual codec based on deep convolutional neural network
  to compress image.
\newblock \emph{{J. Vis. Commun. Image Represent.}}, 63:\penalty0 102589, 2019.

\end{thebibliography}
